\documentclass[12pt,a4,final,twoside]{book}
\usepackage{times}
\usepackage[utf8]{inputenc}
\usepackage{csquotes}
\usepackage[english]{babel}
\usepackage[
bibencoding=utf8,
backend=biber,
style=ieee,
dateabbrev=true,
sorting=none,
url=true
]{biblatex}
\usepackage[T1]{fontenc}
\newcommand{\href}[1]{#1} 

\usepackage{ifthen}
\usepackage{amsmath}
\usepackage{amsfonts}
\usepackage{amssymb}
\usepackage{gensymb}
\usepackage{paralist}
\usepackage[shortlabels]{enumitem}
\usepackage{ragged2e}
\usepackage{hyperref}
\usepackage[normalem]{ulem}
\usepackage[dvipsnames]{xcolor}
\usepackage{graphicx}
\usepackage{epsfig}
\usepackage{booktabs}
\usepackage{multirow}
\usepackage{caption}
\usepackage{subcaption}
\usepackage{comment}
\usepackage{dashrule}

\DeclareMathOperator{\sinc}{sinc}

\newboolean{PrintVersion}
\setboolean{PrintVersion}{false}

\usepackage{amstext} 
\hypersetup{
    plainpages=false,       
    unicode=false,          
    pdftoolbar=true,        
    pdfmenubar=true,        
    pdffitwindow=false,     
    pdfstartview={FitH},    
    pdftitle={Astrophysical\ and\ Cosmological\ Implications\ of Self-interactions\ of\ Ultra-Light\ Dark\ Matter},    
    pdfauthor={Bihag Dave},    
    pdfsubject={Physics},  
    pdfkeywords={Dark matter} {ULDM} {Cosmology}, 
    pdfnewwindow=true,      
    colorlinks=true,        
    linkcolor=blue,         
    citecolor=magenta,        
    filecolor=black,      
    urlcolor=blue           
}
\ifthenelse{\boolean{PrintVersion}}{
\hypersetup{
    citecolor=black,%
    filecolor=black,%
    linkcolor=black,%
    urlcolor=black}
}{} 

\usepackage[a4paper, portrait, margin=1.25in]{geometry}
\usepackage{setspace}
\usepackage{floatrow}
\usepackage[raggedright]{titlesec}
\usepackage{chngcntr}
\usepackage{fancyhdr}
\usepackage[acronym]{glossaries}

\fancypagestyle{publicationpage}{
    \fancyhf{}
    
  \fancyfoot[C]{\thepage}
}

\usepackage[automake,toc,abbreviations]{glossaries-extra} 

\let\origdoublepage\cleardoublepage
\newcommand{\clearemptydoublepage}{%
  \clearpage{\pagestyle{empty}\origdoublepage}}
\let\cleardoublepage\clearemptydoublepage

\begin{document}

\pagestyle{plain}
\pagenumbering{roman}


\begin{titlepage}
        \begin{center}
        \vspace*{1.6cm}
        {\huge{\textbf{Self-Gravitating Scalar Field Configurations, Ultra Light Dark Matter and Galactic Scale Observations}}}
        
        \vspace{1.5cm}
        
        {\fontsize{14}{18}\selectfont\textbf{Bihag Bankimchandra Dave}}

        \vspace{2cm}

        {A dissertation submitted to the faculty of Ahmedabad University in partial fulfilment of the requirements for the degree of}

        \vspace{2cm}

        {\fontsize{16}{18}\selectfont \textbf{Doctor of Philosophy}}

        \vspace{1.5cm}
        \includegraphics[width=0.3\textwidth]{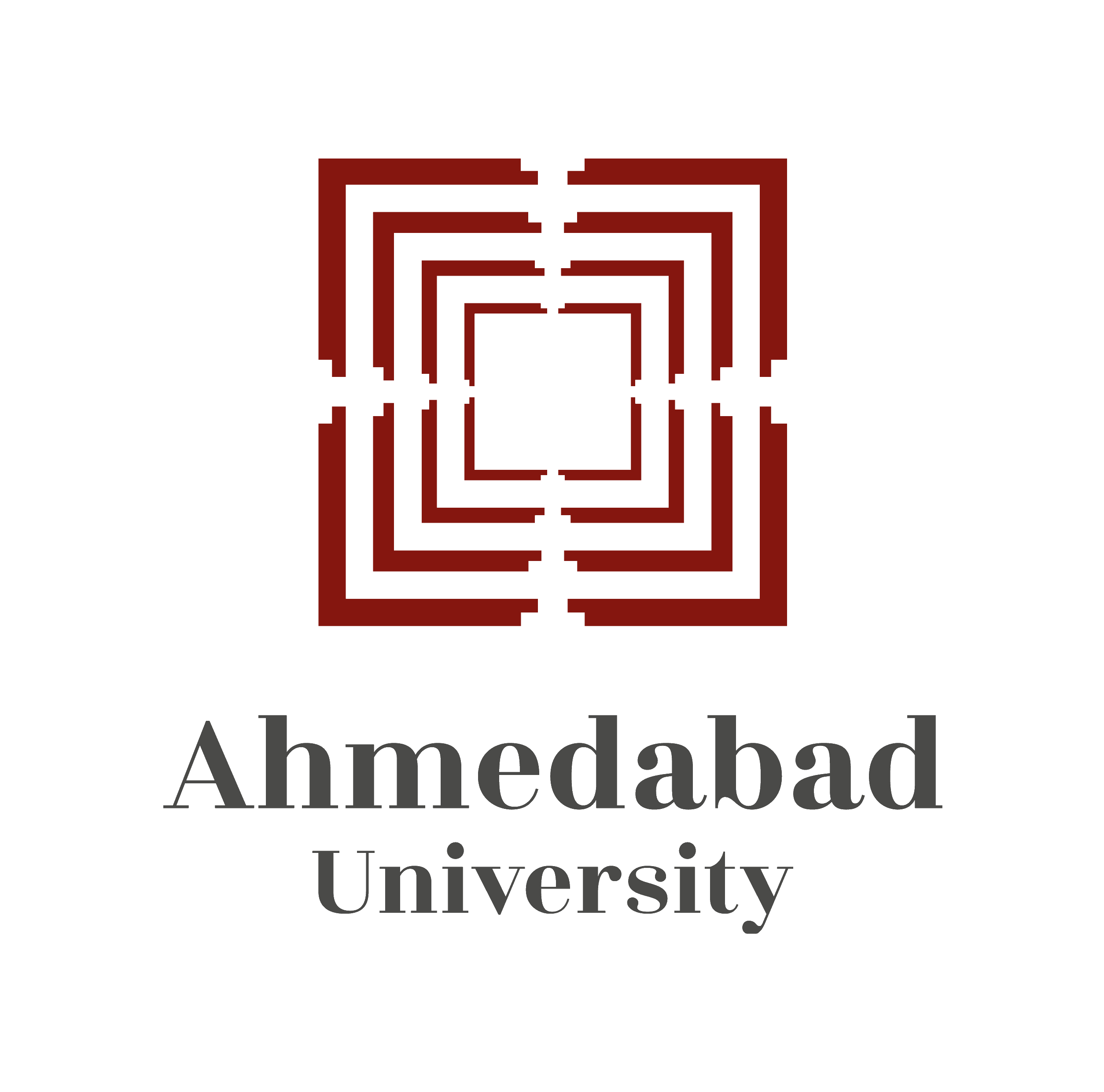}

        \vspace{1cm}

        {September, 2025}

    \end{center}        
\end{titlepage}

\cleardoublepage 
\phantomsection    

\begin{center}
\copyright Bihag Bankimchandra Dave; 2025

All Rights Reserved
\end{center}

\pagestyle{plain}
\setcounter{page}{1}

\cleardoublepage
\phantomsection

 \begin{center}\textbf{Declaration by the candidate}\end{center}

 \noindent
This is to certify that the dissertation titled, “Self-Gravitating Scalar Field Configurations, Ultra Light Dark Matter and Galactic Scale Observations” comprises my original work towards the degree of Doctor of Philosophy at Ahmedabad University and has been completed under the supervision of Dr. Gaurav Goswami. The matter embodied in this dissertation has not been submitted elsewhere for the award of any other degree/diploma. I have not intentionally copied material from any other source such as journals, books, magazines, websites, etc. and due acknowledgement has been made in the text of all material used from other sources.

\vspace{1.5cm}
    
    \flushright {\fontsize{12}{14}\selectfont Bihag Bankimchandra Dave \\ 
    School of Engineering and Applied Science, \\ 
    Ahmedabad University, \\ 
    Commerce Six Roads, Navrangpura, \\ 
    Ahmedabad 380009, India}
    
    
\vspace{0.25cm}
\hrulefill
\vspace{0.5cm}

\begin{center}\textbf{Certificate by the Supervisor}\end{center}

\noindent
\flushleft
It is certified that, to the best of my knowledge, the above statements made by the student are correct.

\vspace{1.5cm}
    
    \flushright {\fontsize{12}{14}\selectfont Dr. Gaurav Goswami\\ 
    Division of Mathematical and Physical Sciences,\\  
    School of Arts and Sciences, \\ 
    Ahmedabad University, \\ 
    Commerce Six Roads, Navrangpura, \\ 
    Ahmedabad 380009, India
    }

    
\cleardoublepage
\phantomsection      

\begin{center}\textbf{Certificate}\end{center}
\justifying
{This is to certify that the dissertation work titled ``Self-Gravitating Scalar Field Configurations, Ultra Light Dark Matter and Galactic Scale Observations'' has been completed by Bihag Bankimchandra Dave (AU1949007) for the degree of Doctor of Philosophy at the School of Engineering and Applied Science of Ahmedabad University under my supervision.}

\vspace{1.2cm}
\flushleft

Dr. Gaurav Goswami\\
(Supervisor)\\ 
Division of Mathematical and Physical Sciences,\\  
School of Arts and Sciences, \\ 
Ahmedabad University, \\ 
Commerce Six Roads, Navrangpura, \\ 
Ahmedabad 380009, India


\hdashrule[0.5ex]{\linewidth}{1pt}{1.5mm}

\flushleft
The members of the Dissertation Advisory Committee were: \\ 

\vspace{0.8cm}


Professor Raghavan Rangarajan \\ 
Division of Mathematical and Physical Sciences,\\ 
School of Arts and Sciences, \\ 
Ahmedabad University, \\ 
Commerce Six Roads, Navrangpura, \\ 
Ahmedabad 380009, India

\vspace{1cm}


Professor Pankaj Joshi \\ 
Division of Mathematical and Physical Sciences,\\  
School of Arts and Sciences, \\ 
Ahmedabad University, \\ 
Commerce Six Roads, Navrangpura, \\ 
Ahmedabad 380009, India

\vspace{1cm}


Professor Anjan Ananda Sen \\ 
Centre for Theoretical Physics, \\ 
Jamia Millia Islamia, \\ 
New Delhi 110025, India \\

    \cleardoublepage
\phantomsection

\addcontentsline{toc}{chapter}{Abstract}
\begin{center}\textbf{Abstract}\end{center}
\justifying
{

We live in an era of data-driven cosmology. Increasingly accurate and precise data from various sources has enabled us to establish the concordance model of cosmology known as the $\Lambda$CDM model. 
However, many fundamental questions about our Universe remain unanswered, such as the mystery of the physical nature of dark matter. 
In this thesis, we will examine the exciting possibility that dark matter consists of spin-zero particles with the smallest possible mass consistent with the existence of dwarf galaxies in the Universe, i.e. $m\sim 10^{-22}\ \text{eV}$. 
There exist many experimental constraints on the couplings of such a new scalar with Standard Model particles.
We however, focus our attention on its coupling to itself, i.e. its self-interactions. In other words, we do not make any assumptions about any non-gravitational interactions of dark matter with known elementary particles, except that they are small enough to be consistent with existing upper limits. 
We adopt a signature-driven approach and ask, what can galactic scale observations say about this class of dark matter scenarios. 

This kind of Ultra Light Dark Matter (ULDM) is expected to form stable self-gravitating configurations, whose macroscopic properties depend on the mass $m$ and self-coupling $\lambda$. 
We explore the observational consequences of describing the inner regions of galactic halos using such self-gravitating scalar field configurations (also called solitons) and ask what values of $\lambda$ can be probed if $m\sim 10^{-22}\ \text{eV}$. 

First, we demonstrate a method to obtain constraints in the $\lambda-m$ plane, using observational upper limits on the amount of mass contained within the central region of a galactic halo containing a supermassive black hole. By obtaining solutions of the Gross-Pitaevskii-Poisson equations, that characterise stationary configurations of the scalar field, we showed that both attractive and repulsive self-interactions with strength $\lambda \sim \pm10^{-96}-10^{-95}$ can be probed by such upper limits. 

Next, we tackled recent constraints on the ULDM scenario with $\lambda = 0$, which stated that ULDM with $m\in\left[10^{-24}\ \text{eV}, 10^{-20}\ \text{eV}\right]$ cannot simultaneously satisfy observed rotation curves of galaxies as well as an empirically expected power law between the mass of the soliton and mass of the halo. 
We found that, for dark matter dominated low surface brightness galaxies from the SPARC catalogue, ULDM with $m = 10^{-22}\ \text{eV}$ and $\lambda \gtrsim 10^{-90}$ can satisfy both observed rotation curves as well as a modified power law relation that includes the effects of self-interactions.

Another interesting consequence of ULDM with $m\sim 10^{-22}\ \text{eV}$ is the tunnelling of dark matter from a satellite dwarf galaxy orbiting around the centre of a large halo over cosmological timescales, due to the tidal effects of the host halo. 
Using the formalism of quasi-stationary solutions, we studied the effects of self-interactions on such a system and found that attractive self-interactions aid the self-gravity of the satellite galaxy against the tidal effects of the halo, extending the lifetime of the satellite. 
On the other hand, repulsive self-interactions do the opposite and shorten the lifetime of satellite galaxies. 
We applied this to the Fornax dwarf spheroidal with a known core mass and orbital period, and found that ULDM with $m = 10^{-22}\ \text{eV}$ and $\lambda \lesssim -2.12\times 10^{-91}$ will enable the dwarf galaxy to survive on cosmological timescales, evading a recent constraint for the same mass but with $\lambda = 0$,
thereby remaining consistent with the ULDM paradigm. 

Additionally, in the context of ULDM, we also explore machine learning models like Artificial Neural Networks (ANNs) to learn from observed data. 
Using simulated rotation curves of several dwarf galaxies, we train neural networks to learn the relationship between the rotation curve and parameters of the dark matter density profile (such as ULDM particle mass $m$, scaling parameter $s$, core-to-envelope transition radius $r_t$ and NFW scale radius $r_s$), along with Baryonic parameters (such as the stellar mass-to-light ratio $\Upsilon_*$). 
We then feed the observed rotation curves of the galaxies and attempt to infer appropriate parameter values along with their uncertainties. 
Here, we explore the importance of noise in the training data, and compare the inferred parameter values and uncertainties obtained using two different methods to those obtained using standard Bayesian methods. 
We find that the trained neural networks can extract parameters that describe observations well for the galaxies we studied.

This thesis is an effort to glean some insight into the nature of dark matter using galactic scale observations. Our work highlights the importance of taking the effects of self-interactions into consideration, as future experiments attempt to detect ULDM. 
Our work also highlights the importance of developing and embracing new techniques to learn from data. 

}
    \cleardoublepage
\phantomsection 

\begin{center}\textbf{Acknowledgements}\end{center}
\justifying

First and foremost, I would like thank my supervisor Dr. Gaurav Goswami. Your guidance, in more ways than one, has shaped not just my approach to research but to life in general. I am very grateful for your mentorship in my journey to become a physicist.
This thesis would not have been possible without your continued support. 

I would also like to thank my collaborators, Prof. Koushik Dutta (IISER Kolkata) and Prof. Sayan Chakraborti (IIT Guwahati) for their help and discussions, especially during the initial stage of my PhD. 
I would also like to thank the members of my Dissertation Advisory Committee: Prof. Raghavan Ranagarajan, Prof. Anjan Ananda Sen and Prof. Pankaj Joshi for their inputs and advice throughout my PhD. 

I would additionally like to thank Prof. Raghavan Ranagarajan and Prof. Amit Nanavati for their guidance regarding research and career throughout the years. 

I want to acknowledge funding from the Department of Science and Technology, Government of India under Indo-Russian call for Joint Proposals (DST/INT/RUS/RSF/P-21) during the initial years of my PhD. 
I am grateful to Ahmedabad University, for providing a tuition waiver as well as financial support in the form of a University Fellowship once external funding was exhausted. I would also like to thank Ahmedabad University and the Dean of Graduate School, Prof. Deepak Kunzru for providing a fantastic working space and environment for all PhD students. 
I would also like to thank Kangna Bagani ma'am, Rahul Bhatiji sir, Isha Ayachit ma'am, and Aarti Jani ma'am for all the help with various PhD related submissions and computational support throughout the years. 

I would also like to acknowledge the use of Param Shavak High Performance Computing System provided by the Gujarat Council on Science and Technology to Ahmedabad University, on which a part of my PhD work was carried out.

I would like thank my cohort, especially Koushiki, Mohit, Ashish and Kishan from School of Arts and Sciences as well as Mahula, Mansi, Jai and Yagnik from School of Engineering and Applied Science for being great colleagues and friends. 
I also want to thank Isha Mahuvakar for the interesting discussions and help with understanding Artificial Neural Networks. 
Finally, I want to thank Jai and Yagnik for their help with tensorflow.

I would like to thank my parents, for being loving, patient and understanding throughout this long journey. Thank you for your support.
Finally, I want to thank my partner Munzerin, for always squashing my doubts and fears. Thank you for keeping me sane.

\vspace{1.25cm}
\flushright
\textbf{Bihag Dave}

\cleardoublepage
\phantomsection    

\renewcommand\contentsname{Table of Contents}
\begin{singlespace}
\tableofcontents
\end{singlespace}
\cleardoublepage
\phantomsection    

\addcontentsline{toc}{chapter}{List of Figures}
\begin{singlespace}
\listoffigures
\end{singlespace}
\cleardoublepage
\phantomsection		

\cleardoublepage
\phantomsection	
\addcontentsline{toc}{chapter}{List of Tables}
\begin{singlespace}
\listoftables
\end{singlespace}
\cleardoublepage
\phantomsection		

\renewcommand*{\abbreviationsname}{List of Acronyms}
\addcontentsline{toc}{chapter}{Abbreviations}
\flushleft {\huge{\textbf{Abbreviations}}}
\justifying
\vspace{0.8cm}
\begin{enumerate}[align=parleft, label={}]
    \item ALP  \dotfill Axion-Like Particle
    \item ANN  \dotfill Artificial Neural Network
    \item BAO  \dotfill Baryon Acoustic Oscillations
    \item BBN  \dotfill Big Bang Nucleosynthesis 
    \item BHSR \dotfill Black Hole Superradiance
    \item CDM  \dotfill Cold Dark Matter
    \item CMB  \dotfill Cosmic Microwave Background
    \item CNN  \dotfill Convolutional Neural Network
    \item DDO  \dotfill David Dunlap Observatory
    \item DE   \dotfill Dark Energy
    \item DM   \dotfill Dark Matter
    \item ESO  \dotfill European Southern Observatory
    \item FDM  \dotfill Fuzzy Dark Matter
    \item GPP  \dotfill Gross-Pitaevskii-Poisson (equations)
    \item GR   \dotfill General Relativity
    \item IAP  \dotfill Internal Adjustable Parameters
    \item IC   \dotfill The Index Catalogue
    \item KGE  \dotfill Klein-Gordon-Einstein (equations)
    \item LITTLE THINGS \dotfill Local Irregulars That Trace Luminosity Extremes, \\
    \vspace{-1cm}\flushright{The HI Nearby Galaxy Survey}
    \item LSB  \dotfill Low Surface Brightness
    \item MACHO \dotfill Massive Astrophysical Compact Halo Object
    \item MCMC \dotfill Markov Chain Monte-Carlo
    \item MOND \dotfill Modified Newtonian Dynamics
    \item NFW  \dotfill Navarro–Frenk–White (density profile)
    \item NGC  \dotfill New General Catalogue (of Nebulae and Clusters of Stars)
    \item MSE  \dotfill Mean-Squared Error
    \item PBH  \dotfill Primordial Black Hole
    \item PNS  \dotfill Primordial Naked Singularities
    \item PTA  \dotfill Pulsar Timing Array
    \item PVC  \dotfill Peak Velocity Condition
    \item QCD  \dotfill Quantum Chromodynamics
    \item ReLU \dotfill Rectified Linear Unit
    \item SFDM \dotfill Scalar Field Dark Matter
    \item SH   \dotfill Soliton-Halo (relation)
    \item SMBH \dotfill Supermassive Black Hole
    \item SM   \dotfill Standard Model (of particle physics)
    \item SP   \dotfill Schrödinger-Poisson (equations)
    \item SPARC \dotfill Spitzer Photometry \& Accurate Rotation Curves
    \item TF   \dotfill Thomas-Fermi
    \item UFD  \dotfill Ultra-Faint Dwarf
    \item UGC  \dotfill The Uppsala General Catalog (of galaxies)
    \item ULA  \dotfill Ultra-Light Axions
    \item ULDM \dotfill Ultra Light Dark Matter
    \item WIMP \dotfill Weakly Interacting Massive Particle
    \item WKB  \dotfill Wentzel–Kramers–Brillouin (approximation)
\end{enumerate}

\cleardoublepage
\phantomsection		
\pagenumbering{arabic}

\justifying
\pagestyle{fancy}
\chapter{Introduction}\label{chpt:introduction}

\section{The observational frontier of cosmology}


Over the last century, our understanding of the Universe has improved dramatically. 
While previously a field dominated by observations in optical wavelengths, observational cosmology today benefits from many windows in the electromagnetic spectrum to the Universe~\cite{Weisskopf_2000, Livio_2003, Bennett_2003, Werner_2004, Perley_2011, Berriman_2014, Clements_2017, Thompson_2022, McElwain_2023}. 
Additionally, our ability to detect cosmic rays (see section~30 of~\cite{PDG_2024}), and recently cosmic neutrinos and gravitational waves \cite{Walter_2008, Abbott_2009, IceCube_2023, LISA_2024} has provided new avenues of observation. 


In this age of data-driven cosmology, the concentrated efforts of theorists and experimentalists working together 
over the last few decades, has established a standard model of cosmology~\cite{Dodelson_2003, Turner_2022}, i.e. the $\Lambda$CDM model. 
The core assumptions of $\Lambda$CDM are: 
\begin{itemize}
    \item Gravity is described by Einstein's general relativity,
    \item The early Universe was homogeneous, isotropic and spatially flat, save for very small scalar metric perturbations, which were adiabatic, Gaussian and had a nearly scale invariant power spectrum,
    \item In addition to the particles in the Standard Model of particle physics, the Universe also contains Dark Matter (DM) and Dark Energy (DE), which dominate the energy density of the Universe.
\end{itemize}

Using the values of the $6$ free parameters obtained from various observations, this model can correctly predict the abundances of light elements in the early Universe, anisotropies of the Cosmic Microwave Background (CMB) sky, and provides a paradigm for understanding the formation of large scale structure in the Universe. 
In particular, observations involving extensive galaxy surveys, supernova data, and precise measurements of CMB anisotropies~\cite{Planck2018, BICEP_2021, ACT_2020, eBOSS_2020, DES_2022}, imply that $\sim 68\%$ of the total energy density consists of dark energy, $\sim 27\%$ consists of dark matter, while the familiar baryonic matter only contributes $\lesssim 5\%$.
The $\Lambda$CDM model is a fantastic phenomenological description of observations, and has largely held its own in face of increasingly accurate and precise observational data. 
However, the most dominant components of the Universe, dark matter and dark energy, are completely unfamiliar to us.
As we will see in section~\ref{sec:DM_evidence}, there is overwhelming observational evidence in favour of the existence of dark matter, 
but the physical nature of this component is largely unknown.
Similarly, while observations of the oldest stars, type Ia supernovae (SN Ia), CMB, baryonic acoustic oscillations (BAO) all point to the existence of dark energy, its fundamental nature is unknown~\cite{Amendola_Tsujikawa_2010}.
In addition to the nature of dark energy, its energy density known from observations is much too small compared to the one expected from short distance physics~\cite{Martin_2012}, and this is dubbed the cosmological constant problem. 
Further limitations of the current picture of cosmology includes a lack of understanding of the origin of baryon asymmetry~\cite{Pereira_2023}, as well as lack of a smoking gun signature of cosmic inflation, which is the generally accepted mechanism to address the flatness and horizon problems~\cite{Achucarro_2022, Ellis_2023}.


In the age of precision cosmology, as datasets get larger and errors get smaller, new issues have cropped up. 
A primary example of this is the recently strengthened Hubble tension as well as the $\sigma_8$ tension~\cite{Di_Valentino_2021, Perivolaropoulos_2022, Abdalla_2022}, which stem from discrepancies between late-time and early-time measurements.
In addition, recent analysis of the BAO measurements from 14 million galaxies as a part of the Dark Energy Survey Instrument (DESI) survey \cite{DESI_2025} has favoured a time evolving equation of state pointing to a dynamical dark energy model.



In addressing the above-mentioned unresolved issues, observational data will be of paramount importance. 
Indeed, as we are expected to acquire ever larger datasets with more accurate observations from ongoing and upcoming missions like LSST,
CMB-S4, DESI, Euclid, etc. \cite{Ivezić_2019, Abazajian_CMBS4_Snowmass2021, DesiCollabVI_2024, Euclid_2024}, confronting theoretical models with observed data using various statistical techniques 
including those based on machine learning techniques 
are expected to play an important role in cosmology and astrophysics. 

Therefore in this thesis, we rely on observations to guide us in tackling one of the primary unresolved issue in cosmology: the physical nature of dark matter. 
In particular, as we shall see in the upcoming sections, this thesis will be based on the following assumptions: 

\begin{itemize}
    \item Dark matter is made of elementary particles, i.e., we don't consider alternative explanations, for instance, Modified Newtonian Dynamics, Primordial Black Holes or Primordial Naked Singularities, 
    \item Dark matter consists of a single species of particles that belong to physics beyond the Standard Model,
    \item Dark matter particles have spin-$0$, and the smallest possible mass consistent with structures in the Universe,
    \item Couplings of the  dark matter particle to Standard Model particles are negligible, i.e. consistent with current experimental constraints. 
\end{itemize}

In the context of these assumptions, we ask whether observations in the central regions of galactic halos can be used to probe the self-interactions of dark matter particles. 
We shall remain completely agnostic about the theoretical model of dark matter, though we shall regard Ultra Light Axions (ULAs) as the benchmark case that we will occasionally compare our results with. 


\section{Astrophysical aspects 
of Dark Matter}\label{sec:dark_matter}

Dark matter is one of the longest standing mysteries in modern cosmology. Over the course of almost a century, various astrophysical observations have  pointed towards a significant fraction of mass within galaxies and clusters to come from a non-baryonic, non-luminous, non-relativistic, collisionless component that interacts mostly gravitationally with itself and Standard Model particles.
We now briefly discuss the evidence for this component from various astrophysical and cosmological observations (see~\cite{Profumo_2019, Safdi_TASI_2023} for details).


\subsection{Evidence for dark matter}\label{sec:DM_evidence}

\subsubsection{Evidence at galactic scales}

Evidence of dark matter at galactic scales comes from the efforts of Vera Rubin and collaborators \cite{Rubin_1970, Rubin_1980}, in the form of rotation curves of galaxies. 
A rotation curve is the circular velocity of stars and gas moving in the gravitational potential of the galaxy.
From Newtonian mechanics, one can relate the circular velocity of any test particle at a radius $r$ from the galactic centre to the mass enclosed within that radius, $M(r)$ by 
\begin{equation}\label{eq:circ_vel_1}
    v(r) = \sqrt{\frac{GM(r)}{r}}\ .
\end{equation}

\begin{figure}[ht]
    \centering
    \includegraphics[width=0.8\linewidth]{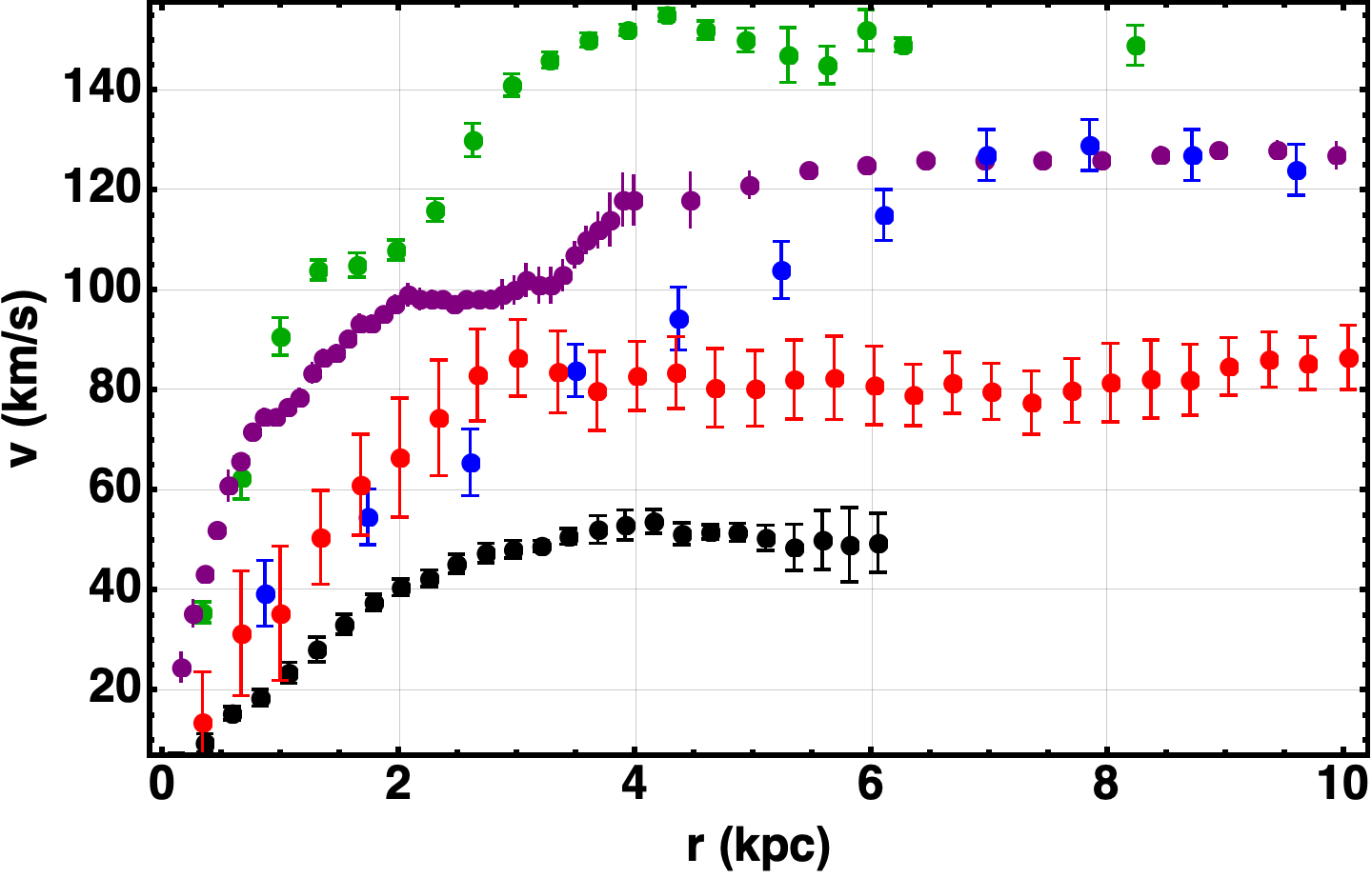}
    \caption[Examples of observed rotation curves]{Observed rotation curves for the following galaxies from the Spitzer Photometry \& Accurate Rotation Curves (SPARC) catalogue~\cite{Lelli_2016}: Black - NGC 2366 (barred irregular dwarf galaxy), Green - NGC 6015 (spiral galaxy), Purple - NGC 2403 (intermediate spiral galaxy), Blue - NGC 4010 (barred spiral galaxy), Red - NGC 2915 (blue dwarf galaxy).}
    \label{fig:rotation_curves_example}
\end{figure}

Since the enclosed mass is just $M(r) = 4\pi\int_0^rdr'r'^2\rho(r')$ (assuming a spherically symmetric density distribution $\rho(r)$), as one goes far from the centre, $\rho(r)$ is expected to be zero and $M(r)$ becomes constant. 
Hence, the rotation curve should feature the so-called Keplerian fall-off where $v(r) \propto r^{-1/2}$. 
However, for most galaxies, it was found that the velocity at large $r$ remained flat (see figure~\ref{fig:rotation_curves_example} for a few examples). 
From eq.~(\ref{eq:circ_vel_1}), this implies a mass distribution that follows $M(r) \propto r$, suggesting the presence of an additional mass surrounding the galaxy, extending far beyond luminous matter.
It is worth noting that decades of simulations and observations suggest that every galaxy forms within a dark matter halo~\cite{Wechsler_2018}, which accounts for a large part of the total mass in a typical galaxy. 

More details about galactic rotation curves and how they can be used to constrain dark matter models are discussed in chapters~\ref{chpt:paper_2} and~\ref{chpt:paper_4}.

\subsubsection{Evidence at cluster scales}

One of the early evidences for the existence of dark matter came in 1930s, when Fritz Zwicky studied the dynamics of the Coma cluster~\cite{Profumo_2019} . 
Using spectral redshift of galaxies within the cluster one can determine their velocities relative to the cluster. 
Using the resultant velocity dispersion of galaxies along with the virial theorem (assuming that the Coma cluster is a stable system), one can then estimate the total dynamical mass of the cluster. 
When Fritz Zwicky compared this mass to the luminous mass of the cluster, he found that the dynamical mass of the cluster was $400$ times larger than the luminous mass, suggesting the presence of additional invisible (dark) mass that did not emit electromagnetic radiation. 
A similar discrepancy was found in the Virgo cluster a few years later \cite{Smith_1936}.
Recent analysis of the galaxies in the Coma cluster~\cite{Lokas_2003} have also shown that most of the mass ($\sim 85\%$) is in the form of dark matter. 

One can also obtain the total mass as well as distribution of mass inside of a cluster using modern techniques like strong and weak gravitational lensing.  This verifies that most matter in galactic clusters is in the form of dark matter~\cite{Kaiser_1993, Vegetti_2023}. 
A prominent example is the Bullet Cluster, which we shall briefly discuss. 
\begin{figure}[h]
    \centering
    \includegraphics[width=0.8\linewidth]{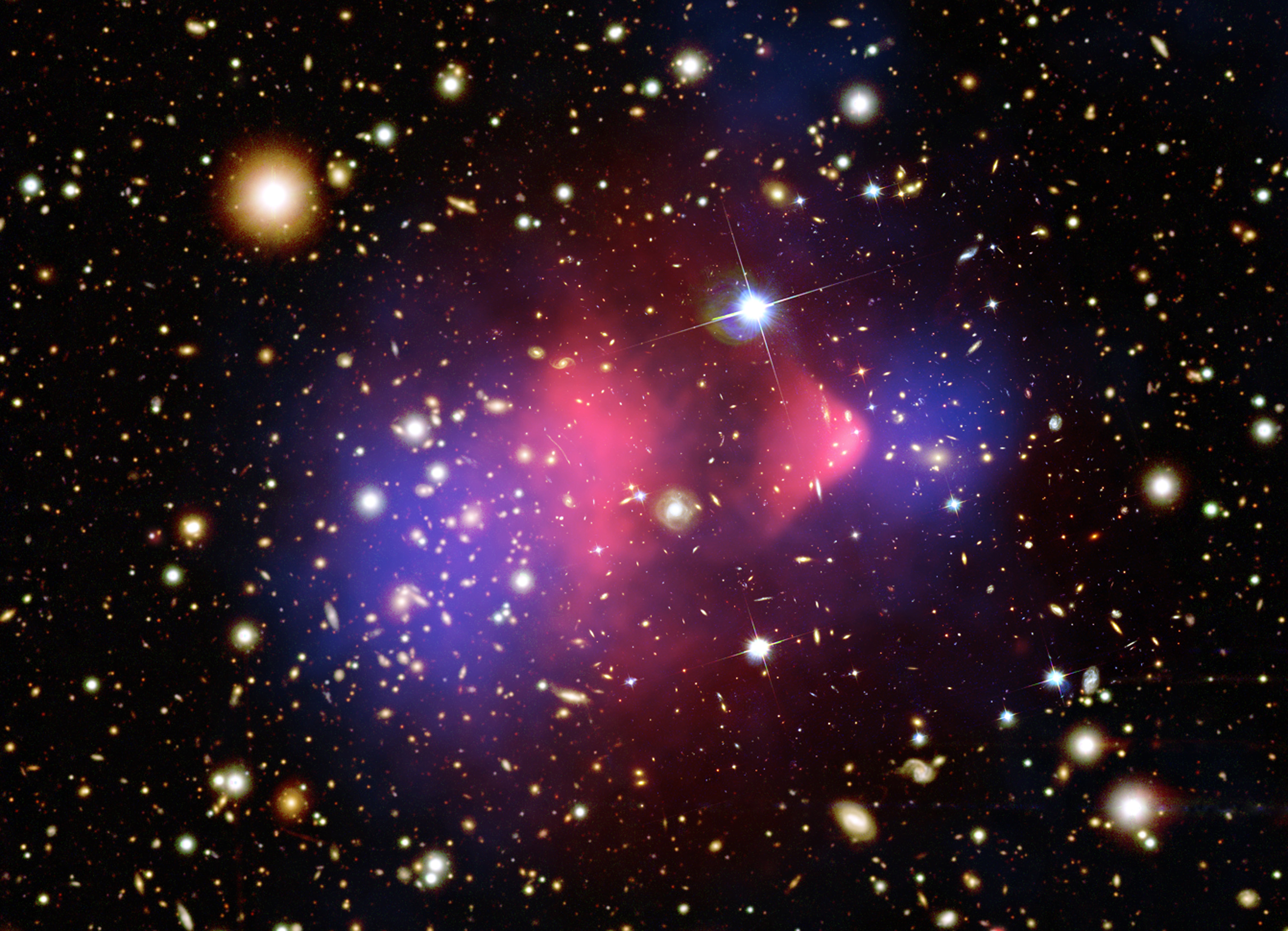}
    \caption[A composite image of the Bullet Cluster]{Credit - X-ray: NASA/CXC/CfA/M.Markevitch, Optical and lensing map: NASA/STScI, Magellan/U.Arizona/D.Clowe, Lensing map: ESO WFI. A composite image of the Bullet Cluster.}
    \label{fig:bullet_cluster}
\end{figure}

The cluster 1E 0657-
56, also known as the ``Bullet Cluster'', is an aftermath of a collision of two clusters. A composite image for the same is shown in figure~\ref{fig:bullet_cluster}. 
The image is a superposition of three different observations: (a) An X-ray image observed from the Chandra telescope of the intracluster gas (pink), (b) An optical image taken by the Magellan and Hubble telescope of the galaxies in the cluster as well as in the background, and (c) Mass distribution in the cluster obtained from a weak gravitational lensing analysis of the background (blue). 

Since most of the baryonic matter in a cluster is in the form of gas, the expectation in the absence of dark matter would be that, the mass distribution obtained from gravitational lensing should overlap the mass distribution of gas.
Note that the hot gas in the cluster interacts during the collision, and slows down due to ram pressure, while the galaxies (i.e., the stellar component) are effectively collisionless. 
However, what is observed is that the peak of the mass distribution obtained form gravitational lensing is displaced away from the X-ray halo as if most of the matter in the cluster was essentially collisionless and interacted only gravitationally~\cite{Clowe_2004, Markevitch_2004}.
Such evidence is available from other merging clusters as well, for instance, MACS J0025.4-1222~\cite{Bradac_2008, Roos_2010}. 
This empirically shows that most of the matter in the cluster is non-luminous, and interacts only gravitationally with itself and with luminous matter in the cluster.

\subsubsection{Evidence at cosmological scales}


In the early Universe, just before photon decoupling, matter and radiation were tightly coupled through Thomson scattering of photons and electrons. 
At the time of decoupling, the temperature of the Universe fell enough for sufficient number of the protons and electrons to form neutral hydrogen and the mean free path of photons became larger than the size of the Universe at the time. 
Now, given some initial density perturbations in the Universe, the inhomogeneities in the temperature that we observe in the remnant radiation (i.e., CMB) are a snapshot of the inhomogeneities in the radiation-baryon fluid when the photons decoupled, around $380000$ years after the big bang. 
Since temperature fluctuations peak around $10^{-4}\ \text{K}$, the density contrast during recombination is approximately $\delta\rho/\rho \sim 10^{-4}$  which occurred when the scale factor $a_{dec} \sim 10^{-3}$, i.e. $1000$ times smaller than it is currently ($a(t_0) \equiv 1$ by definition).

From linear perturbation theory, we know that perturbations in matter (once decoupled from photons) grow as $\delta\rho/\rho \propto a$. 
Hence, not enough time would have passed until present day for these perturbations to form the non-linear structures that we observe in the Universe (i.e. we need $\delta\rho/\rho \gg 1$) \cite{Profumo_2019, PDG_2024}. 
One solution to this is the existence of a dark matter component, which decoupled from the early Universe plasma long before recombination and started forming gravitational wells due to small inhomogeneities. 
Once photons decoupled from baryonic matter, baryons would then fall into the deeper gravitational wells created by dark matter density perturbations, which would explain the more rapid growth of structures expected from observations.

\subsubsection{Evidence for dark matter from early Universe}

As mentioned earlier, at around $z_{dec} \approx 1100$~\footnote{Note that redshift is defined in terms of the scale factor as $1 + z = a(t_0)/a(t)$.}, photons decoupled from the baryons and were free to propagate throughout the Universe uninhibited. 
These Cosmic Microwave Background (CMB) photons that we observe today therefore contain a snapshot of the inhomogeneities in temperature (and hence density) at $z_{dec}$ or $\sim380000$ years after the big bang. 
These inhomogeneities lead to the anisotropies in the CMB and hence the CMB angular power spectrum~\cite{Planck2018}.
The power spectrum features many peaks, representing higher temperature fluctuations at certain angular scales.
The relative heights of these peaks can be explained by the presence of oscillations in the photon-baryon fluid, which manifest because of the attractive force of gravity due to dark matter competing with the repulsive force of the radiation pressure due to relativistic photons. 
Thus, the amount of dark matter relative to the amount of baryons will change the nature of such oscillations, which can be seen by the height and position of peaks in CMB power spectrum~\cite{Dodelson_2003, Balazs_2024}.
Using cosmological (linear) perturbation theory, one can calculate this power spectrum at $z_{dec}$ given some initial conditions and the parameters of a particular cosmological model ($6$ for $\Lambda$CDM). It has been found through various CMB observations~\cite{Bennett_2003, Planck2018} that the present day density of dark matter must be $\Omega_{DM}\sim 0.27$, to explain observed CMB anisotropies. 

The hot big bang model predicts that the light elements $\text{D},\ ^3\text{He},\ ^4\text{He},\ ^7\text{Li}$ that we observe today were formed very early (minutes after the big bang) in the Universe, as temperature of the Universe fell below the binding energy of Deuterium~\cite{Dodelson_2003}.
This resulted in almost all surviving baryons (protons and neutrons) to fuse into these light elements, accounting for all the non-relativistic baryonic matter in Universe. This is called the Big Bang Nucleosynthesis (BBN). 
Observations of metal-poor\footnote{Note that `metals' in astronomy implies all elements heavier than hydrogen and helium.} sources in the Universe has enabled us to constrain these primordial abundances and therefore the energy density of baryons in the Universe~\cite{Cyburt_2016, Schoneberg:2024ifp}. 
Current observations of primordial abundances (for instance~\cite{OMeara_2000}) are consistent with a Universe with a baryon density fraction $\Omega_b = \rho_b/\rho_{crit} \approx 0.05$ (assuming $H_0 = 68\ \text{km}/\text{s}/\text{Mpc}$ for illustration). Here, $\rho_{crit}\equiv 3H_0^2/(8\pi G)$ is called the critical density. 
On the other hand, measurements of the galaxy cluster mass function~\cite{Planck_2015} tells us that the total non-relativistic matter density fraction $\Omega_m = \rho_m/\rho_{crit} \approx 0.3$.
Therefore, simply subtracting the baryon density fraction from the total matter density fraction tells us that most of the non-relativistic matter in the Universe must be non-baryonic.

\subsection{Macroscopic properties of dark matter}\label{sec:DM_macro_nature}


\subsubsection{Cold or warm?}
`Cold' refers to the dark matter component being non-relativistic when it decoupled from the primordial plasma.
It implies that the velocity of DM particles is negligible compared to the speed of light, $v \ll c$ when structure was formed. 
In this limit, the pressure associated with the DM fluid satisfies $P \ll \rho$, where $\rho$ is the energy density of the DM component, implying the equation of state to be $w = P/\rho \approx 0$.
If the velocity of dark matter particles after decoupling is too large, i.e. it is hot, structures at a scale dependent on the velocity of DM will be washed away due to free streaming~\cite{Safdi_TASI_2023}.
Note that there is also an intermediate regime between negligible and quasi-relativistic velocities, where the smoothening out of structure at some length scales (dwarf galaxies or cores of galaxies) is desirable to address some of the small scale issues with $\Lambda$CDM~\cite{Bullock_2017}.
Such models are called Warm Dark Matter, examples for which are sterile neutrinos and gravitinos~\cite{Cirelli_DM_review_2024}.


\subsubsection{Non-luminous}
From the evidence we discussed at cluster scales, it is clear that lensing observations point towards the existence of additional mass in clusters, that does not emit any electromagnetic radiation, nor does it interact with regular matter (for instance hot gas) except gravitationally.
Similarly, observations at galactic scales, such as rotation curves can be explained by introducing a (spherical) dark matter halo within which the galaxy sits and that extends far beyond the galaxy itself. Here as well, the effects of this extra component are considered to be purely gravitational, and it does not interact with or emit any electromagnetic signals. 
This implies that dark matter must be non-luminous, in that it does not couple to electromagnetism~\cite{Profumo_2019, PDG_2024}. 
In the case of particle dark matter, this translates to the DM particle having a small electric charge, and a small electric dipole moment. 

\subsubsection{Collisionless}
Observations of merging clusters can not only infer the mass-distribution of clusters but can also probe the self-interactions of dark matter itself. 
In-fact, one can constrain the self-interaction cross-section of dark matter~\cite{Markevitch_2004, Randall_2008} using the bullet cluster by simulating the merger while including the effects of elastic dark matter scattering. If the self-interactions are too strong, the DM component will lag behind the collisionless galaxies (stellar component), similar to the gas component. 
The lack of separation between the stellar and the dark matter components places stringent constraints on the amount of self-interaction cross-section $\sigma$ of DM, imposing $\sigma/m \lesssim 1\ \text{cm}^2/g$.

\subsection{Beyond particle dark matter}\label{sec:beyond_DM}

Before proceeding further, we note that attempts to explain some astrophysical observations, like galactic rotation curves without invoking the presence of a non-baryonic component have been made, where Newtonian gravity breaks down in the limit of small accelerations \cite{Milgrom_1983, Gentile_2011} (Such models are also called MOdified Newtonian Dynamics or MOND). We shall not explore this further in this thesis. 

There is also a class of dark matter candidates which involves composite macroscopic objects heavier than Planck scale. 
A prominent example is Massive Astrophysical Compact Halo Objects (MACHOs), which are essentially macroscopic objects made of baryons that do not emit light such as large planets, dead stars, stray black holes, etc. But, to account for the late time non-relativistic energy density, MACHOs would require a large baryonic abundance which is ruled out by BBN observations. Hence, at the very least, MACHOs cannot be all of dark matter. 
Current bounds on the amount of dark matter that can be accounted for by MACHOs from gravitational micro-lensing observations is $f < 5\%$ for objects with masses in the range $\left[10^{-11}\ M_\odot, 100\ M_\odot\right]$\cite{Cirelli_DM_review_2024}. Similar bounds exist for heavier objects as well, details for which are discussed in Ref.~\cite{Cirelli_DM_review_2024}. 

Another possibility of macroscopic objects made of baryons are Primordial Black Holes (PBHs). However, to avoid spoiling the light element abundances, these black holes must be created before BBN. Similar constraints to the ones on MACHOs also apply to PBHs since they also rely on gravitational effects. Additional constraints based on PBHs not evaporating until present day, accretion of matter in dense regions, lack of observed gravitational wave signals, etc. exclude much of the allowed parameter space. 
Only a small window $10^{-16}\ M_\odot \lesssim M \lesssim 3\times 10^{-12}\ M_\odot$ where PBHs could account for all of dark matter remains viable \cite{Cirelli_DM_review_2024}.
Recently there have also been attempts to explain dark matter using Primordial Naked Singularities~\cite{Joshi_2024}.

\section{Particle physics nature of dark matter}\label{sec:DM_micro_nature}

If dark matter comprises of particles, then its particle physics nature is unknown. For example, what is the identity of the particle that makes up dark matter? What is its elementary particle mass? What is its intrinsic spin? What is its representation in the internal symmetry group of Standard Model? What is its lifetime? Does it have any couplings to Standard Model particles? What are the values of allowed couplings, etc?
The $\Lambda$CDM model itself incorporates very little information about the microscopic nature of dark matter and 
evidence at both astrophysical and cosmological scales, that we saw in the previous section, sheds light on mostly bulk properties of DM.

\subsection{Dark matter and the Standard Model}


Can particles (elementary or composite) in the Standard Model (SM) describe a cold, non-luminous, collisionless, clustering component, which can act as dark matter?

To answer this, recall that dark matter particle must be sufficiently long-lived, to survive on cosmological timescales.  
On the other hand, all particles in the Standard Model quickly decay into protons, neutrons, neutrinos, photons and electrons.  
We know that pure radiation, i.e. photons, cannot be dark matter since they are relativistic and will wash away structures. 
Similarly, neutrinos are also too hot to be dark matter. 
Since DM is not expected to couple to electromagnetism as we saw in section~\ref{sec:DM_macro_nature}, electrons are out of the picture as well. 
The remaining protons and neutrons, after undergoing big bang nucleosynthesis, account for the light element ($\text{D}$, $^3\text{He}$, $^4\text{He}$, $^7\text{Li}$, etc.) abundances. We saw in section~\ref{sec:DM_evidence} that this fixes the total number of baryons in the Universe at the time of BBN. 
Thus, none of the stable Standard Model particles can be dark matter. 
Therefore, assuming particle nature of dark matter necessitates venturing Beyond the Standard Model (BSM) of particle physics. 

It is worth noting, that there have been speculations that Standard Model particles can form other long-lived configurations which could act as dark matter, but their existence is yet to be established~\cite{Jacobs_2014, Bai_2018}. 



\subsection{Dark matter and physics beyond the Standard Model}


\textbf{Lifetime}: The first important property of the DM particle is that it should be long lived.
Since dark matter plays an important role in the late Universe for structure formation, one requires dark matter particles to survive at cosmological timescales. 
For CDM, this leads to a lower bound on the lifetime of DM particles of $\sim \mathcal{O}(100)\ \text{Gyr}$~\cite{Audren_2014}. 

\noindent\textbf{Mass}: Next is its mass, where there is significant uncertainty. 
The mass of the dark matter particle can span a huge range~\cite{Ferreira_2021, Cirelli_DM_review_2024}, $10^{-22}\ \text{eV} \lesssim m \lesssim 10^{19}\ \text{GeV}$. 
We shall explain the lower limit of this range in section~\ref{sec:lower_limit_m}.
In this range, with a cross-section and mass of the order of electroweak scale $\mathcal{O}(100\ \text{Gev})$, lies an extensively studied class of dark matter models called \textbf{Weakly Interacting Massive Particles (WIMPs)}~\cite{Queiroz_2017}.
There are a plethora of other candidates as well that lie within this range, for example, ultralight scalars and axions (wave dark matter), sterile neutrinos, gravitinos (warm dark matter), WIMPZILLAs, etc.~\cite{Kolb_1998, Boyarsky_2019, Hui_2021, Cirelli_DM_review_2024}.

\noindent\textbf{Spin}: Similarly, the spin of the dark matter particle is unconstrained; in-fact we don't even know if the DM particle is a boson or a fermion. However, as we shall see in section~\ref{sec:wave_DM}, once a DM candidate is assumed to be a fermion, its mass cannot be too low. 
Examples of fermion DM are spin-$1/2$ particles like neutralinos~\cite{Queiroz_2017}, sterile neutrinos~\cite{Boyarsky_2019}, or spin-$3/2$ particles like gravitinos~\cite{Cirelli_DM_review_2024}.
On the other hand, examples of bosonic dark matter include spin-$0$ particles like axions (a pseudo-scalar) and axion-like particles~\cite{Marsh_2016, Hui_2017}, as well as spin-$1$ particles like ultralight vector fields or dark photons~\cite{Fabbrichesi_2020}, and heavy spin-$2$ particles~\cite{Dubovsky_2004, Babichev_2016}.

\noindent\textbf{Couplings}: Another important aspect of the DM candidate involves its couplings. This includes its couplings to SM particles, to itself, and to particles in the Hidden Sector. 
For instance, DM coupling to photons would require it to be electrically charged.
Since electrically charged DM would affect CMB anisotropies by interacting with photons, one can obtain limits on the electric charge for the DM (assuming it is thermally produced) by requiring that it be completely decoupled from the baryon-photon plasma during recombination.
This constrains the electric charge of DM to be $\lesssim 10^{-6}e$, where $e$ is electron charge (see section~27 of \cite{PDG_2024}).
Direct detection experiments probe couplings of DM to Standard Model particles, like axion-photon couplings~\cite{Kimball_book_2023} in case of axions and DM-nucleon or DM-electron cross-sections in case of WIMP dark matter~\cite{Billard_2021}.
Given the sheer number of DM candidates with wildly different physical properties, the exact details of DM's coupling to matter varies wildly although they are still required to be small. 

\noindent\textbf{Hidden Sector}: An intriguing possibility is that DM itself is part of a far larger sector of particles and additional forces, that couple very weakly to Standard Model.
DM could be one of the few stable particles in the Dark Sector similar to how electrons and protons are accompanied by a much larger collection of SM elementary particles~\cite{Abdalla_2020, Cirelli_DM_review_2024}.


\subsubsection{Thermal and non-thermal dark matter}
One way to classify particle dark matter models is by their production mechanism.
If the dark matter was weakly interacting (for instance, a WIMP) with standard model particles, it was in thermal equilibrium with the primordial plasma in the early Universe at some point. Once the interaction rate fell below the Hubble expansion rate, i.e. $\Gamma \ll H$, the dark matter abundance `froze-out' to its current observed value~\cite{Dodelson_2003}. 
WIMPs interacting via the weak force with standard model particles have been a popular candidate for thermally produced dark matter. 
Thermally produced dark matter also imposes a lower bound on the mass of the dark matter particle, $m \gtrsim \text{keV}$~\cite{Cirelli_DM_review_2024} for the case of cold dark matter. 

On the other hand, for candidates that were never in thermal equilibrium with the primordial plasma, more novel mechanisms are required to ensure correct present day abundance. Example for the same are the freeze-in mechanism or initial misalignment (as discussed in section~\ref{sec:production_wave_DM}). 
In this case, as we shall briefly discuss in a later section, the mass of the dark matter particle can be far lighter (ultralight axions) than if it was produced thermally.

\subsubsection{Detection efforts}

There has been a massive effort by the physics community to search for signals of dark matter. These are categorized in two distinct techniques: Direct and indirect detection. 

Direct detection of DM relies on the fact that our solar system lies inside the Milky Way dark matter halo.
This implies that dark matter particles are constantly passing through the earth and terrestrial detectors can potentially pick up signals of DM interacting with a target nucleus or an electron (for the case of WIMPs) or with a strong magnetic field (for the case of axions)~\cite{Billard_2021, Cirelli_DM_review_2024} . 
A few examples of recent direct detection experiments for WIMPs include XENON1T~\cite{XENON_2018} and PandaX-II~\cite{PandaX-II_2017} which are liquid Xenon detectors, DarkSide~\cite{DarkSide_2018} which is a liquid Argon detector, and SuperCDMS~\cite{SuperCDMS_2022}, a Cryogenic detector.  
However, no signal has been observed yet and WIMP DM candidates in the $\text{GeV}$-range and above are getting increasingly constrained~\cite{Billard_2021}.  
Some prominent examples of direct detection experiments are the Axion Dark Matter eXperiment (ADMX)~\cite{ADMX_2023} and Oscillating Resonant Group AxioN (ORGAN)~\cite{McAllister_2017} which are haloscopes that search $\sim \mu\text{eV}$-range axions. 
Other examples include Light-shining-through-the-wall (LSW) experiments (for instance, ALPS at DESY~\cite{Ehret_2010}) which are conducted at laboratory scales using strong magnets to convert photons into axions~\cite{Billard_2021}. 
Yet another example of axion detection experiments includes atomic clocks~\cite{Arvanitaki_2014, VanTilburg_2015}.
However, no signals for axions have been found yet. 

Indirect detection, on the other hand, depends on the detection of excess cosmic rays, gamma rays or neutrinos which are products of DM annihilation or decay~\cite{Cirelli_DM_review_2024}. 
Indirect detection therefore relies on telescopes like the ground-based High Energy Stereoscopic System (H.E.S.S.) and The Cherenkov Telescope Array~\cite{Abdalla_2022, Hofmann_2023}, 
as well as space-based telescopes like Alpha Magnetic Spectrometer (AMS) and
Fermi Gamma Ray Space Telescope (Fermi-LAT)~\cite{Fermi-LAT_2016, Zuccon_2019} which search for DM annihilation signals. As is the case with direct detection, no DM signal has been observed yet~\cite{Arcadi_2024}.

\section{Wave Dark Matter}\label{sec:wave_DM}

While the WIMP paradigm is not completely ruled out, the allowed parameter space has been shrinking rapidly over the years~\cite{Billard_2021, Arcadi_2024}. It is therefore more important than ever to consider alternative scenarios. 
Indeed, a number of promising candidates emerge as we allow the mass of the particle go far below the $\text{GeV}$ scale.
In this section (and in this thesis), we shall consider dark matter with a very small particle mass. 

\subsection{Sufficiently light dark matter must be bosonic}


Consider a dwarf galaxy of radius $R$ and total mass $M$. Suppose it is made of dark matter particles with mass $m$.
The number density of these particles in the galaxy will be $n = N/V = M/m/(4/3\pi R^3)$. 

If these particles are fermions, then this number density implies that the corresponding Fermi energy will be of the order of $E_f \sim \hbar^2(3\pi^2 n)^{2/3}/2m$, which in-turn implies a Fermi velocity of the order of  
\begin{equation}\label{eq:Fermi_velocity}
    v_f = \frac{\hbar}{m}\left(\frac{9\pi}{4}\frac{M}{mR^3}\right)^{1/3}\ . 
\end{equation}
Note that, such a dwarf galaxy will have an escape velocity given by $v_{esc} = \sqrt{2GM/R}$.
Requiring that the Fermi velocity be smaller than the escape velocity, i.e. $v_f \lesssim v_{esc}$, gives us 
\begin{equation}\label{eq:lower_bound_m}
    m \gtrsim \frac{1.257}{M^{1/8}}\left(\frac{\hbar^2}{GR}\right)^{3/8}\ .
\end{equation}
For a typical dwarf galaxy with mass $M \sim 5\times 10^7\ M_\odot$ and radius $R\sim 2.5\ \text{kpc}$, one obtains the lower bound on fermionic dark matter to be $m \gtrsim 100\ \text{eV}$~\cite{Safdi_TASI_2023}.
This is known as the Pauli exclusion principle bound. 
A more detailed argument, incorporating coarse-graining and phase-mixing effects was given by~\cite{Tremaine_1979}, which is also known as the Gunn-Tremaine bound. 

On the other hand, if these particles are bosons, they are not restricted by an exclusion principle and multiple particles can occupy the same quantum state. 
This implies that any dark matter candidate with $m \lesssim 100\ \text{eV}$ necessarily has to be a boson.

\subsection{Sufficiently light bosonic dark matter can be described using classical fields}\label{sec:classical_field_description}
For the case bosonic dark matter with mass $m$ and velocity $v$, the average number of particles in a volume of side deBroglie wavelength $\lambda_{dB} = 2\pi/mv$ in the local Universe is simply \cite{Hui_2017, Cirelli_DM_review_2024}
\begin{equation}
    N_{dB} = \frac{\rho_\odot}{m/\lambda_{dB}^3} \approx \left(\frac{30\ \text{eV}}{m}\right)^4\left(\frac{10^{-3}c}{v}\right)^3\ .
\end{equation}
Here we have used $\rho_\odot = mn_\odot \approx 0.4\ \text{GeV}/\text{cm}^3$~\cite{de_Salas_2021}. 
For smaller masses, i.e. $m \ll 30\ \text{eV}$, $N_{dB}$ is huge number, and in this limit, a collection of DM particles is better described using a classical scalar field (similar to treating a large collection of photons using the classical electromagnetic field). This is why dark matter models involving such small particle masses are often called \textbf{Wave Dark Matter}
~\cite{Hui_2021}, or \textbf{Scalar Field Dark Matter (SFDM)}~\cite{Urena-Lopez_2019, Matos_2000}.

\subsection{Spin-$0$ dark matter}
While very light dark matter can have any integer value of spin, for the rest of this thesis, we shall focus our attention on spin-$0$ particles. 
This is clearly the simplest possibility worth examining, and as we shall see, astrophysical observations can be used to impose constraints on parameters of the Lagrangian of such a scalar. 
Thus, whenever we use the terminology Wave Dark Matter or Ultra Light Dark Matter, we are referring to spin-$0$ particles.

\subsection{Production of wave dark matter}\label{sec:production_wave_DM}

As discussed earlier, such a light scalar cannot be produced thermally, since we require DM to have negligible velocities during structure formation. 
We therefore need an alternative mechanism, an example for which is the misalignment mechanism (see \cite{Marsh_2015, Hui_2017, Kimball_book_2023} for a detailed discussion) to ensure the correct present-day relic density for scalar field dark matter. Consider a real scalar field $\varphi$ with a quadratic potential $U(\varphi) = \frac{1}{2}m^2\varphi^2$, where $m$ is its mass.
The equation of motion of this scalar field in FRW spacetime is then 
\begin{equation}\label{eq:EOM_FRW}
    \ddot{\varphi} + 3H\Dot{\varphi} + m^2\varphi = 0\ ,
\end{equation}
where $H \equiv \Dot{a}(t)/a(t)$, and $a(t)$ is the scale factor. 
Let us now consider the case where the initial value of the field $\varphi$ is not at the minimum of this potential but at some non-zero value $\varphi_i$.
In the early Universe, when $H$ is very large, $H \gg m$ and the Hubble friction term dominates, implying a solution to eq.~(\ref{eq:EOM_FRW}) that is $\varphi = const.$
As the Universe expands and the expansion rate drops, $H < m$, the scalar field rolls towards the minimum and starts to oscillate around it. In this limit, the oscillation amplitude is dampened by the expansion of the Universe and the density of the scalar field, given by $\rho = \frac{1}{2}\Dot{\varphi}^2 + \frac{1}{2}m^2\varphi^2$ goes as $\rho\propto a^{-3}$, save for small oscillations.
This is the expected behaviour for a non-relativistic fluid. 
Assuming the scalar field constitutes all of dark matter, one obtains the current relic density \cite{Ferreira_2021}
\begin{equation}\label{eq:SFDM_relic}
    \Omega_{SFDM} \sim \frac{1}{6}\left(9\Omega_r\right)^{3/4}\left(\frac{m}{H_0}\right)^{1/2}\left(\frac{\varphi_i}{m_{pl}}\right)^2\ ,
\end{equation}
where $H_0$ is the Hubble constant and $m_{pl}$ is Planck mass. 
Requiring that the current relic density agrees with observations, implies $\varphi_i > 10^{14}\ \text{GeV}$ as discussed in \cite{Ferreira_2021}. 
Note that the initial value of field, i.e. $\varphi_i$ is determined by inflationary dynamics in the early universe~\cite{Cirelli_DM_review_2024}.

If the scalar field is an axion-like particle, the full potential is given by~\cite{Hui_2021},
\begin{equation}\label{eq:axion_potential}
    U(\varphi) = \Lambda^4\left[1 - \cos{\left(\frac{\varphi}{f_a}\right)}\right] \ ,
\end{equation}
where $f_a$ is called the axion decay constant.
Expanding the cosine in powers of $\varphi/f_a$ around $\varphi = 0$, yields the following leading order term: $U(\varphi) = \frac{1}{2}\frac{\Lambda^4}{f_a^2}\varphi^2$. 
Defining mass of the axion to be $m_a \equiv \Lambda^2/f_a$, one recovers the quadratic potential that we utilized earlier, and the misalignment mechanism proceeds as normal. 
Note that the initial displacement of the field is now given by $\varphi_i/f_a$ which is also called the misalignment angle, and truncating the Taylor expansion at the quadratic term is valid only for small misalignment angles. 
The expression for the current relic density of axion dark matter can be written as~\cite{Hui_2017}
\begin{equation}\label{eq:axion_relic}
    \Omega_a \sim 0.1\left(\frac{f_a}{10^{17}\ \text{GeV}}\right)^2\left(\frac{m_a}{10^{-22}\ \text{eV}}\right)^{1/2}\ .
\end{equation}
Hence, axions with $m_a\sim 10^{-22}\ \text{eV}$ and a decay constant $f_a\sim 10^{17}\ \text{GeV}$ can indeed fulfil the role of dark matter.

What we have discussed here is just one example of the misalignment mechanism. There are others as well, such as kinetic misalignment, where the scalar field has a non-zero initial velocity \cite{Co_2020, Chang_2020} or trapped misalignment~\cite{Di_Luzio_2020} where the field is trapped in the wrong minimum.

\subsection{Lower limit on particle mass}

\subsubsection{Limit from production mechanism}\label{sec:lower_limit_m}

It is important to realize that the scalar field cannot be arbitrarily light if it is to constitute all of dark matter. 
In-fact, one can put lower bounds on the mass from basic cosmology. 
For instance, from current observed energy density for radiation and non-relativistic matter (which is mostly dark matter), we know that the Universe transitioned from being radiation dominated to being matter dominated at $a_{eq} = \frac{\Omega_{r}}{\Omega_{m}} \sim 3\times 10^{-4}$. 
Since ULDM is produced via the misalignment mechanism, the field must start oscillating (i.e. behaving like DM) at the latest during the matter-radiation equality; using this condition, one can obtain a lower bound on the scalar field mass, $m \gtrsim H(a_{eq}) \approx 10^{-28}\ \text{eV}$ \cite{Kimball_book_2023}.

\subsubsection{Limit from existence of dwarf galaxies}

We have already discussed how Pauli's exclusion principle as well as background cosmology can put lower bounds on the the mass of fermionic as well as bosonic dark matter respectively. 
Now, let us look at how the inhomogeneous Universe, in particular small scale structure, imposes a bound on ultralight bosonic dark matter. A simple example is the existence of dwarf galaxies of size $R\sim 2.5\ \text{kpc}$ and mass $M\sim 5\times 10^7\ M_\odot$. 
Due to its small mass, the effects of the uncertainty principle, i.e. $\Delta x\Delta p \gtrsim 1/2$, for light bosonic dark matter can be felt at galactic scales. 
For instance, given $\Delta x = R$, the uncertainty principle implies that the particle velocity cannot be determined to a precision smaller than 
\begin{equation}
    v \gtrsim 20\ \text{km}/\text{s}\frac{10^{-22}\ \text{eV}}{m}\ .
\end{equation}
For light bosons with $m \lesssim 10^{-22}\ \text{eV}$, $v > v_{esc}$ ($v_{esc}$ is the escape velocity defined below eq.~(\ref{eq:Fermi_velocity})) implying such dwarf galaxies will not be formed. Hence, $m \gtrsim 10^{-22}\ \text{eV}$ is required to be consistent with observed structures. 
One can therefore define a length scale at which effects due to the uncertainty principle or the wave nature will be manifest by the deBroglie wavelength of the scalar field, 
\begin{equation}\label{eq:deBroglie}
    \lambda_{dB} = \frac{2\pi}{mv} \approx \text{kpc}\left(\frac{10^{-22}\ \text{eV}}{m}\right)\left(\frac{100\ \text{km}/\text{s}}{v}\right)\ .
\end{equation}

In-fact, for $m\sim 10^{-22}\ \text{eV}$, such effects will be prevalent at $\mathcal{O}(\text{kpc})$ scales, which will have interesting effects at astrophysical scales.

\subsection{Particle physics of wave dark matter}

Let us now discuss the particle physics properties of scalar fields that could act as dark matter. 
The simplest possibility is that the real scalar field $\varphi$ corresponding to dark matter is a singlet under the Standard Model symmetry group. 
Many extensions of Standard Model involve additional particles, which are in the singlet representation of the Standard Model symmetry group. 
As we noted above, at low energies, the only particles surviving from the Standard Model are photons, neutrinos, electrons, and up and down quarks (which along with gluons will form protons and neutrons). 

Given this information, the only additional terms in the Lagrangian involving $\varphi$ will involve dimension $5$ operators coupling $\varphi$ to the surviving Standard Model particles, of the form (see section~2.2 of~\cite{Damour_2010})
\begin{equation}\label{eq:coupings_SM}
    {\cal L} \supset \frac{\varphi}{\Lambda}\left[-\frac{d_\gamma}{4}F^{\mu\nu}F_{\mu\nu} - \frac{d_g\beta_3}{2}G^A_{\mu\nu}G_A^{\mu\nu} - d_em_e\Bar{e}e - \sum_{q = u,d}d_qm_q\Bar{q}q\right]\ ,
\end{equation}
in addition to the kinetic and potential energy terms of $\varphi$ itself, i.e. 
\begin{equation}\label{eq:lagrangian}
    {\cal L} = -\frac{1}{2}\partial_\mu\varphi\partial^\mu\varphi - \frac{1}{2}m^2\varphi^2 - \frac{\lambda}{4!} \varphi^4 + \cdots \ ,
\end{equation}
where, in eq.~(\ref{eq:coupings_SM}), $F_{\mu\nu}$ is the field strength tensor of the electromagnetic field, $G_{\mu\nu}$ is the field strength tensor of the Gluon field, $e$ is the electron field, $q$ is the quark field (where we sum over only up and down quarks), and the scale $\Lambda$ is of the order of reduced Planck mass $M_{pl}$. 
Note that the dimensionless parameters $d_\gamma,\ d_g,\ d_e,\ d_u,\ d_d$ parametrise the couplings of this additional scalar to the surviving SM particles. 
The current observational constraints and future projections on the values of these parameters, from considerations of stellar cooling, mediation of fifth forces, violation of equivalence principle, time variation of fundamental constants, etc. are summarised in~\cite{Antypas_2022}.
From our discussion in the last section, mass of the scalar field $m$ in eq.~(\ref{eq:lagrangian}), must satisfy $m \ll 100\ \text{eV}$. 
The parameter $\lambda$ characterises the strength of the self-coupling of the scalar field.

In this thesis, we will be interested in constraining the value of $\lambda$ from astrophysical observations. 


\section{Scalar fields and cosmology}

\subsection{Ubiquity of scalar fields in cosmology}
Scalar fields play a prominent in other problems in cosmology. 
For instance, an important example for a model of dark energy involves scalar field, viz., the quintessence model \cite{Tsujikawa_2013}, where an ultralight scalar field $m\sim 10^{-33}\ \text{eV}$ drives the late time accelerated expansion of the Universe.
Similarly, in models of inflation, the rapid expansion of the Universe is driven by a scalar field called inflaton~\cite{Baumann_2009}. 
Novel mechanisms to generate baryon-asymmetry in the Universe like the Affleck-Dine mechanism~\cite{Affleck_1984} utilises a scalar field. 
Scalar fields also arise in models of modified gravity, where extensions to general relativity can involve adding extra scalar fields to GR. For instance, see scalar-tensor theories of modified gravity~\cite{Clifton_2011}. 
As we noted earlier, if scalar fields couple with matter then they also mediate long-range forces. Effects of such fifth-forces have been strongly constrained from experiments and therefore, one could rely on screening mechanisms, to hide such effects at relevant scales~\cite{Hinterbichler_2010, Brax_2021, Burrage_2023}.

\subsection{Axions and ALPs}

The benchmark model for the kind of light scalars that we are interested in are Axion-Like Particles (ALPs) which are cousins of the QCD axion (see section~90 of Ref.~\cite{PDG_2024}).
Recall that the QCD axion is the pseudo-Nambu-Goldstone boson of a global $U(1)$ symmetry called Pecci-Quinn symmetry~\cite{Wilczek_1977, Weinberg_1977}. 
While ALPs could also have a similar origin, 
they may also arise in other ways, such as, as zero-modes of higher dimensional gauge fields compactified on internal manifolds~\cite{Arvanitaki_2010}. 
Even though Axions and ALPs are light scalars, there are reasons to believe that their masses and scalar potentials are insensitive to unknown short distance physics
\footnote{
This is because a NGB enjoys a continuos shift symmetry and no perturbative processes contributes to its potential. The continuous shift symmetry is broken to a discrete shift symmetry by non-perturbative effects but the generated potential is typically small 
(see section~2 of Ref.~\cite{Hui_2017}). 
It is worth mentioning that this issue is still being currently debated~\cite{Dine_2022}. 
}.
As we saw in eq.~(\ref{eq:axion_potential}), axions and ALPs have a scalar potential which goes as a cosine. 
The coupling of axions to SM particles is known to be suppressed by factors of a high energy scale $f_a$, called axion decay constant. 
The pseudo-scalar nature of axions ensures that it does not mediate fifth forces (see appendix~A of Ref.~\cite{Grossman_2025}). 

In this benchmark scenario, the typical value of $\lambda$ that is possible will be discussed in section~\ref{sec:thesis_overview}. 
While we keep this benchmark class of models in mind, in this thesis, we shall remain agnostic about the particle physics aspects of wave dark matter and pursue a signature-driven approach.

\subsection{Extended field configurations}
It is well know that in field theory there are solutions of classical field equations such as kinks, solitons, monopoles, etc. which can have interesting properties (see chapter~92 of~\cite{Srednicki_2007}). 
Similarly, scalar fields coupled to gravity, described by the Klein-Gordon-Einstein (KGE) equations, can form stable, localised configurations, referred to as non-topological solitons (complex scalar fields) or pseudo-solitons (real scalar fields)~\cite{Schunck_2003, Liebling_2012, Cardoso_2019, Visinelli_2021}. 
Such solutions can play an important role in cosmology and astrophysics. 
These self-gravitating scalar field configurations can feature a rich diversity of masses, radii, and stability depending 
on whether the scalar field is real or complex, 
whether effects of self-interactions, repulsive or attractive are taken into account, etc. (see ~\cite{Visinelli_2021} for a detailed exploration of various different objects). 

In the context of astrophysics and cosmology, such configurations can be asteroid-sized, made from QCD axions~\cite{Barranco_2010, Kimball_book_2023}, which can potentially account for dark matter, or they can form exotic compact objects, which can mimic black holes or neutron stars~\cite{Guzman_2009, Macedo_2013, Sennett_2017}.
Other possible objects also include Q-balls, formed by complex scalar fields with attractive self-interaction terms~\cite{Heeck_2020, Ansari_2023}. 
In the non-relativistic limit, scalar fields can form stable configurations of size $\sim\lambda_{dB}$ supported against gravitational collapse by the uncertainty principle. As we shall discuss in chapter~\ref{chpt:numerical_solutions}, such configurations are ground state stationary solutions of the Schrödinger-Poisson equations~\cite{Ruffini_1969, Colpi_1986, Guzman_2004, Chavanis_2011_analytic, Chavanis_2011}. 

For a scalar field with $m\sim 10^{-22}\ \text{eV}$ the size of these field configurations is $\sim\mathcal{O}(\text{kpc})$.
In-fact, recent cosmological simulations involving such scalar field dark matter with $m\sim 10^{-22}\ \text{eV}$ have shown that virialised wave dark matter halos feature a flat density solitonic core of size $\lambda_{dB}$ at the centre of the halo surrounded by a CDM-like envelope~\cite{Schive_Nature_2014, Schive_PRL_2014, Mocz_2017}. 
Additionally, due to the large deBroglie wavelength, simulations also suggest $\mathcal{O}(1)$ density fluctuations at $\text{kpc}$ scales~\cite{Schive_Nature_2014, Mocz_2017, Li_2021}, leading to inference-like patterns in density. 
Therefore, one expects quite unique astrophysical signatures for such a scalar field dark matter candidate.

As we shall see in the next section, these field configurations will play an important role in probing scalar field dark matter.

\section{Thesis overview}\label{sec:thesis_overview}

Bringing things that we discussed in section~\ref{sec:lower_limit_m} together, we know that (a) the existence of dwarf galaxies requires the scalar DM particle to be no lighter than $m\sim 10^{-22}\ \text{eV}$, (b) the uncertainty principle will enable such dark matter to suppress structures below some length scales and (c) large collections of such scalar particles can form stable self-gravitating configurations of size $\lambda_{dB}$. 
Considering these unique features, a dark matter candidate with $m\sim 10^{-22}\ \text{eV}$ with a $\lambda_{dB}\sim \mathcal{O}(\text{kpc})$ will exhibit interesting wave-like behaviour at galactic scales.
Such a dark matter is called Fuzzy Dark Matter (FDM) or Ultra Light Dark Matter (ULDM) in the literature, and it is an attractive alternative to CDM~\cite{Lee_1995, Matos_2000, Hu_2000} to potentially address some of the small scale issues with $\Lambda$CDM (core-cusp problem, missing-satellites problem, and too-big-to-fail problem~\cite{Bullock_2017}) while keeping large scale CDM predictions intact~\cite{Hui_2017, Hui_2021}.
The key feature of this model that we shall be interested in, are the flat density cores at the centres of galactic halos, which are nothing but stable self-gravitating configurations that we briefly discussed in previous section and which we shall discuss in detail in chapter~\ref{chpt:numerical_solutions}.
However, over the last 10 years, stringent constraints have been placed on this model from various astrophysical and cosmological observations, which are summarised in section~\ref{sec:fuzzy_dark_matter}, effectively ruling out $m\sim 10^{-22}\ \text{eV}$.

Of course, as we noted in eq.~(\ref{eq:lagrangian}), the Lagrangian for the scalar field will allow for a $\lambda\varphi^4$ term, representing self-interactions. Here $\lambda$ characterises the strength of the self-interactions, and can be either negative (denoting attractive interactions) or positive (denoting repulsive interactions). Since the potential energy function $U(\bf r)$ in the non-relativistic limit is related to the scattering amplitude $i {\cal M}$ by the relation
\begin{equation}
    U(\textbf{r}) \propto -\int\frac{\mathrm{d}^3\textbf{q}}{(2\pi)^3}e^{i\textbf{q}.\textbf{r}}{\cal M}_{NR}(\textbf{q})\ ,
\end{equation}
where, for 2-2 scattering, for $\lambda \varphi^4$ theory, $i{\cal M}_{NR} = -i\lambda$, the $\lambda\varphi^4$ term in the Lagrangian leads to a potential $U({\bf r}_i, {\bf r}_j)  \propto \lambda \delta^3 ({\bf r}_i - {\bf r}_j)$, which is just contact interaction - repulsive for positive $\lambda$ and attractive for negative $\lambda$. 

It is worth noting that many of the constraints on FDM ignore the self-interactions term (for instance~\cite{Irsic_2017, Hlozek_2018, Safarzadeh_2020, Davies_2020, Dalal_2022, Hertzberg_2023}). 
This is a sound assumption, since in the case of the benchmark ALP model, the self-interaction term manifests as one considers higher order terms in the cosine potential in eq.~(\ref{eq:axion_potential}). 
Here the quartic term $-\Lambda^4\varphi^4/4!f_a^4$ can be written as $\lambda\varphi^4/4!$ by using $m_a = \Lambda^2/f_a$ and defining a dimensionless coupling $\lambda \equiv -\left(m_a/f_a\right)^2$. 
As we saw in eq.~(\ref{eq:axion_relic}), current relic density for such a light particle can account for all dark matter only if $f_a \sim 10^{17}\ \text{GeV}$ which implies $\lambda \sim -10^{-96}$, an incredibly small number. 
As argued in section~IV.D of Ref.~\cite{Chavanis_2021}, in this limit, the self-interactions of FDM will not play a significant role in non-linear regimes.

In this thesis, we however aim to take a signature driven approach, where the mass of the ULDM particle $m$, along with the sign and strength of $\lambda$ probed are dictated by the observations, all the while remaining agnostic regarding the origin of such a $m$ and $\lambda$.
In particular, we attempt to use observational data at astrophysical scales to answer the following questions: (a) What will be the impact of self-interactions (SI) on various observations?, (b) What kind of SI (attractive or repulsive) are preferred by the data, if any? and (c) How strong do SI have to be, for them to be important?
Such an approach has been gaining traction in recent years, where $|\lambda| > 10^{-96}$ can be indeed be probed by various astrophysical and cosmological observations (as we summarise in section~\ref{sec:impact_of_SI}). 
In particular, our interest in this thesis will be on the effect of self-interactions on self-gravitating configurations that can be used to describe the inner regions of galactic halos. We discuss this in chapter~\ref{chpt:numerical_solutions}. 

Cosmology and astrophysics are an observational science, and many of the constraints that we saw earlier are based on analysing and comparing theory predictions to observed data. 
Therefore, it is also important to understand the various methods of learning from data.
In the last chapter of this thesis, we also explore alternative methods (compared to standard Bayesian inference), like likelihood-free inference using neural networks, to learn about the physical nature of DM from astrophysical data like rotation curves. 

The work presented in this thesis, can therefore be split into two broad (albeit closely related) directions:

First, we study the impact of self-interactions on galactic scale observations by considering the following scenarios: (a) Using observational upper limits on the amount of mass contained within some region around the galactic centre, we find that one can impose constraints in the $\lambda-m$ plane, where self-couplings can be as small as $\lambda \sim \pm 10^{-96}$~\cite{Chakrabarti_2022}, (b) requiring that observed galactic rotation curves of dwarf galaxies as well as an empirical soliton-halo relation have to be simultaneously satisfied allows one to probe repulsive self-interactions as small as $\lambda \sim \mathcal{O}(10^{-90})$ while also potentially evading constraints for $\lambda = 0$ for $m\sim10^{-22}\ \text{eV}$~\cite{Dave_2023}, and (c) survival of dwarf satellite galaxies orbiting in the potential of larger halos on cosmological timescales can be used to probe both attractive and repulsive self-couplings as small as $\lambda \sim \pm 10^{-92}$~\cite{Dave_2024}.

Secondly, we explore the use of machine learning techniques, in particular neural networks, to infer parameters of the dark matter density profile (including the ULDM mass $m$ with $\lambda = 0$) along with a Baryonic parameter from observed rotation curve data. 
We did this by training neural networks on simulated rotation curves, to learn the relationship between rotation curves and theory parameters, and finally perform likelihood-free inference from observations. 
We studied the impact of noise in the training set of the neural network as well as compared two different methods of quantifying uncertainty associated with the inferred parameters. 
We found that a neural network trained with noisy training data and a heteroscedastic loss function can infer parameters and uncertainty values that are in agreement with standard Bayesian approach using Markov Chain Monte-Carlo (MCMC)~\cite{Dave_2025}.  
 
The thesis is organised as follows: In chapter~\ref{chpt:numerical_solutions}, we derive the equations of motion for a non-relativistic scalar field, and illustrate how to numerically obtain stationary (and quasi-stationary) solutions. We also study the properties of these scalar field configurations, and the different regimes that manifest due to different self-interaction signs and strengths.
In chapters~\ref{chpt:paper_1},~\ref{chpt:paper_2} and~\ref{chpt:paper_3} we present our work exploring how astrophysical observations can constrain the mass and self-coupling of ULDM. 
In chapter~\ref{chpt:paper_4}, we present our work detailing a neural network approach of obtaining the 'best-fit' parameters describing the dark matter density profile (including ULDM mass $m$ in the $\lambda = 0$ limit) along with uncertainties from observed rotation curves of dwarf galaxies from the SPARC catalogue. 
Finally, in chapter~\ref{chpt:conclusion}, we summarize our work and briefly talk about some potential extensions based on our work that can be pursued in the future. 
Note that throughout the thesis, we shall work with $\hbar = c \equiv 1$ units, except in places where their presence is justified. Here $m_{pl} = \sqrt{\hbar c/G}$ is the Planck mass and $M_{pl} = \sqrt{\hbar c/8\pi G}$ is the reduced Planck mass.


\justifying
\chapter{Self-gravitating scalar field configurations}\label{chpt:numerical_solutions}

Much of the work in this thesis is based on how astrophysical observations can probe the mass and self-interactions of Ultra Light (scalar field) Dark Matter (ULDM). 
To describe structures at the cores of DM halos i.e. at $\mathcal{O}(\text{kpc})$ scales, since DM is cold, we will need to work with the slowly varying part of the real scalar DM field $\varphi$, which will be a complex scalar field $\Psi$. 
In this chapter, we therefore develop the machinery to obtain relevant solutions of the equations of motion for the field $\Psi$ in the presence of self gravitational interactions as well as self-interactions, and discuss the properties of the kind of solutions of these equations which we will need in this thesis.

This chapter is organised as follows: in section~\ref{sec:KGE_to_GPP}, we obtain the equations of motion for the scalar field with self-interactions in the weak field, slowly varying limit - these will turn out to be Gross-Pitaevskii-Poisson (GPP) equations. 
In section~\ref{sec:stationary_states}, we discuss the stationary state (or solitonic) solutions of the GPP system. 
First, we illustrate how to obtain the numerical solution given appropriate boundary conditions, and then, we discuss the scaling symmetry that the GPP system possesses. This scaling symmetry can be exploited to obtain an entire family of solutions from a single numerical solution. 
In section~\ref{sec:mass_radius_curves}, we also discuss different regimes of allowed mass and radius of solitonic solutions depending on the sign and strength of the self-coupling $\lambda$. 
In section~\ref{sec:quasi-stationary} we discuss quasi-stationary solutions of the GPP system, which are required to describe solitons that leak mass over time. This enables us to treat a time-dependent problem in the framework of a time-independent one. 
We end this chapter by talking about the observational status of some ULDM models.
In section~\ref{sec:fuzzy_dark_matter}, we summarize the fuzzy dark matter model (i.e. ULDM without self-interactions) and the current constraints on it. 
Finally, in section~\ref{sec:impact_of_SI}, we briefly summarize how self-interactions of ULDM have been observationally constrained in the recent past.

\section{Equations of motion: from Klein-Gordon-Einstein to Gross-Pitaevskii-Poisson equations}\label{sec:KGE_to_GPP}


In this section, we derive the Gross-Pitaevskii-Poisson system of equations for a classical real scalar field from the Klein-Gordon-Einstein equations. 
The GPP equations describe the non-relativistic part of the scalar field and solutions of which will be used to describe the inner regions of galactic halos. 
Since we are interested in describing structures at galactic scales, we shall ignore the effects due to the expansion of the Universe, and hence work with a static weak-field metric.

\begin{figure}
    \centering
    \includegraphics[width=0.7\linewidth]{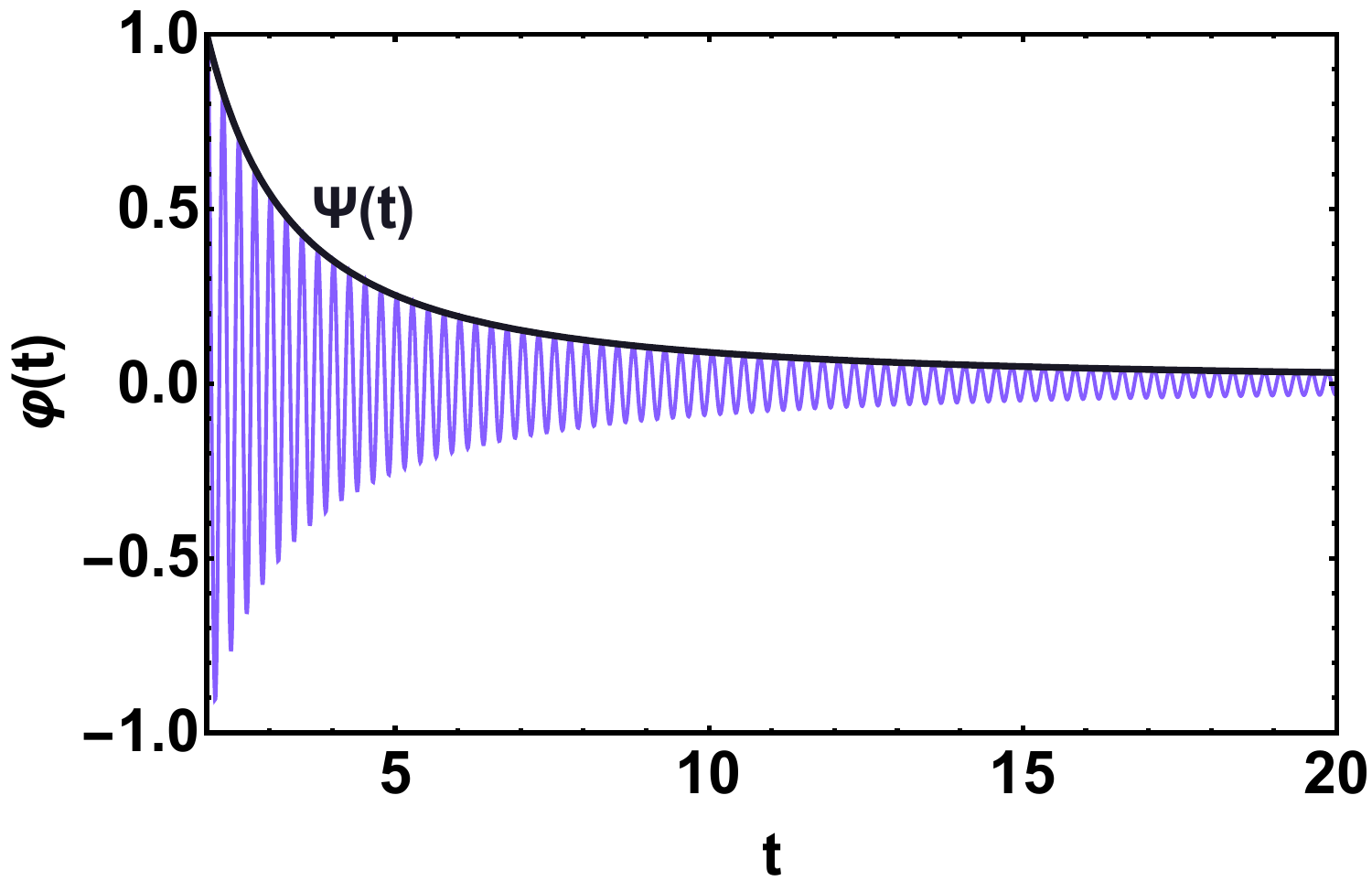}
    \caption[Illustrative schematic to demonstrate slow variation]{A schematic that illustrates a function $\varphi(t)$ undergoing fast oscillations shown in blue (which in our case are captured by the $e^{imt}$ term), along with a slow decay in amplitude with time shown by solid black curve (for our case, this is the scalar field $\Psi(t, \vec{x})$). We are only interested in the gross variation, i.e. time-dependence that is averaged over the fast oscillations.}
    \label{fig:slow_variation}
\end{figure}

Note that for the following derivation, we use an excerpt from our published work~\cite{Chakrabarti_2022}.
We are interested in the classical field theory with action
\begin{equation}
 S = \int d^4 x \sqrt{-g} \left( \frac{M_{\rm pl}^2}{2} R - \frac{1}{2} g^{\alpha \beta} \partial_{\alpha} \varphi \partial_{\beta} \varphi - U(\varphi) \right) \; ,
\end{equation}
where, $U(\varphi) = \frac{m^2 \varphi^2}{2} + \frac{\lambda \varphi^4}{4 !}$. Note that $\lambda$ is the self-interaction strength that we discussed in section~\ref{sec:wave_DM} which can be either attractive ($\lambda < 0$)~\footnote{In case of $\lambda < 0$, while it appears as if the corresponding Hamiltonian will be unbounded from below, it is worth noting that attractive self-interactions are considered in the context of ALP DM, whose full potential is a cosine in eq.~(\ref{eq:axion_potential}). The potential we have used here is an approximation which is valid for small values of $\varphi$.} or repulsive ($\lambda > 0$). If cosmological DM is to be described by the non-relativistic dynamics of the scalar field $\varphi$, we wish to observationally constraint the parameters $m$ and $\lambda$.
Varying the above action w.r.t. ${\varphi}$ will give the equation of motion of the scalar field in any spacetime
\begin{equation} \label{eq:EOM_1}
    \partial_\mu \left[ \sqrt{-g}~ g^{\mu \nu}~ \partial_\nu {\varphi} \right]  = \sqrt{-g} ~U'({\varphi}) \; , 
\end{equation}
where, $U'({\varphi})$ is the derivative of $U$. 
For a regime with weak gravity, the metric takes up the form
\begin{equation} \label{eq:metric}
 ds^2 = - \left(1+ {2 \Phi} \right) (dx^0)^2 + \left(1- {2 \Phi} \right) \delta_{ij} dx^i dx^j \; ,
\end{equation}
where, ${\Phi} \ll 1$ and has only spatial variation. 
Using equation (\ref{eq:EOM_1}) and (\ref{eq:metric}), one finds that
\begin{equation}
    \partial_0^2 {\varphi} - \nabla^2 {\varphi} + U'({\varphi}) = 
    2 \Phi \left[ 2 \partial_0^2 {\varphi} + U'({\varphi}) \right] \; .
\end{equation}
Similarly, we are interested in the situations in which the scalar field dynamics can be adequately described by non-relativistic equations i.e., the scalar field has slow spatial and temporal variation. 
In the absence of gravity and scalar self-interactions, each Fourier mode of the scalar field oscillates with frequency 
${\omega}_{\bf k} = m \left(1 + {\bf k}^2 / m^2 \right)^{1/2}$ and slow variation means that the oscillation frequency is dominated by $\omega_k = m$ save for small corrections.
This suggests that, in order to successfully take the non-relativistic limit, we introduce a complex scalar field $\Psi(t,\vec{x})$ (see figure~\ref{fig:slow_variation} for an illustrative example), defined by
\begin{equation}
 \varphi (t, {\vec x}) = \frac{1}{\sqrt{2} m} \left[ e^{-im t} \Psi (t, {\vec x}) + {\rm c.c.} \right] \; ,
\end{equation} \label{eq:Psi}
here, the new field $\Psi$ captures the dynamics of $\varphi$ over and above the time evolution captured by $e^{-im t}$
and slow variation means that (a) the following hierarchy is maintained:
\begin{equation}
 \Psi ~\gg~ m^{-1} ~ {\dot \Psi} ~\gg~ m^{-2} ~ {\ddot \Psi} \gg \cdots \; ,
\end{equation}
and, (b) when we are interested in time scales large as compared to the Compton time $m^{-1}$, any term which is highly oscillatory (e.g. $e^{-2imt}$ etc) will average out to zero. 
Note that equation (\ref{eq:Psi}) does not define $\Psi$ uniquely and its phase can be freely chosen.

Under this weak gravity and slow variation approximation, equation of motion of the field $\Psi$ and Einstein equations will yield the Gross-Pitaevskii-Poisson (GPP) equations, \cite{Pitaevskii_book, Chavanis_2011}
\begin{eqnarray} 
     i \frac{\partial \Psi}{\partial t} &=&  -\frac{\nabla^2}{2m} \Psi + m \Phi  \Psi +  \frac{\lambda}{8 m^3} |\Psi|^2 \Psi ~+~ \cdots \label{eq:GrossPitaevskii_higher} \\
    \nabla^2 \Phi &=& 4\pi G|\Psi|^2 ~+~ \cdots \; . \label{eq:Poisson_higher}
\end{eqnarray}

In the absence of self-gravity, we do not have eq.~(\ref{eq:Poisson}), the $m \Phi  \Psi$ term in eq.~(\ref{eq:GrossPitaevskii}) will be absent and we recover Gross-Pitaevskii equation (i.e. non-linear Schr\"{o}dinger equation or the limit of Ginzburg-Landau equations in the absence of electromagnetic field). Alternatively, in the absence of the self-interactions, this system reduces to what is usually referred to as the Schr\"{o}dinger-Poisson (SP) system. The ``$\cdots$" in both these equations stand for ``higher-order corrections" (see e.g. \cite{Salehian_2021} for a recent discussion) which we completely ignore for now. We shall have more to say about them in a \ref{sec:scaling_relations}.

As mentioned in the previous chapter, there have been recent efforts to numerically evolve the Schrödinger-Poisson as well as the Gross-Pitaevskii-Poisson system \cite{Schive_Nature_2014, Mocz_2017, Schwabe_2020, Dawoodbhoy_2021, Mocz_2023, Winch_2023} which has enabled us to understand the kinds of structures that can be formed from collections of scalar particles. 
These simulations have shown that the time-averaged density profiles of the inner regions of virialised ULDM halos can be described by the ground state stationary solution of the GPP system. 
Therefore, to describe the inner regions of galactic halos, we shall utilise stationary (as well as quasi-stationary) solutions of the GPP system.

\subsection{A note on the `Quantum' nature of wave dark matter}\label{sec:quantum_nature}
Before proceeding, we should talk about 
a few possible conceptual concerns. Let us first rewrite the Gross-Pitaevskii-Poisson equations with the factors of $\hbar$ and $c$ included:

\begin{eqnarray}
    i \hbar \frac{\partial \Psi}{\partial t} &=&  -\frac{\hbar^2}{2m} \nabla^2 \Psi + m \Phi  \Psi +  \frac{\lambda\hbar^3}{8 m^3c} |\Psi|^2 \Psi\ ,\label{eq:GP_with_factors}\\
    \nabla^2 \Phi &=& 4\pi G|\Psi|^2
\end{eqnarray}

\begin{enumerate}[(a)]
    \item Since the above equations will be used to model DM at galactic scales, factors of $\hbar$ in eq.~(\ref{eq:GP_with_factors}) suggest that there are quantum mechanical effects at galactic length scales. This should not be surprising - the reason atoms are as large as they are is that they are quantum mechanical systems (hence $\hbar$  important) which involve motion of electrons of mass $m_e$ under Coulomb interactions determined by $e$ - the only length scale which can be formed out of these quantities is Bohr radius. Thus, the smallness of atoms (compared to human scales) is not just due to quantum mechanics (i.e. $\hbar$) but also due to the values of $m_e$ and $e$. For ultra-light dark matter, the length scale formed from $m$, $\hbar$ and typical velocity of a particle under the gravitational influence of a galaxy is of Galactic scale. This fact, that quantum mechanical effects are important, is useful in thinking about phenomena such as tunneling which are dealt with in this paper.
    \item In the model of DM we are interested in, since DM particles are Bosons, a lot of them can be in the same (single particle) quantum state - we can thus form coherent states which are well described by classical field theory just like classical electromagnetic waves can be understood in terms of photons. The description of a classical electromagnetic wave doesn't require us to pay attention to the microscopic description in terms of photons. In the same way, under the circumstances of the current problem, we can also describe DM in a galaxy completely in terms of classical field equations.
    \item In particular, in classical Klein-Gordon field theory, $\hbar$ and $m$ don't play any role: the only dimensionful quantities are $c$ and the length scale $L$ in Klein-Gordon equation $( - \partial^2 + L^{-2} ) \varphi = 0$. Working in the convention in which $\varphi$ has dimensions $M^{\frac{1}{2}} T^{-\frac{1}{2}}$, when we introduce a self-interaction term $g_4 \varphi^3$, $g_4$ has dimensions $L^{-2} M^{-1} T^{1}$, we can then express the dimensions of any quantity (e.g. $G$) in terms of $L, c ~{\rm and}~ g_4$. Alternatively, if we wish to include the effect of gravity, we can express all quantities (e.g. $g_4$) in terms of $L, c ~{\rm and}~ G$. In particular, $\Psi$ can be defined from $\varphi$ by an equation which doesn't involve $\hbar$ and $m$ i.e. 
    $\varphi = \sqrt{\frac{c}{2}} \cdot L \cdot \left(e^{-i(ct/L)}\Psi + c.c.\right)$ and
    we can obtain a version of eq. (\ref{eq:GP_with_factors}) which doesn't involve factors of $\hbar$ and $m$ i.e.
    \begin{equation}\label{eq:schrodinger}
        \frac{iL}{c}  \frac{\partial \Psi}{\partial t} = - \frac{L^2}{2} \nabla^2 \Psi + \frac{\Phi_{SG} \Psi}{c^2} + \frac{L^2 G}{8 c^2} \lambda |\Psi|^2\Psi \; , 
    \end{equation}
    which is a purely macroscopic classical description involving $L, c ~{\rm and}~ G$ (note that $\lambda$ is dimensionless). 
\end{enumerate}

One can thus work with either the microscopic quantum mechanical picture (which involves $\hbar$ and $m$) or macroscopic classical picture involving $L$.

\section{Stationary solutions}\label{sec:stationary_states}

Note that much of the text in this section is taken from our published work~\cite{Chakrabarti_2022} and~\cite{Dave_2023}.

As we saw in the last section, that the equations of motions for a real scalar field in the non-relativistic, weak-field limit are just the Gross-Pitaevskii-Poisson equations (ignoring higher order terms), 
\begin{eqnarray} 
     i \frac{\partial \Psi}{\partial t} &=&  -\frac{\nabla^2}{2m} \Psi + m \Phi  \Psi +  \frac{\lambda}{8 m^3} |\Psi|^2 \Psi \ ,\label{eq:GrossPitaevskii} \\
    \nabla^2 \Phi &=& 4\pi G|\Psi|^2 \ ,\label{eq:Poisson}
\end{eqnarray}
where, $\Phi$ is the self-gravitational potential, while $|\Psi|^2$ has the dimensions of mass density.
Note that one can define $\Psi$ in eq.~(\ref{eq:Psi}) with a $\sqrt{m}$ in the denominator which allows $|\Psi|^2$ to be interpreted as number density instead of mass density (see \cite{Schiappacasse_2018} as an example).
We shall see in section~\ref{sec:scaling_symmetry} why we have ignored the higher order terms in the above equations. 

We define \textbf{solitonic solutions} of the system of equations (\ref{eq:GrossPitaevskii}) and (\ref{eq:Poisson}) as those which satisfy the following properties: 

\begin{enumerate}
    \item Stationary in the sense that the field $\Psi$ is of the form $\Psi(t,\vec{x}) = \phi(\vec{x}) e^{-i\gamma t}$ with $\phi$ and $\gamma$ real\footnote{Note that this $\phi$ is different from the relativistic real scalar field $\varphi$ from section~\ref{sec:KGE_to_GPP}}. 
    \item Spherically symmetric $\phi(\vec{x}) = \phi(r)$. 
    \item Spatially localised such that the integral $4\pi \int_0^\infty |\phi|^2r^2dr$ is finite. 
    \item Nodeless, i.e. the ground state or minimum energy solution\footnote{It is worth noting that it was shown that the ground state solution is the only stable solution for this system \cite{Guzman_2006}.}.
    \item Regular everywhere.
\end{enumerate}
For details about such solutions of the GPP and SP system, see \cite{Ruffini_1969, Guzman_2006, Chavanis_2011, Chavanis_2011_analytic, Marsh_2016, Padilla_2021, Davies_2020}.
Using $\Psi(t,\vec{x}) = \phi(r) e^{-i\gamma t}$, one can write the GPP system as coupled time-independent equations,
\begin{eqnarray}
    \frac{1}{2m}\nabla_r^2\phi &=&  m \Phi  \phi - \gamma \phi +  \frac{\lambda}{8 m^3} \phi^2 \phi \label{eq:GrossPitaevskii_TI} \\
    \nabla_r^2 \Phi &=& 4\pi G\phi^2 \ ,\label{eq:Poisson_TI}
\end{eqnarray}
where $\gamma$ is related to the total energy of the system \cite{Marsh_2016} and $\nabla_r^2 = \frac{\mathrm{d}^2}{\mathrm{d}r^2} + \frac{1}{r}\frac{\mathrm{d}}{\mathrm{d}r}$.
Since the solution is stationary, this also implies that the mass density $\rho = |\Psi|^2 = \phi^2$ is time-independent.
Further, for a given field profile $\phi(r)$, one can also define the mass of the soliton contained within a spherical volume of radius $r$ as, 
\begin{equation}
    M(r) = 4\pi G\int_0^r\phi^2(r')r'^2dr'\ . 
\end{equation}
The total mass of the soliton $M_s$ is then obtained by setting the upper limit to $\infty$. 
One can also define a characteristic length scale $R \equiv R_{95}$, which is the radius within which $95\%$ of the total mass is contained (one can define $R_{99}$ similarly).

Now, to solve the coupled system of second order differential equations in eqs.~(\ref{eq:GrossPitaevskii_TI}) and~(\ref{eq:Poisson_TI}), we need four boundary conditions. 
Two boundary conditions can be imposed by requiring that the field and the potential vanish far away from the centre of the core, i.e. $\phi(\infty) = \Phi(\infty) = 0$. 
However, owing to the linearity of the Laplacian in the Poisson equation, one can redefine $\Phi \rightarrow \Phi + C$ where the constant $C$ can be absorbed by $\gamma \rightarrow \gamma - mC$ in the Gross-Pitaevskii equation leaving the system of equations unchanged. 
We therefore use the above redefinition to change boundary condition to $\Phi(0) = 0$. 
On the other hand, requiring regularity at the origin nets us the other two boundary conditions $\Phi'(0) = \phi'(0) = 0$. 

An unknown parameter still remains in the form of $\gamma$, which appears in the phase of $\Psi$. 
To obtain the value of $\gamma$ numerically, we utilize the shooting method \cite{Giordano_Nakanishi}, where we initialize a solution from the origin by specifying a central density $\phi(0)$ and search for a value of $\gamma$ that leads to a nodeless solution satisfying the boundary condition at infinity (i.e. a large enough $r$). 
Note that for a fixed $m$ and $\lambda$, as soon as $\phi(0)$ is chosen, the value of $\gamma$ can be uniquely determined, and the system has only a single solution for the ground state.
If one wants to describe a density profile with a different central density (since different galactic halos have different central densities), the system of equations must be solved again with a new $\phi(0)$ and hence a different $\gamma$. 
We shall see in the next few sections why this is not required. 

First, to make the problem numerically easier to solve on a computer we utilize dimensionless variables, which we can obtain using the scales in the system. 
One way to define dimensionless variables (denoted by a ` $\hat{}$ ') is the following\footnote{See \cite{Eby_2016, Marsh_2016} for different normalization schemes.}: 
\begin{eqnarray}
    \hat{\phi} = \frac{\hbar\sqrt{4\pi G}}{mc^2}\phi, \ \ \hat{\Phi} &=& \frac{\Phi}{c^2}, \ \ \hat{\gamma} = \frac{\gamma}{mc^2}, \label{eq:dimensions1} \\  \hat{r} = \frac{mc}{\hbar}r, \ \ \hat{\lambda} &=& \frac{\lambda}{64\pi}\left(\frac{m_{pl}}{m}\right)^2\ .\label{eq:dimensions2}
\end{eqnarray}
The system of equations now takes the form, 
\begin{eqnarray} 
 \frac{1}{2} {\hat \nabla}_{\hat{r}}^2 {\hat \phi} &=& {\hat \Phi}{\hat \phi} - {\hat \gamma} {\hat \phi} + 2 {\hat \lambda} {\hat \phi}^3 \; , \label{eq:GrossPitaevskii_dimless}\\
{\hat \nabla}_{\hat{r}}^2 {\hat \Phi} &=&{\hat \phi}^2 \; , \label{eq:Poisson_dimless}
\end{eqnarray}
Note that the dimensionless system does not depend on $m$, and hence, $\hat{\gamma}$ now is uniquely determined for fixed values of $\hat{\lambda}$ and $\hat{\phi}(0)$. 

It is easiest to find the solutions of the dimensionless Gross-Pitaevskii-Poisson equations (\ref{eq:GrossPitaevskii_dimless}) and (\ref{eq:Poisson_dimless}) when we impose the additional requirement that ${\hat \phi} ({\hat r} = 0) = 1$. Note that for numerical work, we shall always work with ${\hat \phi} ({\hat r} = 0) = 1$ but the arguments below apply to any ${\cal O}(1)$ value of ${\hat \phi} ({\hat r} = 0)$. 
Such solutions have many nice properties relevant for numerical work since all the relevant quantities are ${\cal O}(1)$ but, as we note below, they are highly unrealistic for the purpose of modelling real solitons. We shall call these solutions ``theoretical solitons". 

\begin{figure}[h]
    \centering
    \includegraphics[width=0.8\linewidth]{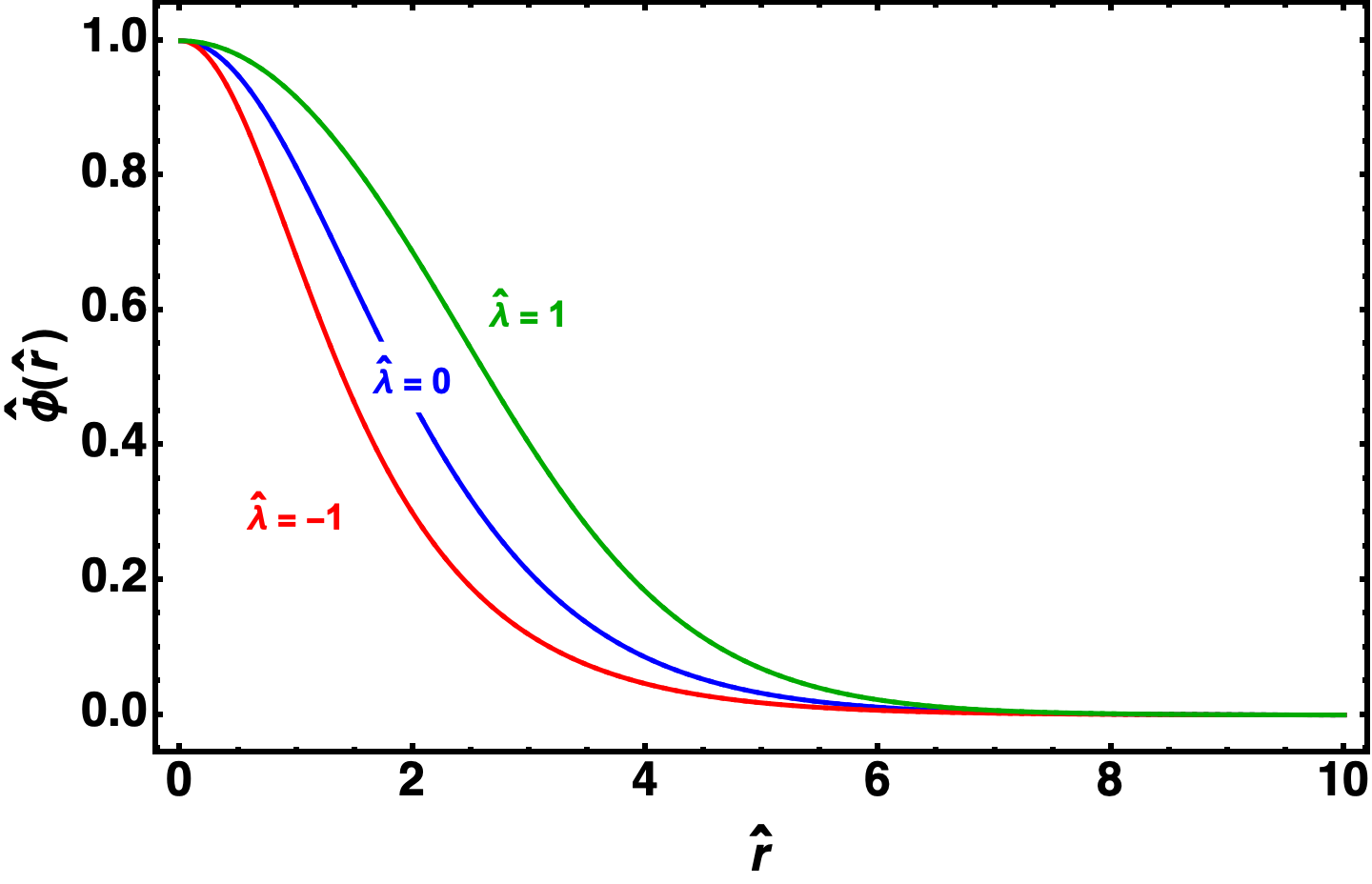}
    \caption[Solitonic solutions for different $\hat{\lambda}$ values]{Example dimensionless ground state solutions (variables defined in eqs.~(\ref{eq:dimensions1}) and~(\ref{eq:dimensions2})) found using the shooting method shown for different values of $\hat{\lambda}$ with the boundary conditions: $\hat{\phi}'(0) = \hat{\Phi}'(0) = 0$, $\hat{\Phi}(0) = 0$ and $\hat{\phi}(0) = 1$.}
    \label{fig:example_soliton}
\end{figure}

The mass of these solitons (from eq.~(\ref{eq:soliton_mass})) can be written as 

\begin{equation}\label{eq:mass_dimless}
    \hat{M}(\hat{r}) = \int_0^{\hat{r}} \hat{\phi}^2(\hat{r})\hat{r}'^2d\hat{r}' = \frac{Gm}{\hbar c}M(r)\ .
\end{equation}
Similarly, one can find $\hat{R}_{95}$ or $\hat{R}_{99}$ as described earlier. 
An example solution is shown in figure~\ref{fig:example_soliton} for $\hat{\lambda} = 0$. 
Since a computer can only evaluate the solution up to a finite value of $\hat{r}$, we denote infinity using a large value, i.e. $\hat{r}_{\text{max}} = 10$ \cite{Davies_2020, Padilla_2021}.
Hence, the boundary condition now implies that the solution $\hat{\phi}$ must vanish at $\hat{r}_{\text{max}}$. 
The dimensionless total mass of the soliton is found to be $\hat{M}_s = \hat{M}_{\hat{r}_{\text{max}}} = 2.0612$. 
We use the same value of initialization and boundary conditions to obtain solutions for $\hat{\lambda}\neq 0$, which are shown by the red (attractive self-interactions with $\hat{\lambda} = -1$) and green (repulsive self-interactions with $\hat{\lambda} = 1$) curves in figure~\ref{fig:example_soliton}.

The corresponding total mass of the solitons is $\hat{M} = 0.9894$ and $\hat{M} = 4.8376$ for the attractive and repulsive case respectively. 
In other words, while all three solutions have the same central densities, the total mass $M_s$ or the total number of particles $N = M_s/m$ is different. 
How does one compare solutions with the same number of particles or the same total mass but with or without self-interactions? 
We answer this, along with the physical significance of the solutions we just obtained in the next few sections.

\subsection{Solutions with $\hat{\phi}(0) = 1$}\label{sec:theoretical_solitons}

As we have mentioned earlier, We solve the system for different values of $\hat{\lambda}$ for a fixed $\hat{\phi}(0) = 1$. 
Adding the dimensions back, where we choose $m = 10^{-22}\ \text{eV}$, 
the definition ${\hat \phi} = \frac{ \sqrt{4 \pi G} \hbar }{m c^2} \phi $ implies that the central density corresponding to $\hat{\phi}(0) = 1$  will be
\begin{equation}
 \rho = 2 \left( \frac{M_{pl}}{m} \right)^2 \frac{m}{(\hbar/mc)^3} \sim 10^{14}
\frac{M_{\odot}}{(pc)^3}  \; .
\end{equation}
Here we have used only the leading contribution to density (i.e. we have ignored contributions of the higher order corrections and scalar self-interactions). This density corresponds to a very large mass within a volume of just $(pc)^3$, hence solutions found by assuming ${\hat \phi}(0) \sim 1$ have too large densities.

A similar argument can be made about the size of the soliton: if the size of the system, in units of $\hbar / mc$, is $\hat L$, then, Poisson's equation tells us that ${\hat \Phi} \sim {\hat L}^2 {\hat \phi}^2$, assuming that all the terms in 
Gross-Pitaevskii equation have the same order of magnitude. This will imply that ${\hat L} \sim 2^{-1/4} {\hat \phi}^{-1/2}$.
If ${\hat \phi} \le 1$, as is the case for theoretical soliton, i.e. if ${\hat \phi}$ is ${\cal O}(1)$, then, ${\hat L}$ will also be ${\cal O}(1)$. Thus, the size of theoretical soliton will be of the order of ULDM / FDM Compton wavelength $\hbar / mc$.
We know that the size of the soliton must be larger than Compton wavelength, in fact it is expected to be of the order of de-Broglie wavelength of the ULDM or FDM particles. 

Since the dimensionless soliton mass $\hat M$ is ${\cal O}(1)$, the gravitational radius of the soliton turns out to be
\begin{equation}
 R_G = \frac{G M}{c^2} = {\hat M} ~ \frac{\hbar}{mc} \sim \frac{\hbar}{mc} \; .
\end{equation}
Thus, the gravitational radius for a soliton with such characteristics is comparable to its size, so, general relativistic effects can not be ignored (even in the absence of a central black hole). 
Real solitons are expected to be much bigger, much less dense and much lighter.

\subsection{Scaling transformations and trustworthy regime}\label{sec:scaling_symmetry}

From the discussion in the previous section, we need a way to obtain solutions with (a) different central densities without having to solve the system again and (b) more realistic central densities, sizes and masses. 
Fortunately, the Gross-Pitaveskii-Poisson system admits a scaling symmetry, i.e. a scaling transformation of the form
\begin{eqnarray}
 {\hat r}  &\rightarrow&   s ~ {\hat r} \; , \label{eq:scaling_length}\\
{\hat \phi}  &\rightarrow& \frac{1}{s^{2}} ~ {\hat \phi} \; , ~{\hat \Phi} \rightarrow \frac{1}{s^{2}}~{\hat \Phi} \; , ~ {\hat \gamma} \rightarrow \frac{1}{s^{2}}~ {\hat \gamma} \; , \label{eq:scaling_Phi}\\ 
{\hat \lambda} &\rightarrow& s^2~ {\hat \lambda} \label{eq:scaling_lambda} \; ,
\end{eqnarray}
causes each term to get scaled by a factor of $1/s^4$. In other words, such a transformation leaves the system of equations invariant.
If $s>1$, such a transformation renders the soliton lighter, bigger and less dense. i.e. a family of solutions for a fixed initial $\hat{\lambda}_{\text{ini}}$. 
Here, we have introduced a subscript `$\text{ini}$' to denote quantities before scaling (or solutions $s = 1$). 
Hence the free parameters of the dimensionful GPP system then are $\{m, \hat{\lambda}_\text{ini}, s\}$.
We also introduce the subscript `$\text{fin}$' to denote quantities after scaling, e.g. $\hat{\lambda}_\text{fin} = s^2\hat{\lambda}_\text{fin}$. 

Note that using the scaling transformations above, higher order terms in Poisson's equation (eq.~(\ref{eq:Poisson_higher})) e.g. ${\hat \lambda} ~ {\hat \phi}^4$, $- 3 {\hat \Phi} {\hat \phi}^2$ as well as those in Gross-Pitaevskii equation (eq.~(\ref{eq:GrossPitaevskii_higher})) will scale by a factor of $1/s^6$ as compared to all the other terms which scale as $1/s^4$, hereby violating the scaling symmetry. 
But, this also implies that for sufficiently large $s$, these terms shall both become negligibly small as compared to the rest of the terms.
Thus, the regime with $s \gg 1$ is highly desirable especially because it is reliably trustworthy. 
Thus, in such a trustworthy regime, all the higher order terms (denoted by ``$\cdots$" in eqs.~(\ref{eq:GrossPitaevskii_higher}) and~(\ref{eq:Poisson_higher}) can be completely ignored. 
Further, we find that from the above scaling relations that the mass in eq.~(\ref{eq:mass_dimless}) scales as 
\begin{equation} \label{eq:scaling_mass}
 {\hat M} \rightarrow \frac{1}{s}~ {\hat M} \; .
\end{equation}
To describe cores of typical dwarf galaxies, with $R \sim 1\ \text{kpc}$ and $M\sim 10^8\ M_\odot$, we require a $s\sim 10^4$, which is sufficiently large enough to solve the GPP system without the higher order terms.

\section{Mass-radius relations}\label{sec:mass_radius_curves}

Solitonic solutions that we described in the previous section are the result of a delicate balance between the outward `quantum pressure' arising from the gradient term in the Gross-Pitaevskii equation, the attractive or repulsive self-interactions of the scalar field, and its self-gravity~\cite{Chavanis_2011_analytic, Chavanis_2011}, leading to a family of allowed masses $M$ and corresponding sizes $R$. Even small values of the self-coupling strength, $\lambda \sim \mathcal{O}(10^{-98})$ can impact the allowed mass and size of solitons.\footnote{This can be easily seen from eq.~(\ref{eq:GrossPitaevskii}), where the self-interaction term is comparable to other terms only when $\hat{\lambda}\sim \mathcal{O}(1)$, which for $m\sim 10^{-22}\ \text{eV}$, gives the self-coupling strength $\lambda \sim \mathcal{O}(10^{-98})$.}

To understand the relation between mass and radius of solitonic solutions, one usually proceeds by using an ansatz \cite{Chavanis_2011_analytic, Schiappacasse_2018} for the form of density profile --- this allows one to write an analytical expression for the energy of the system in terms the soliton mass $M$ and soliton radius $R$. The solutions then correspond to the critical points of the energy. One can then obtain an analytical mass-radius relation for various critical solutions \cite{Chavanis_2011_analytic, Schiappacasse_2018}. 
This approach, while beautiful, relies on assuming a form of the density profile ansatz. It is thus interesting to ask how the expected relationship between the size of the soliton and its mass arises from numerically solving the dimensionless GPP equations. We show that the scaling symmetry of GPP system can be exploited to understand this. 

\begin{figure}[h!]
    \centering
    \includegraphics[width = 0.85\textwidth]{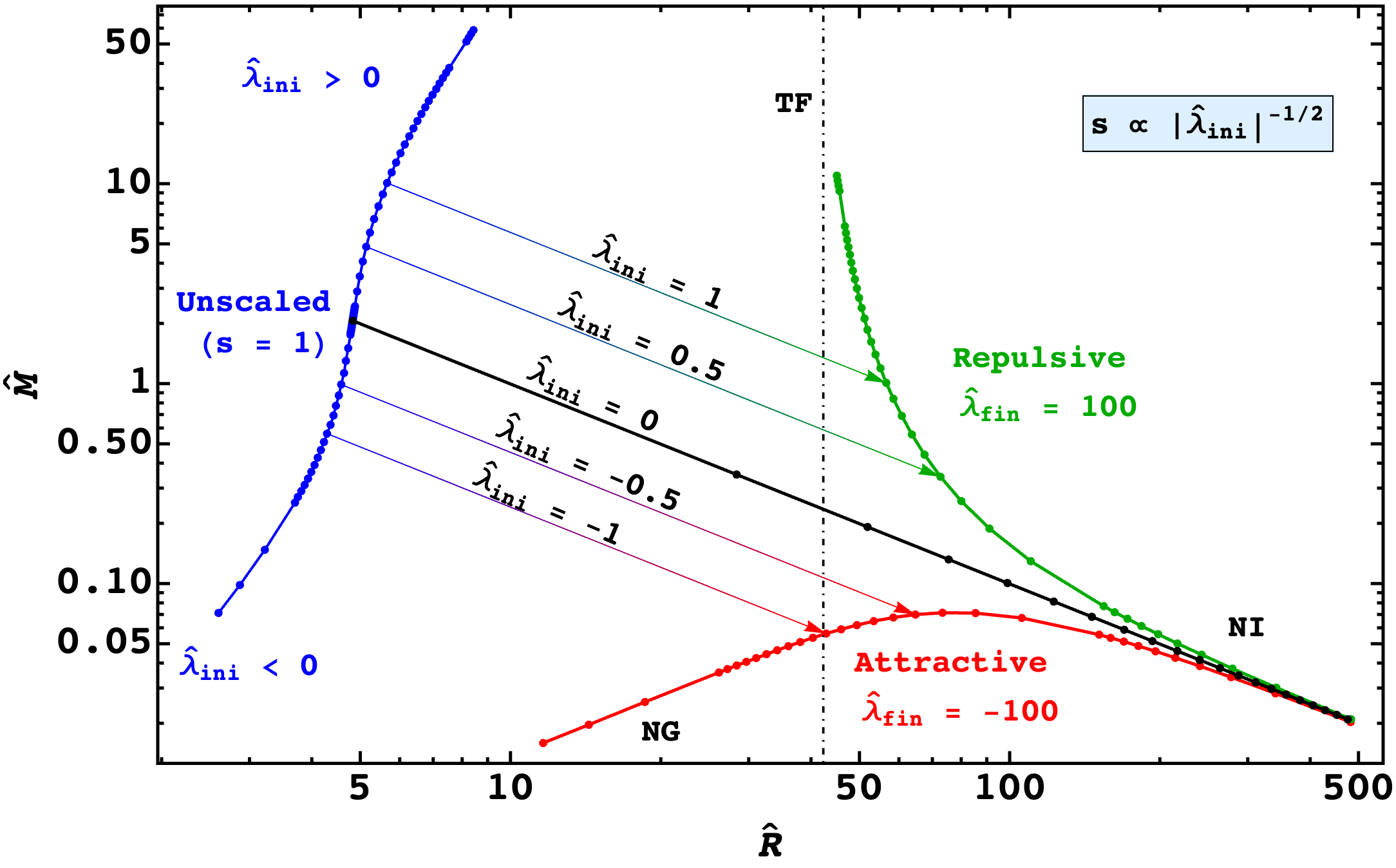}
    \caption[Mass-radius curves for $\lambda = 0$ and a fixed $\hat{\lambda}_\text{fin} = \pm100$]{Blue curve denotes unscaled and dimensionless soliton mass and radius $\left(\hat{M}_\text{ini}, \hat{R}_\text{ini}\right)$ for various values of $\hat{\lambda}_\text{ini}$. Red curve shows the scaled mass-radius curve (dimensionless) for attractive self-interactions, while the green curve does the same for repulsive self-interactions, for a fixed $|\hat{\lambda}_\text{fin}| = 100$. Arrows denote transformation due to scaling from a fixed $s$ to a fixed $\hat{\lambda}_\text{fin}$ curve. NG corresponds to the non-gravitational regime, NI to the non-interaction regime, and TF (vertical dashed line) to the Thomas-Fermi regime. See section~\ref{sec:regimes} for a discussion on different regimes.}
    \label{fig:m_vs_r_scaled}
\end{figure}

\subsection{Mass-radius curves without a density profile ansatz}\label{sec:m_vs_r_procedure}

We begin by solving the GPP system for various choices of $\hat{\lambda}_\text{ini}$, and for each such choice, we calculate the corresponding soliton mass $\hat{M}_\text{ini}$ and radius $\hat{R}_\text{ini} = \hat{R}_{99}$ (see blue curve in figure~\ref{fig:m_vs_r_scaled} marked ``Unscaled"). All the other solutions of GPP system can be obtained from this blue curve by employing scaling transformations as we now argue.

The mass-radius curves for some fixed value of scaled self-interaction strength $|\hat{\lambda}_\text{fin}|$ are shown in figure~\ref{fig:m_vs_r_scaled} for both attractive (red) and repulsive (green) cases. Let us now understand how one gets these curves from the unscaled blue curve. 

For some fixed arbitrary scale value $s$, using eqs.~(\ref{eq:scaling_length}) and~(\ref{eq:scaling_mass}), each point can be scaled to $\left(\hat{R}_\text{fin}, \hat{M}_\text{fin}\right) = \left(s\hat{R}_\text{ini}, \hat{M}_\text{ini}/s\right)$. However, scaling symmetry also implies that each scaled point corresponds to a different value of the scaled self-interaction strength ($\hat{\lambda}_\text{fin} = s^2\hat{\lambda}_\text{ini}$). Since we are interested in the case where $\hat{\lambda}_\text{fin}$ (negative or positive) is fixed, we choose $s$ such that for any $\hat{\lambda}_\text{ini}$,

\begin{equation}\label{eq:lambda_scale}
    s = \sqrt{\frac{\hat{\lambda}_\text{fin}}{\hat{\lambda}_\text{ini}}}\ ,
\end{equation}
where $\hat{\lambda}_\text{fin}$ remains fixed. The corresponding scaled radius and mass of the soliton $\left(\hat{R}_\text{fin}, \hat{M}_\text{fin}\right)$ represent the mass-radius curves for a fixed $\hat{\lambda}_\text{fin}$ as shown in figure~\ref{fig:m_vs_r_scaled} for $\hat{\lambda}_\text{fin} = +100$ (green curve) and $\hat{\lambda}_\text{fin} = -100$ (red curve). Note from eqs.~(\ref{eq:scaling_length}) and~(\ref{eq:scaling_lambda}) that the scaled mass $\hat{M}_\text{fin}$ and radius $\hat{R}_\text{fin}$ change only when the products $\hat{M}_\text{ini}|\hat{\lambda}_\text{ini}|^{1/2}$ and $\hat{R}_\text{ini}|\hat{\lambda}_\text{ini}|^{-1/2}$ vary respectively. This enables one to go from solutions with different $\hat{\lambda}_\text{ini}$ and the same scale ($s = 1$) to solutions with different $s$ values and the same $\hat{\lambda}_\text{fin}$. The final mass-radius curves are consistent with what is obtained by assuming an ansatz for soliton density profile \cite{Chavanis_2011_analytic, Schiappacasse_2018} but can be obtained without making this assumption (see also \cite{Guzman_2006, Chavanis_2011, Padilla_2021}).

An important thing to note is that the choice of $\hat{\lambda}_\text{fin}$ fixes only the location of a mass-radius curve in the $M-R$ plane. It is $\hat{\lambda}_\text{ini}$ that decides the shape of the mass-radius curve, i.e. information about what regime one is in lies with the unscaled dimensionless solutions. This is discussed in greater detail in section~\ref{sec:regimes}. 
We note some interesting features of the mass-radius curves in presence of self-interactions: (a) For attractive self-interactions, there exists a maximum mass $M_{\text{max}}$ for $\hat{\lambda}_\text{ini} = -0.4$; (b) For repulsive self-interactions, for large $\hat{\lambda}_\text{ini} > 0$, $\hat{R}_\text{fin}$ appears to asymptote to a minimum radius (here the system is said to be in the Thomas-Fermi regime). These features are discussed in detail in section~\ref{sec:regimes}. It is also important to note from figure~\ref{fig:m_vs_r_scaled} that for $\lambda < 0$ there exist two radii for the same soliton mass. However, only the larger of the two radii corresponds to a stable solution while the smaller one is unstable under small perturbations \cite{Chavanis_2011, Schiappacasse_2018}. This establishes an upper limit on the amount of attractive self-interactions one can have if one desires a stable solitonic solution at the centre of DM halos. In terms of dimensionless self-coupling strength, this limit is given by $\hat{\lambda}_\text{ini} > -0.4$ (see section~\ref{sec:max_mass}).

\subsection{Different regimes of the mass-radius curves}\label{sec:regimes}

In this section, we shall discuss different regimes of the mass-radius curves in the $M-R$ plane for different signs and strength of self-coupling $\lambda$. First, we shall obtain parametric dependence of $M$ and $R$ by directly comparing terms in the GPP system and then obtain numerical factors from solutions of the system. Note that we can write $\nabla = 1/R$ where $R$ is the characteristic length scale of the system. This also enables us to write the Poisson equation as $\Phi \sim 4\pi G|\psi|^2R^2$. Density can also be written as $\rho = |\psi|^2 \sim 3M/4\pi R^3$. 
\begin{enumerate}
    \item \textbf{The Non-Interacting Regime:} First we consider the case where $\lambda \rightarrow 0$. The self-interaction term is then negligible compared to other terms in eq.~(\ref{eq:GrossPitaevskii}). Comparing the gradient term and self-gravity term (and adding back missing factors of $\hbar$ and $c$) we get,
    \begin{equation}
        R \sim \frac{\hbar^2}{6GMm^2}\ .
    \end{equation}
    When $\hat{\lambda}_\text{ini} = 0$ there are no free parameters in the unscaled system, and there is only one solution with $\hat{M}_\text{ini} = 2.0612$ and $\hat{R}_\text{ini} = 4.822$. When scaling is introduced, $s$ becomes the sole free parameter, leading to a single mass-radius curve (solid black line in figure~(\ref{fig:m_vs_r_scaled})), where each $s$ value leads to a unique point on the curve. Using unscaled values of $\hat{M}_\text{ini}$ and $\hat{R}_\text{ini}$, along with scaling relations in eqs.~(\ref{eq:scaling_length}) and~(\ref{eq:scaling_mass}), and adding back dimensions, one can write, 
    \begin{equation}\label{eq:NI_regime}
        R_\text{fin} = 9.94\frac{\hbar^2}{GM_\text{fin}m^2}\ .
    \end{equation}
    This agrees well with solutions obtained from other numerical work \cite{Chavanis_2011,Chavanis_2011_analytic}. It is straightforward to see from figure~\ref{fig:m_vs_r_scaled} that for $|\hat{\lambda}_\text{ini}| \ll 1$ the mass-radius curve for any $\hat{\lambda}_\text{fin}$ can be approximated by eq.~(\ref{eq:NI_regime}).

    \item \textbf{Thomas-Fermi Regime:} For $\lambda > 0$, consider the case where the outward-acting repulsive self-interactions are balanced by the inward-acting gravity such that the gradient term is negligible. From eq.~(\ref{eq:GrossPitaevskii}) 
    \begin{equation}\label{eq:TF_radius_at_home}
        R^2 \sim \frac{\lambda\hbar^3}{32\pi Gm^4c}\ = \frac{a_s\hbar^2}{Gm^3}\ ,
    \end{equation}    
    where $a_s = \frac{\lambda}{32\pi}\frac{\hbar}{mc}$ is the scattering length for a real scalar field. This is the so-called Thomas-Fermi (TF) approximation. Ref.~\cite{Boehmer_2007} showed that in this limit, the system has an exact solution with density $\rho(r) = \rho_0 \sinc{\left(\pi r/R_{TF}\right)}$ where $R_{TF} = \pi \left(\frac{a_s\hbar^2}{Gm^3}\right)^{1/2}$. Note how $R_{TF}$ depends only on $\lambda$ (or $a_s$) and $m$. In terms of dimensionless variables, eq.~(\ref{eq:TF_radius_at_home}) can be written as $\hat{R} = \sqrt{2\hat{\lambda}_\text{ini}}$.\footnote{One can also write $\hat{R}_{TF} = \pi\sqrt{2\hat{\lambda}_\text{ini}}$.} This allows us to write the gradient term in eq.~(\ref{eq:GrossPitaevskii}) as $\hat{\nabla}^2/2 \sim \hat{R}^{-2}/2 = 1/4\hat{\lambda}_\text{ini}$. The TF approximation is valid when this term is negligible. We define the system to be in the TF regime when $\hat{\lambda}_\text{ini} \geq 2.5$ i.e. when the dimensionless gradient term in the Gross-Pitaevskii equation becomes $\mathcal{O}(0.1)$. To compare with numerical solutions, it is better to define $R_{99}$ for the analytic solution: $R_{99}^{(TF)} = 2.998 \left(\frac{a_s\hbar^2}{Gm^3}\right)^{1/2}$. An estimate for $R_{99}^{(TF)}$ can then be obtained using eqs.~(\ref{eq:scaling_lambda}) and~(\ref{eq:scaling_length}) which in dimensionful form is  
    \begin{equation}\label{eq:R_final}
        R_\text{fin} = \frac{\left(\hat{R}_\text{ini}\hat{\lambda}_\text{ini}^{-1/2}\right)}{\sqrt{2}}\left(\frac{a_s\hbar^2}{Gm^3}\right)^{1/2} = 3.18\left(\frac{a_s\hbar^2}{Gm^3}\right)^{1/2}\ .
    \end{equation}
    Here we have used $\hat{\lambda}_\text{ini} = 3.5$ which is the largest value for which we could solve the GPP equations, with $\hat{R}_\text{ini} = 8.43$. This gives a good approximation of the exact value of $R_{99}^{(TF)}$.

    \item \textbf{Non-Gravitational Regime:} On the other hand, for $\lambda < 0$, when the self-interaction term dominates over the self-gravity term, the equilibrium solutions to the GPP equations are unstable under small perturbations 
    \cite{Chavanis_2011,Guth_2015, Schiappacasse_2018}. Ignoring the self-gravity term in eq.~(\ref{eq:GrossPitaevskii}) and equating the gradient term with the self-interaction term, we get 
    \begin{equation}
        R \sim \frac{3\lambda\hbar M}{16\pi m^2c}\ .
    \end{equation}
    Contrary to the non-interacting regime, here $R$ and $M$ are linearly related, as shown by the left-most part of the red curve in figure~\ref{fig:m_vs_r_scaled}. In its dimensionless form, the relation simply reads $\hat{R} = 12\hat{M}\hat{\lambda}$. Note that for attractive self-interactions, the mass-radius curve transitions from the solitonic regime to the non-gravitational regime after attaining a maximum soliton mass which also separates the stable solutions from unstable ones \cite{Chavanis_2011}. In the next section, we discuss the parametric dependence of this maximum mass and its numerical estimate for a fixed $\hat{\lambda}_\text{fin}$. 
\end{enumerate}

\subsection{Maximum mass for attractive self-interactions}\label{sec:max_mass}

There is no indication of a maximum mass when one solves the GPP equations in its unscaled dimensionless form. However, from figure~\ref{fig:m_vs_r_scaled}, note that fixing the scaled $\hat{\lambda}_\text{fin}$ allows $\hat{M}_\text{fin}$ to indeed attain a maximum value. To find an expression for this maximum mass, we substitute eq.~(\ref{eq:lambda_scale}) in $M_\text{fin} = M_\text{ini}/s$,

\begin{equation}\label{eq:fin_mass}
   M_\text{fin} = 8\sqrt{\pi}\left(\hat{M}_\text{ini}|\hat{\lambda}_\text{ini}|^{1/2}\right)\frac{m_{pl}}{\sqrt{|\lambda_\text{fin}|}}\ .
\end{equation}
Since $\lambda_\text{fin}$ is fixed, $M_\text{fin} = M_{\text{max}}$ only when $\hat{M}_\text{ini}|\hat{\lambda}_\text{ini}|^{1/2}$ attains a maximum. From the unscaled numerical solutions, we find that $\left(\hat{M}_\text{ini}|\hat{\lambda}_\text{ini}|^{1/2}\right)_{\text{max}} = 0.71$, which leads to, for a real scalar field,
\begin{equation}\label{eq:max_mass}
    M_{\text{max}} \approx 10.12\frac{m_{pl}}{\sqrt{|\lambda_\text{fin}|}}\ .
\end{equation}
The form of eq.~(\ref{eq:max_mass}) agrees with what is obtained by \cite{Chavanis_2011, Levkov_2017}. It is worth noting that the maximum mass depends solely on the strength of self-interactions $\lambda$ and not on $m$. 

\subsection{Alternative explanation of regimes}\label{sec:alternative_regimes}

\begin{figure}[h!]
    \centering
    \includegraphics[width=0.8\linewidth]{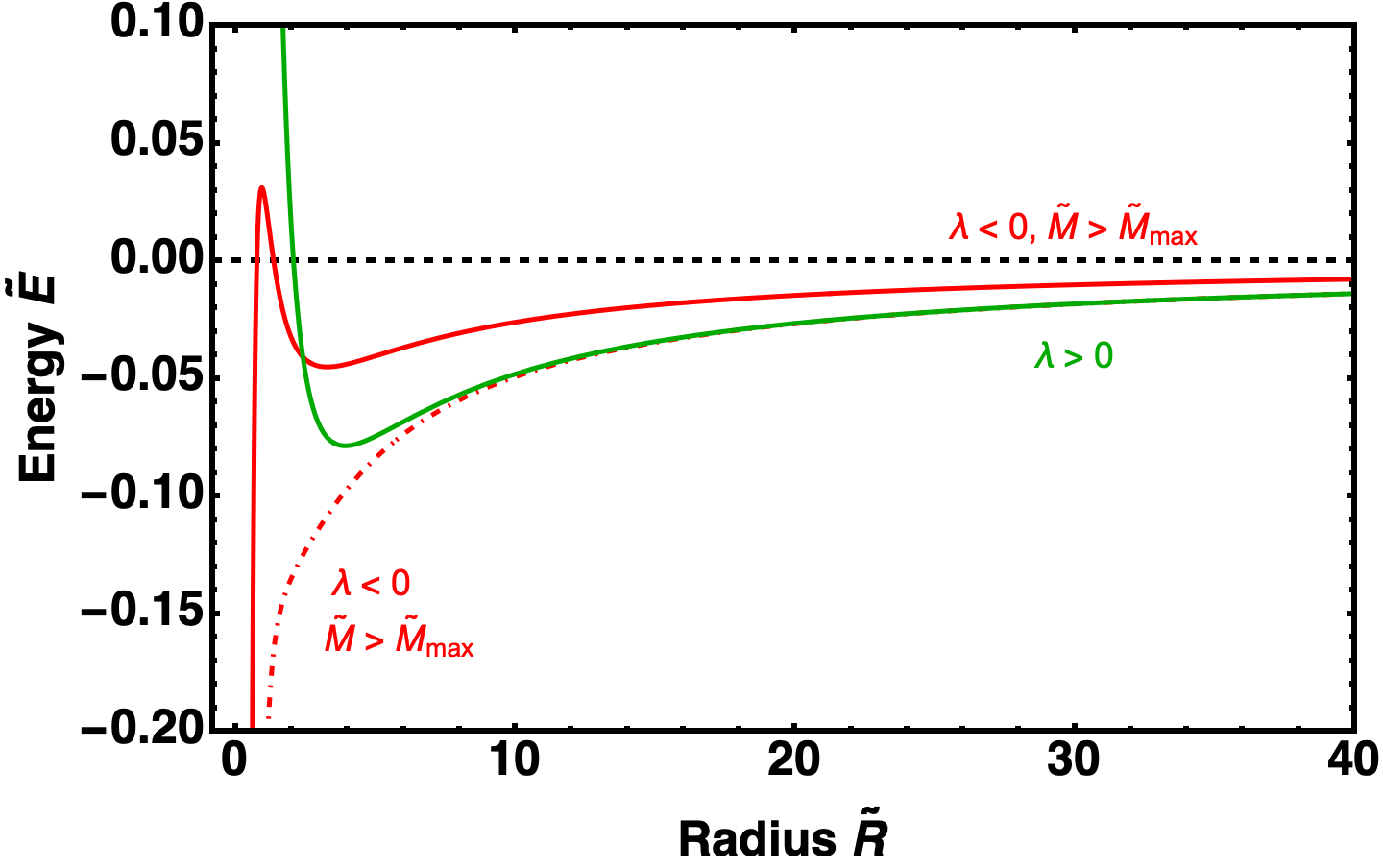}
    \caption[Energy versus radius for the GPP system]{Recreated version of figure~1 in Ref.~\cite{Chavanis_2011}, where we have plotted normalised energy vs. normalised radius (described in section~\ref{sec:alternative_regimes}). Note the presence of a minima for $\lambda > 0$ (green curve) and $\lambda < 0$, when $M < M_{\text{max}}$ (solid red curve), implying a stable solution for both cases. On the other hand, an unstable equilibrium solution (maximum in the solid red curve) is also possible for $\lambda < 0$ (when $M < M_{\text{max}}$). Finally, $\lambda < 0$ (when $M > M_{\text{max}}$) has no equilibrium solutions at all, as shown by the dot-dashed red curve.}
    \label{fig:energy_vs_radius}
\end{figure}

Another way of understanding the different regimes along with the existence of a maximum mass for $\lambda < 0$ is by using the total energy\footnote{One can use the Madelung transformation to write the GPP as equivalent fluid equations \cite{Chavanis_2011} or obtain a non-relativistic Lagrangian which can be then Legendre transformed to a Hamiltonian \cite{Schiappacasse_2018, Delgado_2022}.} of the GPP system, assuming spherical symmetry and an ansatz (for instance, Gaussian~\cite{Chavanis_2011}), 
\begin{equation}\label{eq:energy}
    E = \frac{3}{4}\frac{\hbar^2M}{m^2R^2} + \frac{1}{(2\pi)^{3/2}}\frac{\lambda}{16}\frac{\hbar^3M^2}{m^4R^3c} - \frac{1}{\sqrt{2\pi}}\frac{GM^2}{R}\ .
\end{equation}
Here, the first term corresponds to the Laplacian term (or `quantum pressure') in the Gross-Pitaevskii equation, the second term corresponds to the self-interactions and the third corresponds to the self-gravity of the system. 
For a fixed total mass, the energy can be extremised with respect to the radius, allowing one to obtain the condition for stationary solutions.
Before proceeding, following \cite{Chavanis_2011}, we normalise the energy, mass, and radius by the following: $E_a = GM_a^2/R_a$, where $M_a = \sqrt{32\pi\hbar c/G|\lambda|}$ and $R_a = \sqrt{\lambda\hbar^3/32\pi Gm^4c}$. 
We then plot the normalised energy $\Tilde{E}$ as a function of radius $\Tilde{R}$ in figure~\ref{fig:energy_vs_radius} where each curve corresponds to a fixed mass $\Tilde{M}$ and a fixed sign of $\lambda$. 

In figure~\ref{fig:energy_vs_radius}, we consider the following cases:

\begin{enumerate}
    \item \textbf{$\boldsymbol{\lambda > 0}$}: For repulsive self-interactions, there exists a radius with a clear minimum energy which corresponds to the stationary state. As the mass increases, this minimum will shift towards $R \sim R_{a}$ which is approximately the radius of the soliton in the Thomas-Fermi regime (see the green curve in figure~\ref{fig:m_vs_r_scaled}).  
    Note that the equilibrium is a stable one, i.e. a small perturbation will only lead to the energy coming back to the equilibrium value. 

    \item \textbf{$\boldsymbol{\lambda < 0$, $M < M_{\text{max}}}$}: In the case of attractive self-interactions, for a fixed mass $\Tilde{M}$, there are two values of radius for which the energy is extremised, i.e. there are two equilibrium solutions (as shown by the solid red curve in figure~\ref{fig:energy_vs_radius}). 
    However, note that one of the solutions (the one where the energy is maximised) corresponds to an unstable equilibrium, while the other (the one where energy is minimised) corresponds to a stable equilibrium. 
    These two solutions can be seen in in the M-R plane (figure~\ref{fig:m_vs_r_scaled}) as points on either side of the maximum of the red curve. The unstable (stable) branch in the same figure corresponds to the maximum (minimum) energy solutions for different masses $\Tilde{M}$. 

    To further elucidate the instability of solutions corresponding to the maximum energy is to first recall that from section~\ref{sec:regimes}, balance is maintained between quantum pressure and attractive self-interactions (gravity is ignored). 
    In this limit, in eq.~(\ref{eq:energy}) only the first two terms are relevant. 
    Now, if one increases $\Tilde{R}$ slightly, the self-interaction term $\propto \Tilde{R}^{-3}$ becomes smaller than the quantum pressure term $\propto \Tilde{R}^{-2}$, and since quantum pressure acts outwards, the soliton expands, further increasing the radius and therefore continuing to expand. 
    On other hand, if radius decreases slightly, the self-interaction term $\propto \Tilde{R}^{-3}$ becomes larger than the quantum pressure term $\propto \Tilde{R}^{-2}$, and since $\lambda < 0$, the self-interactions act inwards, reducing the radius of the soliton, which leads to the collapse of the soliton.

    \item \textbf{$\boldsymbol{\lambda < 0$, $M > M_{\text{max}}}$}: 
    In this case, for a fixed $\lambda < 0$, as you increase $\Tilde{M}$, both self-interaction term and the self-gravity term change as $\Tilde{M}^2$, while the quantum pressure term which changes as $\Tilde{M}$. Therefore, the energy as a function of radius is dominated by the negative terms, erasing any extremum above a certain $\Tilde{M}_{\text{max}}$. For $\Tilde{M} > \Tilde{M}_{\text{max}}$, no equilibrium (stationary) solutions of the GPP system exist, as shown by the dot-dashed red curve in figure~\ref{fig:energy_vs_radius}. 
    As we have seen in section~\ref{sec:max_mass}, $M_{\text{max}}$ is given by eq.~(\ref{eq:max_mass}).  
    There are no equivalent points for such a system in the M-R plane (i.e. figure~\ref{fig:m_vs_r_scaled}) since numerical solutions of the GPP system that we obtain are, by definition, stationary solutions.
\end{enumerate}

\section{Quasi-stationary solutions}\label{sec:quasi-stationary}

Having developed the machinery to obtain stationary solitonic solutions, we can now describe time-independent density profiles in the inner regions of galactic halos. 
However, there are certain situations, where stationary solutions do not adequately describe the physics. 
For instance, consider a satellite dwarf galaxy in orbit around the centre of a larger host halo with an angular frequency $\omega$. 
Here, in the rest frame of the satellite galaxy, the self-gravity of the satellite attempts to hold the core together, while the gravitational pull due to the host halo attempts to pull the satellite apart. 
In this case, one can account for the gravitational potential of the host halo, by simulating the full time-dependent system \cite{Du_2018, Glennon_2022}. 
Alternatively, one can continue solving the time-independent system by introducing an external potential that accounts for the effect of the halo. 
To gain intuition for this, we first look at the tidal effects in Newtonian gravity. 
Note that most of the text in this section is an excerpt from our paper~\cite{Dave_2024}.

Consider two point particles near each other falling freely under the gravity of an external, far away object. The acceleration of the relative position, ${\bf r}'$  of the second particle w.r.t. the first particle, is determined by the Hessian matrix of the gravitational potential due to the far away object. Given this acceleration, one can find what is called tidal potential, which is the potential corresponding to the acceleration determined by solving ${\bf a} = - {\bf \nabla} \Phi_{\rm tidal}$ for $\Phi_{\rm tidal}$ i.e. it is given by
\begin{equation}
    \Phi_{\rm tidal} ({\bf r}) -   \Phi_{\rm tidal} ({\bf r}_0) = - \int_{{\bf r}_0}^{\bf r} {\bf a}({\bf r}') \cdot d {\bf r}' \; .
\end{equation}
In order to define this tidal potential unambiguously, we choose ${\bf r}_0$ to be the location of the first particle, which is the origin ${\bf 0}$ and define $\Phi_{\rm tidal} ({\bf r}_0) = 0$. 

As an example, if the far away object has mass $M$ and is located at $(0, 0, d)$ and if the first particle is at the origin and the second particle is at ${\bf r}$, the acceleration of the separation vector ${\bf r}$ is then given by, 
\begin{equation}\label{eq:tidal_acceleration}
    a^i = -\delta^{ij}\left(\frac{\partial^2\Phi}{\partial x^j \partial x^k}\right)_{\bf 0}r^k = -\frac{GM}{d^3}\left(x\hat{x} + y\hat{y} - 2z\hat{z}\right)\ .
\end{equation}
Defining a tidal potential that leads to the acceleration in eq.~(\ref{eq:tidal_acceleration}) such that $\Phi_\text{tidal}({\bf 0}) = 0$, which is given by

\begin{equation}
    \Phi_\text{tidal} = -\frac{GM}{2d^3}\left(2z^2 - x^2 -y^2\right)\ .
\end{equation}
Note that along the line joining the first particle and far away object, this will go as $\Phi_{\rm tidal} \propto - \omega^2 r^2$, where, $\omega^2 = \frac{GM}{d^3}$ from Kepler's third law.

\subsection{Modelling tidal effects in the ULDM paradigm}

Recall that the large deBrogile wavelength of an ultra-light dark matter particle of mass $m \sim 10^{-22}\ \text{eV}$ can lead to wave-like phenomena at galactic scales \cite{Hui_2021}. 
One such intriguing phenomenon is the loss of dark matter from a satellite dwarf galaxy orbiting a larger halo. 
Consider a dwarf satellite galaxy moving in a circular orbit around the centre of a large host halo at a radius $a$ with an angular frequency $\omega = \sqrt{\frac{GM}{a^3}}$, where $M$ is the mass of the halo enclosed within a sphere of radius $a$. We also assume that the mass contained within the radius is close to the total halo mass, i.e. $a$ is such that the satellite is at the outskirts of the halo. 
In this orbit, the self-gravity $\Phi_{SG}$ of the satellite holds it together, while the tidal forces due to the host halo contribute an external potential term.
For $r \ll a$ measured from the centre of the satellite, one can write this external potential as $\Phi_{H} = -\frac{3}{2}m\omega^2r^2$ \cite{Hui_2017, Hertzberg_2023}, which is then included in the time-independent form of the spherically symmetric Schrödinger-Poisson system.  

\begin{figure}[ht]
    \centering
    \includegraphics[width = 0.75\textwidth]{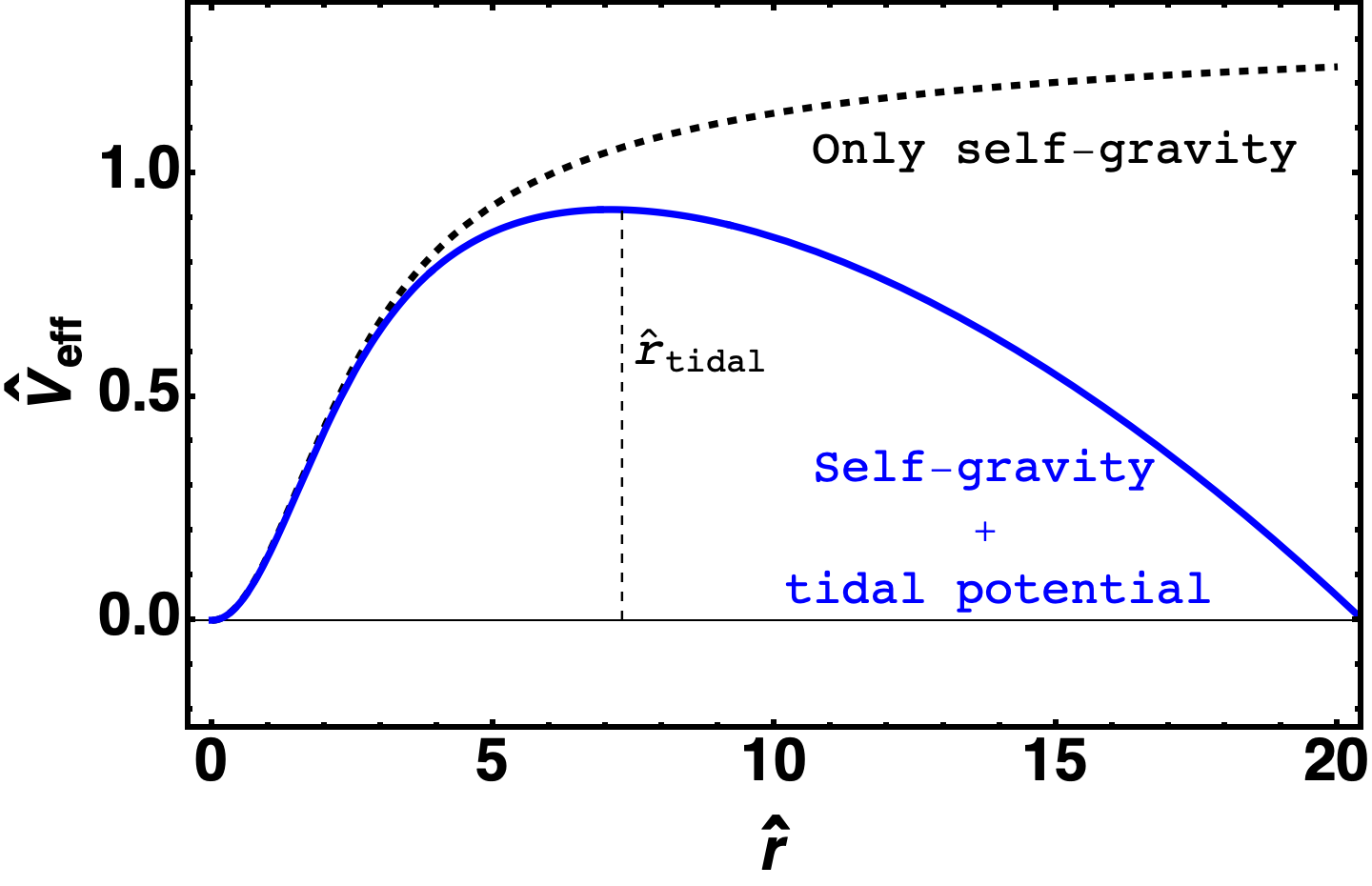}
    \caption[Effective potential in the presence and absence of tidal effects]{Dimensionless effective potential ($\hat{V}_{\text{eff}} = V_{\text{eff}}/c^2$) - as a function of dimensionless distance $\hat{r}$ (see eq.~(\ref{eq:dimensions2})) from the centre of the satellite galaxy - in the absence of a tidal potential (black dashed curve), and in the presence of an external tidal potential (blue curve). In the latter case, the potential peaks at the tidal radius.}
    \label{fig:tidal_potential}
\end{figure}

Note that the tidal effects are in-fact not spherically symmetric. However, as argued in \cite{Hertzberg_2023} one can capture the dominant effect of tidal forces by including a spherically symmetric term $\Phi_H$ (as mentioned above) in the time-independent Schrödinger equation.
One can then define an effective potential $V_\text{eff} \equiv \Phi_{SG} + \Phi_H$, that will possess a local maximum at the so-called tidal radius $r_{\text{tidal}}$ (measured from the origin at the centre of the satellite). The effective potential now resembles a potential barrier (as shown in figure~\ref{fig:tidal_potential}). 

One can model the loss of mass from the satellite by treating it as a tunnelling problem, and looking for quasi-stationary solutions of the following modified time-independent Schrödinger-Poisson system \cite{Hui_2017, Hertzberg_2023}:
\begin{eqnarray}
    \frac{1}{2m}\nabla_r^2\psi &=&  m\Phi_{SG}\psi\ - \frac{3}{2}m\omega^2r^2\psi - \gamma\psi, \label{eq:mod_Shro} \\
    \nabla_r^2\Phi_{SG} &=& 4\pi G|\psi|^2\ , \label{eq:mod_poisson} 
\end{eqnarray}
where recall that $\nabla_r^2 = \frac{\mathrm{d}^2}{\mathrm{d}r^2} + \frac{1}{r}\frac{\mathrm{d}}{\mathrm{d}r}$.
Also, note that here we have used $\Phi_{SG}$ to denote the self-gravity of the scalar field configuration.

Until now, we have only dealt with nodeless stationary state solutions, whose properties, as discussed in section~\ref{sec:stationary_states} bear repeating: 
\begin{inparaenum}[(a)]
  \item is regular everywhere,
  \item has spherical symmetry $\Psi(\Vec{r},t) = \Psi(r,t)$ and $\Phi_{SG}(\Vec{r},t) = \Phi_{SG}(r,t)$,  
  \item is nodeless,
  \item is stationary in the sense that $\Psi(r,t) = e^{-i\gamma t/\hbar}~\phi(r)$ (where $\gamma$ is real and time independent and hence $\phi(r)$ can be taken to be real) and $\Phi_{SG}(r,t) = \Phi_{SG}(r)$, and,
  \item is localized in the sense of being square integrable, i.e. $\int_0^\infty 4\pi r^2|\Psi|^2dr$ is finite.
\end{inparaenum}
As we shall see, when looking for quasi-stationary solutions only the first two properties out of the five will hold good. 

To model the loss of the mass, we first allow the eigenvalue $\gamma$ to be complex, ascribing an exponential time-dependence to the classical field $\Psi$ and consequently to the density $|\Psi|^2$: $e^{-i\gamma t/\hbar} = e^{-i\gamma_R t/\hbar}e^{\gamma_I t/\hbar}$. Requiring $\gamma_I < 0$ ensures that the density decreases with time: $\rho = |\Psi|^2 = e^{-2|\gamma_I|t/\hbar}|\psi|^2$. The solution is no longer stationary as defined in condition (d) above. 
Note that a complex $\gamma$ forces $\psi$ to be complex as well. 

Since matter is now tunnelling through the barrier, the solution will not be zero on the other side of the barrier and will resemble an outgoing wave. 
The form of the outgoing wave can be obtained by using the WKB approximation and taking the large $r$ limit, where the only dominant term will be the tidal potential,

\begin{equation}\label{eq:wkb}
    \lim_{r\rightarrow\infty}\psi(r) = \psi_{WKB}(r) = \frac{A}{r^{3/2}}\exp{\left(i\frac{\sqrt{3}}{2\hbar}m\omega r^2\right)}\ .
\end{equation}
The above equation now serves as a boundary condition that the quasi-stationary solution must match to at some large $r$. However, since the WKB solution is oscillatory, the quasi-stationary solution is also forced to be oscillatory, violating the nodeless condition (c) for stationary solutions.
Further, since $|\psi|^2\propto r^{-3}$ the integral $4\pi\int_0^\infty r^2dr |\psi|^2$ will not converge for $r\rightarrow \infty$ (it will be log-divergent), i.e. the solution will not be localized as described in condition (e). 
The formalism of quasi-stationary solutions described here is often used in quantum mechanics, and can be found in section~134 of \cite{landau1991quantum} and the last chapter of \cite{perelomov1998quantum}. 



\subsection{A note on self-consistency}\label{sec:self-consistency}

It is worth noting that the time-independent treatment of the SP equations in the presence of a tidal potential is only possible if $\Phi_{SG}$ is taken to be time-independent.
Since $\rho(r)$ is no longer time-independent, the source term in the Poisson equation will be $|\psi|^2e^{-2|\gamma_I|t}$. 
This is expected, since the loss of mass over time will alter the density profile of the satellite dwarf and consequently the corresponding self-gravity. 
This was noticed by the authors in \cite{Du_2018} who argued that quasi-stationary solutions of eqs.~(\ref{eq:mod_Shro}) and~(\ref{eq:mod_poisson}) - where the self-gravity is taken to be time-independent - are valid only for initial times $t \rightarrow 0$, i.e. when the loss of density is small enough to not affect the gravitational potential.

It must be noted that one can approximate $\Phi_{SG}$ to be time-independent when $\hbar/\gamma_I^{-1}$ (i.e. the lifetime of the satellite) for a particular object is large compared to the timescales of interest, ensuring that the density does not change appreciably over relevant timescales. We shall see that this is true when the dimensionless angular frequency parameter is small. 
In particular, we find that for a typical orbital frequency of $1.6\ \text{Gyr}^{-1}$, the central density of the dwarf satellite decays to less than $\sim80\%$ its initial value after $1$ ($3$) orbits for $\hat{\omega}_\text{ini} > 0.085$ ($\hat{\omega}_\text{ini} > 0.075$). 
Therefore, for $\hat{\omega}_\text{ini}$ smaller than the above-mentioned values, this description of tunnelling is valid, given all other assumptions.

\subsection{Including self-interactions}
In addition to the presence of a tidal potential, one can also include self-interactions. 
The time-independent spherically symmetric Gross-Pitaevskii-Poisson system can now be written as 

\begin{eqnarray}
   \gamma\phi &=& -\frac{1}{2m}\nabla_r^2\phi + \left(m\Phi_{SG} - \frac{3}{2}m\omega^2r^2 + \frac{\lambda}{8m^3}|\phi|^2\right)\phi\ ,\label{eq:mod_GP} \\ 
   \nabla_r^2\Phi_{SG} &=& 4\pi G|\phi|^2\ , \label{eq:mod_poisson_SI}
\end{eqnarray}
where we have assumed that similar to the no self-interactions case, tidal effects will only lead to an additional $\Phi_H$ term in the Gross-Pitaevskii (GP) equation. It is worth noting that $\Phi_H$ is included as an external potential while $\Phi_{SG}$ must simultaneously satisfy the Poisson equation. One can write eq.~(\ref{eq:mod_GP}) such that it resembles a time-independent Schrödinger equation where one can define an effective potential, $V_\text{eff}$ using the terms in the parenthesis on the RHS, 

\begin{equation}\label{eq:effective_potential}
    V_{\text{eff}} = \Phi_{SG} + \Phi_H + \Phi_{SI}\ .
\end{equation}
Here $\Phi_{SI} = \frac{\lambda}{8m^3}|\phi|^2$. This effective potential will act as the potential barrier for a tunnelling problem. 

Our goal is to solve Eqs (\ref{eq:mod_GP}), (\ref{eq:mod_poisson}) and find how the scalar self-coupling $\lambda$ affects the rate at which DM tunnels out of the satellite galaxy.

\subsection{Numerical solutions}\label{sec:quasi_numerical}

To solve the system numerically, we use the dimensionless variables introduced in eqs.~(\ref{eq:dimensions1}) and~(\ref{eq:dimensions2}), denoted by a ` $\hat{}$ ' over quantities, in terms of $m$ and fundamental constants. 
Here, we consider two additional variables, 

\begin{eqnarray}
    \hat{t} &=& \frac{mc^2}{\hbar}t \label{eq:time} \\  \hat{\omega} &=& \frac{\hbar}{mc^2}\omega\ .\label{eq:omega}
\end{eqnarray}
Recall that $\hat{r}$ is now in units of the Compton wavelength of the scalar field, while $\hat{t}$ is in the units of `Compton time'. The free parameters of the system now are $\hat{\lambda}$ and $\hat{\omega}$ (note that dimensionless version of the system of equations will not depend on $m$ using the re-definitions above). 
When $\hat{\omega} = 0$ the system just reduces to the GPP system, where the $\hat{\gamma}$, $\hat{\phi}$ are real and the boundary conditions for solving the system are described in section~\ref{sec:stationary_states}.
For $\hat{\omega} > 0$, the boundary conditions are different. Since the solution is not a stationary state, $\hat{\gamma} = \hat{\gamma}_R + i\hat{\gamma}_I$. The boundary conditions arising from regularity of the $\hat{\phi}$ and $\hat{\Phi}$ at the origin as well as the initial values of $\hat{\Phi}(0) = 0$ and $\hat{\phi}(0) = 1$ remain unchanged.  
However, for behaviour of $\hat{\phi}$ at large $\hat{r}\rightarrow \infty$, we now have outgoing waves. We impose the BCs at infinity by choosing a large $\hat{r}_m$, and matching the numerical solution as well as its first derivative with the asymptotic solution in eq.~(\ref{eq:wkb}), i.e., $\hat{\phi}(\hat{r}_m) = \hat{\phi}_\text{WKB}(\hat{r}_m)$ and $\hat{\phi}'(\hat{r}_m) = \hat{\phi}_\text{WKB}'(\hat{r}_m)$ (note that this must hold for both real and imaginary parts of $\hat{\phi}$). Similar to the stationary state case, we set $\text{Re}(\hat{\phi}(0)) = 1\ , \text{Im}(\hat{\phi}(0)) = 0$. We look for solutions with $\hat{\gamma}_I < 0$, all of these conditions will uniquely determine the values of $\hat{\gamma}_R$ and $\hat{\gamma}_I$.\footnote{Note that for every solution that satisfies the boundary condition in eq.~(\ref{eq:wkb}), we ensure that the density has negligible `wiggles' and $|\hat{\phi}|^2/|\hat{\phi}_{\text{WKB}}|^2 \propto r^0$ up to a certain accuracy (we require the slope of the above function to be $< 0.05$) around $\hat{r}_m$ to ensure appropriate behaviour.} 
While the solution obtained here is not a stationary state, we continue to call it a soliton for the rest of the section.

\begin{figure}[ht]
    \centering
    \includegraphics[width = 0.8\textwidth]{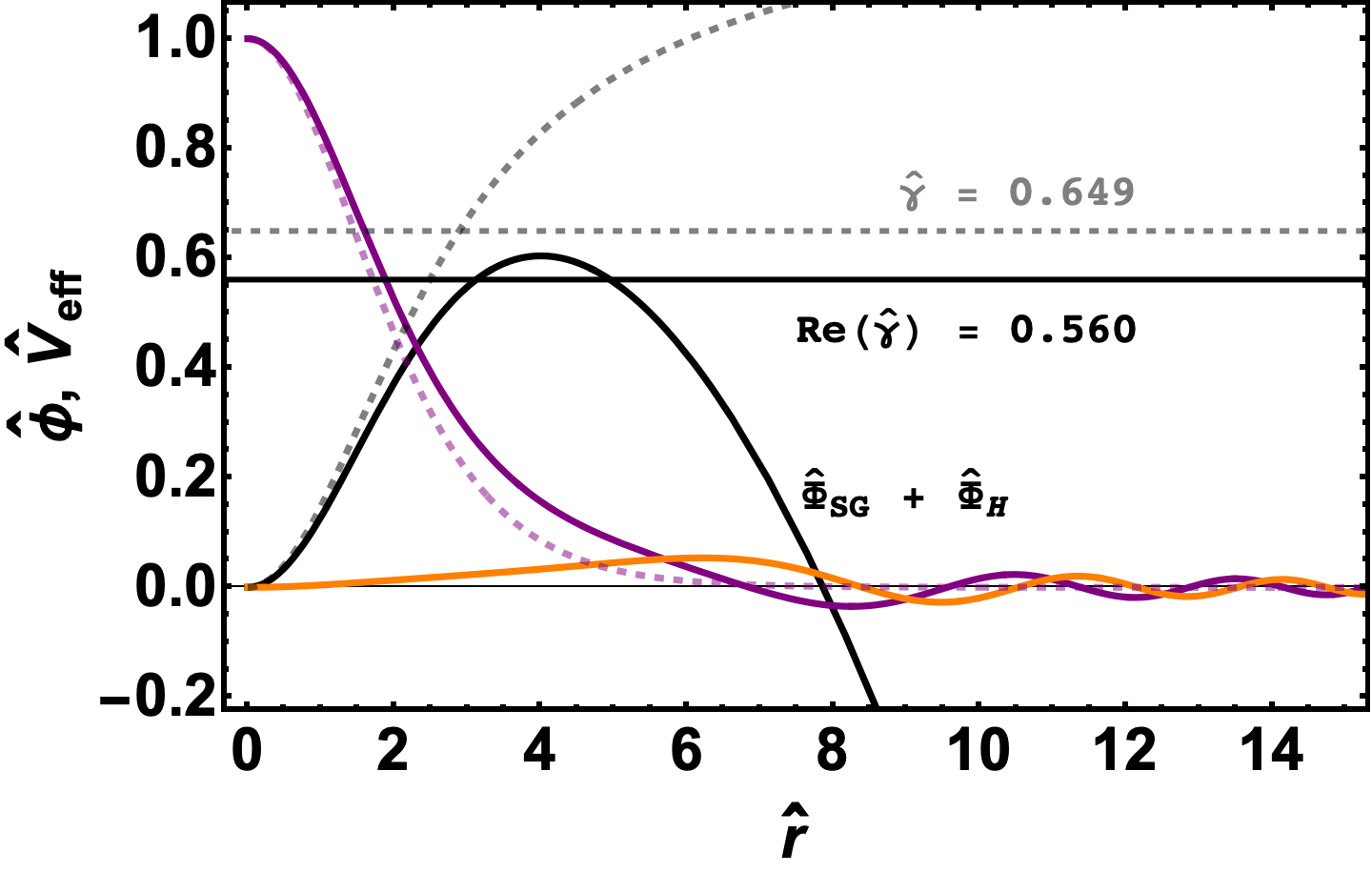}
    \caption[Numerically found quasi-stationary solution for $\hat{\omega} = 0.12$]{An illustrative numerical solution with (solid curves) and without (dashed curves) tidal potential is shown. To highlight the oscillatory nature of the solution in the presence of tidal effects, we have chosen $\hat{\omega} = 0.12$ (however, also see section~\ref{sec:self-consistency} for why we usually will not deal with such large values). 
    The real part of the quasi-stationary solution is shown by the purple solid curve while the imaginary is shown by the orange solid curve. Note that the real part of the eigenvalue for the quasi-stationary $\hat{\gamma}_R$ is smaller than $\hat{\gamma}$ for the stationary state. Also note the presence of a local maximum in the effective potential when a tidal potential is included. All quantities in the plot are dimensionless as defined in eq.~(\ref{eq:dimensions1}) and~(\ref{eq:dimensions2}).}
    \label{fig:example_sol}
\end{figure}

An example solution found using the shooting method \cite{Giordano_Nakanishi} for $\hat{\omega} = 0$ and $\hat{\omega} = 0.12$ with $\hat{\lambda} = 0$ is shown in figure~\ref{fig:example_sol}.

\subsection{Tunnelling rate and lifetime}\label{sec:decay_rates}

Once we have a solution, we have a density profile $\hat{\rho}(\hat{r}) = |\hat{\phi}(\hat{r})|^2$ and can define the mass enclosed in a sphere of radius $\hat{r}$ as
\begin{equation}\label{eq:soliton_mass}
    \hat{M}(\hat{t}, \hat{r}) = \frac{G m}{\hbar c}M(t, r)= \int_0^{\hat{r}}|\hat{\Psi}(\hat{t}, \hat{r}')|^2 \hat{r}'^2 d\hat{r}' = e^{2\hat{\gamma}_I t}\left[\int_0^{\hat{r}}|\hat{\phi}(\hat{r}')|^2 \hat{r}'^2 d\hat{r}'\right]\ , 
\end{equation}
where we have used eqs.~(\ref{eq:dimensions1}) and~(\ref{eq:dimensions2}) and $\hat{\gamma}_I \leq 0$. 

It is important to note that for $\hat{\omega} = 0$, $\hat{\gamma}_I = 0$ and hence mass of the soliton will remain constant over time. On the other hand, for $\hat{\omega} > 0$, from eq.~(\ref{eq:soliton_mass}) and the fact that $\hat{\gamma}_I < 0$ the soliton will lose its mass over time exponentially. We can now define a (dimensionless) decay rate as
\begin{equation}\label{eq:decay_rate}
  \hat{\Gamma} \equiv \frac{-\left(\mathrm{d}\hat{M}/\mathrm{d}\hat{t}\right)}{\hat{M}} = -2\hat{\gamma}_{I}\ ,
\end{equation}
where it is to be noted that $\hat{\gamma}_I$ and hence $\hat{\Gamma}$ are time-independent. Note that this dimensionless decay rate is related to the dimensionful decay rate by the relation $\hat{\Gamma} = \left(\hbar/mc^2\right)\Gamma$. Lifetime of the soliton is just the inverse of the decay rate (or, tunnelling rate) i.e., $\tau \equiv \Gamma^{-1}$.

\subsection{Core radius, core mass and scaling relations}\label{sec:scaling_relations}

In order to calculate any measure of mass of the soliton, one needs to evaluate the integral inside the square brackets in eq.~(\ref{eq:soliton_mass}). Note that one cannot define a total mass (found by replacing the upper limit by $\hat{r}\rightarrow \infty$) for quasi-stationary states, since boundary conditions in eq.~(\ref{eq:wkb}) imply that at large $\hat{r}$, $|\phi|^2\propto r^{-3}$ making the integral in eq.~(\ref{eq:soliton_mass}) log divergent. We can however define some measure of mass by first defining the size of the soliton as the radius at which the density becomes half its central value (also called core radius $\hat{R}_c$). This can then be used to define a soliton mass as the mass contained within core radius $\hat{M}_c = \hat{M}(\hat{R}_c)$ (also called core mass). 

Note that the scaling symmetry that we discussed in section~\ref{sec:scaling_symmetry} remains intact in the presence of the tidal potential. 
One can easily find that this implies the following transformations for the orbital frequency $\hat{\omega}$ and some timescale $\hat{t}$: 
\begin{equation}\label{eq:scaling_omega}
    \hat{\omega}\rightarrow s^{-2}\hat{\omega}, \ \ \ \hat{t}\rightarrow s^2\hat{t}\ .
\end{equation}
Also note that since $\hat{\Gamma}$ has the inverse dimensions of time, $\hat{t}$, it scales as $\hat{\Gamma} \rightarrow s^{-2}\hat{\Gamma}$. 

Until now, in this chapter, we have set up the machinery of solving the GPP system to obtain both stationary and quasi-stationary solutions. 
We are now in a position to confront various galactic scale observations and their ability to probe both the mass and self-coupling of ULDM.
In chapter~\ref{chpt:paper_1}, we look at the observational upper limits on the amount of mass contained within the centre of a galactic halo. 
In chapters~\ref{chpt:paper_2} and~\ref{chpt:paper_4}, we look at rotation curve data from the Spitzer Photometry and Rotation Curve (SPARC) catalogue~\cite{Lelli_2016}. 
In these chapters, we consider stationary state solutions of the GPP system to describe centres of galactic halos. 
Finally, in chapter~\ref{chpt:paper_3}, we utilize quasi-stationary solutions of the GPP system to describe satellite galaxies with observed core mass and orbital frequency in orbit around the centre of a larger host halo.

Before proceeding, let us first briefly  discuss the Fuzzy Dark Matter model, and existing constraints from the literature. We shall also briefly discuss the impact of self-interactions as discussed in the literature. 

\section{Fuzzy Dark Matter}\label{sec:fuzzy_dark_matter}

\subsubsection{Fuzzy dark matter}

Near the turn of the 21st century, it was suggested \cite{Lee_1995, Hu_2000} that a potential DM candidate with an ultralight mass, $m\sim 10^{-22}\ \text{eV}$, can address some of the small scale issues prevalent with CDM.
This is because such Ultra Light DM, due to the uncertainty principle, will suppress structure from forming at $\mathcal{O}(\text{kpc})$ scales, leading to natural solutions to the core-cusp and the missing satellites problem \cite{Bullock_2017, Hui_2017}. 

It is therefore useful at this stage to briefly recall the effect of FDM on linear perturbations. 
For a homogeneous background distribution of scalar field with small linear perturbations, one finds that the growth of perturbations due to gravity can be hindered by the gradient pressure terms arising from the uncertainty principle~\cite{Hui_2021} 

\begin{equation}
    k_J = (16\pi G\Bar{\rho})^{1/4}m^{1/2}\ ,
\end{equation}
where $\Bar{\rho}$ is the background density. The Jeans scale here highlights the relative importance of the pressure due to the uncertainty principle (also called ``Quantum Pressure'') versus gravity. For $k > k_J$, i.e. for length scales smaller than the length scale corresponding to $k_J$, the uncertainty principle dominates, suppressing fluctuations. The Jeans scale is proportional to $m^{1/2}$ implying that as $m$ decreases, the length scales below which perturbations are suppressed becomes larger. 
Hence, there will be a suppression in the matter power spectrum at high $k$ values, which can be a smoking gun signature for FDM~\cite{Hui_2017, Ferreira_2021}. 

Due to its `quantum' nature (as discussed above) at galactic scales, this kind of DM was dubbed Fuzzy Dark Matter (FDM).
At scales larger than $\mathcal{O}(\text{kpc})$, owing to the low density and high inter-particle separation, FDM behaves like CDM (as seen in simulations \cite{Schive_Nature_2014, Schive_PRL_2014, Mocz_2017, Schwabe_2020} due to the so-called Schrödinger-Poisson--Vlasov-Poisson correspondence~\cite{Widrow_1993, Mocz_2018}), leaving the large scale predictions of CDM untouched.
Based on linear perturbation theory and observed CMB anisotropies~\cite{Hlozek_2018, Kimball_book_2023} found that FDM mass must be heavier than $m \gtrsim 10^{-25}\ \text{eV}$. 
Pair with this the ability to address to some of the small-scale issues with CDM (like explaining flat density cores of dwarf spheroidals~\cite{Chen_2016, Gonzalez-Morales_2016yaf}) without invoking baryonic physics, made FDM models very attractive dark matter candidates~\cite{Ferreira_2021, Hui_2021, Kimball_book_2023}. 

Further, a galactic scale deBroglie wavelength also leads to unique observational signatures at these scales, which make it an interesting model to probe with astrophysical observations. 
Examples of unique signatures include flat density cores \cite{Davies_2020, Bar_2022, Zimmermann_2024, Benito_2025, Singh_2025} (we obtained these cores in section~\ref{sec:stationary_states}), suppression of power in the matter power spectrum \cite{Hlozek_2018, Poulin_2018, Liu_2024}, density oscillations and interference \cite{Schive_Nature_2014, Li_2021, Zagorac_2021, Dalal_2022, Amruth_2023, Salasnich_2025}, altered tidal disruption (compared to CDM) \cite{Du_2018, Hertzberg_2023, Glennon_2022, Dave_2024}, etc.

\subsubsection{Testing fuzzy dark matter against observations}

Increased interest in the FDM paradigm has also led to increased scrutiny which, over the last decade has stringently constrained masses in the range $m\in \left[10^{-24}\ \text{eV}, 10^{-20}\ \text{eV}\right]$ from galactic scale observations. 
For instance, an important astrophysical testing ground comes in the form Ultra-Faint Dwarf (UFD) galaxies where along with small stellar masses, the baryonic-feedback effects are also negligible. 
By using observed velocity dispersion of 18 UFDs in the Milky Way Ref.~\cite{Hayashi_2021} found that most galaxies prefer $m \gtrsim 10^{-21}\ \text{eV}$. The mass compatible with Segue I was $m\sim 10^{-19}\ \text{eV}$. 
Similarly, it was found from fitting FDM density profiles to Milky Way UFDs \cite{Safarzadeh_2020} that $m\gtrsim 10^{-21}\ \text{eV}$.
Similar constraints were obtained using stellar heating\footnote{This is the result of $\mathcal{O}(1)$ fluctuations in density of FDM halos due to its large deBroglie wavelength \cite{Schive_Nature_2014, Li_2021, Zagorac_2021}.} of star clusters in UFDs for Eridanus II \cite{Marsh_2019} and Segue 1 and 2 \cite{Dalal_2022} constraining $m \gtrsim 10^{-19}\ \text{eV}$ and $m > 3\times 10^{-19}\ \text{eV}$ respectively. 

Constraints from small-scale structure, which can be probed by observations of Lyman-$\alpha$ flux power spectrum require FDM particle mass to satisfy $m \geq 1.5 \times 10^{-21}\ \text{eV}$ \cite{Irsic_2017, Armengaud_2017}. 
Further, Black hole superradiance (BHSR), a mechanism through which a scalar field cloud can extract energy from a spinning black hole can be used to constrain FDM as well. When applied to observed supermassive black holes, Ref.~\cite{Stott_2018} found that FDM masses $10^{-19}\ \text{eV} \lesssim m \lesssim 10^{-16}\ \text{eV}$ are excluded. 

Rotation curves, which are an important probe for the dark matter distribution in galaxies~\cite{Salucci_2019} prove to be yet another testing ground for FDM. By requiring that galaxies satisfy a core-halo mass relation\footnote{An empirical power law between the mass of the core at the centre of a FDM halo and the mass of the surrounding halo, as observed in simulations~\cite{Schive_Nature_2014, Schive_PRL_2014, Mocz_2017}. See chapter~\ref{chpt:paper_2}. for more details.} as well as observed rotation curves (from the Spitzer Photometry and Accurate Rotation Curves catalogue\cite{Lelli_2016}), \cite{Bar_2018, Bar_2022} ruled out FDM in the entire range of $10^{-24}\ \text{eV} \lesssim m \lesssim 10^{-20}\ \text{eV}$ (see chapter~\ref{chpt:paper_2} for more details). 
On the other hand, Bayesian analysis of rotation curves of dark matter dominated galaxies from the SPARC~\cite{Khelashvili_2023} and LITTLE THINGS~\cite{Banares_Hernandez_2023} catalogues found that no single FDM mass $m$ can describe the rotation curves of all galaxies simultaneously. 

Some other constraints include relating measurements of the environment around SMBHs to FDM mass \cite{Bar_SMBH_2019, Davies_2020, Chiu_2025} (see chapter~\ref{chpt:paper_1} for more details), Pulsar Timing Arrays~\cite{Khmelnitsky_2014, Afzal_2023, Tomaselli_2024}, strong lensing~\cite{Powell_2023}, sub-halo mass function~\cite{Schutz_2020}, tidal effects for ULDM satellites~\cite{Hertzberg_2023} (see chapter~\ref{chpt:paper_3} for more details), density profiles of dwarf galaxies~\cite{Zimmermann_2024}, 21cm signal~\cite{Flitter_2022}, etc.
Note that some of the constraints mentioned in this section become weaker when FDM is allowed to comprise only a fraction of all dark matter (for instance Refs.~\cite{Khmelnitsky_2014, Irsic_2017, Marsh_2019, Bar_2022}). 
Overall, the picture that is being painted, as FDM is confronted with more and more astrophysical observations, is bleak.
It appears that fuzzy dark matter, in the regime where the fuzziness manifests at galactic scales, is ruled out, from comprising of all dark matter. 

\section{Impact of self-interactions}\label{sec:impact_of_SI}

As we saw in section~\ref{sec:KGE_to_GPP} self-interactions of the $\lambda\varphi^4$ kind will lead to an additional $\lambda\varphi^3$ term in the Klein-Gordon equation.  
Due to this, the evolution of the scalar field during the early Universe will be altered. 
Similarly, this additional term will carry over to the Schrödinger-Poisson system in the non-relativistic limit and hence alter structures at the non-linear scales as well. 
In this section, we shall discuss the impact of self-interactions on both the background evolution and summarize the effects on small-scale structure that we discussed in this chapter.

\subsection{Impact on background evolution}

In the presence of repulsive self-interactions (present for moduli fields), the evolution is altered drastically. Here, initially the scalar field will possess an equation of state of $w = -1$ where it will behave like the cosmological constant, i.e. constant density~\cite{Cembranos_2018, Hartman_2021, Lee_2025}. Note that if $\varphi$ is complex, then the initial equation of state will be $w = 1$, i.e. a stiff phase~\cite{Li_2014, Suarez_2017, Foidl_2022}.  
This later transitions to a radiation-like phase where $w = 1/3$, and then finally transitions to a matter-like phase, i.e. $w = 0$, where it behaves like CDM.
Three presence of three different phases in the evolution of the scalar field can alter the expansion history of the Universe depending on the strength of the self-interactions, potentially altering CMB anisotropies, the matter power spectrum and the growth of structure. This was recently explored for the case of a complex light scalar in~\cite{Li_2014, Hartman_2021, Foidl_2022, Yang_2025}. 

On the other hand, for the case of attractive self-interactions, what is usually studied is the axion. In this case, one takes into account the full cosine potential instead of just the quadratic and the quartic term.
It was shown in Ref.~\cite{Cedeno_2017} that this alters the background expansion, where for a smaller decay constant $f_a$, the transition to CDM-like behaviour occurs later and more abruptly.
It will also lead to an enhancement of power at small scales in the matter power spectrum. It is worth noting that a smaller $f_a$ also corresponds to a larger attractive quartic self-coupling since $\lambda = -\left(m_a/f_a\right)^2$. 
Assuming misalignment as the production mechanism, one also needs to consider the higher order terms of the cosine potential if the initial displacement of the axion field is large, i.e. it will `experience' more of the full potential. 
Such models are called extreme axions and can potentially evade some stringent constraints imposed by Lyman-$\alpha$ power spectrum~\cite{Zhang_2017, Leong_2019, Winch_2023}.

\subsubsection{Impact on scalar field configurations and non-linear structure}

As we have seen, the additional $\lambda\varphi^3$ term in the Klein-Gordon equation translates to an additional $\lambda|\Psi|^2\Psi$ term in the Schrödinger equation in the non-relativistic limit.
The Schrödinger equation now resembles the Gross-Pitaevskii equations in eq.~(\ref{eq:GrossPitaevskii}), used to describe Bose-Einstein condensates~\cite{Pitaevskii_book}, while the Poisson equation remains unchanged.
For the new Gross-Pitaevskii-Poisson (GPP) system, solitonic solutions (similar to those for the SP system) do exist~\cite{Guzman_2006, Chavanis_2011, Guth_2015, Brax_2019} albeit with drastically different masses and radii (as have seen in section~\ref{sec:regimes}). 
This is because, equilibrium solutions now have an additional effect at play: self-interactions, which can provide additional support against gravity (in the form of repulsive self-interactions) or aid gravity in further collapse (in the form of attractive self-interactions). 
We looked at these solutions in great detail in this chapter.

For the case of repulsive self-interactions, i.e. $\lambda > 0$, along with the quantum pressure due to the uncertainty principle, self-interactions can also provide support against gravitational collapse~\cite{Khlopov_1985, Colpi_1986, Chavanis_2011, Li_2014, Chavanis_2021}. 
This introduces a new length scale in the system, $R_{TF}\propto\sqrt{\lambda/m^4}$ (see eq.~(\ref{eq:TF_radius_at_home})), which for large enough $\lambda$ can be greater than $\lambda_{dB}$ (also called the Thomas-Fermi limit~\cite{Dawoodbhoy_2021}), significantly deviating from the mass and radius of solitons for the $\lambda = 0$ case. For a fixed $\lambda > 0$, allowed values of mass and radius are shown in figure~\ref{fig:m_vs_r_scaled} by the green curve.

For the case of attractive self-interactions, things are more complicated. 
First, let us note that the presence of attractive self-interactions aid self-gravity in the attempt to collapse the solution. 
Hence the effect on stationary solutions is that the soliton is a smaller and a denser object compared to case of no self-interactions. 
However, including just the quartic self-coupling term $\lambda\varphi^4$ (with $\lambda < 0$) in eq.~(\ref{eq:axion_potential}) without any higher order terms leads to the following cases (as we saw in section~\ref{sec:alternative_regimes}): (a) Soliton solutions exist up to a certain maximum value of soliton mass, (b) soliton solutions with too large a mass do not exist and hence cannot form. 

For case (a), every fixed value of soliton mass corresponds to two solutions with different radius values. The solution with a larger radius value is stable (also called a dilute soliton), while the solution with the smaller radius value is unstable under small perturbations~\cite{Chavanis_2011, Schiappacasse_2018}.
Both the branches are shown in figure~\ref{fig:m_vs_r_scaled} by the red curve. 
For case (b), if a soliton accretes more mass than the allowed maximum mass, it will lead to collapse \cite{Levkov_2017, Chavanis_2018, Mocz_2023}.
In the case where higher order terms from the cosine potential in eq.~(\ref{eq:axion_potential}) are included, one can find stable solitonic solutions even here, albeit for very small radii~\cite{Braaten_2016, Eby_2016b, Schiappacasse_2018}. 
This is because the higher order repulsive terms provide support against the collapse due to the attractive term. 
These are called dense solitons~\cite{Chavanis_2018, Mocz_2023}. 
Thus, including self-interactions can lead to a rich diversity of configurations that deviate significantly form $\lambda = 0$ case.
However, note that when considering attractive self-interactions, we shall not consider the effects of higher order and restrict ourselves to the case of quartic couplings.

Additionally, simulations with $m\sim 10^{-22}\ \text{eV}$ and $\lambda > 0$ show that one can still form ULDM halos with flat density solitonic solutions at the centre surrounded by CDM-like envelopes~\cite{Dawoodbhoy_2021, Foidl_2023, GalazoGarcia_2024}, although the soliton will be larger with a smaller central density.
Similar to the case repulsive self-interactions, simulations with $m\sim 10^{-22}\ \text{eV}$ and $\lambda < 0$~\cite{Mocz_2023, Painter_2024} also feature a flat-density core (smaller and with higher central density than the $\lambda = 0$ case) at the centre, which is surrounded by a CDM-like halo. 
Hence, while similar to the case of no self-interactions, the objects formed in the presence of self-interactions vary a lot in their masses, radii and stability.
Similar to FDM solitons at the centres of DM halos undergoing oscillations and random walks due to interference effects~\cite{Li_2021}, Refs.~\cite{Indjin_2023, Salasnich_2025, Capanelli_2025} explore the impact of self-interactions on such wave interference effects. 
On other hand, Refs.~\cite{Chavanis_2019, Chavanis_2020, Padilla_2021, Chavanis_SH_relation_2019} study the effects of a supermassive black hole at the centre on the mass and radius of the solitonic cores as well as on the core mass-halo mass relation.


\subsection{Probing self-interactions using observations}\label{sec:probing_SI}

The discussion in section~\ref{sec:fuzzy_dark_matter} makes it clear that ULDM with $m\sim 10^{-22}\ \text{eV}$ and $\lambda = 0$ has become increasingly unviable due to stringent constraints. Therefore, there has been a flurry of interest in exploring ULDM with $\lambda \neq 0$, some examples of which we saw in the previous section. 
It is then interesting to ask, can galactic scale observations be used to probe the effect, strength and sign of self-interactions? 


Let us try to understand the magnitude of $\lambda$ that can be probed at astrophysical scales.
We know from previous arguments that the $\mathcal{O}(\text{kpc})$ length scales can be probed by a scalar with $m\sim 10^{-22}\ \text{eV}$. 
Fixing $m= 10^{-22}\ \text{eV}$, let us compare the gradient and self-interaction term in the Gross-Pitaevskii equation in eq.~(\ref{eq:GrossPitaevskii}). 
Now, assuming the typical central region of a galactic halo of size $\sim\text{kpc}$ holds a mass of $\sim 10^8\ M_\odot$, the value of $\lambda$ required to keep the self-interaction term relevant is, 
\begin{equation}
    \lambda \sim 10^{-92}\left(\frac{10^8\ \text{M}_{\odot}}{M}\right)\left(\frac{L}{\text{kpc}}\right)\ .
\end{equation}
This value appears to be extremely small. 
However, it should not be a surprise since it is known that small self-interactions can have non-negligible effects, owing the extremely small particle mass and extremely large occupation number~\cite{Colpi_1986, Guth_2015, Chavanis_2011}. 
Indeed, it was found from simulations~\cite{Glennon_2022} of solitons with self-interactions orbiting in a halo potential that $\lambda\sim \pm 10^{-92}$ lead to $\mathcal{O}(1)$ changes to the lifetime of the soliton under the tidal effects of the halo. 

On the other hand, it is worth noting that for axion DM with $m\sim 10^{-22}\ \text{eV}$ satisfying the current dark matter relic density (see eq.~(\ref{eq:axion_relic})), one can simply use $\lambda = -(m_a/f_a)^2$ with $f_a \sim 10^{17}\ \text{GeV}$, to obtain $\lambda \sim -10^{-96}$, which is 4-5 orders of magnitude smaller than what is probed by galactic scale observations. 
Practically, the effect of such a small value of self-interactions is negligible given the observables of interest~\cite{Chavanis_2021} (also see appendix of~\cite{Bar_SMBH_2019} for an example).

An important example of probing self-interactions of SFDM is using rotation curves of dwarf galaxies~\cite{Bernal_2017, Craciun_2020, Harko_2022, Dawoodbhoy_2021}, where they found that rotation curves can be fit better by SFDM in the TF regime (i.e. large repulsive self-interactions) compared to FDM with no self-interactions. 
In particular, Refs.~\cite{Delgado_2022, Indjin_2025} found that a single value of $m\sim 2.2\times 10^{-22}\ \text{eV}$ and $\lambda\sim 2\times 10^{-90}$ can explain the rotation curves of 17 dark matter dominated galaxies. 


At cosmological scales as well, one can probe very small self-interactions. For instance~\cite{Li_2014, Foidl_2022} found that for a complex scalar field with self-interactions in the TF regime, to ensure that BBN is not disturbed while also imposing that the radiation phase (of the scalar field) does not last too long to spoil CMB anisotropies, for a SFDM mass
$m = 5\times 10^{-21}\ \text{eV}$, repulsive self-interactions must lie in the range $10^{-84} \lesssim \lambda \lesssim 10^{-82.5}$. 
On the other hand, from CMB anisotropies, Ref.~\cite{Cembranos_2018} found that for $m > 10^{-24}\ \text{eV}$, repulsive self-interactions must satisfy $\log_{10}{\lambda} < -91.86 + 4\log_{10}{m/10^{-22}\ \text{eV}}$ for a real scalar field.
Similarly, for axions a large initial displacement of the scalar field will necessarily have to take into account the effect of the quartic term in the cosine potential. 
This, depending on the value of the axion decay constant $f_a$ can evade the stringent constraints on $m\sim 10^{-22}\ \text{eV}$ set by Lyman~$\alpha$ power spectrum~\cite{Desjacques_2018, Leong_2019, Winch_2023}. Such models are called extreme axions.

There has been a growing body of recent work that attempts to constraint self-interactions of ULDM using various other observations as well like gravitational waves~\cite{Dev_2017, Kadota_2024, Boudon_2023, Boudon_2023b}, black hole superradiance~\cite{Baryakhtar_2021, Xie_2025}, neutron stars~\cite{Grippa_2024}, accretion of SFDM into charged black holes~\cite{Ravanal_2023, Ravanal_2024}, etc.

\justifying
\chapter{Constraining ULDM using observational limits on galactic central mass}\label{chpt:paper_1}

\textbf{Based on:} \\S.~Chakrabarti, \textbf{B.~Dave}, K.~Dutta and G.~Goswami, \emph{Constraints on the mass and self-coupling of ultra-light scalar field dark matter using observational limits on galactic central mass}, \href{https://doi.org/10.1088/1475-7516/2022/09/074}{\emph{J. Cosmol. Astropart. Phys.} \textbf{09} (2022) 074} [\href{https://arxiv.org/abs/2202.11081}{arXiv:2202.11081 [astro-ph.CO]}].

\noindent Most of the text in this chapter is taken from our published work, Ref.~\cite{Chakrabarti_2022}.

\section{Motivation}\label{sec:smbh_motivation}

Recently there have been studies \cite{Bar_SMBH_2019, Davies_2020, Pantig_2022, Della_Monica_2023} relating measurements of the dynamical environment of supermassive black holes (SMBH) at galactic centres and ULDM.
Particularly, in \cite{Davies_2020}, the authors studied a novel method to impose observational constraints on the mass $m$ of fuzzy dark matter which works at small ($\sim pc$) scales. 
Using the M87 halo as an example, by requiring that the soliton core obeys a soliton-halo mass relation \cite{Schive_PRL_2014}, is not accreted by the SMBH over cosmological timescales, as well as satisfies an observational upper bound on the amount of mass contained within a region of radius $10\ \text{pc}$, the authors found that mass of the ULDM particle must satisfy $m < 10^{-22.12}\ \text{eV}$. 

The method in Ref.~\cite{Davies_2020} was based on the following considerations:
\begin{itemize}
 \item The core of the DM halo of a galaxy is mathematically modelled by spatially localised, spherically symmetric, stationary solitonic solutions of the Schr\"{o}dinger-Poisson equations.
 \item The existence of an SMBH at the centre of the galaxy can cause the scalar field to accrete into the black hole over a time scale proportional to $m^{-6}$, where, $m$ is the mass of the scalar field; thus, for too large masses, the accretion time will be too small, providing an upper limit on the allowed values of $m$.
 \item In addition, for some galaxies, the upper limit on the amount of DM enclosed within a spherical region of some fixed radius $r_*$ from the centre is observationally known. Certain values of $m$ will produce core profiles which are consistent with these observational results while others will not be. This can provide another upper limit on the allowed values of $m$ which could be lower than the one obtained from accretion time considerations.
\end{itemize}

Since the size and mass of the dark matter core are inversely proportional to $m$, the amount of dark matter mass contained within central regions of the galaxy, which is observationally constrained, will depend on $m$ - thus, observations will constrain $m$. If the scalar field dark matter has self-interactions, for repulsive self-interactions, the core is bigger, while for attractive self-interactions, the core is smaller. This implies that the constraints which the observations of the total mass within central regions of galaxies impose on $m$ will depend on the self-coupling $\lambda$. 
Thus, it is useful to ask whether the approach presented in \cite{Davies_2020} can be extended to impose observational constraints when the scalar field describing DM has non-negligible scalar self-interactions. In other words, can we find the region of $\lambda - m$ plane which gets observationally excluded by such an analysis? 

In section~\ref{sec:GPP_with_SMBH}, we first obtain soliton solutions in the presence of an external potential that characterises the presence of a SMBH at the centre of a halo as a Newtonian point mass.
In section~\ref{sec:m&lambda}, we then apply our procedure to obtain the relevant constraints in the $\lambda-m$ plane and in section~\ref{sec:smbh_results} we discuss the results and finally conclude in section~\ref{sec:smbh_conclusion}.

\section{Gross-Pitaevskii-Poisson equations in the presence of a SMBH}\label{sec:GPP_with_SMBH}

In eqs.~(\ref{eq:GrossPitaevskii}) and (\ref{eq:Poisson}), the only free (i.e. adjustable) parameters of the underlying fundamental theory are $m$ and $\lambda$, which we wish to constrain. In addition, in the presence of a central black hole, the mass of the black hole $M_\bullet$ also acts as a relevant parameter.
Since the solutions we study will serve as models of cores of DM halos, the object specific parameters which are important are:
the mass of the DM halo ($M_{\text{halo}}$), the mass of its core $M$. Since the core of the DM halo will be modelled by a ``soliton" solution (as defined in the last paragraph), $M$ will also be sometimes called the mass of the soliton.
The presence of a black hole will be taken care of by replacing the $m \Phi  \Psi$ term in Gross-Pitaevskii equation (\ref{eq:GrossPitaevskii}) by $m \left( \Phi - \frac{G M_\bullet}{r} \right) \Psi$, where, $M_\bullet$ is the mass of the black hole. 

In order to demonstrate the method, we use the example of M87 (also known as NGC4486), a supergiant elliptical galaxy in the Virgo cluster. It is known to contain a SMBH of mass $ M_{\bullet} = 6.5 (\pm 0.7) \times 10^{9} M_{\odot}$ at its centre. The mass of this SMBH was recently determined by its imaging by the Event Horizon Telescope \cite{EventHorizon_2019}. Following \cite{Hui_2017} and \cite{Davies_2020}, we shall assume that the halo mass of M87 galaxy is $M_{\text{halo}} = 2 \times 10^{14} M_{\odot}$. Similarly, we shall assume that the region within $10 ~{\rm pc}$ of the centre of M87 galactic halo contains a soliton mass not more than $10^{9} M_{\odot}$.

We proceed with the dimensionless variables that were introduced in section~\ref{sec:stationary_states}. In addition, we also introduce $\hat{\alpha}$ to parametrise the ``strength'' of the black hole,
\begin{equation} \label{eq:alpha}
{\hat \alpha} = \frac{GM_\bullet m}{\hbar c} = 4.87 \times 10^{-4} ~ \left(\frac{M_\bullet}{6.5 \times 10^9 ~ M_{\odot}} \right) \left(\frac{m}{10^{-22} ~ {\rm eV} } \right) \; .
\end{equation} 
The dimensionless GPP system now becomes, 
\begin{eqnarray} 
 \frac{1}{2} {\hat \nabla}^2 {\hat \phi} &=& {\hat \Phi}{\hat \phi} - {\hat \gamma} {\hat \phi} - \frac{\hat \alpha}{\hat r} {\hat \phi} + 2 {\hat \lambda} {\hat \phi}^3 \; , \label{eq:GP_dimless_SMBH}\\
{\hat \nabla}^2 {\hat \Phi} &=&{\hat \phi}^2\; , \label{eq:P_dimless_SMBH}
\end{eqnarray}

We impose the same boundary conditions as we did in section~\ref{sec:stationary_states}, i.e. ${\hat \phi}({\hat r}=\infty) = 0,~ {\hat \phi} ' ({\hat r} = 0) = 0, {\hat \Phi}({\hat r} = 0 ) = 0,~ {\hat \Phi} ' ({\hat r} = 0) = 0$.
At this stage, the parameters $\hat \alpha$ and ${\hat \lambda}$ are free and their values are to be chosen before we can proceed to solve the system. This is not true about $\hat \gamma$ $-$ for a chosen $\hat \alpha$ and ${\hat \lambda}$, the value of the parameter $\hat \gamma$ is determined by the boundary conditions and the requirement of having no nodes in the desired solution. 

\begin{figure}[h!]
  \includegraphics[width = 0.92\textwidth]{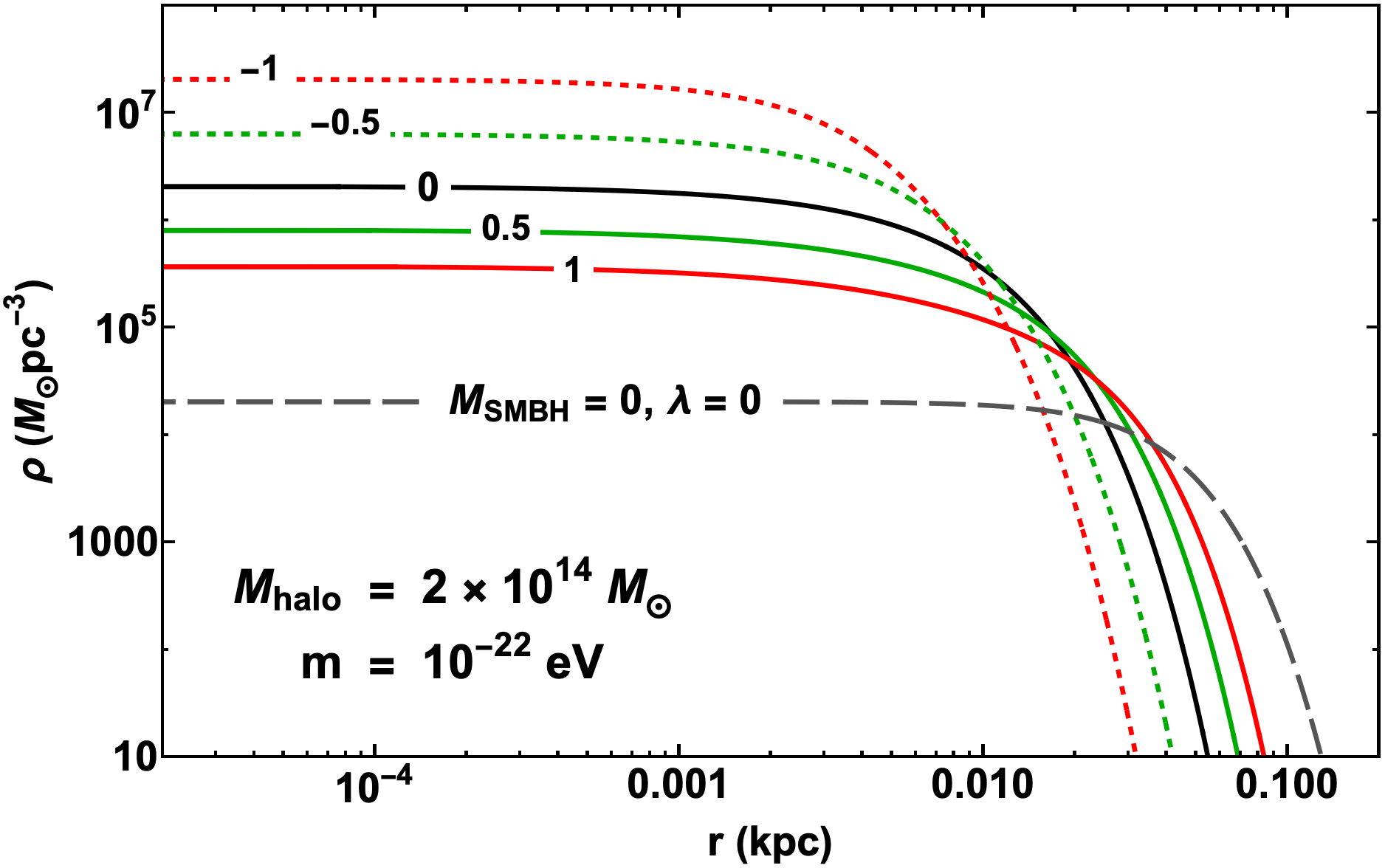}
  \caption[Effects of self-interaction as well as a SMBH on soliton density profiles]{
Density profiles for the core of M87 halo for fixed value of Dark Matter mass, $m$, and various values of self-coupling $\lambda$. The black dashed curve at the bottom corresponds to a halo without a SMBH at the centre while assuming no self-interactions of DM. The other curves show the density profile but in the presence of a SMBH at the centre. The label on each curve is the value of dimensionless $\hat{\lambda}_\text{ini}$ (defined in eq.~(\ref{eq:dimensions2})). 
The various ${\hat \lambda}_\text{ini}$ values correspond to the following  $\lambda$ values (top to bottom): $-6.2 \times 10^{-95} $, $-5.6 \times 10^{-95}$, $0$, $1.6 \times 10^{-94}$, $4.6 \times 10^{-94}$. 
Note that the top and bottom red curves are solely for an illustrative purpose, since we do not consider the regime where self-interactions are more important than self-gravity as discussed in section~\ref{sec:typical}.
It can be seen that repulsive self-interactions expand the soliton while attractive self-interactions compress it.} 
  \label{fig:densityprofile}
\end{figure}

Thus, once $\hat \alpha$ and ${\hat \lambda}$ are specified, the system of equations can be readily solved by using the familiar shooting method, see e.g. \cite{Giordano_Nakanishi} and \cite{Davies_2020}. 
The density profiles obtained from such solutions are shown in figure \ref{fig:densityprofile}. 
In section~\ref{sec:scaling}, we discuss which values of $\hat \alpha$ and ${\hat \lambda}$ need to be used in the solution of eqs. (\ref{eq:GP_dimless_SMBH}) and (\ref{eq:P_dimless_SMBH}) in order to obtain realistic solitonic solutions such as those shown in figure \ref{fig:densityprofile}.

Given any solution of the system, one can calculate the dimensionless mass of a spherically symmetric soliton i.e. $\hat{M}$ using eq.~(\ref{eq:mass_dimless}).

In section~\ref{sec:theoretical_solitons}, we saw that typical size of solitons with where the central density is initialized to be $\hat{\phi}(0) = 1$, will be $\mathcal{O}(m^{-1})$. 
Similarly, in the presence of central BH: If ${\hat \alpha} \ll 1$, then, the ${\hat \Phi}{\hat \phi}$ term will still dominate over 
$\frac{\hat \alpha}{\hat r} {\hat \phi}$ term in Gross-Pitaevskii equation and the size $\hat L$ will still be ${\cal O}(1)$. 
When the $\frac{\hat \alpha}{\hat r} {\hat \phi}$ term dominates, the size will be ${\hat L} \sim 1 / {\hat \alpha}$.
But, this happens only when ${\hat \alpha}$ is at least ${\cal O}(1)$, so, the dimensionless size $\hat L$ will still remain ${\cal O}(1)$.

Using eq.~(\ref{eq:mass_dimless}) and a similar reasoning it is possible to convince oneself that the dimensionless mass $\hat M$ will be ${\cal O}(1)$. This corresponds to a soliton mass of the order of $10^{12} ~M_{\odot}$ which is much larger than the corresponding value for e.g. M87 i.e. $M \lesssim 10^9 M_{\odot}$.

\subsection{The amount of scaling}

The scaling symmetry that we discussed in section~\ref{sec:scaling_symmetry} remains intact in the presence of the extra potential term for the point-mass SMBH, if one scales $\hat{\alpha}$ as 
\begin{equation}\label{eq:alpha_scaling}
    \hat{\alpha} \rightarrow \frac{1}{s}\hat{\alpha}\ .
\end{equation}

For theoretical solitons, having fixed the values of $\hat \alpha$ and ${\hat \lambda}$, once the solutions ${\hat \phi}({\hat r})$ and ${\hat \Phi}({\hat r})$ are obtained, one can readily find the dimensionless soliton mass, $\hat M$, using eq.~(\ref{eq:mass_dimless}).
As we saw in the previous section as well as in section~\ref{sec:theoretical_solitons}, for the theoretical soliton, for typical realistic values of ${\hat \alpha}$ and ${\hat \lambda}$ of interest to us, we find that the mass of the soliton ${\hat M}$ is typically ${\cal O}(1)$ which is far bigger than the typically observed values (since ${\hat \phi}(0) = 1$ corresponds to too large a central density).

One thus needs to do the following:
\begin{itemize}
 \item start with some initial values ${\hat \alpha} = {\hat \alpha}_\text{ini}$ and ${\hat \lambda} = {\hat \lambda}_\text{ini}$, while solving eqs.~(\ref{eq:GP_dimless_SMBH}) and (\ref{eq:P_dimless_SMBH}) and obtain the theoretical solitons,
 \item use the scaling transformations to turn theoretical solitons into real solitons whose parameters take realistic values. 

\end{itemize}
In other words, the scaling transformation parameter ``$s$'' needs to be chosen to ensure that $\hat M$ of the real soliton as well as the corresponding $\hat \alpha$ are equal to the observed values. The way to do this is explained in detail in 
appendices \ref{sec:scaling}
and 
\ref{sec:parameter}.
Thus, the correct value of scaling parameter $s$ will be such that 
\begin{equation} \label{eq:s}
s = \frac{{\hat M}_\text{ini}}{{\hat M}_{\text{emp}}} = \frac{{\hat \alpha}_\text{ini}}{{\hat \alpha}_{\text{emp}}} \; ,
\end{equation}
where, ${\hat M}_\text{ini}$ and ${\hat \alpha}_\text{ini}$ are defined in the appendix \ref{sec:parameter}.
Since the soliton size scales as $s$, the ratio of the physical size $L$ of the actual soliton and $m^{-1}$, the size of ``theoretical" soliton, will also be related by
\begin{equation} \label{eq:s2}
s = \frac{L}{m^{-1}} \; .
\end{equation}
It is also worth noting that since ${\hat M}_\text{ini}$ is large compared to ${\hat M}_{\text{emp}}$, we must have $s \gg 1$ so that ${\hat \alpha}_{\text{trans}} \ll {\hat \alpha}_\text{ini}$.

\subsection{Typical values of interest}
\label{sec:typical}

At this stage, it is useful to compare the strength of gravity and scalar self-interactions.
Using Poisson's equation (\ref{eq:Poisson}), it is easy to see that, in Gross-Pitaevskii equation (\ref{eq:GrossPitaevskii}), the ratio of gravitational term $m \Phi \Psi$ and the self-interaction term $\frac{\lambda}{8m^3} |\Psi|^2 \Psi$ will be 
\begin{equation}
 \frac{\rm gravitation}{\rm scalar~ interaction} = \frac{4}{\lambda} \left( \frac{m}{M_{pl}} \right)^2 \left( \frac{L}{m^{-1}} \right)^2 \; ,
\end{equation}
i.e. as the size of the object increases, gravity becomes far more important than self-interactions.
This is expected since the quartic self-interactions of the scalar field are contact interactions while gravity is a long range force.
Now using eqs.~(\ref{eq:s2}) and (\ref{eq:lambda}), this ratio is also the same as
\begin{equation}
 \frac{\rm gravitation}{\rm scalar~ interaction} = \frac{1}{2 {\hat \lambda}_\text{ini}} \; ,
\end{equation}
so that if gravity always dominates over scalar self-interactions then we should have 
\begin{equation}
  - \frac{1}{2} \le {\hat \lambda}_\text{ini} \le \frac{1}{2} \; . 
\end{equation}
In this work, we will not focus on the case in which self-interactions dominate over gravity, hence, the range of ${\hat \lambda}_\text{ini}$ of interest to us is given by the above equation.
Needless to say, if gravity is negligibly small, we go from the domain of Gross-Pitaevskii-Poisson equations to the domain of non-linear Schr\"{o}dinger equation.
Since we are going to be interested in Ultra Light Dark Matter, we work with the following range of $m$
\begin{equation}
10^{-24}~{\rm eV} \le m \le 10^{-21}~{\rm eV} \; . 
\end{equation}
As stated, for the purpose of this work, we restrict our attention to the case of M87 galaxy, so we shall keep the parameter $M_{\text{halo}}$ to the value $2 \times 10^{14} M_{\odot}$.

\section{Imposing constraints in $m - \lambda$ plane}
\label{sec:constraints}

Having developed the machinery to probe the self-coupling, we now go back to the method proposed by Davies and Mocz \cite{Davies_2020} to impose constraints on the mass of the fuzzy dark matter and find out what it can teach us about the joint constraints on the mass $m$ and the scalar self-coupling $\lambda$.

As discussed earlier, the first step involved in the method studied by \cite{Davies_2020} is to model the soliton using stationary, spherically symmetric solutions of Schr\"{o}dinger-Poisson equations. For the case of our interest, this step involved modelling the soliton using similar solutions of Gross-Pitaevskii-Poisson equations. The machinery outlined in the previous section was developed to obtain solitonic solutions of Gross-Pitaevskii-Poisson equations for any value of $m$, ${\hat \lambda}_\text{ini}$ and for a galaxy with halo mass $M_{\text{halo}}$.

The next step involves the realisation that the mass of the ultra light dark matter will be constrained simply because of considerations based on the accretion of the scalar field into the central black hole. Again, following \cite{Davies_2020}, we shall simply remind the reader that in the presence of a SMBH at the centre of the scalar field soliton, the scalar field will accrete into the black hole. 
According to the calculations of \cite{Barranco_2011} and \cite{Barranco_2017}, the corresponding accretion time  will be proportional to $t_{\text{acc}} \propto M_{\bullet}^{-5} m^{-6}$ and hence, for a given black hole mass there is always a value of ULDM mass, $m$ such that, for all masses larger than that, the accretion time is too small compared to cosmological time scales. This consideration alone imposes a constraint on the mass of the DM - specifically, it will provide an upper limit on $m$. 
The calculations of \cite{Barranco_2011} and \cite{Barranco_2017} do not deal with a scalar field with self interactions, but, since this aspect is not the main point of our work, we shall demonstrate all our results by simply assuming that the accretion based constraints hold good even when the scalar field has non-negligible scalar self-interactions. 

The third step involves calculating the DM mass within some central region of a galactic halo core using the solutions we already have. Needless to say, this mass will depend on the free parameters such as $m$ and $\lambda$ and comparison with observations will then exclude some regions in the $\lambda - m$ parameter space. 

Keeping this in mind, we now turn to the calculation of the DM mass within a central region of some observable size.

\subsection{The amount of centrally concentrated dark matter for a soliton core}
\label{sec:M&D}

Using the solitonic solutions we have obtained (see figure~\ref{fig:densityprofile}), we need to calculate the DM mass, $M_{r < r_*}$, within a spherical region of some fixed radius, say $r_{*}$, from the centre of the galactic halo. Note that $M_{r < r_*}$ could be calculated irrespective of whether there is SMBH at the centre of the galaxy or not.
The dimensionless variable specifying the radius of the region will be
 \begin{equation} \label{eq:rhatstd}
 {\hat r}_* = \frac{r_*}{\hbar / mc} \; .
\end{equation} 

As we saw in section~\ref{sec:GPP_with_SMBH}, it is assumed that for M87, a soliton mass of only $10^9~M_{\odot}$ can be contained within a distance of 10 pc of the SMBH. In other words, $r_*$ is taken to be 10 pc - we shall continue to work with this number.
The discussions in the last section suggest that there must exist a scaling parameter $s$ which takes us from the theoretical soliton to the real soliton. 
When we use the empirical relations to find the value of $s$, it is found that the value of $s$ in the presence of black hole is smaller than the value of $s$ in the absence of black hole (though both of them are larger than 1) i.e. the presence of a central black hole squeezes the soliton, as was found in \cite{Davies_2020} and also can be seen in figure~\ref{fig:densityprofile}.

In section~\ref{sec:GPP_with_SMBH}, we had argued that the unscaled size of the soliton i.e. the size of theoretical soliton is ${\cal O}(m^{-1})$. Note that the size of the soliton could be defined more precisely e.g. one could define the distance within which 95\% of the mass of the soliton exists to be the size of the soliton. 
Using any such definition of the size of the soliton, if the unscaled dimensionless size of the soliton is ${\hat r}_{\rm sol}$, then, eq.~(\ref{eq:s2}) suggests that the dimensionless size of the soliton after scaling will be $ s {\hat r}_{\rm sol}$. 

Now, there are two possibilities
\begin{itemize}
 \item $ s {\hat r}_{\rm sol} > {\hat r}_* $ i.e. only a fraction of the soliton is inside the region of interest. In this case, the mass within this region should be
\begin{equation} \label{eq:M_limited}
M_{r < r_*} = \left( \frac{\hbar c}{G m} \right) \int_0^{{\hat r}_*} d{\hat r} ~ {\hat r}^2 {\hat \phi}^2 \; .
\end{equation}
where, it is to be noted that the quantities used inside the integral are scaled.
 \item $s {\hat r}_{\rm sol} < {\hat r}_* $ i.e. the entire soliton is inside the region of interest. In this case, the mass within this region is
\begin{equation} \label{eq:M_full}
M_{r < r_*} = \left( \frac{\hbar c}{G m} \right) \int_0^{s {\hat r}_{\rm sol}} d{\hat r} ~ {\hat r}^2 {\hat \phi}^2 \; ,
\end{equation}
where, again the quantities used inside the integral are scaled. Since the entire soliton is well inside the region of interest, at $s {\hat r}_{\rm sol}$ the field takes up very small values and so if we replace the upper limit in the integral in the last expression by $\infty$, the error in evaluating it is expected to be very small. 
In this approximation, $M_{r < r_*}$ will be the same as the total mass of the soliton itself (recall the discussion just before eq.~(\ref{eq:mass_dimless})).
\end{itemize} 
This is how the quantity $M_{r < r_*}$ can be evaluated by solving eqs.~(\ref{eq:GP_dimless_SMBH}) and (\ref{eq:P_dimless_SMBH}), using the machinery developed in the last section and chapter~\ref{chpt:numerical_solutions}. Let us suppose that for a given galaxy, the halo mass $M_{\text{halo}}$ and black hole mass $M_{\bullet}$ are known, then, in principle, we could find out $M_{r < r_*}$ for a particular choice of $m$ and $\lambda$. 

\begin{figure}[h!]
\centering
  \includegraphics[width = 0.8\textwidth]{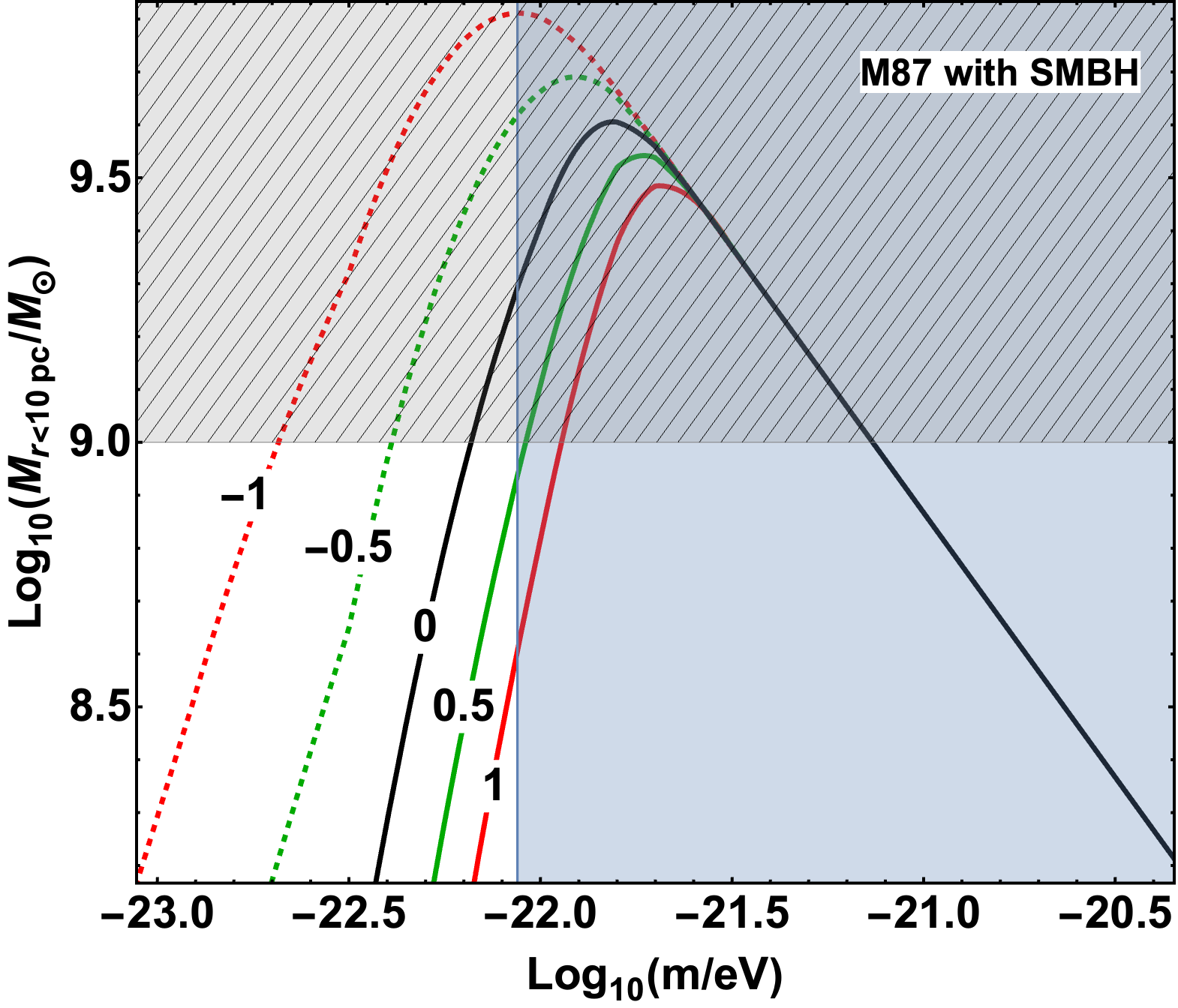}
  \caption[Mass contained within a region of radius $10\ \text{pc}$ as a function of $\lambda$ and $m$]{The curves represent the mass within a distance of 10 pc of the centre of M87 galaxy ($M_{\text{halo}} = 2\times 10^{14} ~ M_{\odot}$) against values of mass of Ultra Light Dark Matter i.e. $m$ and for a few different values of ${\hat \lambda}_\text{ini}$ in the presence of a central SMBH at the centre of the soliton. The shaded and marked regions are excluded from observations, see text for details. In figure~\ref{M87WithBH},  the solid curve labelled ``0" corresponds to the results of \cite{Davies_2020}. Recall that the top and bottom red curves are again shown for an illustrative purpose, since we do not consider the regime where self-interactions are more important than self-gravity as discussed in section~\ref{sec:typical}.}
  \label{M87WithBH}
\end{figure}

To begin with, we proceed in the following manner - we fix ${\hat \lambda}_\text{ini}$ and $m$, obtain the density profile (such as in figure~\ref{fig:densityprofile}) and find $M_{r < r_*}$. Then, we repeat this process for different values of $m$. It is in this manner that each curve in figure~\ref{M87WithBH} has been obtained. It is worth noting that though all points on a given curve in figure~\ref{M87WithBH} have the same ${\hat \lambda}_\text{ini}$, each point on a given curve in figure~\ref{M87WithBH} corresponds to a different value of $\lambda$ since the corresponding value of $m$ is different (see eq.~(\ref{eq:dimensions2})).
Towards the end of appendix~\ref{sec:parameter}, we note that each curve in figure~\ref{fig:densityprofile} has different value of the scaling parameter $s$. 
In comparison, in figure~\ref{M87WithBH}, the value of the scaling parameter $s$ is not only different for different curves, it is also different for different points on the same curve, in this context, see appendix~\ref{sec:m&lambda} for more details.

Thus, for any given DM halo (i.e. fixed $M_{\text{halo}}$ and $M_{\bullet}$), one can obtain a plot such as figure \ref{M87WithBH}.
We have explained the origin of the shapes of various curves in figure~\ref{M87WithBH} in appendix~\ref{sec:Shape1}. 
The only noteworthy facts at this stage are (a) for any object (such as M87) and for any choice of ${\hat \lambda}_\text{ini}$, one can obtain a plot of 
$M_{r < r_*}$ against $m$, (b) due to accretion time considerations and due to observational limits on mass within $r_*$ of the centre of the halo, there will be regions of these plots which are observationally disallowed,
(c) the combinations of $m$ and ${\hat \lambda}_\text{ini}$ which correspond to the portion of the curves that lie within either of the shaded regions will also get observationally ruled out.

\section{Results}\label{sec:smbh_results}

In figure~\ref{M87WithBH}, note that $m$ values in the shaded region (i.e. those to the right of the vertical line) will be excluded based on accretion time considerations. Similarly, the values of $M_{r < r_*}$ above the horizontal line, which are shaded by slanted lines, are ruled out as the horizontal line represents the observationally inferred limit on the mass within the central 10 pc region of the centre of the galaxy. This rules out the combinations of $m$ and ${\hat \lambda}_\text{ini}$ which correspond to the portion of the curves which lie within either of these regions.
It can be seen that the upper limit on allowed $m$ based on central mass considerations is smaller than the upper limit on $m$ obtained from accretion time considerations. This corresponds to the row with first entry ``0.0" in table~\ref{table:constraints}. 

It is easy to see from figure~\ref{M87WithBH} as well as table~\ref{table:constraints} that when the scalar self-interactions are allowed, these constraints change - the sign and strength of ${\hat \lambda}_\text{ini}$ determine the amount of this change.
For sufficiently large positive values of ${\hat \lambda}_\text{ini}$, there may be no lowering of the allowed upper limit on $m$ from that found from accretion considerations. On the other hand, for negative values of ${\hat \lambda}_\text{ini}$, the lowering of the allowed upper limit on $m$ is more than the lowering for the case considered in \cite{Davies_2020} i.e. the upper limits on $m$ become more stringent.
 
\begin{table}\centering
\begin{tabular}{|cr|c|}
\hline
~~~~~~~${\hat \lambda}_\text{ini}$ &~~~~~ & excluded $m$ \\ 
\hline
\hline
 &~~~~~& \\
~~~~~-1.0 &~~~~~ & $ m \geq 10^{-22.68} $ eV\\
~~~~~-0.5 &~~~~~& $ m \geq 10^{-22.38} $ eV \\ 
~~~~~0.0 &~~~~~& $m > 10^{-22.18}$ eV\\
~~~~~0.5 &~~~~~ & $m > 10^{-22.06}$ eV\\
~~~~~1.0 &~~~~~ & $m > 10^{-22.06}$ eV\\
\hline
\end{tabular}
\caption[Excluded values of $m$ for different $\hat{\lambda}$ values]{The excluded values of ULDM mass $m$, obtained by the method outlined, get modified in the presence of scalar field DM self-interactions.}
\label{table:constraints}
\end{table}

\subsection{Allowed regions in $\lambda - m$ plane}

\begin{figure}[h]
\centering
  \includegraphics[width=0.8\linewidth]{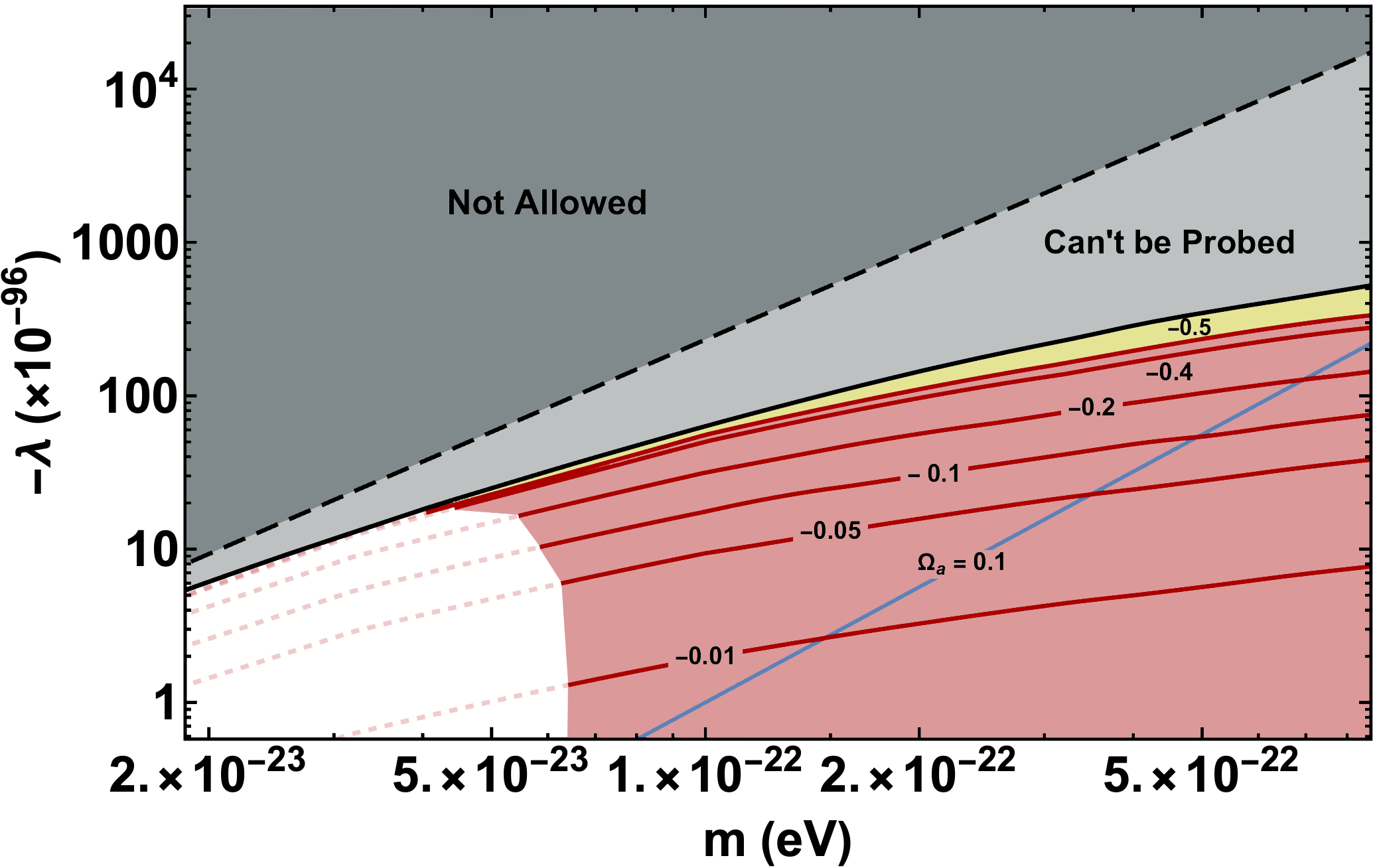}
  \caption[Constraints in $\lambda-m$ plane for attractive self-interactions]{For a given galaxy (such as M87), for every choice of ${\hat \lambda}_\text{ini}$ of interest, we obtain a curve in 
  $\lambda - m$ plane. In this case, we show the results for negative $\lambda$ i.e. attractive self-interactions. The solid blue line marked ``$\Omega_a = 0.1$" corresponds to the situation in which all of DM consists of Ultra-Light Axions. Note that we have only shown results for a few negative values of ${\hat \lambda}_\text{ini}$. The dashed line corresponds to eq.~\ref{lambda_constraint}. See main text for discussion about the various regions such as the inaccessible region in this plot.}
  \label{Excluded_-}
\end{figure}

\begin{figure}[h]
\centering
  \includegraphics[width=0.8\linewidth]{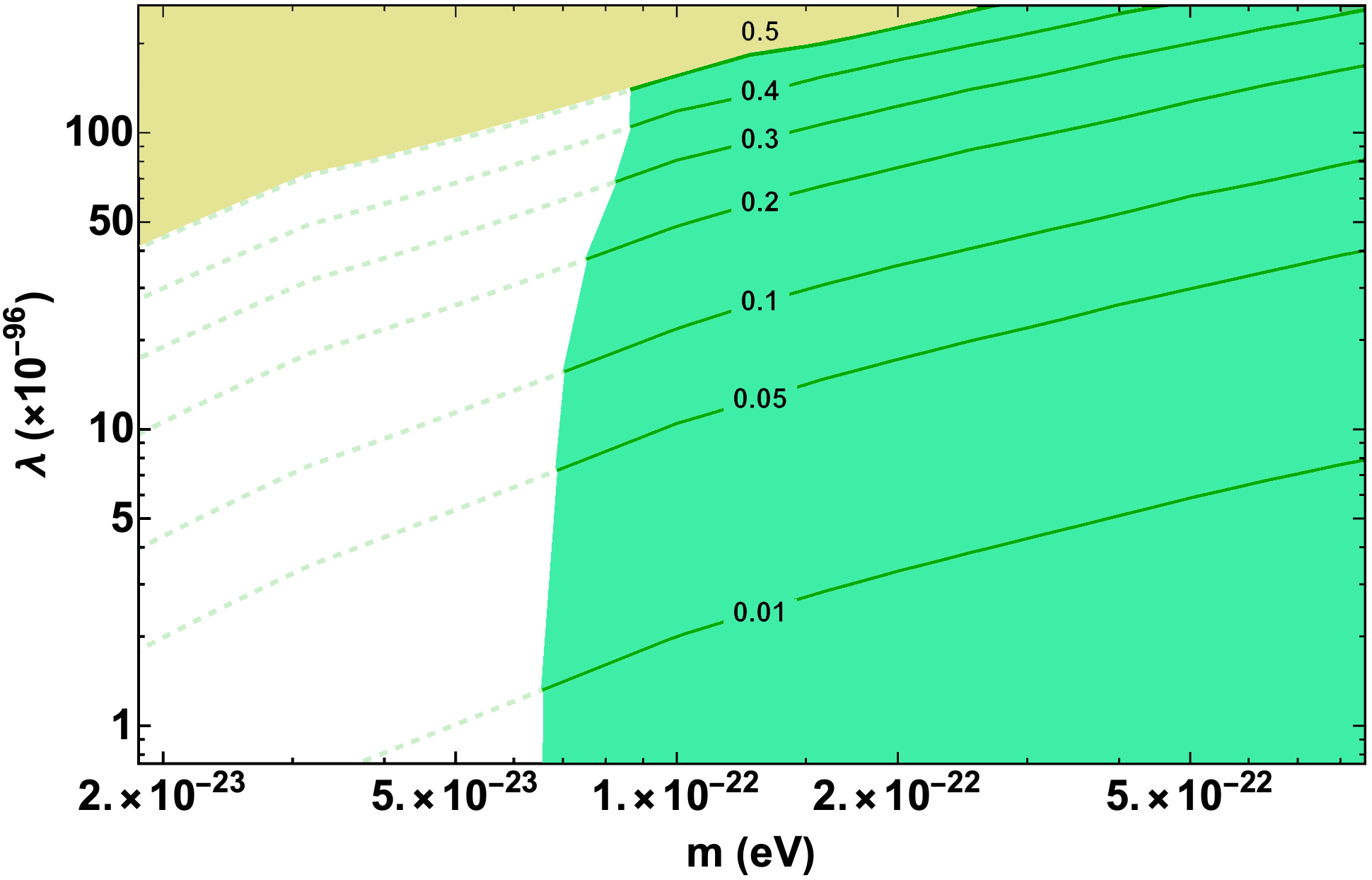}
  \caption[Constraints in $\lambda-m$ plane for repulsive self-interactions]{The same as the previous figure but for positive self-interactions. Note that, for the same $|{\hat \lambda}_\text{ini}|$, the curves in this plot are higher as compared to figure \ref{Excluded_-} and there is no inaccessible region (see the main text for details). }
  \label{Excluded_+}
\label{fig:test}
\end{figure}

We found the impact of ${\cal O}(1)$ values of ${\hat \lambda}_\text{ini}$ on the mass within central region of the soliton core. 
Using the constraints on $m$ and ${\hat \lambda}_\text{ini}$ and using eq.~(\ref{eq:lambda}) and eq.~(\ref{eq:dimensions2}), we can find the constraints in $\lambda - m$ plane, i.e. we can identify which regions of the $\lambda - m$ parameter space get excluded by the data.
Every curve in figure~\ref{M87WithBH} is for a fixed object (i.e. fixed $M_{\text{halo}}$) and for a fixed value of ${\hat \lambda}_\text{ini}$. Corresponding to each such curve in figure~\ref{M87WithBH}, eq.~(\ref{eq:lambda}) and eq.~(\ref{eq:dimensions2}) [or, see eq.~(\ref{eq:lambda0})] suggest that there will be a curve in $\lambda - m$ plane, as is shown in figures~\ref{Excluded_-} and \ref{Excluded_+}. 
Note that figure~\ref{Excluded_-} is obtained for the case of negative (i.e. attractive) scalar self-interactions while figure~\ref{Excluded_+} is obtained for the case of positive (i.e. repulsive) scalar self-interactions. 
Note that the origin of the shapes of the curves in these two  figures is explained in appendix~\ref{sec:m&lambda}.

Since a part of each curve in figure~\ref{M87WithBH} gets excluded, a portion of each of the curves in figures~\ref{Excluded_-} and \ref{Excluded_+} also gets excluded - the excluded portion of each curve is shown as a solid curve while the allowed portion is shown as a dashed curve.
When we consider a whole family of curves corresponding to different values of ${\hat \lambda}_\text{ini}$, we end up excluding regions in $\lambda - m$ plane.

From figures~\ref{Excluded_-} and \ref{Excluded_+}, we can see that a large region in $\lambda-m$ plane gets excluded by the observational constraints on the mass within 10 pc of the centre of M87 galaxy and accretion time constraints. Note that, for $m \sim 10^{-22}$ eV, the typical values of $\lambda$ which we probe are $\lambda \sim - 10^{-96}$, which correspond to a scattering length of $a_s = \frac{\hbar}{mc}\frac{\lambda}{32\pi} \sim 10^{-82}$ m. 
From eq.~\eqref{eq:dimensions2}, it can be seen that, for $m \sim10^{-22}$ eV, $\hat \lambda$ of ${\cal O}(1)$ corresponds to $\lambda$ of the order of $10^{-98}$, but, scaling transformations, with scaling parameter $s \sim 100$ (see figure~\ref{s_m} in appendix~\ref{sec:m&lambda}), through eq.~(\ref{eq:lambda}) causes the probed values of $\lambda$ to be two orders of magnitude higher (see also, eq.~(\ref{eq:lambda0})).

As is argued in appendix~\ref{sec:m&lambda}, for a given $|{\hat \lambda}_\text{ini}|$, the value of the scaling parameter $s$ for positive interactions is always going to be greater than the value of the scaling parameter $s$ for negative interactions. This, along with eq.~(\ref{eq:lambda}) and eq.~(\ref{eq:dimensions2}) [or, see eq.~(\ref{eq:lambda0})], explains why, for the same $|{\hat \lambda}_\text{ini}|$, the curves in figure~\ref{Excluded_+} are higher than the corresponding curves in figure~\ref{Excluded_-}.
Again, as is argued in appendix~\ref{sec:m&lambda}, for negative interactions, for any given object (i.e. DM halo), there will be regions in the $\lambda-m$ plane which can not be probed using the method we have presented. For M87 observations, this region in figure~\ref{Excluded_-} is above the topmost solid black curve (labelled `Can't be Probed') and is shaded gray. However, see discussion below eq.~(\ref{lambda_constraint}) and appendix~\ref{app:max_mass_smbh} to understand the physical intuition behind this region. 

Note that in figure~\ref{Excluded_-}, we only show results assuming that ${\hat \lambda}_\text{ini}$ is within the range $- \frac{1}{2} \le {\hat \lambda}_\text{ini} < 0 $ while in figure~\ref{Excluded_+}, we show results assuming that ${\hat \lambda}_\text{ini}$ is within the range $ \frac{1}{2} \ge {\hat \lambda}_\text{ini} > 0 $. 
As we saw in section~\ref{sec:typical}, this range corresponds to gravity being stronger than the scalar self-interactions.
If we attempt to obtain similar curves for ${\hat \lambda}_\text{ini} > \frac{1}{2}$, they will simply lie above the curve labelled ``$0.5$" in figure~\ref{Excluded_+}. This is why the yellow shaded portion above the curve marked ``$0.5$" in figure~\ref{Excluded_+} is the region in which gravity becomes weaker than repulsive scalar self-interactions.
On the other hand, for reasons discussed in appendix~\ref{sec:m&lambda}, for attractive self-interactions, as we attempt to obtain similar curves for ${\hat \lambda}_\text{ini} < - \frac{1}{2}$, some part of the curves could be below the curve labelled ``$-0.5$" in figure~\ref{Excluded_-}.

\subsection{Probing mass and self-coupling of ULAs}\label{sec:mass_lambda_ULA}

For a canonically normalised field describing an axion or Axion-Like Particle (ALP), \cite{Marsh_2016}, the potential has a residual discrete shift symmetry, when all the higher harmonics could be ignored, the potential, from eq.(~\ref{eq:axion_potential}), is of the form
\begin{equation}
 U(\varphi) = m_a^2 f_a^2 \left[ 1 - \cos \left( \frac{\varphi}{f} \right) \right] \; ,
\end{equation}
where, $f$ is the ``decay constant" of the axion or ALP and using this potential, the quartic self coupling shall be given by
$\lambda_a = - \left( \frac{m_a}{f_a} \right)^2 $.
For axions with $f \sim {\cal O}(10^{17})~ {\rm GeV}$ and $m_a \sim 10^{-22}~ {\rm eV}$ (hence called Ultra-light axion or ULA) one finds that $\lambda_a \sim - 10^{-96}$, which is a dramatically small negative number.
Furthermore, by calculating the relic abundance of axion DM one finds that, if all of DM consists of axions, the corresponding density is $\Omega_a \sim 0.1\left( \frac{f_a}{10^{17} {\rm GeV} } \right)^2 \left( \frac{m}{10^{-22} {\rm eV} } \right)^{1/2}$ \cite{Hui_2017}
where, $\Omega_a = \rho_a / \rho_{\rm crit}$, which implies that 
\begin{equation}\label{eq:axion_imp}
\Omega_a \sim 0.1\left(\frac{m}{10^{-22}\ \text{eV}}\right)^{5/2} \left( \frac{10^{-96}}{ -\lambda } \right) \; .
\end{equation}
Thus, in $\log \lambda - \log m $ plane, fixed $\Omega_a$ (such as 0.1) corresponds to a line with slope 5/2. This is shown by the solid blue line labelled ``$\Omega_a = 0.1$" in figure~\ref{Excluded_-}. Needless to say, since the self-coupling of the axion is negative, there will be no axion line in figure~\ref{Excluded_+}.

For sufficiently large strength of attractive self-interactions, gravity and self-interactions can no longer be balanced by ``quantum pressure" and hence there is no stable soliton solution possible \cite{Levkov_2017, Chavanis_2011_analytic}. This implies that, for a stable soliton, there is a maximum possible value of soliton mass, denoted by $M_{\text{max}}$ (as discussed in section~\ref{sec:max_mass}), given by \cite{Chavanis_2011_analytic, Levkov_2017}
\begin{equation}\label{critical_mass}
    M_{\text{max}}
    \approx 10.2 \frac{m_{\text{pl}}}{(-\lambda)^{1/2}} \; .
\end{equation}
For any galaxy, the empirical mass of the soliton is given in the dimensionful form as $M_{\text{emp}} = \frac{\hbar c \hat{M}_{\text{emp}}}{G m}$, where $\hat{M}_{\text{emp}}$ is given by eq.~(\ref{eq:memp}). Rearranging  eq.~(\ref{critical_mass}), one can obtain the maximum $\lambda$ allowed such that the corresponding soliton mass is always less than $M_{\text{max}}$. For M87, with $M_{\text{halo}} = 2\times 10^{14}\ M_\odot$ we get
\begin{equation}\label{lambda_constraint}
    - \lambda \lesssim 2.34\times 10^{-94}\left(\frac{m}{10^{-22}\ \text{eV}}\right)^2 \; .
\end{equation}
The dashed line in figure~\ref{Excluded_-} corresponds to the value of $- \lambda$ above which there can be no soliton (stationary) solution. 
At this point it is worth noting that the solid black line in the figure actually corresponds to eq.~(\ref{lambda_constraint}) but obtained in the presence of a black hole. Hence, there is no `Can't be Probed' region in the figure, only the `Not Allowed' region, i.e., there are no stationary state solutions for the values of $\lambda$ and $m$ in the light gray region. 
This is further explained in detail in appendix~\ref{app:max_mass_smbh}.
It is worth noting that the constraints we obtain for attractive self interactions are for allowed range of values of the self-coupling.

Finally, using eq.~(\ref{eq:lambda}) and eq.~(\ref{eq:dimensions2}) [or, see eq.~(\ref{eq:lambda0})], one can easily show that, for ${\hat \lambda}_\text{ini}$ of ${\cal O}(1)$, the scaling parameter $s \sim {\cal O}(10)$. Notice that too small values of $s$, smaller than 10 or so, will take us out of the domain of applicability of our formalism. In particular, when $s$ becomes ${\cal O}(1)$, general relativistic effects can no longer be neglected and we have to include all the higher order terms we ignored in reliably modelling the solitons (see section~\ref{sec:theoretical_solitons}). 

But since the scaling parameter does depend on the mass of scalar $m$, ${\hat \lambda}_\text{ini}$, as well as the galaxy itself (through e.g.~$M_{\text{halo}}$), it is expected that there will be some galaxies for which the mass and couplings of an ultra light axion-like scalar field which forms all of DM will be probed. In this context, we would like to point the reader to appendix~\ref{sec:m&lambda} especially to figure~\ref{s_m}. 
For the case of M87 galaxy, the corresponding results are shown in figure~\ref{Excluded_-} and suggest that if all of dark matter is made of ultra light axions, then axions with mass $m \sim {\cal O}(10^{-22})~{\rm eV}$ are ruled out by the observational constraints on the mass within 10 pc of the centre of M87 galaxy and accretion time constraints. Figure~\ref{Excluded_-} also suggests that if ULAs form all of DM, its mass has to be less than
$ \sim 6 \times 10^{-23}$ eV.

\section{Summary and discussion}
\label{sec:smbh_conclusion}

In the standard model of cosmology, Dark Matter is conjectured to consist of non-relativistic (hence cold) particles, undergoing classical dynamics and with mean free path large compared to the length scales of interest (i.e. collisionless). 
There exist observations at sufficiently small length scales, e.g. galactic scales, which could be interpreted to imply that this picture of dark matter may be inaccurate or even invalid. Given the fact that this is inferred from observations at galactic scales, it is sensible to find ways to test models of DM at those scales. In particular, its worth asking whether observations of the myriads of galaxies can teach us something about the fundamental parameters in the Lagrangian of DM.

With this context in mind, we began by indicating how the Gross-Pitaevskii-Poisson equations follow from the weak field, slow variation and non-relativistic limit of a classical field theory with a canonical self-interacting scalar field and Einstein gravity. We then assumed that the astrophysical DM is in fact the self-interacting classical scalar field in this theory. We then modelled the cores of DM halos by solving the Gross-Pitaevskii-Poisson equations in section~\ref{sec:GPP_with_SMBH} and looked for stationary, spherically symmetric, nodeless solutions. 

This was done by first finding the solutions with an additional normalisation condition ${\hat \phi}(0) \sim 1$. It was then argued that such solutions were both physically unrealistic and untrustworthy (which we called ``theoretical solitons"). We then used the scaling symmetry in the Gross-Pitaevskii-Poisson equations to obtain realistic, trustworthy solutions from the theoretical solitons. In section~\ref{sec:scaling} and section~\ref{sec:parameter}, we described the strategy used to obtain the amount of scaling which is required to arrive at a model of the DM halo core for M87 galaxy.

In addition, the DM halo of interest may have a central black hole (BH) present - we model it as a point particle in Newtonian gravity. 
In the absence of a central black hole or scalar self-interactions, the density profile of the scalar field has a universal form \cite{Schive_Nature_2014}. In \cite{Davies_2020}, it was shown that the presence of a central supermassive black hole squeezes the soliton. 
We found that the presence of scalar self-interactions leads to many interesting possibilities; e.g. the presence of self interactions could squeeze the soliton if the interactions are attractive ($\lambda < 0$) and stretch it if the self interactions are repulsive ($\lambda > 0$). This can be easily seen in figure~\ref{fig:densityprofile}.

Needless to say, if the self-interaction strength is too low, it will have no observable effects, but, if it is too high, the structure of DM halos will be governed not by gravity but by self-interactions alone - this is a scenario which we have not explored in this work. When we impose this additional requirement that scalar self-interactions do not overtake gravity - this implies that the parameter $|{\hat \lambda}_\text{ini}|$ should be less than $1/2$ (see section~\ref{sec:typical}), and we get a curve in the parameter plane such that all constraints of interest are below this curve (the curve marked $-0.5$ in figure~\ref{Excluded_-} and the curve marked $+0.5$ in figure~\ref{Excluded_+}).
As we argued in appendix \ref{sec:m&lambda}, for a given value of $|{\hat \lambda}_\text{ini}|$, the probed $\lambda$ obtained from eq.~(\ref{eq:lambda0}) for attractive self-interactions will be more than the probed $\lambda$ for negative interactions. We also found that, for every halo, if the scalar self-interactions are attractive, there will be regions in $\lambda-m$ plane, which can never be probed by the method presented in this paper. 

Using the dependence of accretion time of scalar field into super massive black holes on $m$ and using the upper limits on the amount of DM concentrated near the central black holes, \cite{Davies_2020} placed constraints on the mass of Ultra Light DM particles. In the presence of self-interactions of the scalar field, an analysis, presented in section~\ref{sec:M&D}, which was a modification of the one in \cite{Davies_2020}, implies that the lower limit on the mass $m$ depends upon the sign and strength of the self-interactions (see figure~\ref{M87WithBH} as well as table~\ref{table:constraints}). 
This consideration is important while comparing the limits on $m$ obtained by the methods of \cite{Davies_2020} with the limits on $m$ obtained elsewhere in the literature.

Before closing, it is important to note a few caveats and sources of errors and uncertainties. 
This method is obviously currently limited by various uncertainties associated with (a) theoretical modelling of accretion of scalar field, (b) observational limits on the masses of SMBH and the amount of DM contained within the central region etc. Furthermore, as we noted in appendix~\ref{sec:scaling} the empirical relations used to obtain the amount of scaling need not be applicable for many galaxies. We have used them to illustrate the ideas we present here. As discussed in the beginning of section~\ref{sec:constraints}, the accretion time considerations we used are based on the results of accretion modelling for a scalar field without self-interactions. Similarly, as we noted in section~\ref{sec:M&D}, we have not been committed to any specific definition of the exact size of theoretical soliton which we took to be ${\cal O}(m^{-1})$ i.e. there is an ${\cal O}(1)$ uncertainty in the size of theoretical soliton. Throughout this paper, we illustrated all our ideas using the numerical estimates for M87 galaxy as discussed in section~\ref{sec:GPP_with_SMBH}. These numbers (e.g. the mass of soliton) have uncertainties associated with them and this can cause the actual numbers in our plots to change considerably.

Some of the problems due to these uncertainties are expected to be ameliorated in the near future e.g. the theoretical modelling of the accretion process, the observational limits on the DM mass within the central region of galaxies, the observational limits on the mass of super-massive black holes etc are all expected to improve. 
Needless to say, the region of the $\lambda-m$ plane we constrain has been explored by other approaches \cite{Fan_2016,Cembranos_2018,Suarez_2017,Chavanis_2021,Desjacques_2018,Urena-Lopez_2019_cosh,Delgado_2022}. Our primary focus here has been developing the method and finding the constraints using an illustrative set of parameter values. With that in mind, we 
can conclude that (a) for every galaxy, there will be a region in the $\lambda - m$ parameter plane which will get excluded by accretion time considerations and estimates of mass within central regions of a galaxy; (b) by including the effects of self-interactions of the ultra light scalar DM under consideration, the values of self-coupling which can be probed is extremely small i.e. ${\cal O}(10^{-96})$ - i.e. of the order of those corresponding to ultra-light axions with Planckian decay constants;
(c) the probed values of self-coupling for positive self-interactions are higher than those of negative self-interactions,
(d) our analysis suggests that if axions form all of DM, then, axionic DM with $m \sim {\cal O}(10^{-22)}~{\rm eV}$ is ruled out by central mass estimates for M87 galaxy - its mass has to be less than $ \sim 6 \times 10^{-23}$ eV (e) for negative self-interactions, for any given DM halo, there will be regions in $\lambda - m$ plane which can never be probed using the methods described in this chapter.
\justifying
\chapter{Self-interactions of ULDM to the rescue?}\label{chpt:paper_2}

\textbf{Based on:} \\\textbf{B.~Dave} and G.~Goswami, \emph{Self-interactions of ULDM to the rescue?}, \href{https://doi.org/10.1088/1475-7516/2023/07/015}{\emph{J. Cosmol. Astropart. Phys.} \textbf{07} (2023) 015} [\href{https://arxiv.org/abs/2304.04463}{arXiv:2304.04463 [astro-ph.CO]}]

\noindent Note that most of the text in this chapter is an excerpt from our published work, Ref.~\cite{Dave_2023}.

\section{Observations and Motivation}\label{sec:rc_observations}

Galactic rotation curves i.e. orbital velocity of stars and gas as a function of distance from the centres of galaxies are an important probe of the matter (visible and dark) distribution in said galaxies \cite{Carroll_Ostlie_2017, Sofue_2001}. 
In general one can obtain the circular velocity of a test particle in an orbit of radius $r$ in the gravitational potential of a spherically symmetric distribution of matter using
\begin{equation}\label{eq:circ_vel}
v(r) = \sqrt{\frac{GM(r)}{r}} = \sqrt{\frac{4\pi G\int_0^r\rho(r')r'^2dr'}{r}}\ .
\end{equation}
The total observed velocity can be split into various components corresponding to different distributions of matter in the galaxy: (a) stellar disk ($V_d$), (b) stellar bulge ($V_b$), (c) gas ($V_g$) and (d) dark matter ($V_{DM}$). Hence, a typical observed rotation curve for a galaxy from, for instance the Spitzer Photometry \& Accurate Rotation Curves (SPARC) catalogue \cite{Lelli_2016}, can be written as
\begin{equation}\label{eq:observed_vel}
    V_{obs} = \sqrt{V_{DM}^2 + V_{g}|V_{g}| + \Upsilon_d V_{d}|V_{d}| + \Upsilon_b V_{b}|V_{b}|}\ .
\end{equation}
Note that here contributions from the disk and bulge can be tuned using the stellar mass-to-light ratios $\Upsilon_d$ and $\Upsilon_b$ respectively. Usually, observed rotation curves exhibit a velocity that increases in the inner region, and then flattens as one goes further away from the centre. The inner regions of large galaxies are well-explained by the sizeable amount of baryonic matter contained near the centre. On the other hand, dark matter is required to explain the flat rotation curves at large $r$ where baryonic contribution is very little. DM-only simulations \cite{Navarro_1996} suggest a density profile for DM that goes like $r^{-1}$ in the inner region and $r^{-3}$ at large $r$. This is the well-known Navarro-Frenk-White (NFW) profile given by

\begin{equation}
    \rho_{NFW}(r) = \frac{\rho_s}{\frac{r}{r_s}\left(1 + \frac{r}{r_s}\right)^2}\ ,
\end{equation}
where $\rho_s$ and $r_s$ are parameters of the profile. These parameters can be chosen such that the corresponding velocity curve exhibits a flat portion at a desired scale. The NFW velocity curve also attains a maximum at $\sim 2.16r_s$. For small $r$, using eq.~(\ref{eq:circ_vel}) one can see that NFW profile implies an increasing velocity where $v\propto \sqrt{r}$. However, for many low mass and low surface brightness (LSB) galaxies where the baryonic contribution is thought to be small even at small radius, observed velocities in the inner regions ($\sim \mathcal{O}(1)\ \text{kpc}$) point to a more slowly increasing velocity curve, $v \propto r$. This is the manifestation of the well-known core-cusp problem \cite{Bullock_2017}. 

FDM resolves this issue by considering the inner regions to be described by stable solutions of the Schrodinger-Poisson equations. These core-like structures have flat density profiles and are called solitons. Independent numerical simulations \cite{Schive_Nature_2014, Schive_PRL_2014, Mocz_2017, Mina_2022} have confirmed a core-halo structure for FDM, where the inner region is described by the FDM core (also called a soliton) while further from the centre, it behaves like CDM. Here, the total density profile can be written as, 
\begin{equation}\label{eq:tot_dens}
    \rho(r) = \Theta{(r_t - r)}\rho_{SFDM}(r) + \Theta{(r - r_t)}\rho_{NFW}(r)\ ,
\end{equation}
where imposing the continuity of density at $r_t$ implies $\rho_{SFDM}(r_t) = \rho_{NFW}(r_t)$. Note that this fixes one of the parameters of the NFW profile ($\rho_s$) leaving two free parameters for the outer envelope: $\{r_t, r_s\}$.

\subsection{Soliton-Halo (SH) relations}\label{sec:SH_relations_obs_curves}

Simulations in \cite{Schive_Nature_2014, Schive_PRL_2014} also obtained a power-law relationship between mass of the soliton $M_{SH}$ and mass of the halo $M_h$\footnote{We remind the reader that halo mass is often defined as $M_{h} = \frac{4\pi}{3}(200\rho_{c})R_{200}^3$, where $R_{200}$ is the radius at which the average density of the mass contained is $200$ times the critical density ($\rho_c$) of the Universe.} of the form $M_{SH} \propto M_h^{1/3}$, or, more precisely \cite{Schive_PRL_2014}
\begin{equation}\label{eq:Schive_SH}
\left(\frac{M_{SH} }{10^9\ M_\odot}\right) = 1.4\left(\frac{M_h}{10^{12}\ M_\odot}\right)^{1/3}\left(\frac{m}{10^{-22}\ \text{eV}}\right)^{-1}\ .
\end{equation}
The simulations report a scatter of roughly a factor of $2$ in this soliton-halo (SH) relation. It is important to note that other simulations \cite{Mocz_2017, Mina_2022} have reported a different power law between soliton mass and halo mass: $M_{SH} \propto M_h^{0.556}$: this disagreement could partly be the result of different merger histories and tidal stripping \cite{Chan_2022}. 

Therefore, if simulations in \cite{Schive_Nature_2014, Schive_PRL_2014} are correct, then along with describing observed rotation curves, FDM is expected to satisfy the SH relation in eq.~(\ref{eq:Schive_SH}). In other words, such relations can be considered to be a sharp prediction of the FDM paradigm.


\subsection{Rotation curves and SH relations for FDM}\label{sec:FDM_constraints}

Recently, using all 175 galaxies in the SPARC catalogue, Ref.~\cite{Bar_2022} reported that, for FDM, the SH relation in eq.~(\ref{eq:Schive_SH}) is not consistent with the observed rotation curves. 
In this context, we direct the reader to figure~1 of Ref.~\cite{Bar_2022} as well as figure~\ref{fig:LSB_no_inter} of this work. 
The details of the procedure followed in Ref.~\cite{Bar_2022} which are relevant for our purpose are discussed in section~\ref{sec:imposing_mod_SH_relations}. For now, we just highlight the following: the soliton masses allowed by the rotation curves data were much smaller than the corresponding soliton masses expected from the SH relation obtained from eq.~(\ref{eq:Schive_SH}) for $m\in \left[10^{-24}\ \text{eV}, 10^{-20}\ \text{eV}\right]$.\footnote{Authors in \cite{Bar_2019} that the SH relation in eq.~(\ref{eq:Schive_SH}) was equivalent to the ratio of kinetic energy and total mass being roughly the same for the soliton and halo: $(K/M)_{sol} \approx (K/M)_{halo}$. In particular, Ref.~\cite{Bar_2018} showed that for $m\in\left[10^{-22}\ \text{eV}, 10^{-21}\ \text{eV}\right]$, if a soliton with mass $M_s$ is expected to satisfy the SH relation, then the corresponding velocity curve significantly overshoots observed velocity in the inner regions of dark matter dominated galaxies.}

Hence, the SH relation for an ultra-light scalar field with no self-interactions seems to be incompatible with observed rotation curves. This along with other constraints mentioned in section~\ref{sec:rc_observations} potentially rules out FDM being a significant fraction of all dark matter.

\subsection{SFDM self-interactions}
\label{sec:selfinter}

Let us begin by recalling from chapter~\ref{chpt:introduction} some basics about the quartic (i.e. $\lambda\varphi^4$-type) self-interaction term in the Lagrangian of a scalar field. In the non-relativistic limit (relevant to cold DM), this self-interaction leads to an inter-particle interaction potential energy function of the form $U = U_0 ~ \delta^3 ({\bf x}_i - {\bf x}_j)$ i.e. it is a contact interaction. Depending on the sign of self coupling, this interaction could be attractive ($\lambda < 0$) or repulsive ($\lambda > 0$). 

Scalar field dark matter (SFDM) with attractive self-interactions is well motivated if one considers axions, where Taylor expansion of the cosine potential will lead to a quartic self-interaction term with $\lambda < 0$ \cite{Marsh_2016}. On the other hand, repulsive ($\lambda > 0$) self-interactions are expected from e.g. moduli fields ubiquitous in theories of high energy physics. 

If ultra-light axions (ULAs) are to comprise all of DM, the self-interaction strength must be of the order $\sim 10^{-96}$ for mass $\sim 10^{-22}\ \text{eV}$ \cite{Hui_2017}. Recent simulations in \cite{Mocz_2023} focus on the impact of such attractive self-interactions on cosmological structure formation. In fact, it is well known that even very small self-interactions can dramatically change the resultant stable configuration \cite{Colpi_1986}. The effect of self-interactions on the mass and radius of solitons is apparent even in the Newtonian-limit, as \cite{Chavanis_2011, Chavanis_2011_analytic} demonstrated for both attractive and repulsive self-interactions. 

Thus, there are very good reasons to consider SFDM with small but non-negligible self-interactions. Before proceeding, we note that SFDM with self interactions, in particular in the Thomas-Fermi (TF) regime (see discussion in section~\ref{sec:regimes}), has been constrained in the past in various ways e.g. by looking at cosmological evolution and structure formation \cite{Li_2014, Foidl_2022}, and even using rotation curves \cite{Bernal_2017, Craciun_2020, Harko_2022, Dawoodbhoy_2021, Delgado_2022}. See also \cite{Khlopov_1985, Dev_2017, Banerjee_2023, Davoudiasl_2023, Fox_2023, Feng_2022} for some selected references that consider scalar fields with self-interactions in this context.

\subsection{SH relations for SFDM with self-coupling}\label{sec:SFDM_SH_relations}

Other parameters being fixed, the mass of a soliton in the presence of self-interactions gets changed. 
This suggests that, in the presence of self-interactions, the corresponding soliton-halo relation will take up a form different from eq.~(\ref{eq:Schive_SH}) which is not expected to be valid when $\lambda \neq 0$. Recently, \cite{Chavanis_SH_relation_2019, Padilla_2021} arrived at the corresponding SH relation, which takes up the following form (see also section~V of Ref.~\cite{Chavanis_2021})
\begin{equation}\label{eq:modified_SH}
\left(\frac{M_{SH} }{10^9\ M_\odot}\right) = 1.4\left(\frac{M_h}{10^{12}\ M_\odot}\right)^{1/3}\left(\frac{m}{10^{-22}\ \text{eV}}\right)^{-1}\sqrt{1 + (1.16\times 10^{-7})\hat{\Lambda}\left(\frac{M_h}{10^{12}\ M_\odot}\right)^{2/3}}\ ,
\end{equation}
where the $\hat{\Lambda}$ in the above equation is proportional to the self coupling $\lambda$ of the scalar.\footnote{
More precisely, $\hat{\Lambda}$ is the same as $\frac{\lambda}{4} \left( \frac{M_{pl}}{m} \right)^2$ which is the same as $2 (s^2\hat{\lambda}_\text{ini})$ in our notation introduced in section~\ref{sec:stationary_states}.} The origin of the numerical factor in front of $\hat{\Lambda}$ can be understood from the discussion above eq.~(80) in \cite{Padilla_2021}. Note that in the absence of self-interactions the SH relation reduces to eq.~(\ref{eq:Schive_SH}).
While eq.~(\ref{eq:modified_SH}) is valid for both attractive and repulsive self-interactions, for $\lambda$ that is too negative, this SH relation will no longer be applicable (see also section~\ref{sec:why_no_att_inter}). 

\subsection{Velocity curve for a soliton}

We have already discussed in section~\ref{sec:KGE_to_GPP} that the equations of motion for a scalar field $\varphi$ with self-interactions in the Newtonian, slow varying limit are the Gross-Pitaevskii-Poisson (GPP) equations in eq.~(\ref{eq:GrossPitaevskii_higher}) and~(\ref{eq:Poisson_higher}), where one defines a complex scalar field $\Psi$ in eq.~(\ref{eq:Psi}) as the slowly varying part of the real scalar $\varphi$.
As we have done in chapter~\ref{chpt:paper_1}, we shall describe the inner region of dark matter halos using the spherically symmetric, ground state stationary solutions of the GPP system (albeit without the central SMBH). 
We have noted that the free parameters of the system for such a solution are mass of the ULDM particle, $m$, dimensionless\footnote{We use hatted variables to denote dimensionless quantities variables. See  section~\ref{sec:stationary_states} for details.} unscaled self-interaction strength $\hat{\lambda}_\text{ini}$ and the scaling parameter $s$, related to the total mass of the soliton solution\footnote{See section~\ref{sec:scaling_symmetry} for details about scaling.}.
Once we solve the system (see chapter~\ref{chpt:numerical_solutions} for details) numerically, we obtain the density profile $\hat{\rho}(\hat{r}) = \hat{\phi}^2$, where the complex field in the GPP system is written as $\hat{\Psi} = e^{-i\hat{\gamma}\hat{t}}\hat{\phi}(\hat{r})$.

\begin{figure}[ht]\captionsetup[subfigure]{font=footnotesize}
    \centering
    \includegraphics[width = 0.6\textwidth]{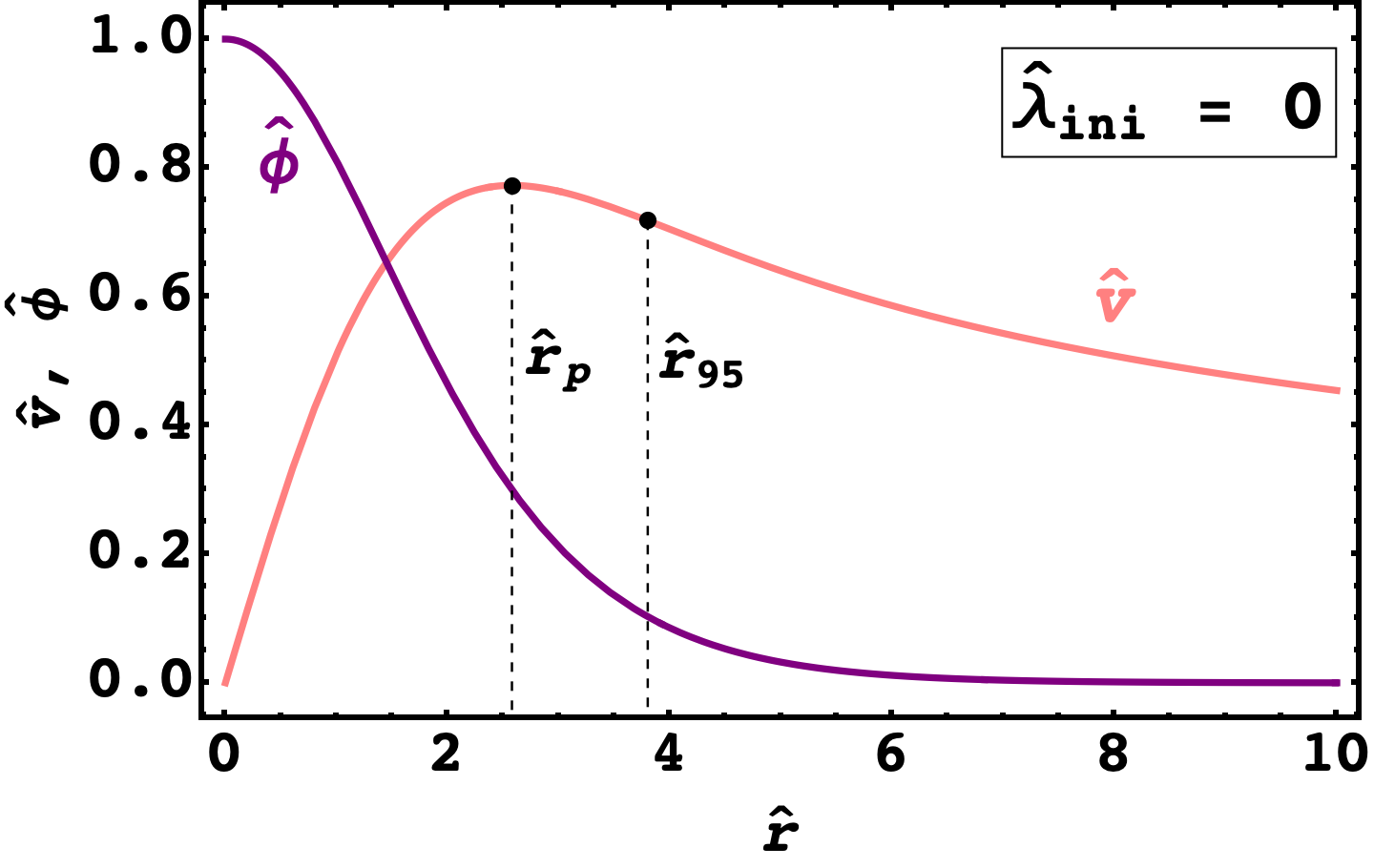}
    \caption[Theoretical soliton solution along with velocity curve]{Dimensionless solution satisfying the boundary condition $\hat{\phi}(\infty) = 0$ for $\hat{\lambda}_\text{ini} = 0$ is shown by the purple curve, while the corresponding velocity curve is shown by the pink curve. Note that $\hat{r}_{p}$ is the radius at which the velocity peaks. Recall that the dimensionless variables are defined in eq.~(\ref{eq:dimensions1}) and~(\ref{eq:dimensions2}).}
    \label{fig:density_velocity}
\end{figure}

Once density profile $\hat{\rho}(\hat{r})$ is obtained from the solution, one can also find the corresponding velocity curve from eq~(\ref{eq:circ_vel}). It is easy to see that $\hat{v} = v/c$ (velocity curve for $\hat{\lambda}_\text{ini} = 0$ is shown by the pink curve in figure~\ref{fig:density_velocity}). Note that scaling symmetry implies that velocity scales as $\hat{v}\rightarrow \hat{v}/s$. 

\section{Self-interacting ULDM, rotation curves and soliton-halo relations}\label{sec:soliton_properties}

Before proceeding, note that we fix DM particle mass to the fiducial value of $m = 10^{-22}\ \text{eV}$ unless mentioned otherwise and hence focus our attention on the effect of (a) varying DM self-coupling $\lambda$ (parameterised by the dimensionless quantity ${\hat \lambda}_\text{ini}$ defined in eq.~(\ref{eq:dimensions2})), and, (b) varying the total number of DM particles forming the soliton acting as the core of the DM halo of a given galaxy (this is parameterised by soliton mass $M_s$ or scale $s$). 

For a particular DM species with a fixed physical $m$ and $\lambda$, as we consider various soliton solutions with different total masses, the size of the soliton is different. In section~\ref{sec:mass_radius_curves} we arrive at some important results about the connection between the mass of the soliton and its radius. 
Since we eventually need to satisfy rotation curves, in section~\ref{sec:impact_of_parameters} we shall briefly look at the impact of the free parameters $\{\hat{\lambda}_\text{ini}, s\}$ on the circular velocity (i.e. rotation curves). Finally, in section~\ref{sec:imposing_mod_SH_relations} we shall check the compatibility of the modified SH relation in eq.~(\ref{eq:modified_SH}) with observed rotation curves for solitons formed from SFDM with self-interactions.

\subsection{Impact of parameters on rotation curves}\label{sec:impact_of_parameters}

Every combination of the free parameters $\{m, \hat{\lambda}_\text{ini}, s\}$ will lead to a unique density profile and a unique corresponding circular velocity profile. As $m$ is fixed, it is useful to ask how varying other two free parameters $\hat{\lambda}_\text{ini}$ and $s$, affects the velocity curves of solitons.

\begin{figure}[ht]\captionsetup[subfigure]{font=footnotesize}
    \centering
    \includegraphics[width = 1\textwidth]{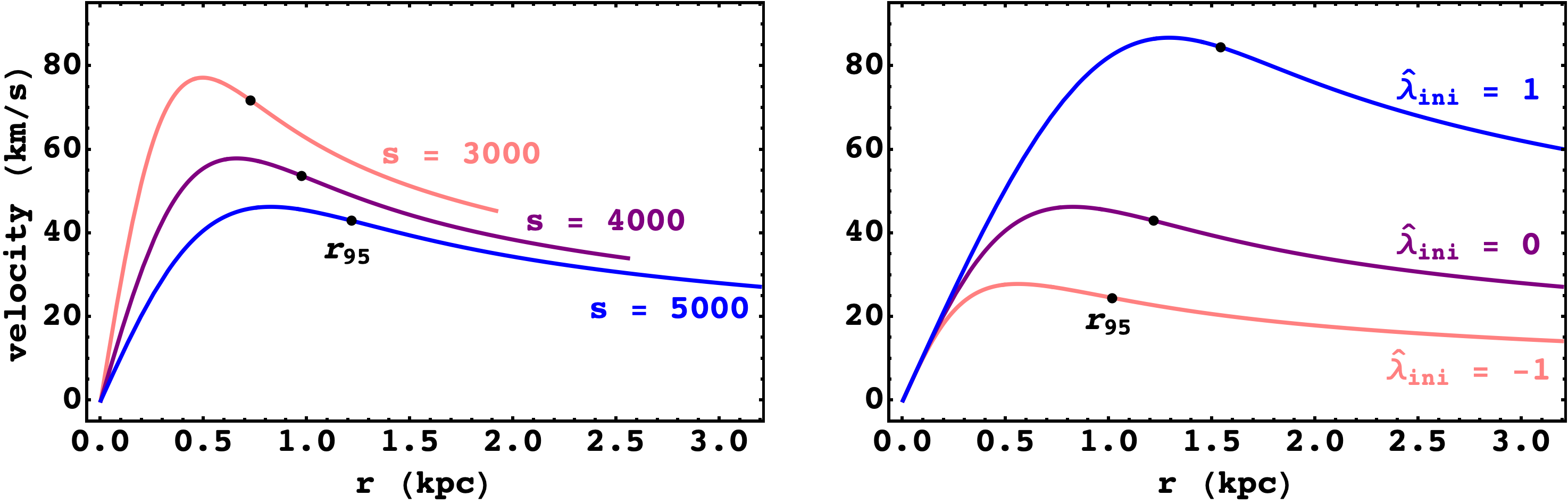}
    \caption[Impact of $s$ and $\hat{\lambda}_\text{ini}$ parameters on velocity curves]{The left panel demonstrates how larger scale values lead to larger cores but smaller peak velocities, and vice-versa ($\hat{\lambda}_\text{ini} = 0$ is fixed). The right panel shows the impact of changing the self-interaction strength. For $\hat{\lambda}_\text{ini} > 0$, peak velocity and size of the core increases, while for $\hat{\lambda}_\text{ini} < 0$ peak velocity decreases and the core gets smaller ($s = 5000$ is fixed).
    Note that while we have shown the velocity curve for $\hat{\lambda}_{\text{ini}} = -1$, any solution with $\hat{\lambda}_{\text{ini}} < -0.4$ will lead to a soliton that is not stable under small perturbations. This limits the impact attractive self-interactions can have on theoretical rotation curves (see discussion in section~\ref{sec:why_no_att_inter} for more details).}
    \label{fig:impact_of_parameters}
\end{figure}
\begin{enumerate} 
    \item \textbf{Impact of $s$:}
    Scaling symmetry implies that $v \rightarrow v/s$ and $r \rightarrow r\cdot s$. Therefore, an increase in $s$ leads to the stretching of the $r$-axis, while squeezing the $v$-axis, leading to a larger soliton but a smaller peak velocity. This effect is shown in the left panel of figure~\ref{fig:impact_of_parameters}. Note that $M_s$ scales in the same way as $v$, implying that for a fixed $m$ and $\lambda$ a smaller peak velocity corresponds to a lighter soliton and vice-versa.
    
    \item \textbf{Impact of $\hat{\lambda}_\text{ini}$:}
    As $\hat{\lambda}_\text{ini}$ increases, the corresponding $\hat{R}_\text{ini}$ and $\hat{M}_\text{ini}$ also increase (see blue curve in figure~\ref{fig:m_vs_r_scaled}). Hence, for a fixed $m$ and $s$, increasing $\hat{\lambda}_\text{ini}$ will stretch both the $r$ and $v$ axes leading to a larger peak velocity and $r_{95}$. The opposite is true when $\hat{\lambda}_\text{ini}$ decreases. The effect can then be described as the stretching and squeezing in roughly the direction of the slope of the linear region of the velocity curve (see right panel of figure~\ref{fig:impact_of_parameters}). Note that the squeezing effect of $\hat{\lambda}_\text{ini} = -1$ is much smaller than the expanding effect of $\hat{\lambda}_\text{ini} = 1$.
\end{enumerate}

\subsection{Confronting observed rotation curves}\label{sec:analysis_SPARC}
Armed with the knowledge of how different parameters impact soliton velocity curves, we can now confront observations. We consider low surface brightness (LSB) galaxies from the Spitzer Photometry \& Accurate Rotation Curves (SPARC) catalogue \cite{Lelli_2016}, which hosts surface photometry at $3.6\ \mu m$ and HI/H$\alpha$ rotation curves for 175 galaxies. In this section, we probe the compatibility of the modified SH relation in eq.~(\ref{eq:modified_SH}) with observed rotation curves.  

Before proceeding, we ensure that we are dealing with good quality rotation curves by eliminating galaxies with quality flag $Q = 3$. This removes galaxies with large asymmetries and non-circular motions. Since LSB galaxies are characterized by a low effective surface brightness ($B_{\text{eff}}$), we only keep galaxies with $\log(B_{\text{eff}}) \leq 1.5\ L_\odot/\text{pc}^2$ \cite{Khelashvili_2023}. This leaves us with a sample of 56 galaxies. Note that galaxies in our sample are bulgeless i.e. $V_b = 0$ in eq.~(\ref{eq:observed_vel}) at all radii for every galaxy.

\subsection{Rotation curves and SH relation for repulsive self-interactions}\label{sec:imposing_mod_SH_relations}

In this section, we check the compatibility of the modified SH relation in eq.~(\ref{eq:modified_SH}) with observed rotation curves of the sample of LSB dwarf galaxies from the SPARC catalogue. We carry out a procedure similar to the one performed in Ref~\cite{Bar_2022}, however here instead of varying scalar field mass $m$, we keep $m = 10^{-22}\ \text{eV}$ fixed and vary the dimensionless self-interaction strength $\hat{\lambda}_\text{ini} \geq 0$. For the sake completeness we also show the results for a varying $m$ with no self-interactions $\lambda = 0$ (for comparison with the results of \cite{Bar_2022}) and $\lambda \neq 0$ (in section~\ref{sec:impact_of_mass}). We also talk briefly about the effect of negative self-interactions in section~\ref{sec:why_no_att_inter}.

\begin{figure}[h!]\captionsetup[subfigure]{font = {footnotesize}}
    \centering
    \begin{subfigure}[t]{0.45\textwidth}
        \centering
        \includegraphics[width = \textwidth, height=2in]{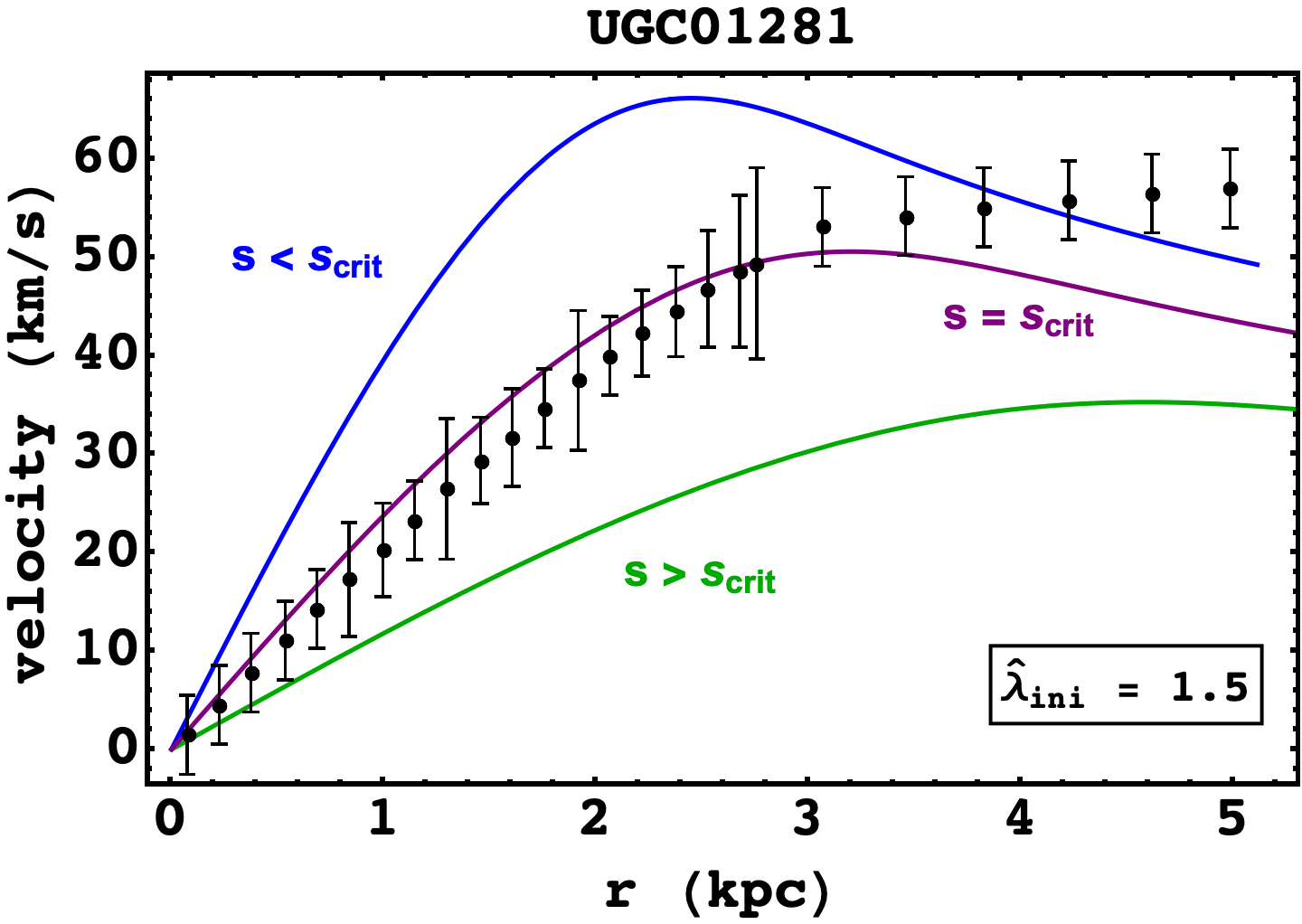}
        \caption{Black dots with error bars are the observed velocity values for `UGC 1281'. For a fixed $\hat{\lambda}_\text{ini} = 1.5$, as we increase $M_s$ (by decreasing $s$), the slope of the soliton-only velocity curve increases until it overshoots (blue curve) the observed velocity. Soliton masses are allowed (purple and green curves) if they fit the observed velocities or undershoot them at any radius. In case of undershooting, it is expected that the background components can be tuned to fit observations and hence are allowed.}
        \label{fig:eg_exclusion_procedure}
    \end{subfigure}
    \hfill
    \begin{subfigure}[t]{0.45\textwidth}
        \centering
        \includegraphics[width = \textwidth, height=2in]{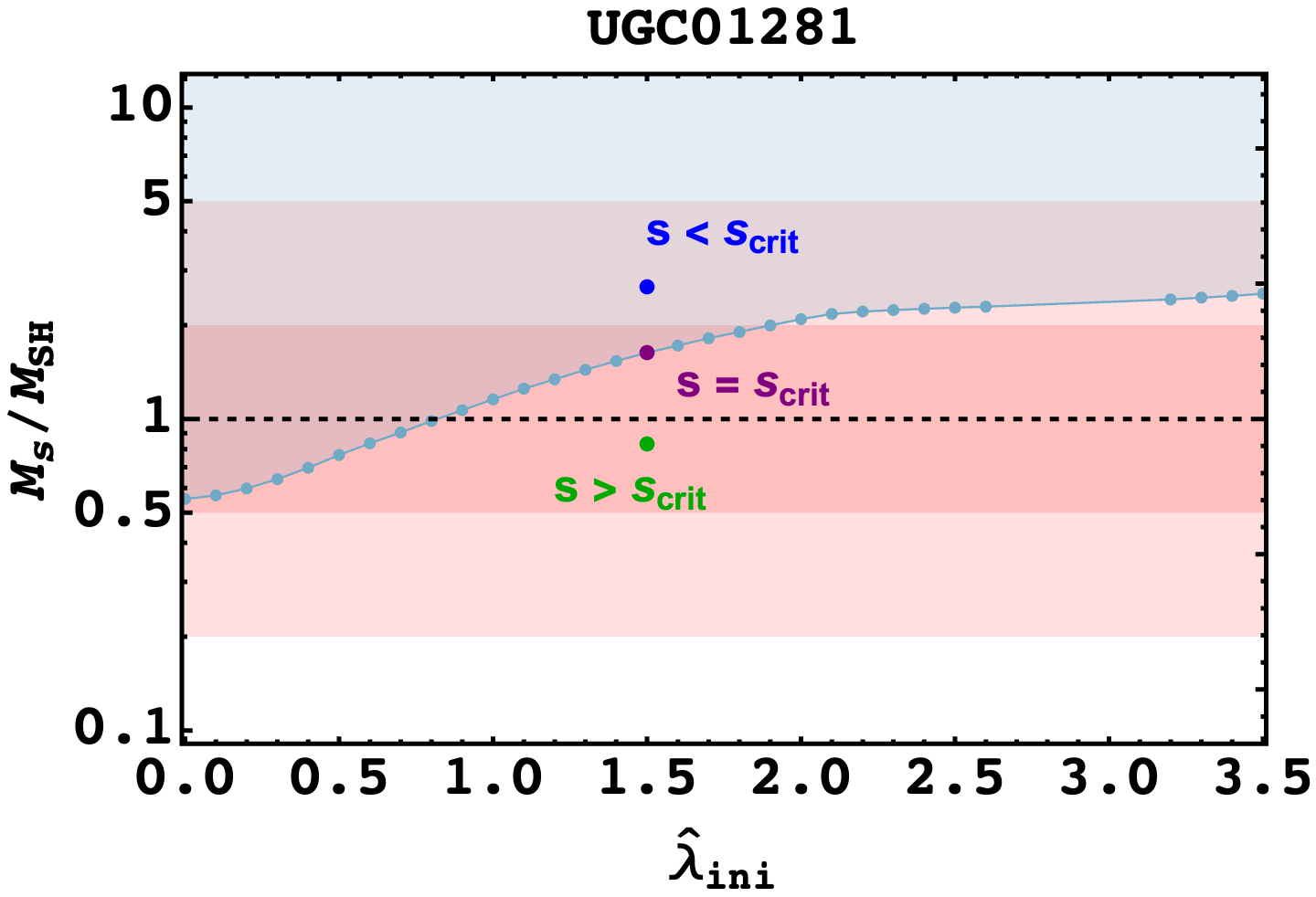}
        \caption{The maximum allowed $M_s/M_{SH}$ for various values of $\hat{\lambda}_\text{ini}$ are shown by the blue dots while the curve connecting them marks the boundary of the shaded (excluded) region. Solitons with $M_s/M_{SH}$ in the shaded region with lead to a velocity curve that overshoots the observed velocity. The three cases shown in figure~\ref{fig:eg_exclusion_procedure} are shown by dots of the corresponding colours. The dark and light shaded pink regions correspond to a scatter of 2 and 5 from the SH relation in eq.~(\ref{eq:modified_SH}), while the dashed line corresponds to $M_s = M_{SH}$.}
        \label{fig:eg_exclusion_region}
    \end{subfigure}
    \caption[Demonstration of procedure for UGC 1281]{A demonstration of the procedure detailed in this section and the resulting exclusion region for the galaxy `UGC 1281'.}
    \label{fig:eg_exclusion}
\end{figure}

To illustrate the procedure, we shall use the example of the galaxy `UGC 1281' whose observed rotation curve is shown in figure~\ref{fig:eg_exclusion_procedure} using data points. These data points correspond to $V_{obs}$ while the solid curves in figure~\ref{fig:eg_exclusion_procedure} correspond to $V_{DM}$ in eq.~(\ref{eq:observed_vel}). The dark matter velocity $V_{DM}$ can be obtained from eq.~(\ref{eq:circ_vel}) for the density profile given by eq.~(\ref{eq:tot_dens}) in which $\rho_{SFDM}$ can be obtained from the numerical solution of GPP equations. If $V_{DM}$ happens to be smaller than $V_{obs}$, the other terms on the RHS could be such that eq.~(\ref{eq:observed_vel}) is still satisfied and the corresponding model parameters leading to said $V_{DM}$ will be allowed. On the other hand, if $V_{DM}$ is larger than $V_{obs}$, eq.~(\ref{eq:observed_vel}) will not be satisfied and the corresponding model parameters will be ruled out. Note that for $r > r_t$, the density profile $\rho_{NFW}$ will be determined by NFW parameters which we assume can be adjusted to ensure that $V_{DM}$ is not larger than $V_{obs}$.

Since $m$ is fixed, free parameters of the system are $\{\hat{\lambda}_\text{ini}, s\}$. Now for a fixed $\hat{\lambda}_\text{ini}$, we have seen in figure~\ref{fig:impact_of_parameters} that a larger $s$ corresponds to a soliton velocity curve with a smaller slope in the inner region and a smaller peak velocity. Hence, for a large enough value of scale $s$ the soliton velocity curve does not overshoot the observed velocity at any point (see the green curve in figure~\ref{fig:eg_exclusion_procedure}). As $s$ decreases, the corresponding soliton mass increases (since $M_s \rightarrow M_s/s$) along with the slope of the inner region of the soliton velocity curve and its peak velocity. For a small enough value of $s$, $V_{DM}$ overshoots $V_{obs}$, as shown by the blue curve in figure~\ref{fig:eg_exclusion_procedure}. The smallest value of scale that does not cause $V_{DM}$ to overshoot $V_{obs}$ (as shown by the purple curve in figure~\ref{fig:eg_exclusion_procedure}) is denoted by $s_{\text{crit}}$. The soliton mass corresponding to $s_{\text{crit}}$ is the largest soliton mass $M_s^{\text{crit}}$ allowed by the data. For galaxies in our sample, the typical values of $s_{\text{crit}} \sim \mathcal{O}(10^4)$.

Before proceeding further, we make the following assumptions:
\begin{enumerate}
    \item We set the following overshooting condition: At an ith observed radius we calculate $\chi^2_i = \frac{(v_i^{\text{pred}} - v_i^{obs})^2}{\sigma_i^2}$, where $v_i^{\text{pred}}$ is the predicted velocity, $v_i^{obs}$ is the observed velocity and $\sigma_i$ is the uncertainty at that radius. We exclude a soliton if for any ith observed radius bin, both $v_i^{\text{pred}} > v_i^{obs}$ and $\chi_i^2 > 1$ are satisfied.\footnote{We have verified that the results shown here only change slightly when one uses $\chi^2_i > 3$ instead.}
    
    \item For $\lambda = 0$ there is an analytical expression for density profile \cite{Schive_Nature_2014} (also called the Schive profile) that can be evaluated at an arbitrary radius. However, for $\lambda \neq 0$ the density profile is evaluated numerically. Since the numerical solution is calculated only till the dimensionless distance $\hat{r}_{\text{max}}$, we only consider observed data points up to which a scaled soliton solution can be calculated ($r_{\text{max}} \propto m^{-1}s\hat{r}_{\text{max}}$).\footnote{However, it is expected for the density to keep falling for $\hat{r} > \hat{r}_{\text{max}}$, implying fall-off in velocity as well. Therefore if the velocity does not overshoot for any radius covered by the numerically solved part, it will not overshoot for the rest of the rotation curve as well. This is because the galaxies in our sample do not exhibit a fall-off in observed velocity at large $r$.}
    
    \item We assume that the total halo mass is the same as the CDM halo mass and does not change in the presence of a soliton at the centre \cite{Robles_2018, Dawoodbhoy_2021}. This implies that for a given galaxy, $M_h$ in eq.~(\ref{eq:modified_SH}) can be fixed to the best-fit value obtained in \cite{Li_2020}. 
\end{enumerate}

\noindent \textbf{\emph{Absence of self-interactions}}
\vspace{0.25cm}

\begin{figure}[h!]
    \centering
    \includegraphics[width = \textwidth]{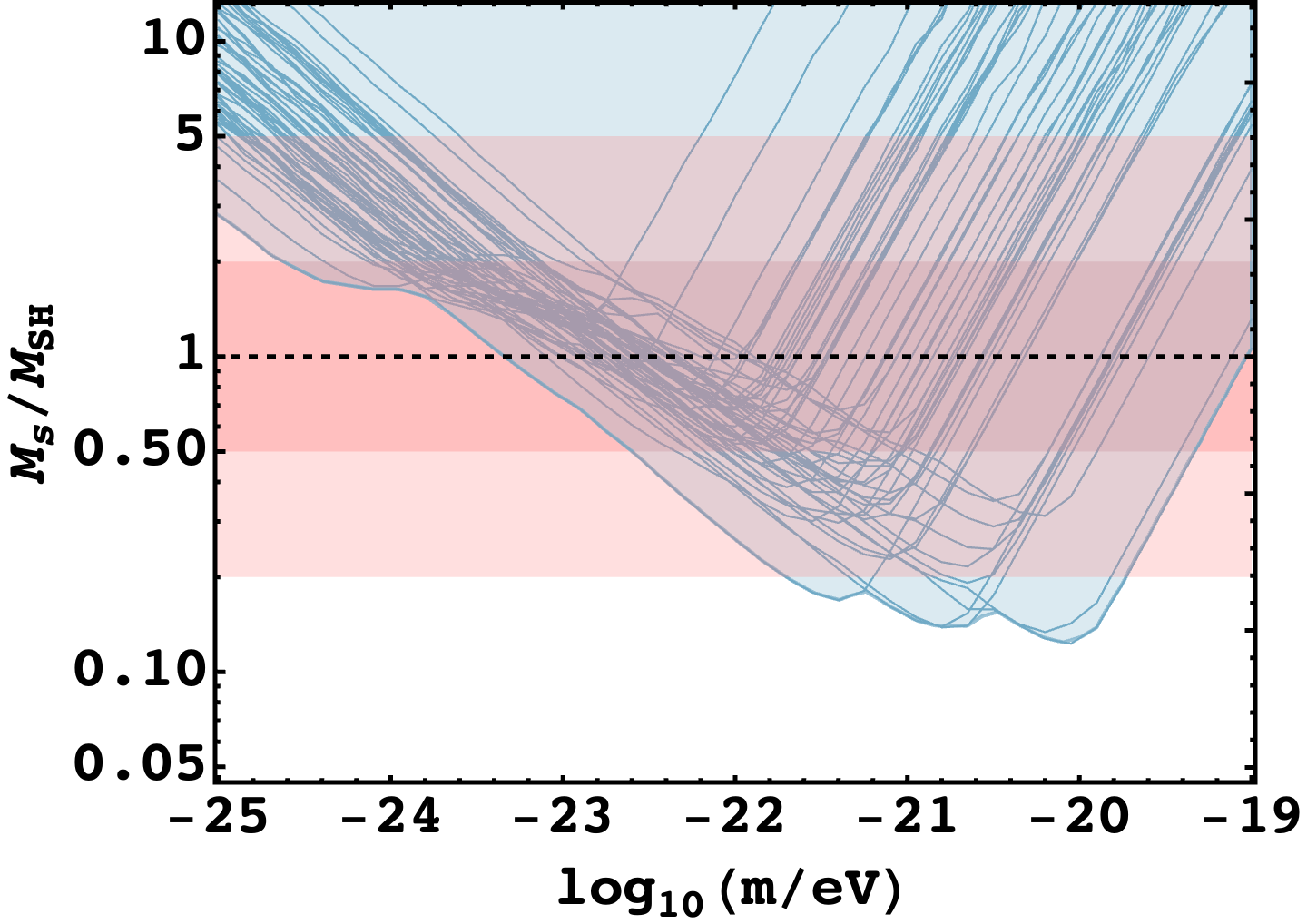}
    \caption[Constraint on $m$ from 56 LSB galaxies for $\lambda = 0$]{Here, $\hat{\lambda}_\text{ini} = 0$ is fixed and $m$ is allowed to vary. The dark and light shaded pink regions correspond to a scatter of factor of 2 and 5 from the SH relation in eq.~(\ref{eq:Schive_SH}) respectively.}
    \label{fig:LSB_no_inter}
\end{figure}

\noindent In the special case of SFDM with no self-interactions ($\hat{\lambda}_\text{ini} = 0$) i.e. FDM, the only free parameter is $s$ (when $m$ is fixed). Furthermore, soliton mass expected from soliton halo relation i.e. $M_{SH}$, is given by eq.~(\ref{eq:Schive_SH}) and hence is independent of $s$. This implies that smaller values of the ratio $M_s / M_{SH}$ corresponding to $s > s_{\text{crit}}$ will be allowed by rotation curves. Ref.~\cite{Bar_2022} found that for many galaxies the values of $M_s$ allowed by rotation curves are smaller than $\sim 0.5M_{SH}$ when $m$ is allowed to vary within the range $\left[10^{-24}\ \text{eV}, 10^{-20}\ \text{eV}\right]$ implying that the SH relation is not satisfied (assuming a scatter of a factor of 2 in the SH relation eq.~(\ref{eq:Schive_SH})) for solitons that are allowed by rotation curves. To verify this, we first conducted an analysis for $\hat{\lambda}_\text{ini} = 0$ and varied $m$ in the range $\left[10^{-25}\ \text{eV}, 10^{-19}\ \text{eV}\right]$ for the 56 LSB galaxies in our sample. The results are shown in figure~\ref{fig:LSB_no_inter}. The asymptotic dependence of $M_s/M_{SH}$ on $m$ here is consistent with what is expected from the work done in Ref.~\cite{Bar_2022} (i.e. $M_s/M_{SH} \propto m^{-1/2}$ for small $m$ and $M_s/M_{SH} \propto m$ for large $m$). The galaxy that imposes the strongest constraint for $m = 10^{-22}\ \text{eV}$ is `IC 2574', where all soliton masses with $M_s/M_{SH} \gtrsim 0.2$ are excluded.

However it is worth noting that while the general idea of the exclusion is the same as in Ref.~\cite{Bar_2022}, our approach is slightly different. For instance, as discussed earlier, overshooting is defined when $\chi^2_i$ at any single radius bin exceeds 1. Further, we also utilize the exact SH relation in eq.~(\ref{eq:modified_SH}) (which reduces to eq.~(\ref{eq:Schive_SH}) when $\hat{\lambda}_\text{ini} = 0$) which requires an input of $M_h$. This, along with the assumptions given earlier suggest that our results are not expected to be exactly identical.   

\vspace{0.5cm}
\noindent \textbf{\emph{Presence of repulsive self-interactions}}
\vspace{0.25cm}

\noindent Let us now investigate what happens in the presence of self-interactions. For fixed values of $M_h$, $m$ and $\hat{\lambda}_\text{ini}$, every value of $s$ will also correspond to a different soliton mass $M_{SH}$ expected from the SH relation in eq.~(\ref{eq:modified_SH}), since $M_{SH} \propto \sqrt{1 + As^2}$ ($A$ is just some numerical factor). As $s$ decreases, while soliton mass $M_s$ increases, $M_{SH}$ decreases, causing the ratio $M_s/M_{SH}$ to be larger. Hence for every $\hat{\lambda}_\text{ini}$, all $M_s/M_{SH} > M_s^{\text{crit}}/M_{SH}^{\text{crit}}$ correspond to $s < s_{\text{crit}}$. As discussed earlier, this causes $V_{DM} > V_{obs}$ and the corresponding $M_s/M_{SH}$ is excluded. The solid blue curve in figure~\ref{fig:eg_exclusion_region} is the ratio $M_s^{\text{crit}}/M_{SH}^{\text{crit}}$ for different $\hat{\lambda}_\text{ini}$, where the filled region above it represents the excluded soliton masses. Note that we do not expect the SH relation to be satisfied exactly. Here, the dark and light shaded pink regions in figure~\ref{fig:eg_exclusion_region} represent a scatter from eq.~(\ref{eq:modified_SH}) of factors of 2 and 5 respectively. As $\hat{\lambda}_\text{ini}$ increases, more of the region around $M_s = M_{SH}$ is allowed by observed rotation curves. This demonstrates that the requirement of satisfying SH relation as well as observed rotation curves can impose constraints on self-coupling of ultra-light scalar field dark matter.

The rise of the boundary curve in figure~\ref{fig:eg_exclusion_region} can be understood in the following manner: We know from the unscaled curve in figure~\ref{fig:m_vs_r_scaled} that as $\hat{\lambda}_\text{ini}$ increases $\hat{M}_\text{ini}$ also increases. Due to the shape of the inner region of the observed rotation curve, we also find that $s^{\text{crit}}$ increases with $\hat{\lambda}_\text{ini}$. From our earlier discussion, a larger $s^{\text{crit}}$ implies a smaller $M_s^{\text{crit}}$ and a larger $M_{SH}^{\text{crit}}$ leading to an overall smaller $M_s^{\text{crit}}/M_{SH}^{\text{crit}}$. However, the amount of increase in $\hat{M}_\text{ini}$ outweighs the decrease in $M_s^{\text{crit}}/M_{SH}^{\text{crit}}$ which leads to the rise of the boundary curve. 

\begin{figure}[h!]\captionsetup[subfigure]{font=footnotesize}
    \centering
    \includegraphics[width = \textwidth]{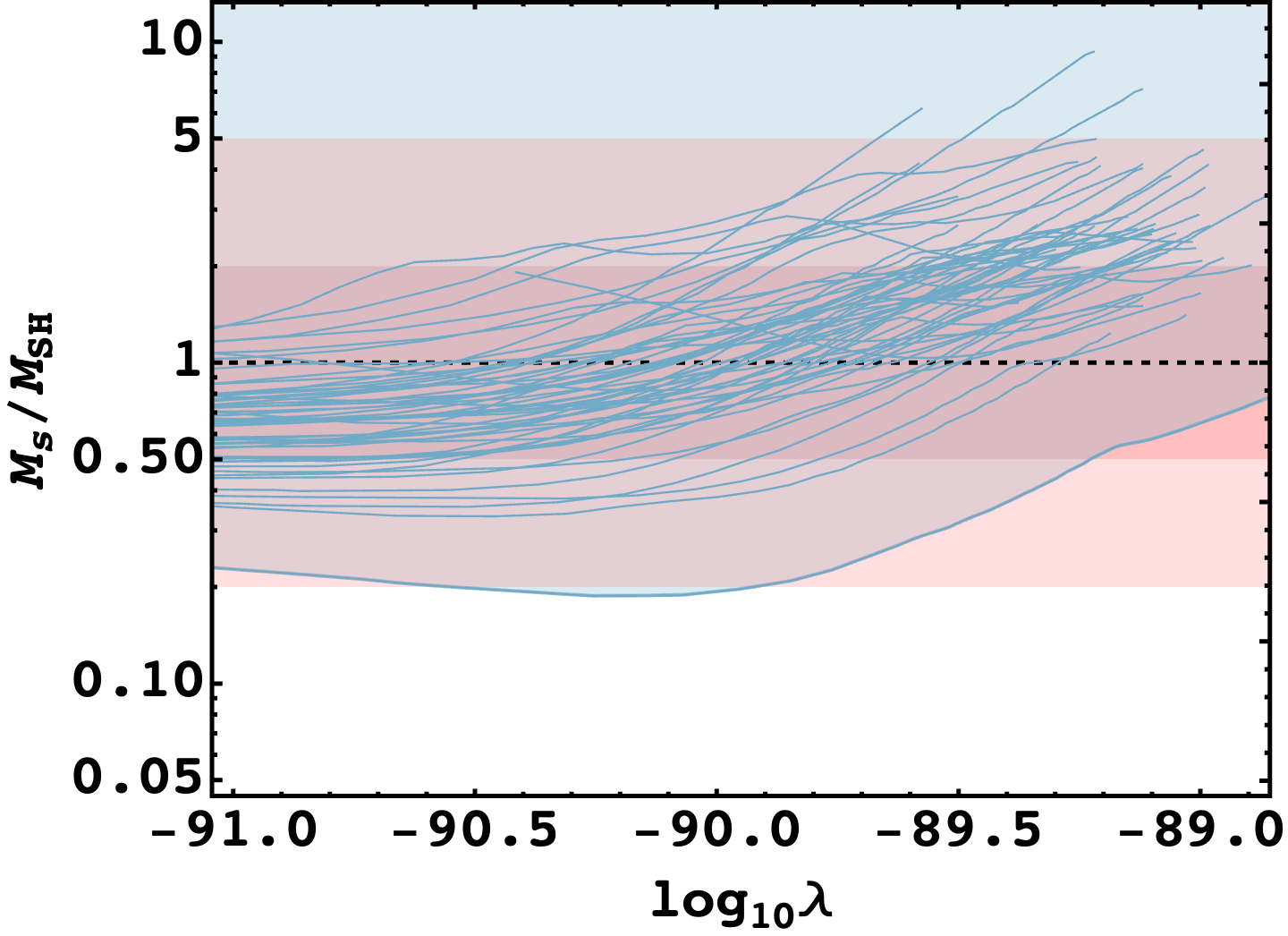}
    \caption[Constraint on $\lambda > 0$ from 56 LSB galaxies for $m = 10^{-22}\ \text{eV}$]{Here, $m = 10^{-22}\ \text{eV}$ is fixed and $\hat{\lambda}_\text{ini}$ is allowed to vary. The dark and light shaded pink regions correspond to a scatter of factor of 2 and 5 from eq.~(\ref{eq:modified_SH}) respectively.}
    \label{fig:LSB_with_inter}
\end{figure}

We repeat the above procedure for the remaining 55 galaxies in our sample and obtain figure~\ref{fig:LSB_with_inter} (note that in the horizontal axis we use $\lambda = 8s^2\hat{\lambda}_\text{ini}(M_{pl}/m)^2$). For all galaxies, the boundary between the shaded and un-shaded region is pushed upwards as $\lambda$ increases. The strongest constraints are imposed by the galaxy `IC 2574' where, for $\lambda \sim 10^{-91}$, the ratio $M^{\text{crit}}_s/M^{\text{crit}}_{SH} \sim 0.2$. As $\lambda$ increases the boundary of the excluded region is pushed upwards. This means that for large repulsive self-interactions $\lambda > \mathcal{O}(10^{-90})$, a larger region that satisfies the SH relation is allowed by rotation curves as shown in figure~\ref{fig:LSB_with_inter}. The dark and light shaded pink regions correspond to a scatter of a factor of 2 and 5 around $M_s = M_{SH}$ respectively. Note that eq.~(\ref{eq:Schive_SH}) is obtained from simulations, while eq.~(\ref{eq:modified_SH}) is its extension in the presence of self-interactions and hence we do not have an estimate for the scatter in the relation. 

Also note that some of the blue curves in figure~\ref{fig:LSB_with_inter} have a larger minimum $\lambda$. This is because the numerical solution has a finite size, requiring a larger $\hat{\lambda}_\text{ini}$ for the numerically calculated soliton to be large enough to reach the first observed radius bin. This leads to a larger minimum $\lambda$. On the other hand, for $\hat{\lambda}_\text{ini} = 3.5$ (the largest value for which we obtained a numerical solution), every galaxy will allow a different value of $s_{\text{crit}}$, leading to different values of maximum $\lambda$. For some galaxies the combination of these two effects lead to a smaller curve in figure~\ref{fig:LSB_with_inter}.

It must be noted that we have made assumptions in the beginning of this section, relaxing which could change the results of our analysis. From section~\ref{sec:regimes}, $\hat{\lambda}_\text{ini} > 2.5$ implies that we are in the Thomas-Fermi (TF) regime. This is close to the values of $\hat{\lambda}_\text{ini}$ required to push the boundary upwards sufficiently for many galaxies. For SFDM in the TF regime (SFDM-TF), there already exist constraints on $\lambda/m^4$ from background evolution for a complex scalar field \cite{Li_2014, Dawoodbhoy_2021}. In particular, requirement of a timely transition from radiation to matter domination from CMB power spectrum imposes an upper limit on $\lambda/m^4$: For $m = 10^{-22}\ \text{eV}$ this corresponds to $\lambda \leq 10^{-89.38}$ for a real scalar field, which can further constrain the value of $\lambda$ we can allow in our analysis.

\subsection{Impact of attractive self-interactions}\label{sec:why_no_att_inter}

As discussed in section \ref{sec:thesis_overview}, axions can have attractive self interactions corresponding to a negative $\lambda$. However, as we have mentioned in section~\ref{sec:mass_radius_curves}, too strong attractive self-interactions lead to solutions that are unstable under small perturbations. The transition from stable to unstable solutions occurs at $\hat{\lambda}_\text{ini} = -0.4$ (see discussion in section~\ref{sec:max_mass}). This sets the allowed range of $\hat{\lambda}_\text{ini}$ to be between 0 and -0.4. 

Also note that the second term in the square-root in the SH relation in eq.~(\ref{eq:modified_SH}) will be negative for attractive self-interactions. For a fixed value of $M_h$, a large enough $s^2\hat{\lambda}_\text{ini}$ will lead to an imaginary $M_{SH}$ which is unphysical. Therefore, given the small range of allowed $\hat{\lambda}_\text{ini}$ and the form of modified SH relation in eq.~(\ref{eq:modified_SH}), the presence of attractive self-interactions is not expected to improve constraints from the analysis carried out in this chapter. 

We conducted a similar analysis to the one we did for repulsive self-interactions in section~\ref{sec:imposing_mod_SH_relations} and found that within the allowed values of $\hat{\lambda}_\text{ini}$ the boundary of excluded region is not altered a lot for most galaxies in our sample. This is also seen from the exercise in section~\ref{sec:PVC} where, we compare velocity curves for different self-interaction strengths (attractive and repulsive) but the same peak velocity. From figure~\ref{fig:eg_slope_peak} note that the velocity curve for the strongest allowed attractive self-interaction strength, i.e. $\hat{\lambda}_\text{ini} = -0.4$ (red curve) is not very different from the velocity curve for no self-interaction case. 

\subsection{Impact of changing scalar field mass}\label{sec:impact_of_mass}
\begin{figure}[h!]\captionsetup[subfigure]{font=footnotesize}
    \centering
    \includegraphics[width = 0.6\textwidth]{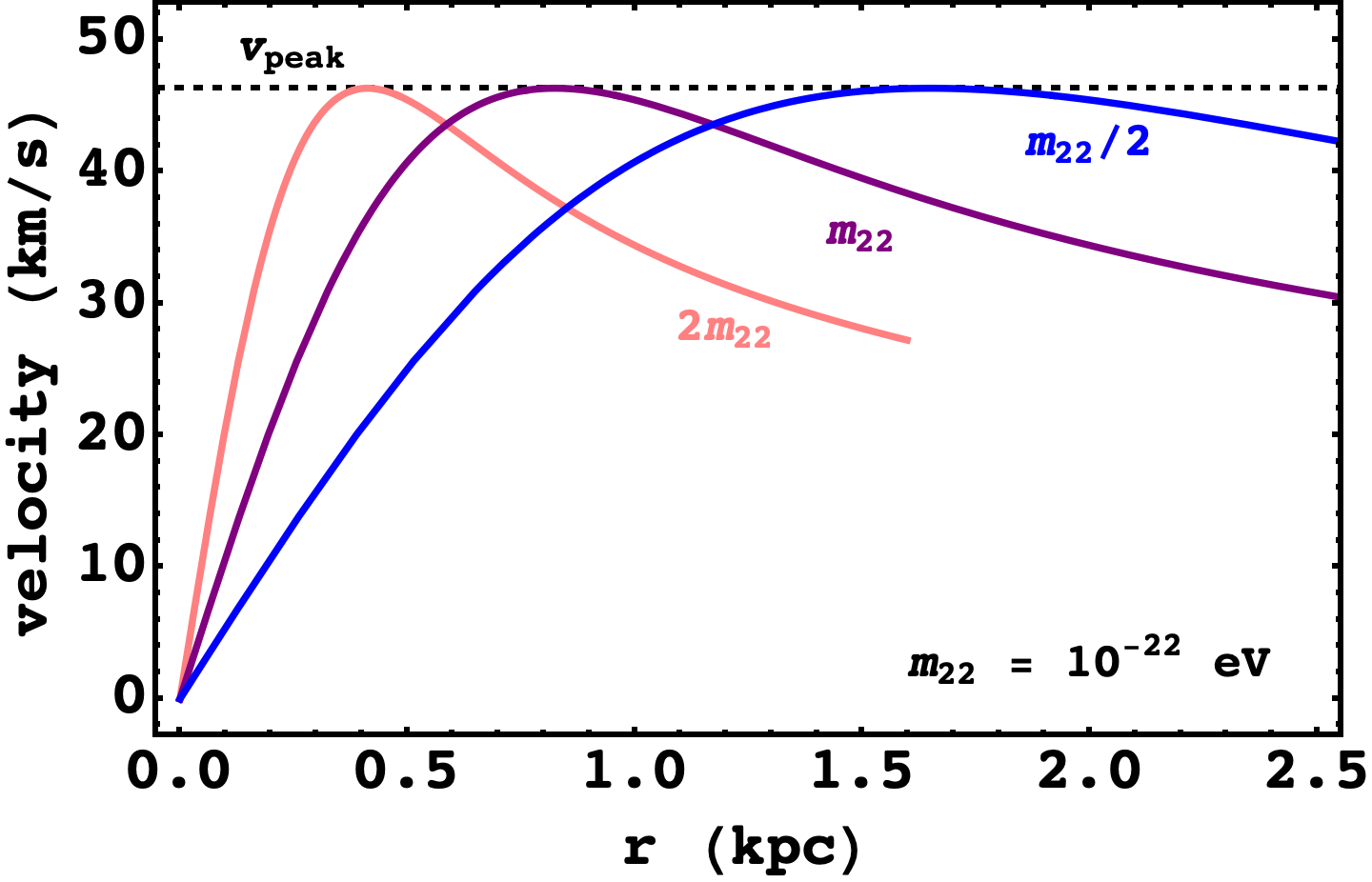}
    \caption[Impact of parameter $m$ on velocity curves]{Velocity curves for different values of $m$ are plotted using different colours. Here values of $s = 5000$ and $\hat{\lambda}_\text{ini} = 0$ are fixed. The dashed horizontal lines denotes the peak velocity of each curve which remains unaffected by a change in $m$.}
    \label{fig:mass_effect}
\end{figure}

It is worth noting that in figure~\ref{fig:LSB_with_inter} we have kept $m$ fixed at its fiducial value of $10^{-22}\ \text{eV}$. It is then important to ask how changing $m$ changes (a) the velocity curves and (b) the behaviour of the boundary in figure~\ref{fig:LSB_with_inter}. The effect of changing $m$ on the velocity curves is shown in figure~\ref{fig:mass_effect}.

\begin{figure}[h!]\captionsetup[subfigure]{font = {footnotesize}}
    \centering
    \begin{subfigure}[t]{0.45\textwidth}
        \centering
        \includegraphics[width = \textwidth, height=2in]{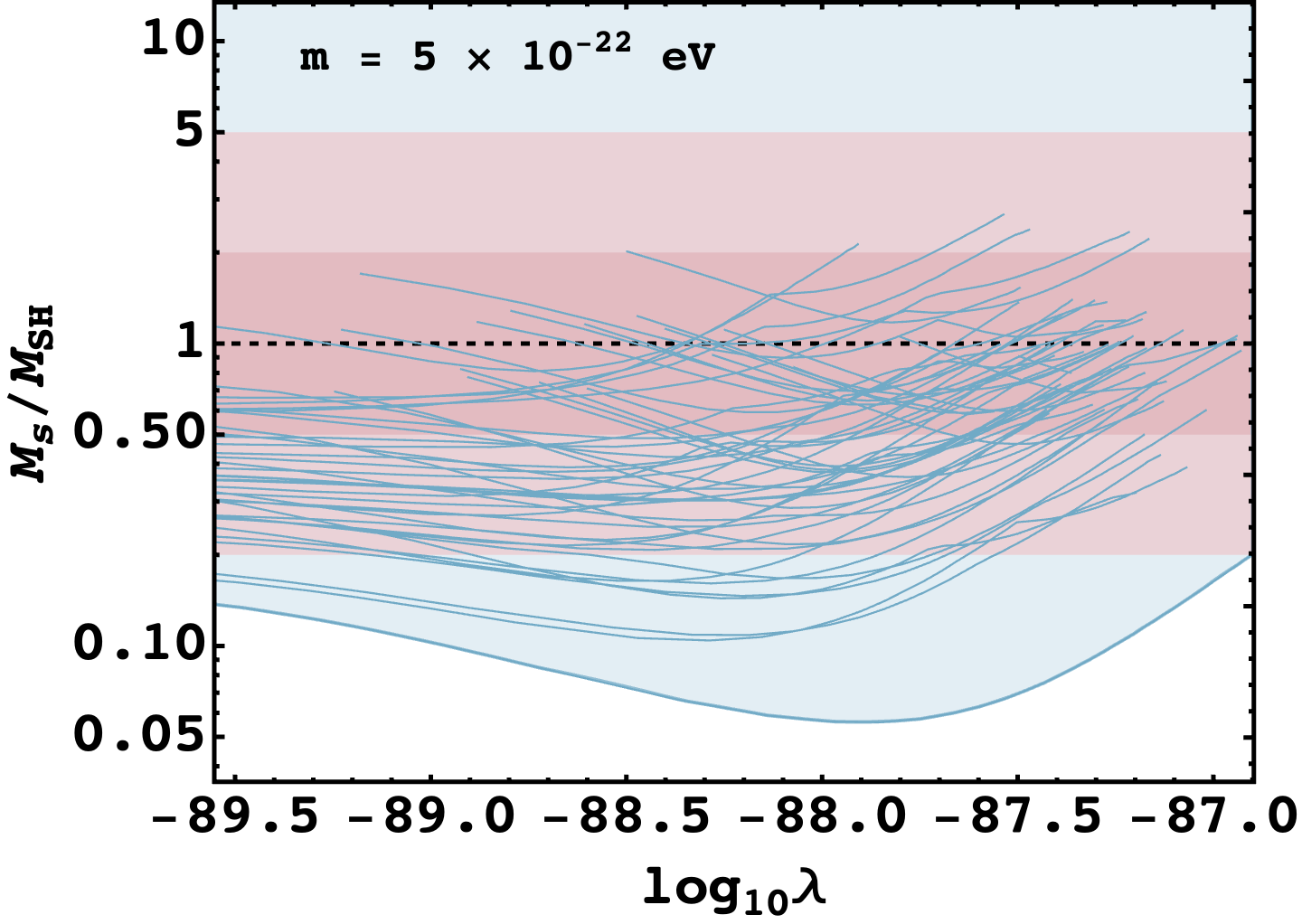}
        \caption{Here, $m = 5\times 10^{-22}\ \text{eV}$ is fixed and $\hat{\lambda}_\text{ini}$ is allowed to vary. The dark and light shaded pink regions correspond to a scatter of factor of 2 and 5 from eq.~(\ref{eq:Schive_SH}) respectively.}
        \label{fig:LSB_522}
    \end{subfigure}
    \hfill
    \begin{subfigure}[t]{0.45\textwidth}
        \centering
        \includegraphics[width = \textwidth, height=2in]{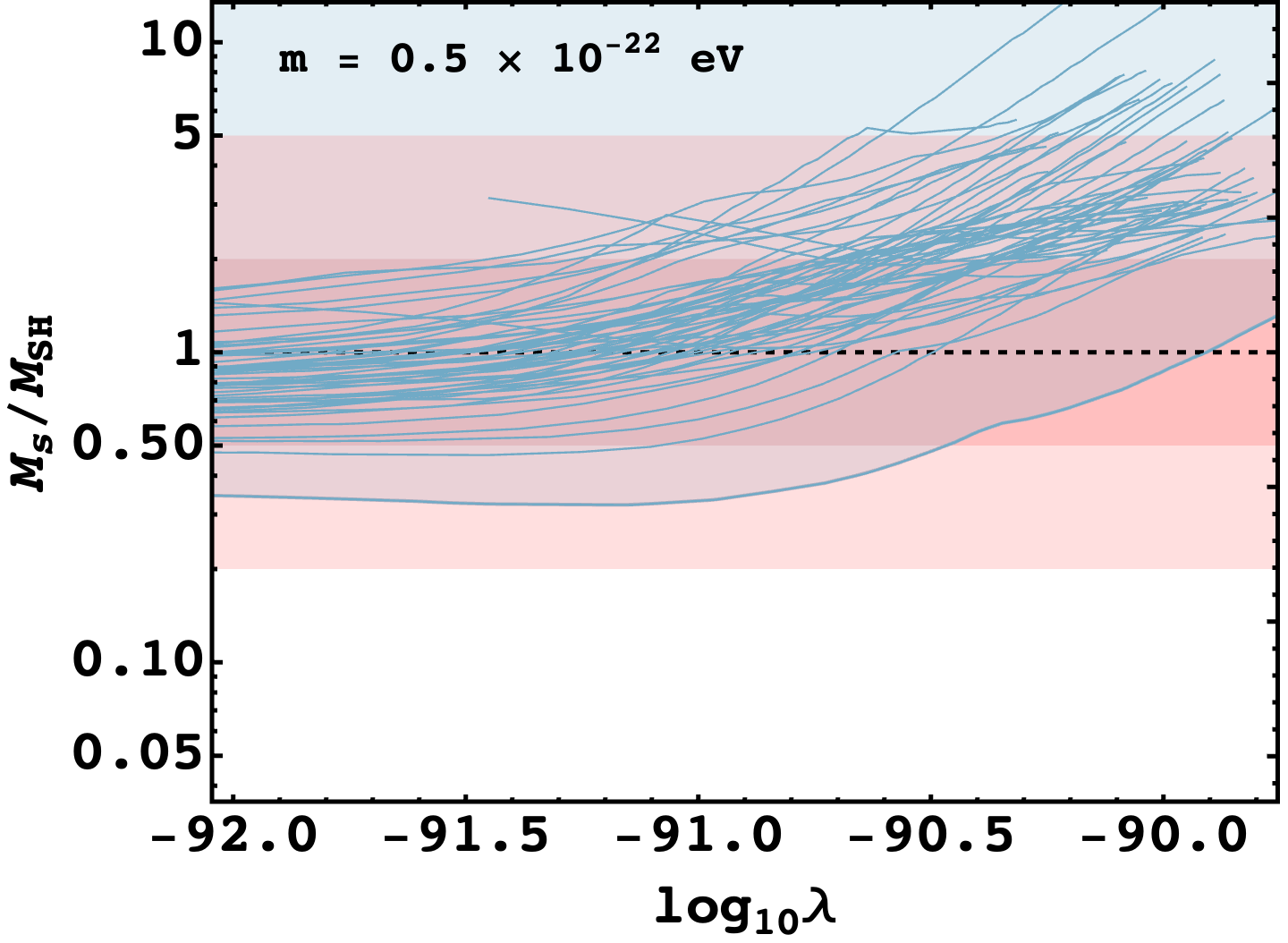}
        \caption{Here, $m = 0.5\times 10^{-22}\ \text{eV}$ is fixed and $\hat{\lambda}_\text{ini}$ is allowed to vary. The dark and light shaded pink regions correspond to a scatter of factor of 2 and 5 from eq.~(\ref{eq:modified_SH}) respectively.}
        \label{fig:LSB_523}
    \end{subfigure}
    \caption[Impact of changing $m$ on constraints for $\lambda > 0$]{Plotting $M_s^{\text{crit}}/M_{SH}^{\text{crit}}$ for a different fixed values of $m$ for 56 LSB galaxies from SPARC, where soliton masses in blue region are excluded by the data.}
    \label{fig:LSB_diff_m}
\end{figure}
From figure~\ref{fig:LSB_no_inter} it is also evident that compared to $m = 10^{-22}\ \text{eV}$, FDM masses in the range $10^{-22}$ -- $10^{-20}\ \text{eV}$ are more constrained while $m < 10^{-22}\ \text{eV}$ are less constrained by the LSB sample used. It is then important to verify whether the upwards movement of boundary occurs even when one considers a different $m$. To demonstrate this, we consider two values of scalar field mass ($m = 0.5\times 10^{-22}\ \text{eV}$ and $m = 5\times 10^{-22}\ \text{eV}$) and repeat the procedure. We find that the general behaviour of the boundary is similar to that for $m = 10^{-22}\ \text{eV}$ (see figure~\ref{fig:LSB_diff_m}). Note that since $\lambda = 64\pi\hat{\lambda}\left(\frac{m}{m_{pl}}\right)^2$, a larger value of $m$ will probe larger values of $\lambda$ for the same $\hat{\lambda}$, while the opposite is true for smaller values of $m$.

\subsection{Peak velocity condition}\label{sec:PVC}

Authors in \cite{Bar_2018} showed that for a FDM ($\lambda = 0$) core surrounded by a NFW halo, the SH relation in eq.~(\ref{eq:Schive_SH}) is equivalent to the soliton peak velocity being approximately equal to the halo peak velocity. This is also called the `velocity dispersion tracing' as seen in \cite{Chavanis_SH_relation_2019, Chavanis_2021, Padilla_2021}. If this `peak velocity condition' (PVC) is imposed, the authors found that for $m \in \left(10^{-22}\ \text{eV}, 10^{-21}\ \text{eV}\right)$ FDM over-predicts velocities in the inner region for dark matter dominated galaxies. In other words, FDM velocity curves can either obtain the correct peak velocity or the observed slope of the inner region but cannot obtain both simultaneously. 

It is then natural to ask: Can self-interactions help? We try to answer this question in this section. 
Before proceeding, tt is important to note that while PVC was obtained for eq.~(\ref{eq:Schive_SH}), we assume that satisfying PVC is also equivalent to satisfying the SH relation in eq.~(\ref{eq:modified_SH}). This may not be true in general since the expression for total energy of the soliton will have extra terms due to self-interactions.

The `peak velocity condition' (PVC) is used as a proxy for SH relation in eq.~(\ref{eq:Schive_SH}) \cite{Bar_2018} where the peak velocity in the halo turns out to be roughly equal to the peak velocity in the soliton. Without obtaining halo fits, one can estimate halo peak velocity for a galaxy as the DM velocity ($V_{DM}$) measured in the flat region of the rotation curve sufficiently far from the centre. We estimate this as $V_{DM}$ obtained at the farthest observed radius bin \cite{Bar_2018}.

For this section, in addition to the LSB condition, we also consider dwarf galaxies with halo masses $M_h \sim 10^{10}\ M_\odot$, with maximum velocities $v_{\text{max}} \sim 80\ \text{km/s}$ bringing our number down to 36 galaxies. This constraint comes from the limitation of our numerical approach where solutions for $\hat{\lambda}_\text{ini} > 3.5$ are difficult to obtain due to the fine-tuned initial guesses required for $\hat{\gamma}$. We further note that not all galaxies in the sample have flat rotation curves at farthest measured radii, i.e. there are no measurements far enough from the center to establish a halo peak velocity. Hence, we shall discard galaxies with no established $V_{flat}$ (provided by the SPARC catalogue) for this analysis, leaving us with 17 galaxies. We first fix $m = 10^{-22}\ \text{eV}$ to focus on the effect of self-interactions. We vary $\{\hat{\lambda}_\text{ini}, s\}$ and minimize reduced $\chi^2$ \footnote{Reduced $\chi^2$ is defined as $\chi^2_{\text{red}} = \frac{1}{N-K}\sum_i^N\frac{(v_i^{\text{pred}} - v_i^{\text{pred}})^2}{\sigma_i^2}$ where $N$ is the number of data points and $K$ is the number of free parameters.} for all 17 galaxies while also requiring that the peak velocity condition is satisfied within some scatter. First, we make the following assumptions:

\begin{enumerate}    
    \item SPARC database also provides contribution of baryonic components like gas, disk and bulge to the total observed velocity for every galaxy. Using the best fit value $\Upsilon_d = 0.5$ \cite{Lelli_2016}, we obtain DM-only rotation curves $V_{DM}(r)$ from eq.~(\ref{eq:observed_vel}). 

    \item We are only interested in whether SFDM can describe inner regions of galaxies and hence do not try to fit the NFW envelope. We assume that at some $r = r_t$, the NFW profile takes over and free parameters $\{r_t, r_s\}$ can be adjusted to fit the rest of the rotation curve. 

    \item We also assume that $r_t = r_{95}$, ensuring that NFW envelope takes over sufficiently far from the radius ($r_p$) at which the soliton velocity peaks (see figure~\ref{fig:density_velocity}). Hence, we consider data-points where the corresponding radius bin $r_{obs} \leq r_{95}$. Given that we have two free parameters $\{\hat{\lambda}_\text{ini}, s\}$, we only allow parameters for which the number of data-points within $r_{95}$ is at least 3. 

    \item Soliton and halo peak velocities are denoted by $v_p^{(s)}$ and $v_p^{(h)}$ respectively. We allow for a scatter of a factor of 2 from $v_p^{(s)} = v_p^{(h)}$ which is the allowed scatter in eq.~(\ref{eq:Schive_SH}) \cite{Schive_PRL_2014, Bar_2018}. Hence, only parameters satisfying $0.5 < v_p^{(s)}/v_p^{(h)} < 2$ are considered.
\end{enumerate}
To illustrate the effect of imposing PVC on SFDM solitons with attractive or repulsive self-interactions, we first consider the velocity curve for the galaxy `UGC 1281' for which $v_p^{(h)} = 51.5\ \text{km/s}$.

\begin{figure}[h]\captionsetup[subfigure]{font=footnotesize}
    \centering
    \includegraphics[width = 0.6\textwidth]{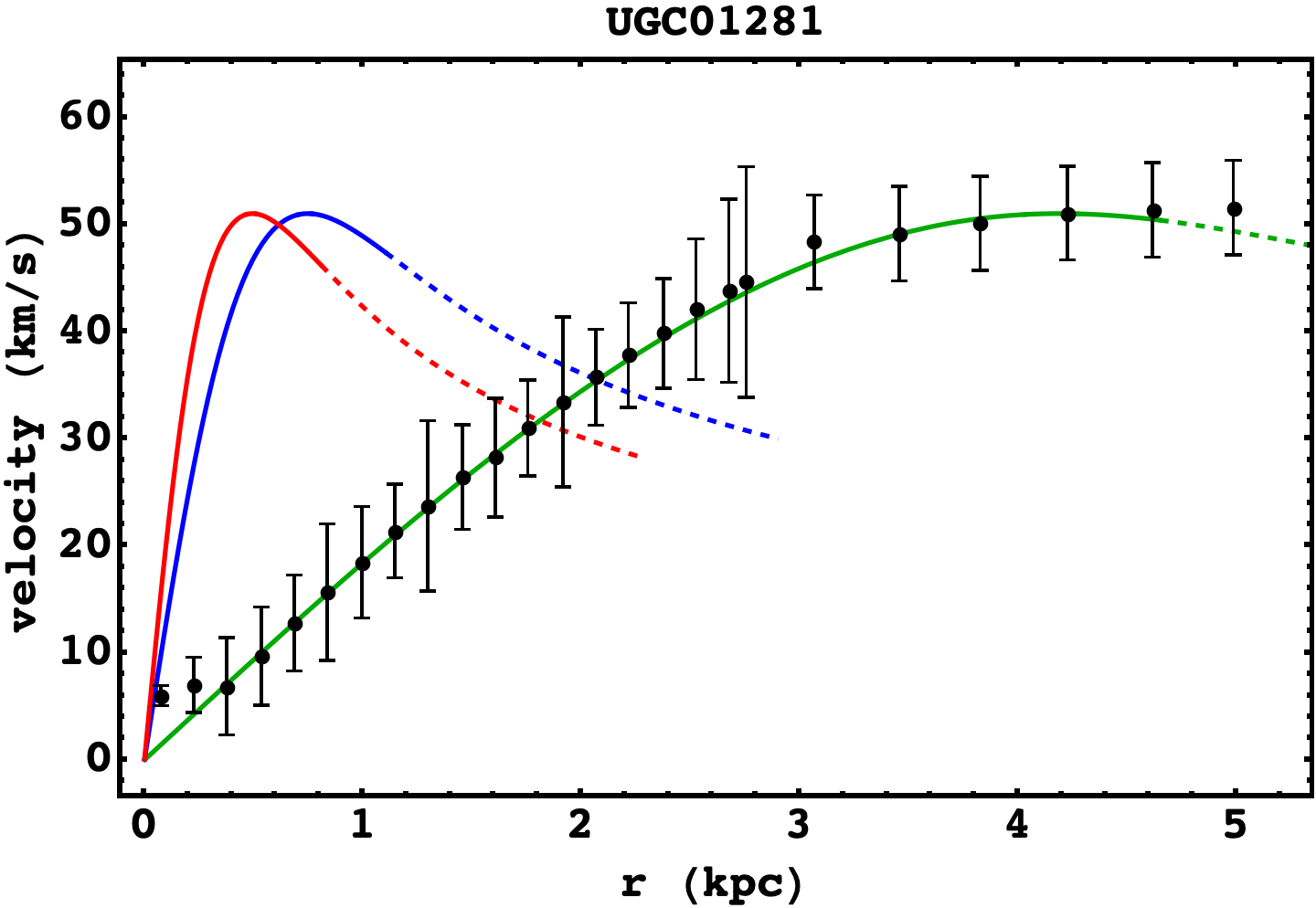}
    \caption[Imposing peak velocity condition for different $\lambda$ for UGC 1281]{The green curve represents a soliton with $\hat{\lambda}_\text{ini} = 2$ while the blue and red curves correspond to $\hat{\lambda}_\text{ini} = 0$ and $\hat{\lambda}_\text{ini} = -0.4$ respectively. Dashed curves correspond to $r > r_{95}$. The corresponding scaled self-coupling strength is $\lambda = 3.98\times 10^{-90}$ for repulsive self-interactions.}
    \label{fig:eg_slope_peak}
\end{figure}

\begin{figure}[h]
    \centering
    \includegraphics[width = 0.8\textwidth]{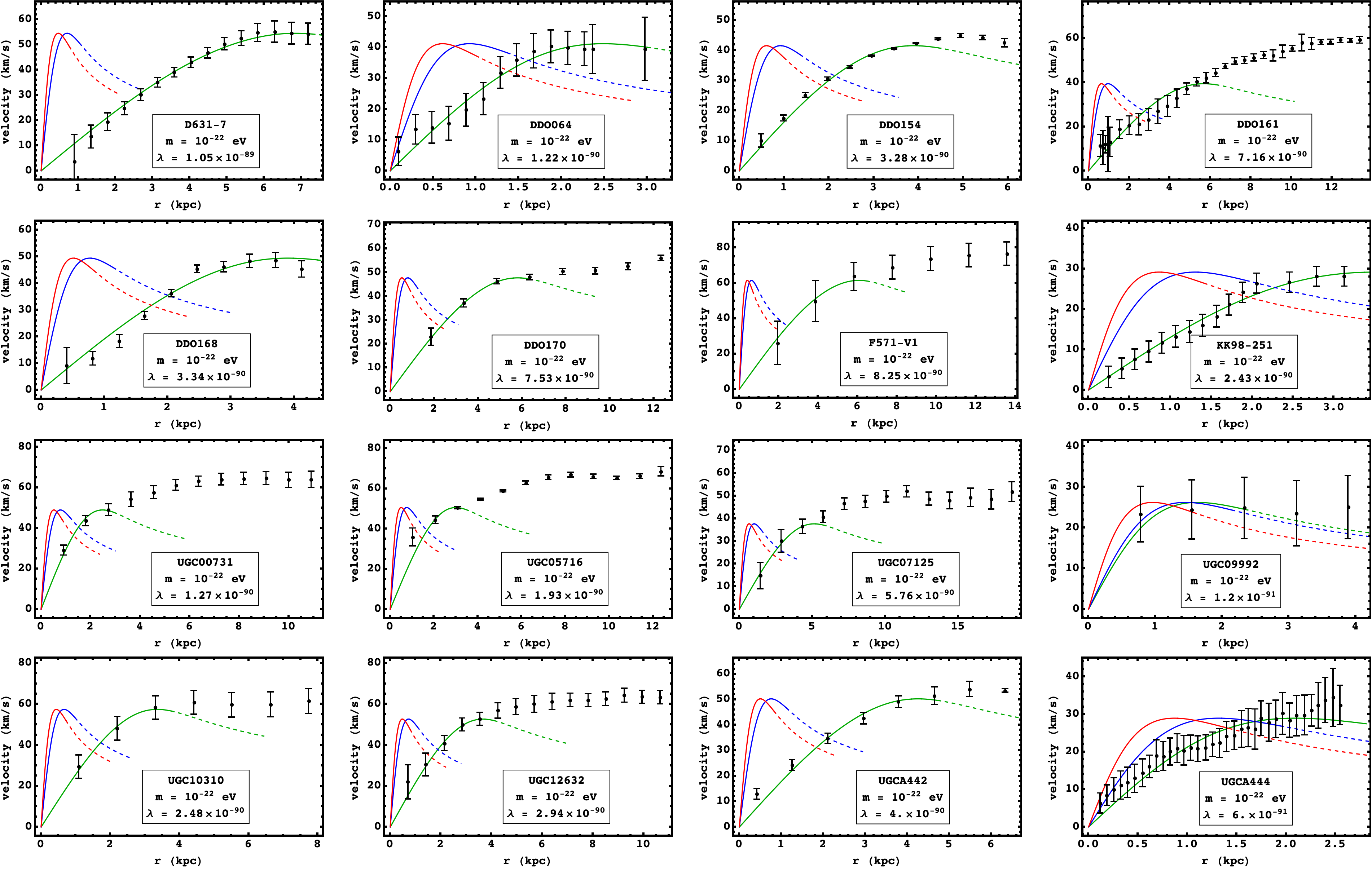}
    \caption[Peak velocity condition imposed on the remaining 16 galaxies for different $\lambda$ values]{Remaining 16 galaxies from LSB galaxy sample. Blue curves represent solitons with no self-interactions, while red and green are solitons with attractive ($\hat{\lambda}_\text{ini} = -0.4$) and repulsive self-interactions respectively. For all curves, $50\%$ deviation from PVC is allowed. The dashed part of the soliton curves represents $r > r_{95}$. See main text for discussion.}
    \label{fig:LSB_PVC}
\end{figure}

\begin{enumerate}
    \item \textbf{No Self-Interactions:} The blue curve in figure~\ref{fig:eg_slope_peak} corresponds to the velocity curve of a soliton with $\lambda = 0$ and $m = 10^{-22}\ \text{eV}$. Here the only free parameter is the scale $s$, which is chosen to satisfy $v_p^{(s)} = 51\ \text{km/s} = v_p^{(h)}$. The slope of the resultant soliton velocity curve is quite steeper than the slope of the observed curve. From the discussion in section~\ref{sec:impact_of_parameters}, it is clear that if one tries to reduce the slope by increasing $s$, peak velocity will consequently be smaller. Hence, for $\lambda = 0$, one cannot satisfy the observed slope and peak simultaneously, which is in agreement with the results of \cite{Bar_2018}.

    \item \textbf{Attractive Self-Interactions:} Now, we consider $\lambda < 0$ and allow $\hat{\lambda}_\text{ini}$ to vary. For every $\hat{\lambda}_\text{ini}$ one can choose a scale $s$ such that $v_p^{(s)} \approx v_p^{(h)}$. However, since we require a stable soliton, we cannot consider any $\hat{\lambda}_\text{ini} < -0.4$ (see section~\ref{sec:mass_radius_curves}), limiting the effect of attractive self-interactions. Moreover, as discussed in section~\ref{sec:impact_of_parameters} presence of attractive self-interactions squeezes the velocity curve in the direction of the slope requiring a smaller $s$ value to attain the same soliton peak velocity as the $\lambda = 0$ case. This leads to an even steeper slope, worsening the prediction (see red curve in figure~\ref{fig:eg_slope_peak}).
    
    \item \textbf{Repulsive Self-Interactions:} Finally, we allow $\hat{\lambda}_\text{ini} > 0$ to vary and find that a large $\hat{\lambda}_\text{ini} = 2$ allows one to satisfy the peak velocity condition without over-predicting velocities in the inner region. One can choose an $s$ to satisfy the observed slope in the inner region, while $\hat{\lambda}_\text{ini}$ can be varied to obtain the correct peak velocity. This is because a larger $\hat{\lambda}_\text{ini}$ will lead to a larger soliton, which alters the peak velocity while keeping the slope of the inner region roughly unchanged (see section~\ref{sec:impact_of_parameters}). For `UGC 1281' $\hat{\lambda}_\text{ini} = 2$ and $s = 11942$ satisfies the observed slope and peak simultaneously. The corresponding scaled value of self-coupling is $\lambda = s^2\lambda_\text{ini} = 3.98\times 10^{-90}$. 
\end{enumerate}
Velocity curves corresponding to parameters with minimum $\chi^2_{\text{red}}$ for the remaining 16 galaxies are shown in figure~\ref{fig:LSB_PVC} by the green curves. Note that dimensionless self-coupling $\hat{\lambda}_\text{ini}$ is allowed to vary within the range $\left[-0.4, 3.5\right]$ ensuring stable solutions (for attractive self-interactions).

For all galaxies, SFDM with repulsive self-interactions better describes the slopes of inner regions. Curves for attractive self-interactions using $\hat{\lambda}_\text{ini} = -0.4$ (red) and no self-interactions (blue) with the same soliton peak velocity as the best-fit case are also shown for comparison. `DDO 161' exhibits the highest deviation from PVC ($v_p^{(s)} \sim v_p^{(h)}/2$) for the least $\chi^2_\nu$ value, while describing only a part of the linearly increasing region. On the other hand, in many cases velocity curves of solitons with $\hat{\lambda}_\text{ini} > 0$ trace out a large portion of the observed rotation curve. Attractive self-interactions however, already constrained to be small to ensure stability, do not appear to be well probed by the rotation curve data presented here. Even with their limited effect on velocity curves, they seem to fare worse, predicting steeper slopes than even the no self-interactions case for the same peak velocity. 

For repulsive self-interactions, we find that large values of self-coupling ($\hat{\lambda}_\text{ini} \geq 1$) are preferred for many galaxies. The values of $\hat{\lambda}_\text{ini}$ seems to suggest that the solitons are close to the Thomas-Fermi regime (see section~\ref{sec:regimes}). However, given the diversity of slopes and sizes of inner regions, it appears to be unlikely that a single value of $\lambda$ will fit all rotation curves. The order of magnitude of self-interaction strength is $\mathcal{O}(10^{-90})$, with an average value $\lambda \sim 3.92\times 10^{-90}$. We stress that we do not attempt to conduct a full parameter estimation and perform only a crude analysis to study the effects of self-interactions for a fixed $m = 10^{-22}\ \text{eV}$. An important caveat here is that we do not allow for the baryonic contribution to change ($\Upsilon_d = 0.5$ is fixed) or fit the outer envelope using the NFW profile. One can potentially tune these components appropriately to obtain best fits for a fixed $\lambda$, which is left for future work.

\section{Summary and discussion}\label{sec:summary}

In order to learn about the nature of Dark Matter, the spin, mass, couplings and other fundamental properties of DM particles need to be uncovered. We considered spinless DM particles which are ultra light ($m \sim 10^{-22}$ eV). 
What could be the self-coupling strength of these particles? For axions with mass $m_a$ and decay constant $f_a$, the self coupling is suppressed and is given by $\lambda_a = - \left( \frac{m_a}{f_a} \right)^2$ and for ULAs forming dark matter, turns out to be $- {\cal O}(10^{-96})$. For other particles, the self-coupling could be positive and much larger. 

Impact of self-interacting scalar field dark matter (SFDM) on galactic rotation curves has been studied extensively in recent years. For instance in \cite{Craciun_2020, Harko_2022} authors test the SFDM model (with a complex scalar) in the TF regime against rotation curves from SPARC. They also consider additional contributions from the global rotation of the halo, random confining potentials and baryonic matter distribution. Authors in \cite{Dawoodbhoy_2021} also work in the TF regime. They show that an inner core described by SFDM-TF surrounded by a NFW envelope fits high mass dwarf galaxies ($M_h \sim 10^{11}\ M_\odot$) better than CDM or FDM ($\lambda = 0$) for $m = 0.8\times 10^{-22}\ \text{eV}$. On the other hand, authors in \cite{Delgado_2022} use a Gaussian ansatz for $\lambda > 0$ and try to fit the inner regions of 17 bulgeless galaxies from the SPARC catalogue. They obtain a best-fit value of $\lambda \sim 2\times 10^{-90}$ and $m = 2.2\times 10^{-22}\ \text{eV}$. In this paper we did not assume a particular approximation (e.g. the TF approximation) or an ansatz to estimate density profiles. We deal directly with the numerical solutions of Gross-Pitaevskii-Poisson (GPP) equations. 

Recently \cite{Bar_2018, Bar_2022}, it has been argued that if FDM in the mass-range $10^{-24}\ \text{eV} \leq m \leq 10^{-20}\ \text{eV}$ is to be allowed by rotation curves from the SPARC database, it cannot also satisfy the soliton-halo relation in eq.~(\ref{eq:Schive_SH}) at the same time (see figure~1 of \cite{Bar_2022}). In section~\ref{sec:analysis_SPARC}, we obtained a similar result for a smaller sample of LSB galaxies from the SPARC database, see figure~\ref{fig:LSB_no_inter}. Later in the same section we conducted an analysis similar to the one in \cite{Bar_2022} but with two key differences (along with a few other minor ones): (a) Instead of varying over a range of FDM masses, we fixed $m = 10^{-22}\ \text{eV}$ and allowed the self-interaction strength $\lambda$ to vary, and, (b) we used a modified SH relation, eq.~(\ref{eq:modified_SH}), which takes into account the impact of self-interactions. We found that SFDM with $m =10^{-22}\ \text{eV}$ and $\lambda \gtrsim 10^{-90}$ can in-fact be allowed by rotation curves while simultaneously satisfying the modified SH relation within a smaller scatter than before. The upward trend of the boundary in the figure~\ref{fig:LSB_with_inter} is indicative of this effect (see section~\ref{sec:imposing_mod_SH_relations} for a detailed discussion). Note that from the analysis in section~\ref{sec:PVC} we found similar values of $m$ and $\lambda$ can satisfy the `peak velocity condition' within a scatter of a factor of 2 for a sub-sample of LSB galaxies. 
We also note that our results are in agreement with those in the appendix~E of Ref.~\cite{Chavanis_2021}. We also briefly discussed why attractive self-interactions are not expected to play a big role in altering these constraints in section~\ref{sec:why_no_att_inter}.

The work in this chapter motivates a full parameter search in the $\lambda-m$ parameter space which is left for future work. It will be interesting to see how constraints are altered when all 175 galaxies are taken into account, along with their baryonic contribution and parameters of the NFW envelope. 
\justifying
\chapter{Self-interactions of ULDM to the rescue? - Part 2}\label{chpt:paper_3}

\textbf{Based on:} \\\textbf{B.~Dave} and G.~Goswami, \emph{ULDM self-interactions, tidal effects and tunnelling out of satellite galaxies}, \href{https://doi.org/10.1088/1475-7516/2024/02/044}{\emph{J. Cosmol. Astropart. Phys.} \textbf{02} (2024) 044} [\href{https://arxiv.org/abs/2310.19664}{arXiv:2310.19664 [astro-ph.CO]}]

\noindent Note that most of the text in this chapter is an excerpt from our published work, Ref.~\cite{Dave_2024}.

\section{Satellite galaxies in the presence of a tidal potential}

Until now, we have looked at astrophysical observations that have only required us to consider stationary solutions of the GPP system, like rotation curves or the amount of mass contained within some central region. 
However, there could be situations in which such solutions are no longer useful, for instance, when the core of the galaxy loses its mass over time. 
For an astrophysical example of such a situation, consider a satellite galaxy in a circular orbit around the centre of a larger host DM halo. In this system, self-gravity of the satellite holds it together, while tidal forces due to the host halo contribute a tidal potential that leads to a local maximum in the effective potential ($V_{\text{eff}} = \Phi_{SG} + \Phi_H$, where $\Phi_{SG}$ is the self-gravity while $\Phi_H$ is the tidal potential due to the host halo) corresponding to a tidal radius $r_{tidal}$ (measured from the centre of the satellite galaxy)~\cite{Hertzberg_2023}. 
This was discussed in section~\ref{sec:quasi-stationary}, and can be seen in figure~\ref{fig:tidal_potential} by the solid blue curve. 

In such a situation, if DM was described by particle-like CDM, DM contained within the tidal radius ($r < r_{tidal}$ in the rest frame of the satellite) will be safe from tidal disruption indefinitely. 
On the other hand, wave-like dark matter (owing to its large deBroglie wavelength in eq.~(\ref{eq:deBroglie})) inside the tidal radius will tunnel out through the potential barrier over time, leading to the eventual tidal disruption of the satellite~\cite{Hui_2017}, which can potentially address the ‘Missing Satellites’ problem of CDM. 
If one finds that the rate of tunnelling for an observed satellite is too high to survive on cosmological timescales, one can impose constraints on the parameters of ULDM~\cite{Du_2018, Hertzberg_2023}. 

\begin{figure}[h]
    \centering
    \includegraphics[width = 0.8\textwidth]{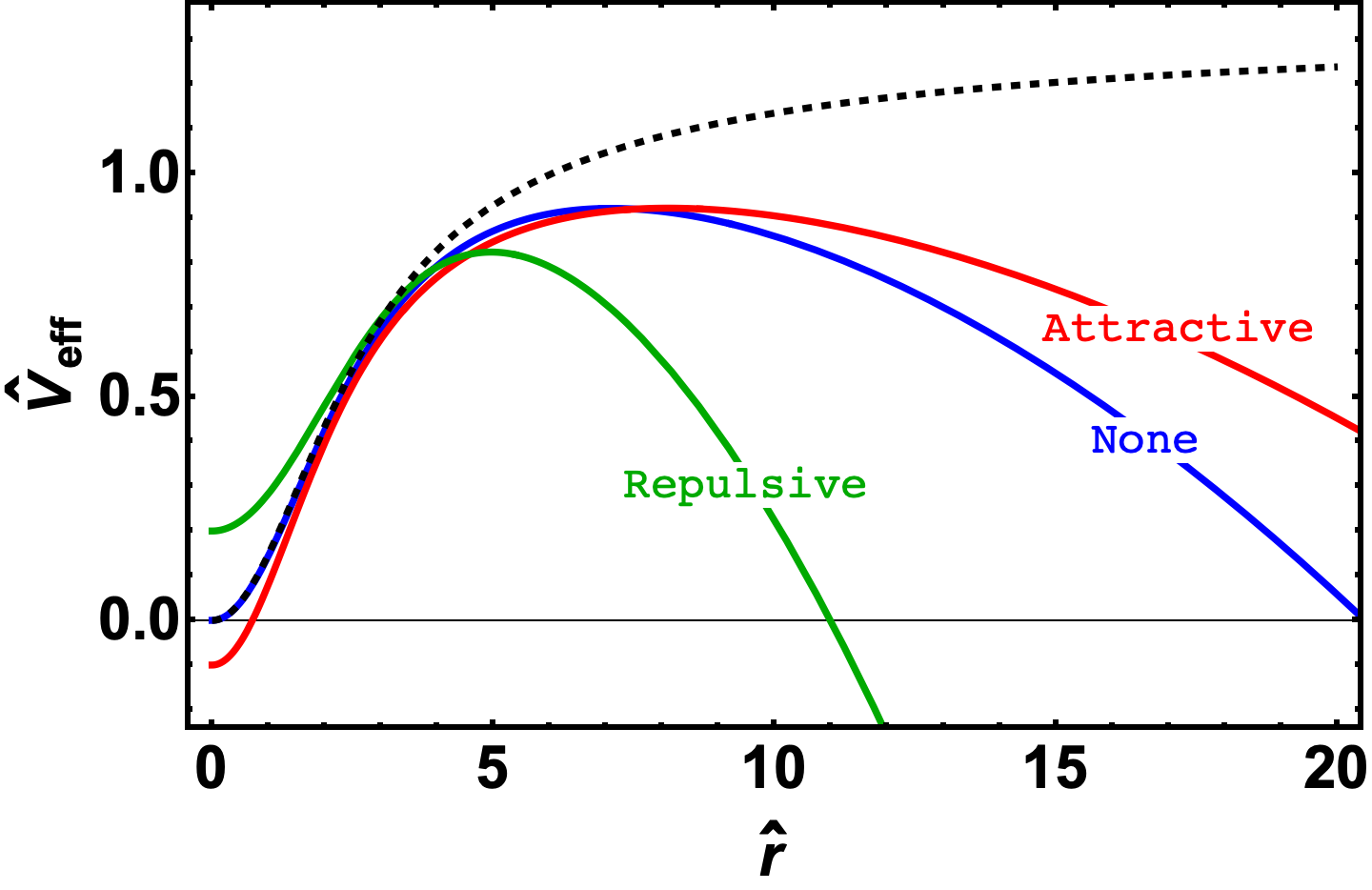}
    \caption[Impact of self-interactions on the effective potential]{For an object with a fixed mass and orbital period (used in section~\ref{sec:fixed_Mc}), effective potential (dimensionless) as defined in section~\ref{sec:quasi-stationary} is plotted for different values of self-coupling $\lambda$ as a function of dimensionless radius $\hat{r}$ (see eq.~(\ref{eq:dimensions2})). 
    Note that the potential barrier shrinks in the presence of repulsive self-interactions (green), while it stretches when self-interactions are attractive (red).}
    \label{fig:SI_barrier}
\end{figure}

In particular Ref.~\cite{Hertzberg_2023} suggested that given the observed core mass and orbital frequency of the Fornax dwarf spheroidal,  the Milky Way satellite could not have survived until present day if $m = 10^{-22}\ \text{eV}$. In fact, the allowed mass values of for FDM lie within the narrow range $2\times 10^{-22}\ \text{eV} \lesssim m \lesssim 6\times 10^{-22}\ \text{eV}$. 
This and similar analyses of the survival of Milky Way satellites~\cite{Du_2018} based on full simulations of the SP system, rely on the assumption that $\lambda = 0$. 
Therefore, it is interesting to explore the impact of self-interactions on these constraints. 

In this chapter we attempt to understand the effect of self-interactions on the solutions of the GPP system in the presence of a tidal potential. 
Using the formalism of quasi-stationary solutions (that we discussed in section~\ref{sec:quasi-stationary}), we shall see that for an object with a known core mass and orbital period, attractive (repulsive) self-interactions lead to a wider (narrower) potential barrier which in-turn leads to a longer (shorter)-lived satellite. 
We illustrate this in figure~\ref{fig:SI_barrier}. 

Further, we find that for ULDM with $m = 10^{-22}\ \text{eV}$, and the systems we are interested in, very small self-interactions $|\lambda|\sim 10^{-91}$ can have drastic impact on the lifetimes of satellite galaxies. 
Indeed, as we shall see in section~\ref{sec:saving_ULDM}, the observed core mass and orbital period of the Fornax dwarf spheroidal is consistent in the ULDM paradigm with $m\sim 10^{-22}\ \text{eV}$ as long as $\lambda < -2.12\times 10^{-91}$. 

Note that most of the text in this chapter is an excerpt from our published work~\cite{Dave_2024}.

\section{Quasi-stationary solutions}
We have discussed in detail (see section~\ref{sec:quasi-stationary} in chapter~\ref{chpt:numerical_solutions}) the equations of motion and their quasi-stationary solutions that can describe the density profiles of such satellite galaxies. 

We rewrite them for clarity, where in the rest frame of the satellite galaxy, the equations of motion are just,
\begin{eqnarray}
   \gamma\phi &=& -\frac{1}{2m}\nabla_r^2\phi + \left(m\Phi_{SG} - \frac{3}{2}m\omega^2r^2 + \frac{\lambda}{8m^3}|\phi|^2\right)\phi\ ,\label{eq:mod_GP_again} \\ 
   \nabla_r^2\Phi_{SG} &=& 4\pi G|\phi|^2\ , \label{eq:mod_poisson_SI_again}
\end{eqnarray}
where $\omega$ is the orbital frequency of the satellite galaxy, while $\Phi_{SG}$ is the gravitational potential due to the self-gravity of ULDM in the satellite. 
Recall from section~\ref{sec:quasi_numerical} that we allow $\gamma \in \mathbb{C}$ (which force $\phi$ to also have a complex part) and set outgoing wave boundary conditions far from the centre, which we obtained using the WKB approximation in eq.~(\ref{eq:wkb}). 
The resultant density profile $\rho(r) = |\phi|^2$ therefore will pick up $e^{-2|\gamma_I|t}$ time-dependence where $\gamma_I < 0$ and one can define the tunnelling rate of matter leaking through the potential barrier by eq.~(\ref{eq:decay_rate}). 
In its dimensionless form, the solutions of the resulting system of equations can be parametrised by $\hat{\omega}_\text{ini}$ and $\hat{\lambda}_\text{ini}$ when $\hat{\phi}(0) = 1$ is fixed. 
An example solution is shown in figure~\ref{fig:example_sol}. 

Once we have obtained the quasi-stationary solutions, we must utilise the scaling relations from sections~\ref{sec:scaling_symmetry} and~\ref{sec:scaling_relations} to then describe physical objects. 
Before doing so, in this section, we shall first understand how the decay rates obtained numerically changes with a change in the unscaled parameters $\hat{\omega}_\text{ini}$ and $\hat{\lambda}_\text{ini}$.
We shall also compare our results to results from Ref.~\cite{Hui_2017, Hertzberg_2023} for the case of $\lambda = 0$.

\subsection{Decay rates vs. $\rho_c/\rho_H$}\label{sec:comparison}

The results of recent work \cite{Hui_2017, Du_2018, Hertzberg_2023} suggests that for FDM, the rate of decay or mass loss depends only on the ratio of the central density of the satellite galaxy $\rho_c$ and the average density of the host halo within the orbital radius $\rho_H$. In-fact, they obtain a fitting function from numerical solutions that suggests that the decay rate exponentially decreases with an increasing $\rho_c/\rho_H$: $\Gamma \propto e^{-\rho_c/\rho_H}$. 

In order to compare our solutions with previous work, we first note that for a fixed $\hat{\lambda}_\text{ini}$ a different $\hat{\omega}_\text{ini}$ will lead to a different solution. Now using scaling relations in eqs.~(\ref{eq:scaling_Phi}) and~(\ref{eq:scaling_omega}), it is straightforward to see that $\hat{\rho}_{0, \text{fin}}/\hat{\omega}_{\text{fin}}^2 = \hat{\rho}_{0, \text{ini}}/\hat{\omega}_{\text{ini}}^2$ where $\hat{\rho}_{0,\text{ini}}$ and $\hat{\rho}_{0,\text{ini}}$ are the unscaled and scaled values of density at the origin, i.e. the central density of the satellite galaxy. If the satellite is orbiting the halo centre at a radius $a$ with orbital period $T$, one can define the average density of the host halo in a sphere of radius $a$ as $\rho_H = \frac{3M_{enc}}{4\pi a^3} = \frac{3\omega^2}{4\pi G}$ (where we have used Kepler's third law and $\omega = 2\pi/T$ is the scaled orbital frequency of the satellite). Denoting the scaled central density as $\rho_c \equiv \rho_{0,\text{fin}}$ one can write the ratio of the scaled dimensionful central density and average halo density as, 
\begin{equation}\label{eq:density_ratio}
    \frac{\rho_c}{\rho_H} = \frac{\hat{\rho}_{0,\text{fin}}}{3\hat{\omega}_\text{fin}^2} = \frac{1}{3}\frac{s^{-4}\hat{\rho}_{0,\text{ini}}}{s^{-4}\hat{\omega}_\text{ini}^2} = \frac{1}{3\hat{\omega}_\text{ini}^2}\ ,
\end{equation}
where we have $\hat{\rho}_{0,\text{ini}} = |\hat{\phi}(0)|^2 = 1$ is fixed in our numerical solutions. Hence, changing $\hat{\omega}_\text{ini}$ implies changing the ratio of central density and average halo density.

Now, for different values of $\hat{\omega}_\text{ini}$, one can always choose scale values $s$ such that either $\hat{\rho}_{0,\text{fin}}$ or $\hat{\omega}_\text{fin}$ remain fixed. Hence a changing $\rho_c/\rho_H$ can imply either changing central density or a changing the orbital frequency. A higher $\rho_c/\rho_H$ can therefore imply that the satellite has a higher central density or a smaller orbital frequency (i.e. a larger orbital radius if $M_{enc}$ is fixed). 

\subsection{Recovering the case of negligible DM self-interactions}

We obtain decay rates for different values of $\hat{\omega}_{\text{ini}}$ by solving the system of equations in eqs.~(\ref{eq:mod_GP}) and~(\ref{eq:mod_poisson}) numerically. We then plot decay rates $\Gamma/(\omega/2\pi)$ versus the ratio of central density of the satellite and the average density of the host halo $\rho_c/\rho_H$ in figure~\ref{fig:decay_vs_ratio} (blue curve). We then compare our solutions with the fitting function obtained by \cite{Hertzberg_2023} (black curve) and find that the decay rates obtained by us agree well with it. 

In the context of potential barriers, one can understand why a larger $\hat{\omega}_\text{ini}$ leads to a larger decay rate using potential barriers from figure~\ref{fig:approx_pot_barrier} (approximate potential barriers). It is clear that a larger $\hat{\omega}_\text{ini}$ implies a narrower potential barrier which in the tunnelling picture should make it easier for DM to tunnel out, leading to a shorter lifetime, while a smaller $\hat{\omega}_\text{ini}$ implies a wider barrier and hence a longer lifetime.

\subsection{Core mass and lifetime}

Now, recall from section~\ref{sec:decay_rates} that we can define a core radius $\hat{R}_c$ as the distance at which the density becomes $1/2$ its central value, and hence define the mass of the core as $\hat{M}_{\text{ini}}\equiv\hat{M} (\hat{R}_c)$ using eq.~(\ref{eq:soliton_mass}). Using scaling relations in eqs.~(\ref{eq:scaling_Phi})and~(\ref{eq:scaling_mass}), one can write the central density in terms of scaled core mass $M_c \equiv M_{\text{fin}} = s^{-1}\frac{\hbar c}{Gm}\hat{M}_\text{ini}$ as, 
\begin{equation}\label{eq:mass_density}
   M_c^4  \propto \frac{\hbar^6}{G^3m^6}\rho_c\ .
\end{equation}
Here, for a fixed $\omega_{\text{fin}}$ if the scaled core mass $M_c$ is larger, then $\rho_c/\rho_H$ is also larger and consequently the lifetime of the soliton is longer. In other words, heavier satellites survive for longer than lighter satellites. It was suggested in \cite{Hui_2017, Hertzberg_2023} that in general for a typical dwarf satellite, $\rho_c/\rho_H \gtrsim 70$ is required for it to survive till present day in orbit around a host halo.

\subsection{The impact of self-interactions of DM: a first look}

\begin{figure}[ht]
    \centering
    \includegraphics[width = 0.7\textwidth]{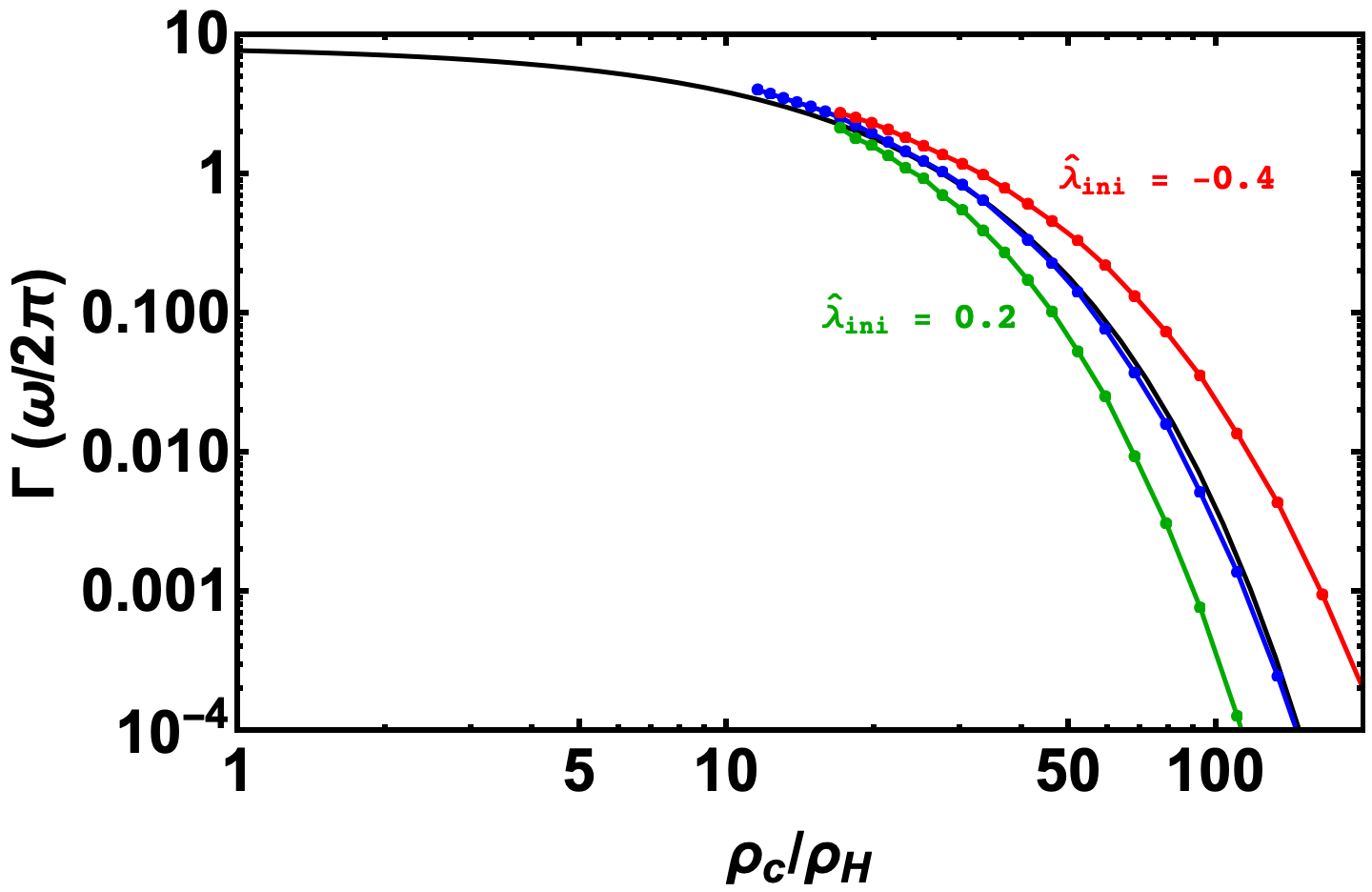}
    \caption[Decay-rates versus ratio of soliton central density and average halo density]{We have plotted decay rates (in the units of $T^{-1} = \omega/2\pi$) vs. the ratio of central density of the satellite and average density in the host halo. The solid black line is the fitting function obtained in \cite{Hertzberg_2023} for $\lambda = 0$. The blue curve are the decay rates obtained using our numerical implementation while red and green curves are decay rates for a fixed unscaled dimensionless attractive and repulsive self-interaction strength $\hat{\lambda}_\text{ini} = -0.4, 0.2$ respectively. It is worth noting that (a) for all three cases, we could not obtain reliable solutions for larger values of $\hat{\omega}_{\text{ini}}$ (as $\rho_c/\rho_H = 1/3\hat{\omega}_\text{ini}^2$), and (b) for larger values of $\hat{\omega}_\text{ini}$ the tunnelling approximation will not be valid for sufficiently long as discussed in section~\ref{sec:self-consistency}.}
    \label{fig:decay_vs_ratio}
\end{figure}

This procedure can be repeated in the presence of self-interactions: for two values, $\hat{\lambda}_\text{ini} = -0.4$, and $\hat{\lambda}_\text{ini} = 0.2$, we show the corresponding decay rates in figure~\ref{fig:decay_vs_ratio} as the red and green curves respectively. 
It is also worth noting that for the case of attractive self-interactions, we only look at solutions corresponding to $\hat{\lambda}_\text{ini} > -0.4$. This is because in the absence of tidal effects, $\hat{\lambda}_\text{ini} < -0.4$ corresponds to the unstable branch of solutions (or the so-called non-gravitational regime) \cite{Dave_2023, Chavanis_2011_analytic}. 

We see that just like the case of no interactions, a larger $\hat{\omega}_\text{ini}$ leads to a larger decay rate and a smaller lifetime. However, we also see that for a fixed $\hat{\omega}_\text{ini}$ attractive self-interactions appear to have a higher decay rate than repulsive self-interactions. This suggests that attractive self-interactions lead to a satellite galaxy surviving for a shorter time before losing most of its mass to tidal effects while repulsive self-interactions allow the satellite to survive for longer. This result bears a closer inspection, which we shall carry out in section~\ref{sec:fixed_rhoC}.

To understand exactly how these solutions translate to describing realistic objects, we must employ scaling relations, as we shall see in the next section~\ref{sec:fixed_rhoC}.

\section{Impact of self-interactions}\label{sec:SI_impact}
In this section we shall look at how one can use scaling relations to model realistic objects and analyse the impact of self-interactions in two different ways. In section~\ref{sec:fixed_rhoC} we shall look at the case where the scaled central density is fixed along with $\omega_\text{fin}$, and allow $\lambda$ to vary. Next, in section~\ref{sec:fixed_Mc} we shall look at objects with fixed mass (in this case, core mass) $M_\text{fin}$ and orbital frequency $\omega_\text{fin}$ and see how self-interactions alter corresponding life-times. Finally in section~\ref{sec:saving_ULDM} we shall look under what conditions will self-interactions allow objects to survive that are ruled out for the $\lambda = 0$ case.

Recall that the numerical solutions we obtain are for $s = 1$. However, to model realistic objects we need to obtain an appropriate value of scale which at least satisfies $s \gtrsim 10$ (this describes the `trustworthy regime' where the effects of general relativity will not be important \cite{Chakrabarti_2022}). 

We can estimate the order of magnitude of $s$ required by considering the typical orbital period of a satellite to be $T = 2\pi/\omega \sim \mathcal{O}(1)\ \text{Gyr}$ (For instance, $T \approx 1.6\ \text{Gyr}$ and $T \approx 3.9\ \text{Gyr}$ for UMi and Fornax respectively \cite{Hertzberg_2023}). However, for $m = 10^{-22}\ \text{eV}$, $\hat{\omega}_{\text{ini}} \sim 0.05$ implies an unscaled time period of $T_{\text{ini}} = 2\pi/\omega_{\text{ini}} \sim \mathcal{O}(10)\ \text{yr}$ which is far smaller than what is observed. The required scale value can then be obtained from eq.~(\ref{eq:scaling_omega}) as $s \sim \mathcal{O}(10^4)$. Additionally, for an estimate of the central density or the core mass of a satellite, obtaining the correct $s$ that scales $\hat{\omega}_\text{ini}$ and $\hat{M}_\text{ini}$ (or $\hat{\rho}_{c, \text{ini}}$) to required values is more involved, as we shall see in the next sections. 


\subsection{Objects with known central density and orbital period}\label{sec:fixed_rhoC}

Suppose the known quantities for a satellite galaxy are the central DM density and the orbital period around the host halo at the centre. We denote these as $\rho_{c,\text{fin}}$ and $T_\text{fin} = 2\pi/\omega_{\text{fin}}$ respectively. In our numerical solutions, the central density without scaling is fixed since $\hat{\rho}_{c,\text{ini}} = |\hat{\phi}(0)|^2 = 1$ for all values of $\hat{\omega}_\text{ini}$ and $\hat{\lambda}_\text{ini}$. 

\subsubsection{The procedure}

To find the value of scale $s$ that satisfies observed values $\rho_{c, \text{fin}}$ and $\omega_\text{fin}$, we employ the following procedure:

\begin{enumerate}
    \item Fix dimensionless quantities $\hat{\rho}_{c,\text{fin}}$ and $\hat{\omega}_{\text{fin}}$ using the observed values.
    
    \item Using scaling relations in eq.~(\ref{eq:scaling_Phi}), obtain scale value $s = (\hat{\rho}_{c, \text{fin}})^{-1/4}$ (recall that $\hat{\rho}_{c,\text{ini}} = 1$). 

    \item Using $s$ and eq.~(\ref{eq:scaling_omega}), obtain the unscaled orbital frequency such that $\hat{\omega}_{\text{ini}} = s^2\hat{\omega}_{\text{fin}}$ is satisfied. 

    \item For a fixed $\hat{\lambda}_{\text{ini}}$ and the obtained $\hat{\omega}_{\text{ini}}$ solve the dimensionless form of eqs.~(\ref{eq:mod_GP}) and~(\ref{eq:mod_poisson}) numerically to obtain the unscaled decay rate $\hat{\Gamma}_{\text{ini}} = 2|\hat{\gamma}_I|$. Scale the decay rate using $s$ to obtain the lifetime of the soliton $\tau = s^2\frac{\hbar}{mc^2}\hat{\Gamma}_\text{ini}^{-1}$. 
\end{enumerate}

\begin{figure}[ht]
    \centering
    \includegraphics[width = 0.7\textwidth]{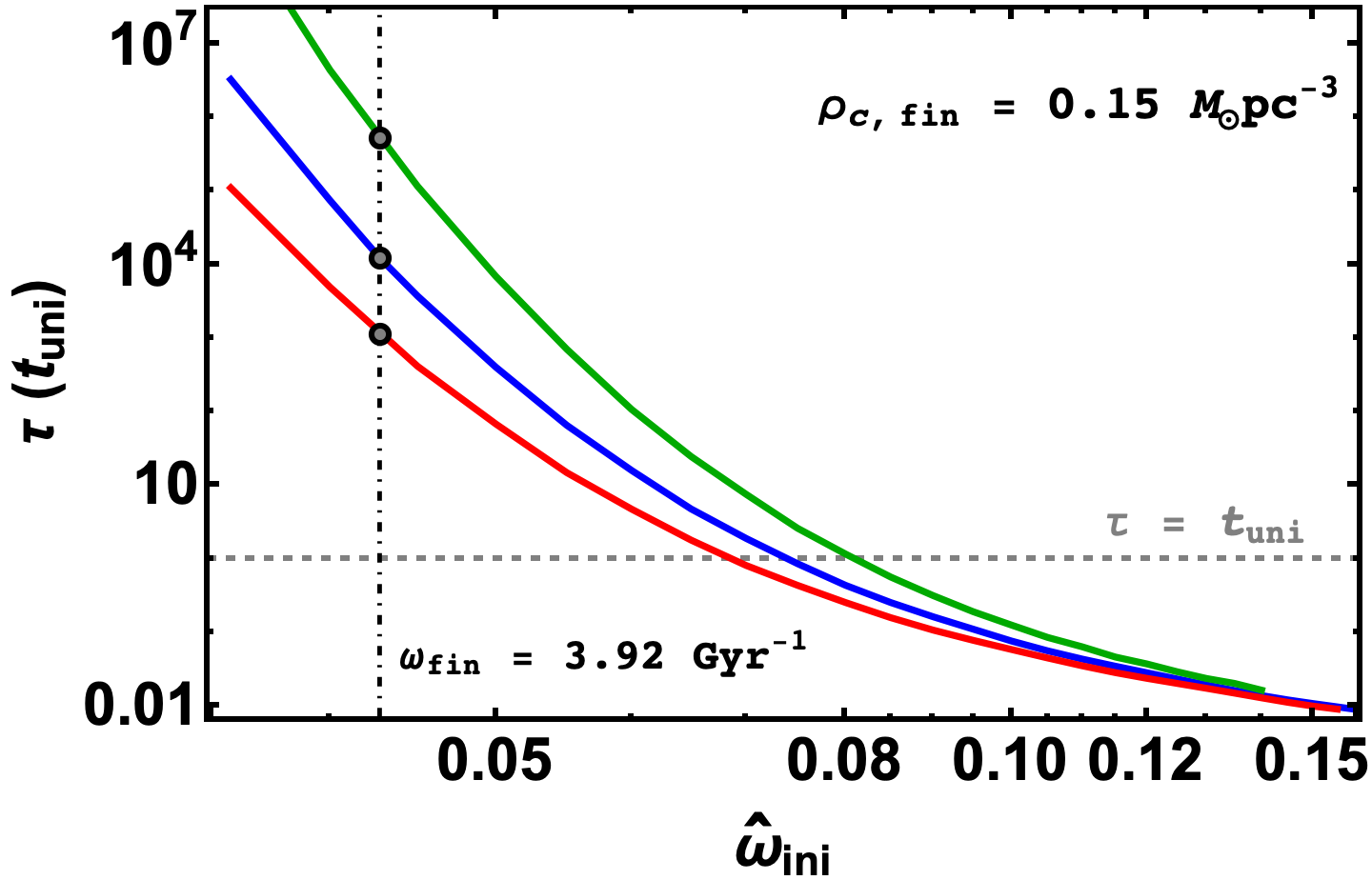}
    \caption[Lifetime versus $\hat{\omega}_\text{ini}$ for different $\hat{\lambda}_\text{ini}$ values]{Scaled lifetimes for $\hat{\lambda}_\text{ini} = -0.2, 0, 0.2$ are shown by the red, blue and green curves respectively. Here $s = 7225$ is taken to satisfy a scaled $\rho = 0.15\ \text{M}_\odot\text{pc}^{-3}$. The black dots correspond to lifetimes when the scaled time period is $2\pi/\omega_{\text{fin}} = 1.6\ \text{Gyr}$, i.e. time period of the UMi dwarf galaxy.}
    \label{fig:lifetime_fixed_rho}
\end{figure}

\subsubsection{The absence and presence of self-interactions}
 
To demonstrate the above procedure, we take an example of the Ursa Minor (UMi) spheroidal dwarf with central density $\rho_c = 0.15\ \text{M}_\odot \text{pc}^{-3}$ and $T = 1.6\ \text{Gyr}$ \cite{Hertzberg_2023}. The scale $s$ and $\hat{\omega}_{\text{ini}}$ that satisfy observed values are $s = 7725$ and $\hat{\omega}_{\text{ini}} = 0.0427$. We work initially in the FDM regime, where $\hat{\lambda}_\text{ini} = 0$. The eigenvalue is found to be $\hat{\gamma} = 0.639 - i(3.19\times 10^{-8})$. The scaled decay rate $\Gamma$ and lifetime $\tau$ are then,
\begin{equation}
    \Gamma \approx 0.8\times 10^{-4}\ \text{t}_{\text{uni}}^{-1} \ \  , \ \ \tau \approx 1.23\times 10^4\ \text{t}_{\text{uni}}\ ,
\end{equation}
where, $\text{t}_{\text{uni}} = 13.8\ \text{Gyr}$ is the age of the Universe for the standard $\Lambda$CDM model \cite{Planck2018}. The above decay rate agrees well with the one obtained by Ref.~\cite{Hertzberg_2023}. The corresponding lifetime of the satellite is far larger than the age of the universe, allowing it to survive until present time. Also note that the scaled core mass of the satellite is $M_c = 9.31\times 10^7\ M_\odot$. 

Now we turn on self-interactions, and choose $\hat{\lambda}_\text{ini} = \pm 0.2$, along with $s = 7225$ and $\hat{\omega}_\text{ini} = 0.0427$. Figure~\ref{fig:lifetime_fixed_rho} plots the lifetime of an object with the above-mentioned properties as a function of $\hat{\omega}_{\text{ini}}$. It is worth noting that the values of $s$ and $\hat{\omega}_\text{ini}$ will be fixed independently of the value of $\hat{\lambda}_\text{ini}$ since $s$ depends on $\hat{\rho}_{0,\text{ini}}$ which is fixed to $1$ for all values of $\hat{\omega}_\text{ini}$ and $\hat{\lambda}_\text{ini}$. The corresponding scaled self-coupling strength is $\lambda = s^2\hat{\lambda}_\text{ini} = \pm 1.41\times 10^{-91}$. The lifetime turns out to be $\tau \approx 5.17\times 10^5\ \text{t}_\text{uni}$ for repulsive self-interactions and $\tau \approx 1.15\times 10^3 \ \text{t}_\text{uni}$ for attractive self-interactions. This corresponds to a larger decay rate for attractive self-interactions while a smaller one for repulsive self-interactions compared to the no self-interactions case which agrees with the what we have seen in figure~\ref{fig:decay_vs_ratio}.

It is worth noting that the above results agree qualitatively with what we obtain from the approximate approach in appendix~\ref{app:approx_appraoch}, where we solve the system without the tidal potential, construct an effective potential with the tidal potential and use WKB approximation inside the barrier to estimate the decay-rate (similar to the $\alpha$-decay problem in quantum mechanics). In the approximate approach, while the numbers do not agree with the numerical solutions, the lifetime does decrease (increase) as one goes from no self-interactions to attractive (repulsive) self-interactions. In particular, note the widening (narrowing) of the barrier for $\lambda > 0$ ($\lambda < 0$) in figure~\ref{fig:wkb_fixed_rhoc} indicating that in the presence of repulsive self-interactions DM is less likely to tunnel out while the presence of attractive self-interactions increases the transmission probability.

\subsubsection{A discrepancy}\label{sec:discrepancy}

Intuitively as well as from simulations \cite{Glennon_2022} one expects attractive self-interactions to increase the lifetime of the satellite. In stark contrast, in the results above, we find that attractive self-interactions lead to a smaller lifetime while repulsive self-interactions lead to the satellite galaxy surviving for longer. What exactly is happening here?

\begin{figure}[ht]
    \centering
    \includegraphics[width = 0.65\textwidth]{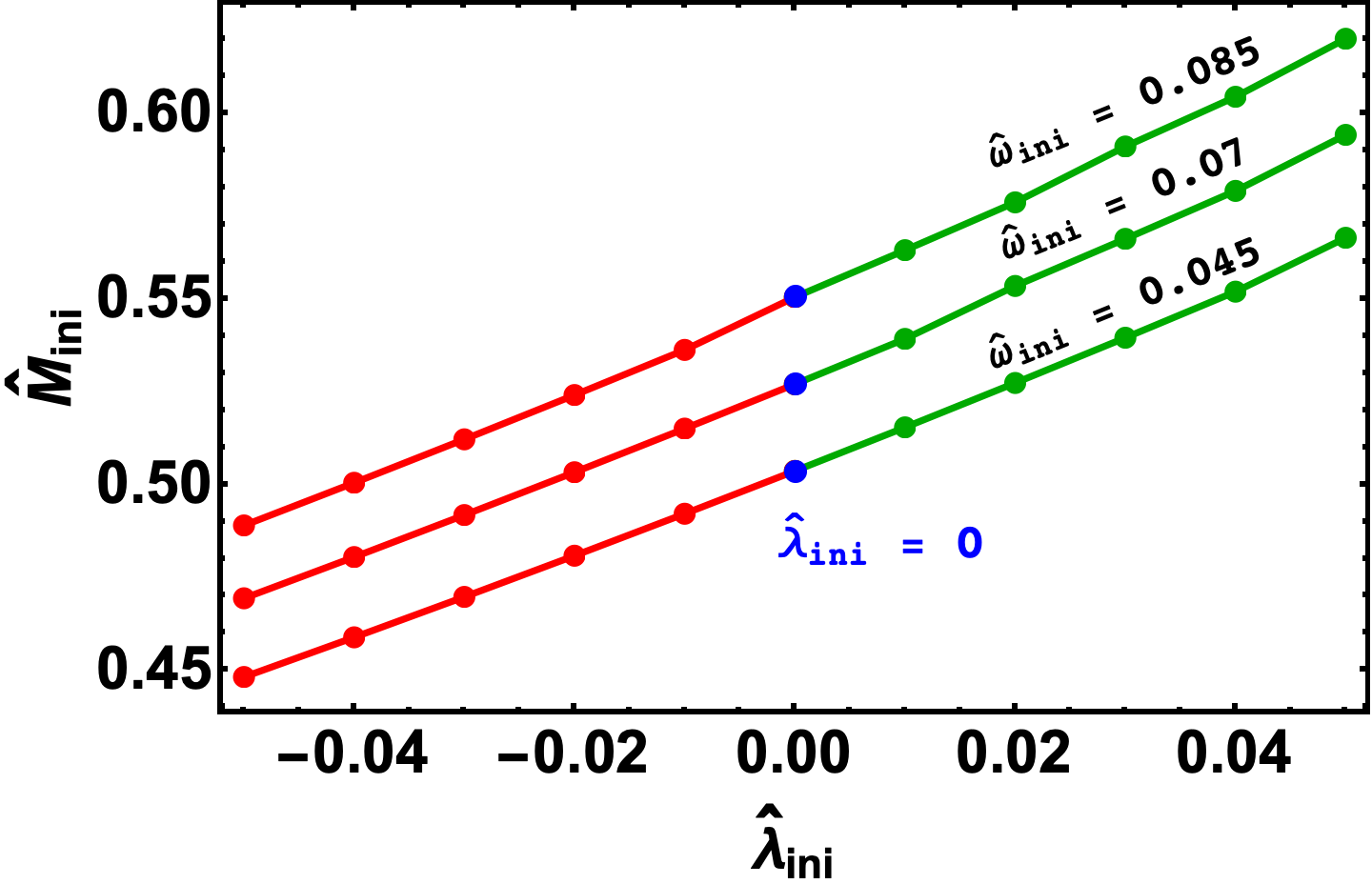}
    \caption[$\hat{M}_\text{ini}$ versus $\hat{\lambda}_\text{ini}$ for different values $\hat{\omega}_\text{ini}$]{Here we plot the unscaled dimensionless core mass $\hat{M}_\text{ini}$ versus unscaled dimensionless self-interaction strength $\hat{\lambda}_\text{ini}$ for different values of $\hat{\omega}_\text{ini}$. Core mass is defined as the mass contained within a sphere of radius $\hat{R}_{1/2}$, where $\hat{R}_{1/2}$ is the distance at which density becomes half its central value}
    \label{fig:mhat_vs_lambda}
\end{figure}

To understand this discrepancy, we note that for $\hat{\lambda}_{\text{ini}} > 0$ ($\hat{\lambda}_\text{ini} < 0$) any integrated mass $\hat{M}_\text{ini}$ is found to be larger (smaller) than the $\hat{\lambda}_\text{ini} = 0$ case (for a fixed value of $\hat{\omega}_\text{ini}$), as shown in figure~\ref{fig:mhat_vs_lambda}.  

As we have mentioned earlier, for a fixed scaled central density, the scale value $s$ remains the same for different values of $\hat{\lambda}_\text{ini}$. This implies that similar to unscaled core masses, scaled core masses will also be larger for repulsive self-interactions and smaller for attractive self-interactions. In other words, when we fix the scaled central densities for different values of $\hat{\lambda}_\text{ini}$, the scaling relations ensure that we are looking at objects with different scaled soliton masses. Further, as we have seen from eq.~(\ref{eq:mass_density}), a large core mass for a fixed orbital frequency leads to a smaller decay rate and a longer lifetime, and vice-versa. This reasoning also explains the effect of self-interactions on the curves in figure~\ref{fig:decay_vs_ratio}, since $s = 1$ is the same for all values of $\hat{\omega}_\text{ini}$ for a fixed $\hat{\lambda}_\text{ini}$.

To study the effects of self-interactions for objects with the same core mass, we must use a different implementation of the scaling relations, as we shall discuss in the next section.

\subsection{Objects with known core mass and orbital period}\label{sec:fixed_Mc}

Before proceeding, it is important to note that one can use various definitions of the mass of the satellite once we have a density profile. For instance, one can use tidal mass $\hat{M}_\text{tidal}$ which is defined as the mass enclosed within the tidal radius $\hat{r}_\text{tidal}$. 
In our analysis, we use core mass, i.e. the mass contained within the radius at which density becomes half its central value (as defined in section~\ref{sec:decay_rates}) which can be uniquely found for every choice of $\hat{\lambda}_\text{ini}$ and $\hat{\omega}_\text{ini}$. 

\subsubsection{The procedure}

In order to study the effects of only self-interactions, we consider an object with a fixed $M_\text{fin}$ and $\omega_{\text{fin}}$. Then, for every $\hat{\lambda}_{\text{ini}}$, we require a scale value $s$ such that $\hat{M}_{\text{fin}} = s^{-1}\hat{M}_{\text{ini}}$ and $\hat{\omega}_\text{fin} = s^{-2}\hat{\omega}_{\text{ini}}$ both are satisfied. The procedure for this is as follows:

\begin{enumerate}
    \item Fix $\hat{M}_{\text{fin}}$ and $\hat{\omega}_{\text{fin}}$ from observations. 
    
    \item Fix the value of $\hat{\lambda}_{\text{ini}}$. For a chosen value of $\hat{\omega}_{\text{ini}}$ solve the dimensionless form of eqs.~(\ref{eq:mod_GP}) and~(\ref{eq:mod_poisson}). 

    \item Use $\hat{M}_{\text{fin}} = s^{-1}\hat{M}_{\text{ini}}$ to get a value of $s$, and check if the condition $\hat{\omega}_{\text{fin}} = s^{-2}\hat{\omega}_{\text{ini}}$ is satisfied within some desired accuracy ($5\%$ for our analysis). 

    \item If $\hat{\omega}_\text{fin} \neq s^{-2}\hat{\omega}_\text{ini}$, repeat steps $2$ and $3$ for a different value of $\hat{\omega}_{\text{ini}}$. Once the condition is satisfied for some $\hat{\omega}_\text{ini}$, using the corresponding numerical solution obtain $\hat{\Gamma}_{\text{ini}} = 2|\hat{\gamma}_I|$, and consequently obtain the scaled lifetime as $\tau = s^2\frac{\hbar}{mc^2}\hat{\Gamma}_\text{ini}^{-1}$. 
\end{enumerate}

A satellite with observed core mass $M_{\text{fin}}$ and orbital frequency $\omega_{\text{fin}}$ can be described by ULDM only if $\tau > t_\text{uni}$ where $t_\text{uni}$ is the age of the universe. 

\subsubsection{The absence and presence of self-interactions}\label{sec:absence_and_presence}

As a demonstration we have carried out the above procedure for a satellite whose core mass and orbital period are $M_{\text{fin}} = 9.1\times 10^7\ \text{M}_\odot$ and $T_{\text{fin}} =2\pi/\omega_\text{fin} = 1.6\ \text{Gyr}$ respectively (similar to values for UMi dwarf spheroidal from section~\ref{sec:fixed_rhoC}).

\begin{figure}[ht]\captionsetup[subfigure]{font = {footnotesize}}
    \centering
    \begin{subfigure}[t]{0.45\textwidth}
        \centering
        \includegraphics[width = \textwidth, height=2in]{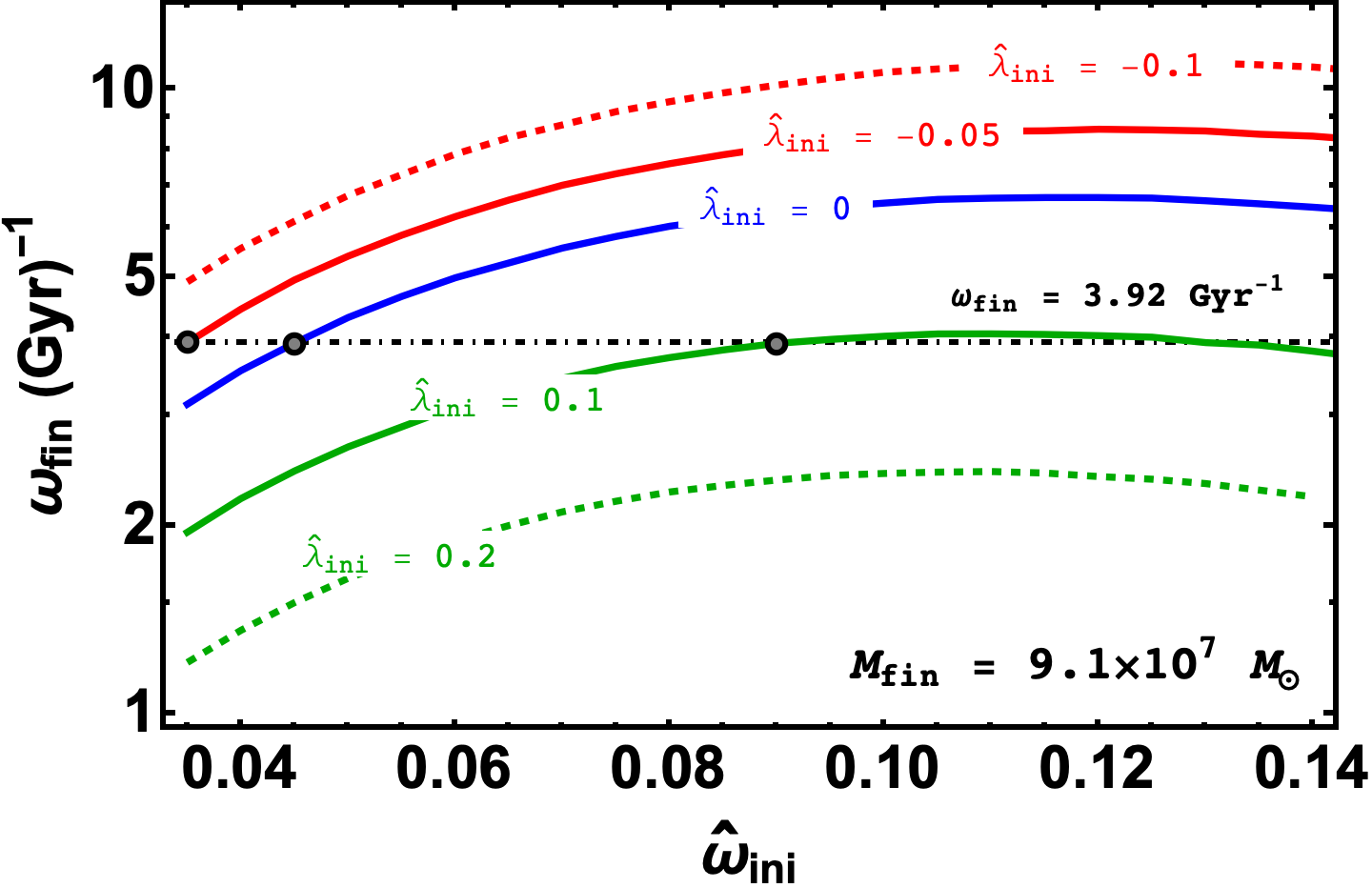}
        \caption{Multiple possible values of $\omega_\text{fin}$ for a fixed $M_\text{fin}$ are shown. Each colored curve corresponds to a different $\hat{\lambda}_\text{ini}$. The horizontal gray line represents a fixed $\omega_\text{fin} = 3.92\ \text{Gyr}^{-1}.$ }
        \label{fig:fixed_core_mass_omega}
    \end{subfigure}
    \hfill
    \begin{subfigure}[t]{0.45\textwidth}
        \centering
        \includegraphics[width = \textwidth, height=2in]{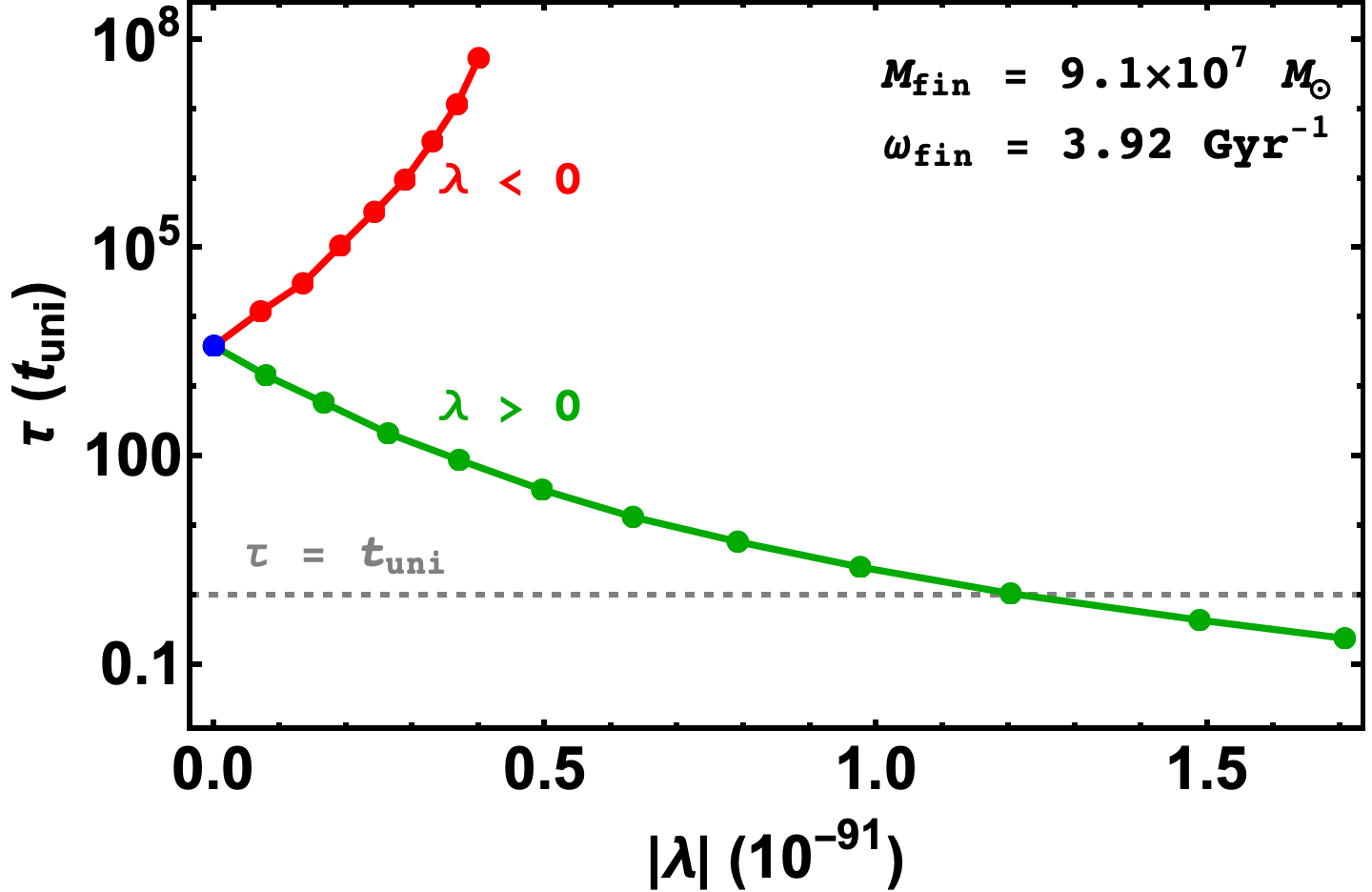}
        \caption{Lifetime in the units of the age of the universe is plotted for both attractive (red) and repulsive (green) self-interactions. The blue dot corresponds to the lifetime when $\lambda = 0$.}
        \label{fig:lifetime_fixed_core_mass}
    \end{subfigure}
    \caption[Procedure for obtaining lifetimes for a fixed $M_\text{fin}$ and $\omega_\text{fin}$ for different $\lambda$]{Left panel illustrates how one can model an object with a fixed core mass $M_\text{fin} = 9.1\times 10^7\ M_\odot$ and $\hat{\omega}_\text{fin} = 3.92\ \text{Gyr}^{-1}$. We plot how the lifetime changes in the presence of self-interactions for the same object in the right panel.}
    \label{fig:fixed_core_mass}
\end{figure}

It is important to note that not every combination of chosen values of $M_\text{fin}$ and $\omega_\text{fin}$ is allowed if one requires that both the scaling relations $M_\text{fin} = s^{-1}M_\text{ini}$ and $\omega_\text{fin} = s^{-2}\omega_\text{ini}$ are satisfied. As shown in figure~\ref{fig:fixed_core_mass_omega}, we find that for a fixed $\hat{\lambda}_{\text{ini}}$ and a chosen  $M_\text{fin}$, there is a maximum scaled $\omega_{\text{fin}}$ that can be obtained no matter which $\hat{\omega}_{\text{ini}}$ we start with. As a result, for the object with the above mentioned core mass and orbital frequency, self-coupling strengths with $\hat{\lambda}_{\text{ini}} > 0.1$ are not allowed, which is shown in figure~\ref{fig:fixed_core_mass_omega}. 

Following the procedure above, for the $\hat{\lambda}_\text{ini}$ values that can give rise to the observed values of $M_\text{fin}$ and $\omega_\text{fin}$, we have also plotted the lifetime of the satellite against $|\lambda|$ in figure~\ref{fig:lifetime_fixed_core_mass}. We find that for $\lambda = 0$, the lifetime of the satellite is $\tau \approx 3922\ t_\text{uni}$, which is already well above the minimum allowed lifetime $\tau = t_\text{uni}$. In the presence of attractive self-interactions, the lifetime increases as $|\lambda|$ increases shown by the red curve in figure~\ref{fig:lifetime_fixed_core_mass}. 
On the other hand, for ever larger repulsive self-interactions, lifetime of the satellite keeps on decreasing (shown by the green curve), until for $\lambda \gtrsim 1.2 \times 10^{-91}$,  $\tau < t_\text{uni}$ i.e., the satellite no longer survives until present day. Hence, if an object with $M_{\text{fin}} = 9.1\times 10^7 \ \text{M}_{\odot}$ and $\omega = 3.92\ \text{Gyr}^{-1}$ is observed, it will rule out $\lambda \gtrsim 1.2\times 10^{-91}$, imposing an upper limit on the strength of the self-coupling $\lambda$.

\subsubsection{Intuition for the effects of self-interactions}\label{sec:intuition}

As we have discussed in the previous section, the effect of self-interactions can be understood by looking at solutions for an object with a fixed core mass. Note that for stationary state solutions, we know that if soliton mass is fixed, central density is larger (as the soliton itself is squeezed) when self-interactions are attractive, while it is smaller (as the soliton expands) for repulsive self-interactions (see figure~1 in \cite{Chakrabarti_2022}). This is similar to the situation we have for the tunnelling problem, albeit for a leaky soliton. For instance, the object considered in figure~\ref{fig:fixed_core_mass}, has the orbital frequency $\omega_\text{fin} = 3.92\ \text{Gyr}^{-1}$ and core mass $M_\text{fin} = 9.1\times 10^7\ \text{M}_\odot$. The scaled central density for this object when $\lambda = 0$ turns out to be $\rho_c \approx 0.13\ \text{M}_\odot/\text{pc}^{3}$. On other hand, for $\lambda = -2.88\times 10^{-92}$, the central density is $\rho_c = 0.22\ \text{M}_\odot/\text{pc}^{3}$ while for $\lambda = 1.48\times 10^{-91}$, $\rho_c \approx 0.03\  \text{M}_\odot/\text{pc}^{3}$. From the discussion in section~\ref{sec:comparison} we also know that a larger (smaller) central density implies a larger (smaller) lifetime, which explains the effect of attractive and repulsive self-interactions.

Further, we have plotted potential barriers for three values of $\hat{\lambda}_\text{ini} = \{-0.05, 0, 0.1\}$ for the above mentioned object in figure~\ref{fig:SI_barrier} to confirm our intuition. To satisfy scaled quantities, the value of $\hat{\omega}_\text{ini}$ required is smaller (larger) for attractive (repulsive) self-interactions, leading to a wider (narrower) potential barrier. Hence, in the presence of attractive self-interactions, the satellite survives for longer while for repulsive self-interactions, the opposite is true.

We find that the approximate approach in appendix~\ref{app:approx_appraoch} holds up qualitatively for satellites with constant core masses as well. As seen in figure~\ref{fig:wkb_fixed_core_mass}, the approximate potential barriers exhibit the same behaviour as the numerically solved barriers in figure~\ref{fig:SI_barrier}. This is because even in the absence of tidal effects a $\hat{\lambda}_{\text{ini}} < 0$ ($\hat{\lambda}_{\text{ini}} > 0$) will lead to a smaller (larger) initial core mass. Hence, one requires a different $\hat{\omega}_{\text{ini}}$ for different $\hat{\lambda}_\text{ini}$ to obtain the same core mass. 
However, it is worth noting that the numbers obtained from the approximate approach may not be accurate since the system is not numerically solved in the presence of the tidal potential (see appendix~\ref{app:approx_appraoch} for more details).

\subsection{Saving ULDM?}\label{sec:saving_ULDM}

From the discussion in the previous sections it is clear that repulsive self-interactions make the satellite even more susceptible to tidal disruption, thereby decreasing lifetime. In this section, we shall discuss how attractive self-interactions can help the case of ULDM for a fixed $m$, by looking at two different objects. 

\begin{figure}[ht]\captionsetup[subfigure]{font = {footnotesize}}
    \centering
    \begin{subfigure}[t]{0.45\linewidth}
        \centering
        \includegraphics[width = \textwidth, height=2in]{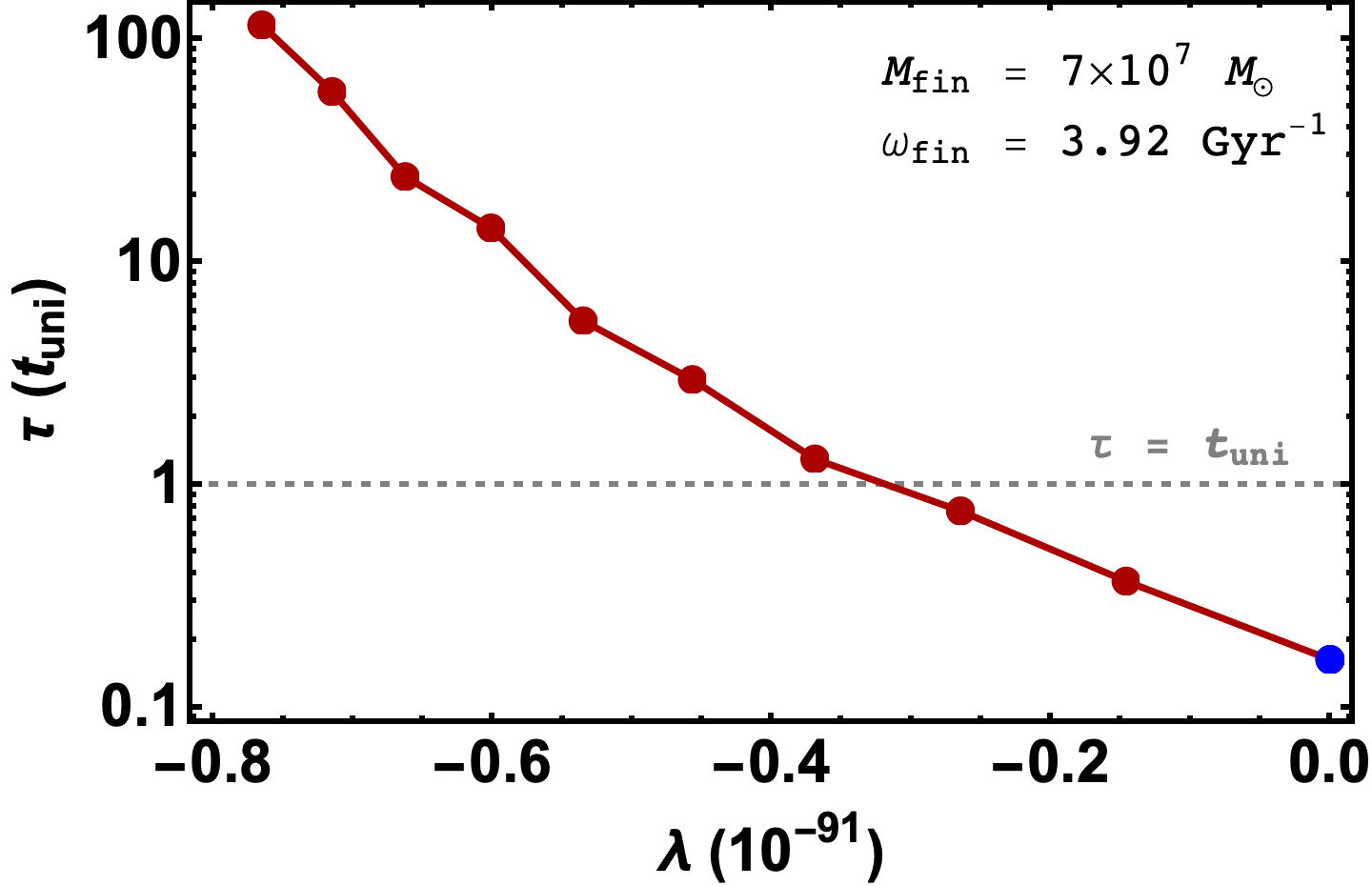}
        \caption{If this hypothetical object is to be described within the ULDM paradigm with $m =10^{-22}\ \text{eV}$, attractive self-interactions with $\lambda \lesssim -0.32\times 10^{-91}$ are required for it survive until present day.This can be used to impose constraints on self-interactions for a fixed $m$, as we see in the right panel}
        \label{fig:fixed_Mc_att_lambda}
    \end{subfigure}
    \hfill
    \begin{subfigure}[t]{0.45\linewidth}
        \centering
        \includegraphics[width = \textwidth, height=2in]{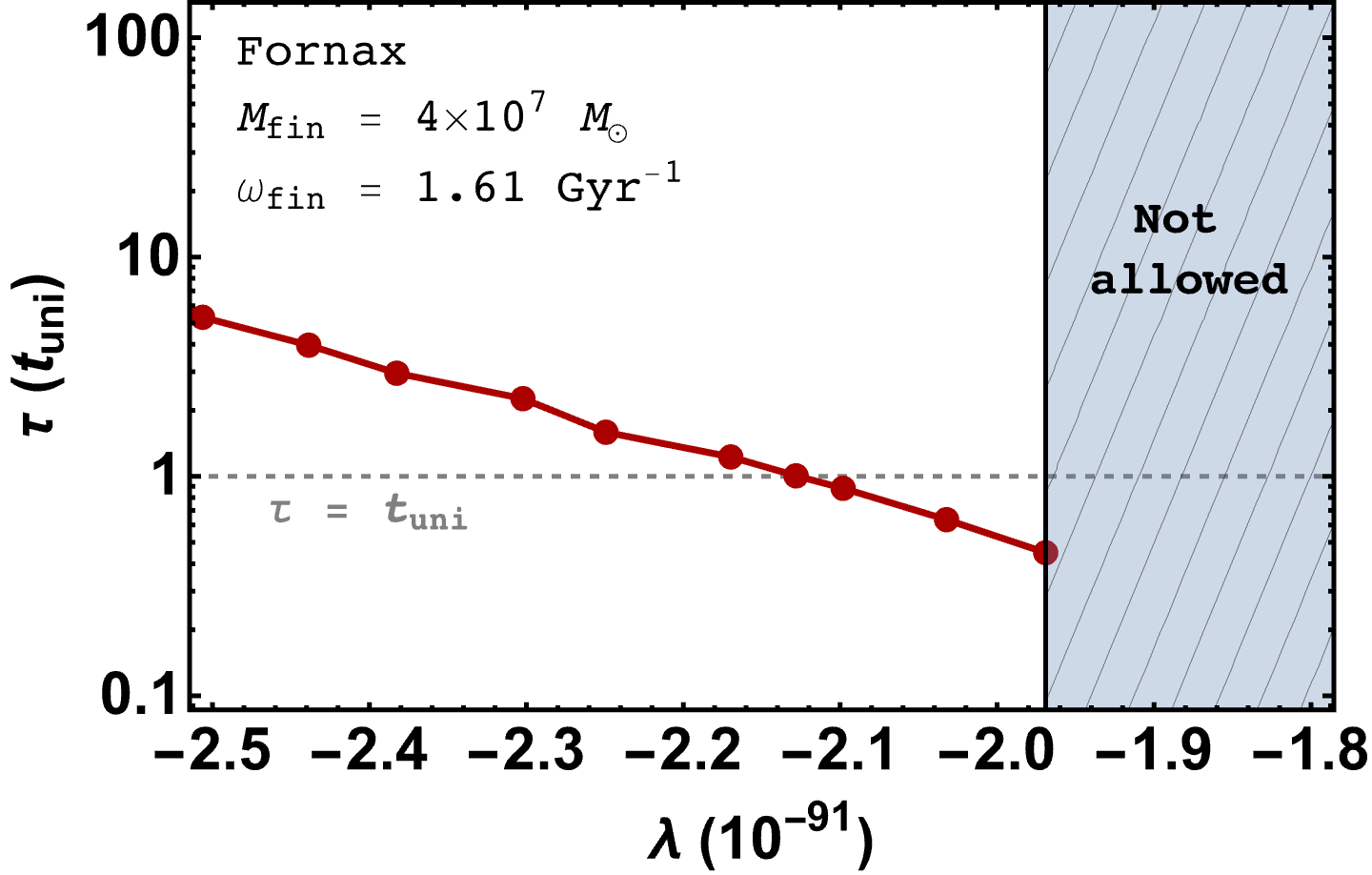}
        \caption{ULDM with $m = 10^{-22}\ \text{eV}$ and $\lambda > -1.9\times 10^{-91}$ cannot model the Fornax dwarf galaxy. Further, for $-2.12\times 10^{-91} < \lambda < -1.97\times 10^{-91}$ while we can model it, it will not survive until present day. For $\lambda < -2.12\times 10^{-91}$ however, its lifetime will be larger than the age of the universe.}
        \label{fig:saving_fornax}
    \end{subfigure}
    \caption[Lifetime of galaxy versus strength of attractive self-interactions]{Lifetime of the satellite galaxy versus strength of attractive self-interactions are plotted for a hypothetical object (left panel) and Fornax dwarf galaxy (right panel).}
    \label{fig:saving_ULDM}
\end{figure}

Firstly, let us consider an object with orbital period $T_{\text{fin}} = 1.6\ \text{Gyr}$ and a core mass $M_\text{{fin}} = 7\times 10^7\ \text{M}_\odot$. For this object, if we follow the procedure from the previous section, we find that in the FDM regime (i.e. $\lambda = 0$) the  lifetime of the satellite is smaller than the age of universe: $\tau \approx 0.16\ t_\text{uni}$. Here, repulsive self-interactions are only going to make things worse, by reducing the lifetime even further. However, from figure~\ref{fig:lifetime_fixed_core_mass}, it is clear that attractive self-interactions should help and we find that they indeed do, as shown in figure~\ref{fig:fixed_Mc_att_lambda}. 

As $\lambda$ becomes more negative, the lifetime of the satellite increases until $\lambda \lesssim -0.32\times 10^{-91}$, when $\tau \gtrsim t_\text{uni}$ and the object can survive until present day. Hence, if an object with core mass $\sim 7\times 10^7\ \text{M}_\odot$ and orbital frequency $\sim 3.92\ \text{Gyr}^{-1}$ is observed, while $\lambda = 0$ will be ruled out, $\lambda < 0$ can still account for the survival of the satellite. 

We then look at another object, viz. the Fornax dwarf spheroidal, whose core mass and orbital period are $M_\text{fin} = 4\times 10^7\ M_\odot$ and $T_\text{fin} = 2\pi/\omega_\text{fin} = 3.9\ \text{Gyr}$ \cite{Hertzberg_2023}. For $m = 10^{-22}\ \text{eV}$ the $\hat{\lambda}_\text{ini} = 0$ case cannot model Fornax (following a similar argument to the case for $\hat{\lambda}_\text{ini} > 0.1$ for the object in figure~\ref{fig:fixed_core_mass_omega}).
This should be compared with results in \cite{Hertzberg_2023} where it is claimed that the existence of Fornax dwarf galaxy implies that $m = 10^{-22}\ \text{eV}$ is ruled out for FDM. 

From figure~\ref{fig:saving_fornax}, one can see that Fornax can be described only if the self-interaction strength $\lambda \approx -1.97\times 10^{-91}$. However, the corresponding scaled lifetime of Fornax will be smaller than the age of the universe. As we consider increasingly negative $\lambda$ we again see that the lifetime increases until, for $\lambda \lesssim -2.12\times 10^{-91}$, the lifetime becomes larger than the age of the universe, allowing Fornax to survive until present day. This implies that if ULDM mass is to be $m = 10^{-22}\ \text{eV}$ then the attractive self-interactions $\lambda \lesssim -2.12\times 10^{-91}$ are necessary to explain the survival of Fornax until present day. 

Also note the slight wobbly behaviour of the curves in figure~\ref{fig:saving_ULDM}. Our understanding is that this could be because: (a) for all $\hat{\lambda}_{\text{ini}}$ and $\hat{\omega}_\text{ini}$ values, we have fixed $\hat{r}_{m}$ (the distance at which we match the numerical solution with the WKB one - see discussion in section~\ref{sec:quasi_numerical} to be $55$, (b) the numerical solution becomes highly oscillatory at large $\hat{r}$ (see eq.~(\ref{eq:wkb})). We also found that this wobble appears to occur at larger values of $\hat{\lambda}_\text{ini}$ (which require smaller values of $\hat{\omega}_\text{ini}$ to describe objects with the same scaled core mass).

Hence, we have demonstrated that one can impose constraints on the strength of attractive self-interactions of ULDM from observations of DM core masses and orbital periods of satellite galaxies.

\subsection{Comparison with simulations}
Recently, the effects of self-interactions have been studied by carrying out full time-dependent simulations in \cite{Glennon_2022}. Here, authors simulate solitons with total mass $M_s$ in the narrow range $1.04 \times  10^8\ M_\odot - 1.13\times 10^8\ M_\odot$ orbiting at the outskirts of a halo ($M_h\sim 10^{11}\ M_\odot$) in the presence of attractive and repulsive self-interactions with $\lambda \approx \pm 2\times 10^{-92}$. 
Note that here the authors deal with a full time-dependent problem, and simulate the evolution of the solitons for multiple orbits. 
They find that going from $\lambda = 0$ to $\lambda = \pm 2.4\times 10^{-92}$ can increase or decrease disruption time by up to a factor of $\sim 1.5$. 

Using quasi-stationary solutions, we find that for core mass $M_{\text{fin}} = 7\times 10^7\ M_\odot$ (which is not the total soliton mass) and $\omega_\text{fin} = 3.92\ \text{Gyr}^{-1}$, lifetime of the satellite galaxy will be larger by a factor of $\sim 4$ for $\lambda = -2.4\times 10^{-92}$ compared to $\lambda = 0$ (see figure~\ref{fig:fixed_Mc_att_lambda}).

Further, for a larger core mass $M_\text{fin} = 9.1\times 10^7\ M_\odot$, we find that the effects of SI are larger, and the lifetime of the satellite increases by two orders of magnitude as one goes from $\lambda = 0$ to $\lambda = -2.4\times 10^{-92}$.
On the other hand, going from no self-interactions to $\lambda = 2.4\times 10^{-92}$ decreases the lifetime by an order of magnitude (see figure~\ref{fig:lifetime_fixed_core_mass} for curves considering both attractive and repulsive self-interactions). Hence, our results agree with those of \cite{Glennon_2022} : attractive self-interactions lead to longer-lived satellite galaxies compared to the absence of self-interactions, while repulsive self-interactions lead to shorter-lived satellites. However, the extent of the effect of $\lambda\neq 0$ differs. 

It should be noted that apart from working with quasi-stationary solutions there are some key differences in our treatment of the problem compared to Ref.~\cite{Glennon_2022}:
\begin{itemize}
    \item In our approach, we cannot reliably define a total soliton mass since the integral in eq.~(\ref{eq:soliton_mass}) does not converge at large $r$. Hence, it is not clear if we are looking at similar soliton masses to \cite{Glennon_2022}. The orbital periods (assuming a circular orbit) that the authors consider are also greater than what we can probe for the core masses that we consider. This is because in our method, the scaling relations in eqs.~(\ref{eq:scaling_mass}) and~(\ref{eq:scaling_omega}) restrict the combinations of $M_\text{fin}$ and $\omega_\text{fin}$ that can be probed. 
    \item In \cite{Glennon_2022}, the authors initialize the solitons with a fixed $\rho_c/\rho_H \approx 50$ for all $\lambda$ values. However, in our work, we utilize scaling relations to describe realistic solutions which requires different $\hat{\omega}_\text{ini}$ values to probe a fixed scaled core mass and orbital period for different $\lambda$ values. This will lead to a different ratio $\rho_c/\rho_H$ for every $\lambda$ value which can explain the larger effect of self-interactions in our case.  
\end{itemize}

\section{Summary and discussion}

Dark Matter (DM) could be composed of elementary or composite particles of a single species. Since there are no elementary or composite particles in the standard model of particle physics with properties expected from DM, it is important to look for theories beyond the standard model. Thus, astrophysical and cosmological observations of DM could teach us something about new physics.

If the particles constituting DM are Bosons, for sufficiently small particle masses, the number density is allowed to be so large that a classical wave description is valid (see \cite{Allali_2020, Allali_2021, Allali_GR_2021, Herdeiro_2022, Eberhardt_2023, Chakrabarty_2021, Eberhardt_SCF_2022, Eberhardt_2022, Alcubierre_2023, Proukakis_2023, Proukakis_BEC_2023} and references therein).
This so-called Ultra Light Dark Matter (ULDM) or Fuzzy Dark Matter (FDM) (in the absence of self-interactions) will lead to unique wave-like observational signatures at galactic scales while behaving like Cold Dark Matter (CDM) at larger scales \cite{Matos_2000, Lee_1995, Hu_2000, Schive_Nature_2014} (also see \cite{Ferreira_2021, Hui_2021, Kimball_book_2023} for recent reviews), and can potentially solve small-scale issues with CDM \cite{Bullock_2017}. However in recent years, much of the relevant mass range for ULDM with no self-interactions $\left[10^{-24}\ \text{eV}, 10^{-20}\ \text{eV}\right]$ has been severely constrained using various astrophysical and cosmological observations. See for instance constraints from rotation curves \cite{Bar_2022, Khelashvili_2023, Banares_Hernandez_2023}, Lyman-$\alpha$ data \cite{Irsic_2017, Kobayashi_2017, Rogers_2021}, CMB and Large Scale Structure \cite{Hlozek_2018, Lague_2022}, dynamical heating of stars in Ultra-Faint Dwarf galaxies \cite{Dalal_2022}, as well as recent results from NANOGrav and Pulsar Timing Arrays \cite{Afzal_2023, Smarra_2023}. 

There is however a possible out, in the form of non-negligible self-interactions. For ULDM, many of the constraints mentioned earlier are obtained in the FDM regime, i.e. the effect of self-interactions is taken to be negligible. As we have mentioned earlier, presence of extremely small self-interactions can drastically impact solutions of the Gross-Pitaevskii-Poisson system, which can potentially alter the above-mentioned constraints. For instance, the stringent constraints imposed by Lyman-$\alpha$ data can potentially be relaxed if the effects of self-interactions are taken into account (see discussion below eq.~(3.45) in section~3.2.3 in \cite{Kimball_book_2023}, section~5.1.2 in \cite{Ferreira_2021} as well as \cite{Zhang_2017, Leong_2019}). Indeed, there has been a growing interest in recent years to study and constrain self-interactions of scalar field dark matter \cite{Li_2014, Dev_2017, Suarez_2017, Desjacques_2018, Cembranos_2018, Delgado_2022, Chakrabarti_2022, Mocz_2023, Dave_2023, Ravanal_2023, Kadota_2024, Budker_2023, Boudon_2023, Winch_2023}. 

In this chapter, we studied the effects of self-interactions for a system consisting a satellite dwarf galaxy orbiting the centre of a host halo at a distance $a$ with an orbital period $T = 2\pi/\omega$ where $\omega$ is the orbital frequency of the satellite. As shown by \cite{Hertzberg_2023}, the tidal effects due to the halo will lead to an external potential of the form $-\frac{3}{2}m\omega^2r^2$ in the time-independent Schrödinger-Poisson system. Note that the authors consider only the $\ell = 0$ mode since higher order modes contribute very little for small values of $\omega$. It is also important to note that the factor of $\frac{3}{2}$ is valid only if the satellite is orbiting far enough away from the halo centre. Assuming that one can introduce a similar term to the Gross-Pitaevskii-Poisson equations (i.e. equations of motion for a classical scalar field with self-interactions), we obtain a time-independent system given by eqs.~(\ref{eq:mod_GP}) and~(\ref{eq:mod_poisson}). 

One can solve this system by treating it as a quantum mechanical tunnelling problem as authors in \cite{Hui_2017, Du_2018, Hertzberg_2023} have done albeit for $\lambda = 0$. To model tunnelling in the time-independent approximation, one requires quasi-stationary states which arise as a consequence of outgoing wave boundary conditions describing a scattering-like state (given by eq.~(\ref{eq:wkb}) at a large distance from the origin viz. the centre of the satellite in our case). Note that this forces the eigenvalue $\gamma$ to be complex where $\text{Im}(\gamma) \equiv \gamma_I$ characterises the rate of tunnelling while its inverse gives us the timescale over which the satellite loses its mass through tunnelling. To solve the system, we looked for the correct values of $\gamma_R$ and $\gamma_I$ such that the solution matches with the outgoing wave solution at large $r$ (see section~\ref{sec:quasi_numerical} for details). 

An important point worth noting here is that our approach attempts to solve a time-dependent problem using a time-independent description, in the sense that mass of DM in the satellite decays exponentially, i.e. $M(t) = M_0e^{-\Gamma t/\hbar}$ where $\Gamma$ is constant. However, detailed modelling \cite{Du_2018} based on explicit solution of time-dependent SP system indicates that the time-dependence of the mass of DM in the satellite could be different from exponential. It is worth noting that the effects of self-interactions have been considered in recent simulations \cite{Glennon_2022} where the authors tackle a full time-dependent problem. The results we have arrived at by working with the exponential model agree well with the results of their simulations, in that attractive (repulsive) self-interactions lead to a longer (shorter) lifetime for a satellite dwarf orbiting the host halo. 

We have then solved the modified time-independent GPP system numerically to obtain decay rates for various values of dimensionless $\hat{\lambda}_\text{ini}$ and $\hat{\omega}_\text{ini}$ for a fixed $m = 10^{-22}\ \text{eV}$. Using scaling relations in eqs.~(\ref{eq:scaling_mass}) and~(\ref{eq:scaling_omega}) to describe satellite galaxies with typical core masses and orbital periods, we found that for a satellite with core mass $M_c \sim 10^8\ M_\odot$, and orbital period $T\sim \mathcal{O}(\text{Gyr})$, self-couplings as small as $\lambda \gtrsim \pm 10^{-92}$ can have a substantial effect on its lifetime (see section~\ref{sec:SI_impact} for details). We looked at an object with a given core mass and orbital period and found that as we increase the strength of repulsive self-interactions, the lifetime of the satellite becomes shorter while the opposite is true for attractive self-interactions (see figure~\ref{fig:fixed_core_mass}).

Thus, one can place upper limits on repulsive self-interactions by requiring that the satellite survive at least as long as the age of the universe (see section~\ref{sec:fixed_Mc}). This is because repulsive self-interactions are responsible for a lower central density and a narrower potential barrier leading to short-lived satellite compared to no self-interactions. On other hand, we found that attractive self-interactions which lead to higher central densities and wider potential barriers, allow for a satellite to survive for longer compared to the case of no self-interactions (see figure~\ref{fig:SI_barrier} and the discussion in section~\ref{sec:intuition}). Assuming no DM self-interactions, if an object of known core mass and orbital period has a lifetime much less than the age of the universe, then introduction of sufficiently strong attractive self-interactions will make its lifetime large enough to survive until present day (see figure~\ref{fig:fixed_Mc_att_lambda}). This can severely modify the conclusions of \cite{Hertzberg_2023} which assumes negligible self-interactions. 

In chapter~\ref{chpt:paper_2}, based on~\cite{Dave_2023}, we had found that (a) repulsive self-interactions with $\lambda \gtrsim 10^{-91}$ can explain observed rotation curves for dark matter dominated galaxies as well as satisfy an appropriate soliton-halo relation, and, (b) including attractive self-interactions did not have the same interesting effect. On the contrary, in this chapter, we found that consideration of lifetimes of satellite galaxies prefers attractive self-couplings of comparable strengths. 

Ultra light scalars like the ones considered in this chapter could turn up in many ultraviolet completions of the standard model (see e.g. \cite{Cicoli_2022, Hamaide_2022, Cicoli_2023, Gendler_2023kjt} for some recent papers and references therein). If observations can be used to constrain the self-couplings of such scalars, this can go a long way towards uncovering physics which operates at ultra-short distances. It is well-known that axions will have a negative quartic self-coupling - in particular, ultra light axions can have extremely small self-couplings. Thus the results of this chapter are intriguing, given recent interest in axion dark matter \cite{Amruth_2023}.
\justifying
\chapter{Learning from rotation curves: a neural network approach}\label{chpt:paper_4}

\textbf{Based on:} \\\textbf{B.~Dave} and G.~Goswami, \emph{Learning from galactic rotation curves: a neural network approach} [\href{https://arxiv.org/abs/2412.03547}{arXiv:2412.03547 [astro-ph.CO]}], (\textbf{Under review})

\noindent Note that most of the text in this chapter is an excerpt from our preprint, Ref.~\cite{Dave_2025}. 
\vspace{1em}

\noindent In the preceding chapters, we investigated how self-interactions of the type $\lambda\varphi^4$ can alter the phenomenology of ULDM at galactic scales and consequently, constraints on the model from astrophysical observations. 
Rotation curves, as discussed in chapter~\ref{chpt:paper_2}, are one such important set of observations. 
In particular, rotation curves of dwarf galaxies have been used extensively to impose constraints on ULDM models in the presence as well as absence of self-interactions (see~\cite{Bernal_2017, Delgado_2022, Khelashvili_2023, Banares_Hernandez_2023} for more details).

In this chapter, however, noting the emergence of the very interesting use-cases of machine learning techniques for parameter estimation, emulation and non-parametric reconstruction, we shall shift focus and explore a novel way of extracting information from data, i.e., using neural networks. 
In particular, we focus on parameter estimation using neural networks, where we shall use rotation curves as the observed data, to which we would like to fit parameters describing a dark matter density profile as well as Baryonic contribution. 
The dark matter density profile will have a core-halo structure as discussed in section~\ref{sec:rc_observations} where the core is described by a ULDM soliton with $\lambda = 0$.
The goal of the work carried out in this chapter is to understand and evaluate the viability of neural networks in the above context and compare our results with standard Bayesian inference.

\section{\label{sec:ANN_motivation}Introduction}

In the last few decades, large observational data sets at both astrophysical and cosmological scales have enabled us to place ever-stringent constraints on many exciting new physics ideas such as dark matter, dark energy, inflation, etc. (see for instance \cite{PDG_2024, Planck2018}). 
In the near-future, we expect even larger data sets with more accurate observations from various upcoming experiments like LSST, CMB-S4, DESI, etc. \cite{Ivezić_2019, Abazajian_CMBS4_Snowmass2021, DesiCollabVI_2024}. 
At the same time, in the last decade or so, advances in computer hardware, and especially parallel computing, have led to a renewed interest in machine learning techniques. In particular, deep learning using Artificial Neural Networks (ANNs), Convolutional Neural Networks (CNNs), transformers, etc. \cite{Mehta_2019, Alzubaidi_2021, Vaswani_2023}, has proved to be very useful in extracting information from data. 

These novel tools and techniques are currently being applied to a wide range of astrophysical and cosmological datasets \cite{Graff_2012, Wang_2020, Wang_ECoPANN_2020, Fluri_2021, Ho_2021, Gomez-Vargas_2021, Pal_2023, Pal_ANN-CMB_2023, Pal_CNN-CMB_2023, Shah_2024, GarciaArroyo_2024, Pal_2024, Hagimoto_2024, Artola_2024, Garuda_2024, Nerin_2024}. 
For instance, Refs.~\cite{Graff_2012, Gomez-Vargas_2021} use ANNs to compute likelihood in Bayesian inference to reduce computational time. 
On the other hand, Ref.~\cite{Wang_2020} carries out non-parametric reconstruction of the Hubble parameter as a function of redshift while  Ref.~\cite{GarciaArroyo_2024} does so for galactic rotational velocity as a function of radius.
Ref.~\cite{Wang_2020} finds that if one conducts Bayesian inference using Markov Chain Monte-Carlo (MCMC) on the reconstructed data, the resultant cosmological parameter posteriors are in agreement to those obtained from observed data. 
Deep learning has also been applied to the reconstruction of CMB B-modes\cite{Pal_2024}, as well as to reconstruct full CMB spectra from partial sky data \cite{Pal_ANN-CMB_2023, Pal_CNN-CMB_2023}.
Neural networks have also been used to estimate parameters from observational data like $H(z)$ data \cite{Pal_2023}, CMB angular power spectrum \cite{Wang_ECoPANN_2020}, CMB birefringence maps \cite{Hagimoto_2024} as well as Lyman-$\alpha$ spectra \cite{Artola_2024}.

In this chapter, we use neural networks to extract model parameters from galactic rotation curves from the Spitzer Photometry \& Accurate Rotation Curves (SPARC) catalog \cite{Lelli_2016}. 
Recall that rotation curves  are pivotal in tracing the mass distribution of dark matter and baryons in galaxies, and are an important test of dark matter models \cite{Salucci_2019, Profumo_2019}.
We consider dark matter to comprise of a spin-zero particle with mass $m\sim 10^{-22}\ \text{eV}$ (called Ultra-Light Dark Matter (ULDM)), whose large deBroglie wavelength leads to the formation of a flat density core surrounded by a cold dark matter-like envelope \cite{Schive_Nature_2014, Schive_PRL_2014} (see also \cite{Ferreira_2021, Hui_2021, Kimball_book_2023}).
We thus have the following five free parameters: mass of the dark matter particle $m\ (\text{eV})$, along with galaxy specific parameters such as the scaling parameter $s$, which characterizes the dark matter core, core-to-envelope transition radius $r_t\ (\text{kpc})$ and NFW scale radius $r_s\ (\text{kpc})$ which characterize the surrounding halo. The effect of Baryons is parameterized by the stellar mass-to-light ratio $\Upsilon_*\ (M_\odot/L_\odot)$, which tunes contribution from the stellar disk (see Section~\ref{sec:model_and_data} for details). 

For a model with the above parameters, we ask the following: for a chosen galaxy, given the observed rotation curve [i.e. observed values of velocities $V_{obs}(r)$ for some finite number ($N_{obs}$) of radius values along with their uncertainties $\sigma(r)$], what can we say about the values and uncertainties of parameters $m$, $s$, $r_t$, $r_s$ and $\Upsilon_*$?

The usual approach to answer this question involves Bayesian inference, where given some prior distribution of parameters and a likelihood function, one can obtain the posterior distribution of parameters using Bayes' theorem.
MCMC methods \cite{McKay_book_2023} are then used to sample from this posterior which in-turn gives the best-fit parameters along with confidence intervals (see \cite{Bernal_2017, Delgado_2022, Khelashvili_2023, Banares_Hernandez_2023} for some recent work on constraining ULDM parameters using rotation curves). 
In the context of our problem, given the $N_{obs}$ values of rotational velocity, for the case of uniform priors, this problem is equivalent to the problem of finding regions in the five dimensional parameter space in which the likelihood function is large. 

Since the five parameters are estimated from $N_{obs}$ values of rotational velocity, it is interesting to ask whether there could be a well defined function from $\mathbb{R}^{N_{obs}}$ to $\mathbb{R}^5$ which, when fed the rotation curve (i.e. a point in $\mathbb{R}^{N_{obs}}$), gives the ``best fit" parameters (i.e. a point in $\mathbb{R}^5$). 

We explore whether, using simulated rotation curves, one can train a neural network to approximate this function. 
Typically, $N_{obs}\sim \mathcal{O}(15)$ for the galaxies we consider, while the number of parameters we want to infer is $5$ (or $10$ if uncertainties are also inferred). 
If the neural network has $2$ hidden layers with $200$ neurons each, it will have $\sim (40-50)\times 10^3$ internal adjustable parameters (called weights and biases).
To fix these internal parameters, we need to train the neural network. 
For training, we use simulated rotation curves whose parameter values are already known. 
We generate training data for $7$ dwarf galaxies from the SPARC catalog \cite{Lelli_2016} and train a different neural network for each galaxy. 
The size of the training data, i.e., the number of known pairs of rotation curves and parameters in our work is $\sim 10^5$. 
The details of neural networks used and their architectures are discussed in Sections~\ref{sec:ANN_basics} and~\ref{sec:architecture} respectively. 

To test our trained neural networks, we use the observed rotation curves for the $7$ galaxies as input and infer parameter values in Section~\ref{sec:noiseless_case}.
Then, in Section~\ref{sec:noisy_case}, we explore the effect of noise in the training data on performance of the neural network during parameter inference, and find that the including noise improves point-estimates of parameters when confronted with observed rotation curves, i.e., the rotation curve obtained from these parameter values agree well with the observed rotation curves. 
In Section~\ref{sec:uncertainties}, we also utilize two different ways of obtaining uncertainties in the model parameters: during inference (as carried out in \cite{Wang_ECoPANN_2020}), or during training (following the work in \cite{Pal_2023}). 
Finally, we compare the parameter point-estimates and uncertainties obtained using our approach to those obtained using MCMC in Section~\ref{sec:mcmc_comparison}.
We conclude in Section~\ref{sec:discussion}.

\section{\label{sec:rot_curves_ANN}Rotation curves and artificial neural networks}

\subsection{Model and data\label{sec:model_and_data}}

Galactic rotation curves, i.e. orbital velocity of stars and gas as a function of distance from the centres of galaxies are an important probe of the matter (visible and dark) distribution in said galaxies \cite{Salucci_2019}. 

The total gravitational potential of the galaxy includes contribution from both baryonic (disk, bulge, gas) and dark matter components, allowing one to split the total velocity\cite{Lelli_2016}:  
\begin{equation}\label{eq:vel_components}
    V_{obs} = \sqrt{V_{DM}^2 + V_{g}|V_g| + \Upsilon_d V_{d}|V_d| + \Upsilon_b V_{b}|V_b|}\ ,
\end{equation} 
where $V_d$, $V_b$, and $V_g$ are contributions from the stellar disk, bulge and gas components, while $V_{DM}$ is the dark matter contribution.
Contributions from the stellar disk and bulge can be further tuned by $\Upsilon_d$ and $\Upsilon_b$, i.e. the disk and bulge mass-to-light ratios respectively, which are free parameters. 
Baryonic velocities, i.e. $V_d$, $V_g$, and $V_b$ can be obtained by fitting relevant density profiles to observed surface brightness profiles \cite{Lelli_2016}.
For galaxies without a bulge, $V_b = 0$ at all radius values, and $\Upsilon_* \equiv \Upsilon_d$ is the only free parameter. 

In this chapter, as we have done throughout the thesis, we consider dark matter to comprise of ultralight spin-zero scalars, with $m\sim 10^{-22}\ \text{eV}$. 
Due to the large deBroglie wavelength $\mathcal{O}(1\ \text{kpc})$, simulations suggest that dark matter halos in the ULDM paradigm have a core-halo structure where, the inner regions of the halo are described by flat density cores \cite{Schive_Nature_2014}. 
These cores are stationary state solutions of the Schrödinger-Poisson system of equations.
In the outer regions, beyond a transition radius, ULDM behaves like CDM and the corresponding density profile can be described by the well known Navarro-Frenk-White (NFW) profile \cite{Navarro_1996}. 
Hence, for a galactic halo, the total dark matter density profile can be written as
\begin{equation}\label{eq:uldm_nfw}
    \rho_{DM}(r) = \rho_{ULDM}\Theta(r_t - r) + \rho_{NFW}\Theta(r - r_t)\ ,
\end{equation}
where $r_t$ is the transition radius. 
In the absence of self-interactions\footnote{Recall from section~\ref{sec:stationary_states} that we solve the GPP system for fixed values of  $\hat{\lambda}_{\text{ini}}$ using the shooting method. Since one has to choose the initial choice of $\hat{\gamma}$ carefully for every $\hat{\lambda}_{\text{ini}}$, we have only solved the system for a finite number of discrete values of $\hat{\lambda}_\text{ini}$.
Hence we leave considering one parameter to be discrete while others are continuous to future work and stick to the $\lambda = 0$ case.}, instead of solving for the stationary state solution, $\rho_{ULDM}$ can also be described by the following fitting function \cite{Schive_Nature_2014}

\begin{equation}\label{eq:schive_profile}
    \rho_{ULDM}(r) \simeq \frac{0.019\times (m/10^{-22}\ \text{eV})^{-2}(r_c/\text{kpc})^{-4}}{\left[1 + 0.091\times (r/r_c)^2\right]^8}\ M_\odot/\text{pc}^3\ ,
\end{equation}
where $r_c$ is defined as the radius at which the density becomes half its central value, and is given by 
\begin{equation}\label{eq:core_radius}
    r_c = 0.8242\left(\frac{s}{10^4}\right)\left(\frac{m}{10^{-22}\ \text{eV}}\right)^{-1}\ \text{kpc}\ .
\end{equation}
Note that the free parameters here are the ULDM particle mass $m$ and the scale parameter $s$. The scale parameter allows one to describe solitonic solutions of different masses and radii \cite{Chakrabarti_2022, Dave_2023}.

The NFW density profile, obtained form CDM-only simulations \cite{Navarro_1996} is given by 
\begin{equation}\label{eq:nfw}
    \rho_{NFW}(r) = \frac{\rho_s}{\frac{r}{r_s}\left(1 + \frac{r}{r_s}\right)^2}\ M_\odot/pc^3.
\end{equation}
Here $\rho_s$ and $r_s$ are halo-specific parameters. 
Since we impose continuity at the transition radius, i.e. $\rho_{ULDM}(r_t) = \rho_{NFW}(r_t)$, one can eliminate $\rho_s$ and describe the NFW part of the profile using only $r_t$ and $r_s$.

The circular velocity of a test particle moving under the influence of the spherically symmetric density profile in Eq.~(\ref{eq:uldm_nfw}), is simply 
\begin{equation}\label{eq:circ_vel_2}
    v(r) = \sqrt{\frac{GM(r)}{r}} = \sqrt{\frac{4\pi G\int_0^r\rho_{DM}(r')r'^2dr'}{r}}\ .
\end{equation}
Hence, the free parameters that characterize the rotation curve are: mass of the ULDM particle $m$, the scaling parameter $s$, the radius at which ULDM transitions to NFW $r_t$, the scale radius of the NFW profile $r_s$ and the stellar mass-to-light ratio $\Upsilon_*$, i.e. 
\begin{equation}\label{eq:params}
    {\bf P} = \left(m, s, r_t, r_s, \Upsilon_*\right)\ .
\end{equation}

\subsubsection{Observed Rotation Curves}\label{sec:RCs}

In this chapter, we utilize observed rotation curves from the SPARC catalog which hosts
high quality HI/H$\alpha$ rotation curves for 175 galaxies \cite{Lelli_2016}. 
The SPARC catalog has been utilized previously to constrain ULDM parameters \cite{Bernal_2017, Bar_2022, Delgado_2022, Khelashvili_2023} as well as models of modified gravity \cite{de_Almeida_2018,Bhatia_2024} and obtain bounds on the cosmological constant \cite{Benisty_2024}.

Since ULDM affects dark matter distribution in the inner regions of galaxies, one must look at galaxies where baryons are not the dominant component even at small radii, or dark matter dominated galaxies. 
To study ULDM, authors in Ref.~\cite{Delgado_2022} chose $17$ dark matter dominated dwarf galaxies from the SPARC catalog with well-defined inner regions.
In this chapter, we choose a subset of $7$ galaxies from the sample of $17$. 
It is important to note that, along with observed rotation curves, the SPARC catalog provides values for $V_d$ and $V_g$ by fitting relevant stellar density profiles. We shall utilize these values directly in eq.~(\ref{eq:vel_components}), while allowing $\Upsilon_*$ to vary. 

At this point, as discussed in Section~\ref{sec:ANN_motivation}, we note that we are trying to approximate a function that can take an observed rotation curve as input and infer model parameters in Eq.~(\ref{eq:params}) as well as their uncertainties as output. 
To understand how a neural network can do this, we must briefly discuss the ingredients involved in defining and training an artificial neural network in the following sub-section.

\subsection{Artificial neural networks}\label{sec:ANNs}

\subsubsection{Basics of neural networks}\label{sec:ANN_basics}

Neural networks (NNs) are an important tool in supervised machine learning, that can approximate complex relationships between some input ${\bf x}$ and output ${\bf y}$. 
This is done by looking at a set of examples called `training data' consisting of known pairs of inputs (also called features) and corresponding outputs (also called targets). 
Note that the output ${\bf y}$ can either be a vector of continuous values (in case of regression), or categories from a finite set (in case of classification). 
For a set of $I$ known input-output pairs $\{{\bf x}_i, {\bf y}_i\}_{i = 1}^I$, a feed-forward neural network is simply a function ${\bf f}$ of the input ${\bf x}$ parameterized by ${\bf \Omega}$, 
\begin{equation}\label{eq:ANN_func}
    {\bf f} = {\bf f}({\bf x}; {\bf \Omega})\ .
\end{equation}
Here, ${\bf \Omega}$ are the internal adjustable parameters (IAPs) of the neural network. 
To understand how to construct such a function, we first look at the fundamental building block of a neural network, the neuron. 
For an input vector ${\bf x}\in \mathbb{R}^{N_0}$, the output of a neuron is defined as 
\begin{equation}\label{eq:neuron}
    v = a({\boldsymbol \omega}\cdot {\bf x} + \beta)\ .
\end{equation}
Here, components of ${\boldsymbol \omega}$ are called weights and $\beta$ is called bias. 
The function $a(z)$ is called the activation function, which imparts a non-linearity to the transformed input. 
The choice of the activation function depends on the kind of problem at hand. 
For instance, in the case of classification problems where the required output is discreet, the sigmoid function, $a(z) = (1+e^{-z})^{-1}$ is useful since the output is contained between $0$ and $1$. For regression problems, where the required output is continuous, the rectified linear unit (ReLU), $a(z) = \mathrm{max}(z, 0)$ can be used. 

One can also define a layer (often called a hidden layer) of $N$ neurons, where the input for each neuron is the same albeit with different weights and biases. 
The output of the layer can be written as a vector of size $N$, where each component is given by Eq.~(\ref{eq:neuron}) with different weights and biases, 
\begin{equation}
    {\bf v} =  {\bf a}\left({\boldsymbol \omega}{\bf x} + {\boldsymbol \beta}\right)\ .
\end{equation}
In this case, ${\boldsymbol \omega}$ denotes a matrix of size $N\times N_0$ while $\beta$ is a $N\times 1$ column matrix. 
Also note that activation function vector ${\bf a}$ is applied element-wise. 

A neural network can have multiple such layers, where the output of one layer acts as the input for the next one; in particular when for all hidden layers, the output of each neuron in the current layer acts as an input to every neuron in the next layer, it is called a fully connected neural network. 
For a neural network with $L$ layers, we use the index $j$ to keep track of which layer we are talking about, where $1 \leq j \leq L$. The number of neurons in the $j$th layer is denoted by $N_j$.

In this case, output of the $j$th layer with $N_j$ neurons, is defined to be ${\bf v}_j$,
\begin{equation}\label{eq:layer}
  {\bf v}_j = {\bf a}\left({\boldsymbol \omega_j}{\bf v}_{j-1} + {\boldsymbol \beta}_j\right)\ . 
\end{equation}
Here, $j = 1, 2, ..., L$, while ${\boldsymbol \omega}_j$ and ${\boldsymbol \beta}_j$ are the weights and biases for the $j$th layer. 
For ease of notation, one can denote weights and biases for all layers by $\boldsymbol{\Omega}$. 

The final output of a neural network, can be written as a composition of multiple functions ${\bf f}({\bf x}, {\bf \Omega}) = {\bf v}_L({\bf v}_{L-1}(...{\bf v}_1({\bf x}, {\bf \Omega})))$, where $L$ is the total number of layers in the neural network \cite{Strang_book_2018}. 

It must be noted that one can use $j = 0$ to denote the input layer, which is just the input vector ${\bf x}$. For ANNs, no transformations occur at this layer.
One can also define $j = L+1$ as the output layer, whose size will be the same as ${\bf y}$. 
Unlike the input layer, going from $j=L$ to $j = L+1$ does involve a transformation similar to Eq.~(\ref{eq:neuron}), but with $a(z) = z$. 

The universal approximation theorem \cite{Hornik_1990} shows that a neural network with a single hidden layer can approximate any non-linear function with a finite number of neurons.
However, it is often found that utilizing multiple layers, i.e. deep neural networks are easier to train and generalize better than shallow ones \cite{Prince_book_2023}. 
This is not understood well and is an active area of research \cite{Mhaskar_2017, Meir_2023}.

\begin{figure}[h]
    \centering
    \includegraphics[width=0.7\textwidth]{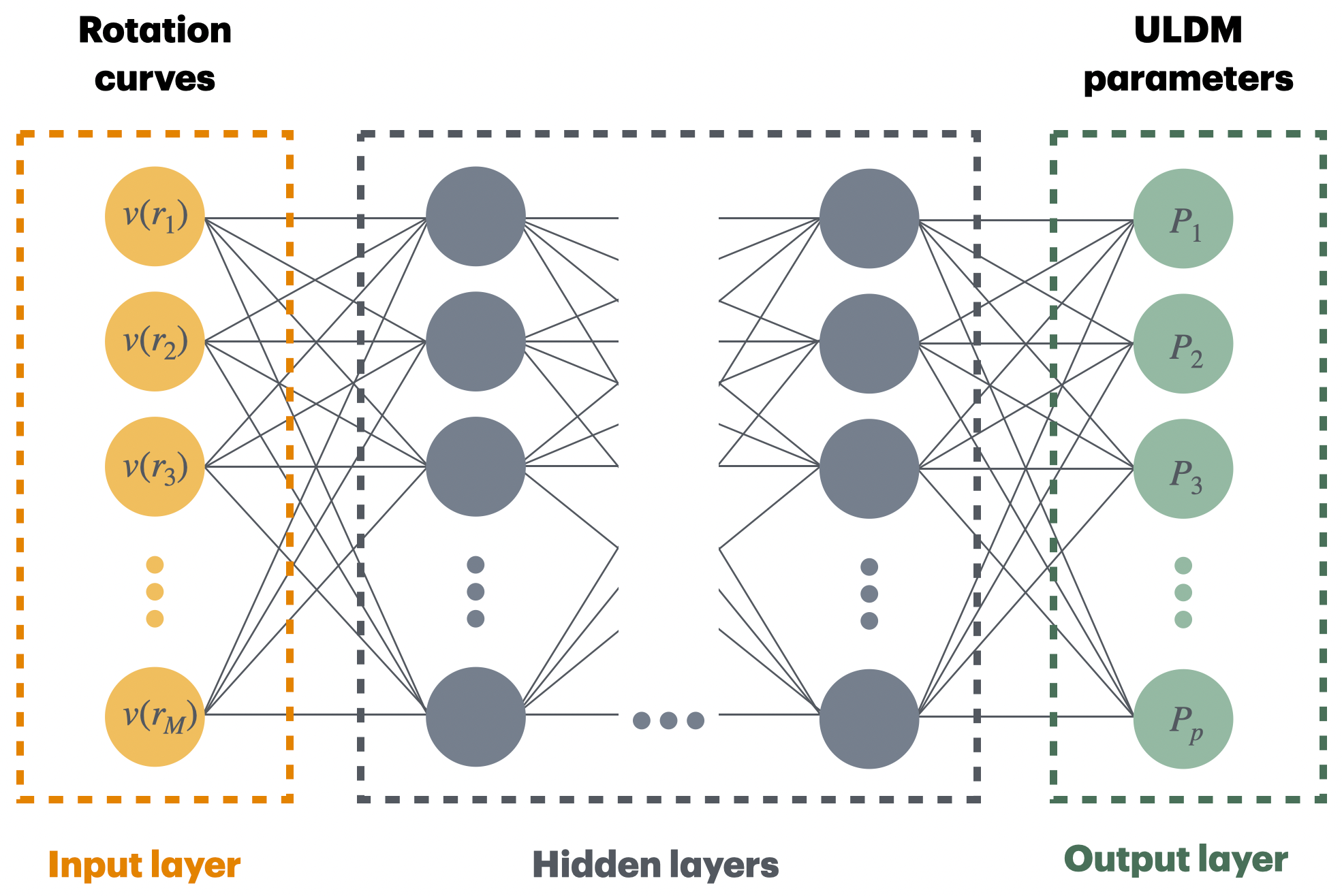}
    \caption[Neural network schematic]{\justifying A schematic of the neural network that we want to construct, where given a rotation curve of dimension $M$ for a galaxy as input, one can obtain the corresponding $p$ ULDM parameters. In our work, while the number of parameters to predict will be $5$ (Eq.~(\ref{eq:params})), the number of output neurons $p = 5$ or $p = 10$ depending on the loss-function used.}
    \label{fig:ANN_cartoon}
\end{figure}

In our case, the neural network of interest is shown in Fig.~\ref{fig:ANN_cartoon}, where the velocities for $N_{obs}$ radius values are in the input layer, while the model parameters in Eq.~(\ref{eq:params}) are in the output layer, and we allow for more than $1$ hidden layers. 
It is worth noting, that for the case of a heteroscedastic loss function, as we shall see in Section~\ref{sec:hetero_uncertainty}, the size of the output layer will be doubled. 
This is because we shall require the uncertainties in the parameter values to be learned during training itself. 

It is also important to note that the number of observed data points in the rotation curves of different galaxies will not be the same, implying that the size of the input layer will be different for each galaxy. 
Further, the range of radius values for which there are observed velocities will also be different for each galaxy. 
Due to this, one must define a different neural network for every galaxy in our sample, i.e., we train a total of $7$ neural networks for every case in Section~\ref{sec:MSE_predictions} and~\ref{sec:uncertainties}.

\subsubsection{Training a neural network}\label{sec:ANNs_training}

The goal of training a neural network is to find the optimum values of the IAPs or weights and biases ${\bf \Omega}$, such that for an input from the training data, the neural network output ${\bf f}$ (also called predicted or inferred output) is close to the target output ${\bf y}$ (also called ground truth). 
To quantify this closeness, one employs a loss function $\mathcal{L}({\boldsymbol \Omega})$, which assigns a real number to each pair of predicted and target output. 
Training thus involves finding a set of ${\boldsymbol \Omega}$ such that $\mathcal{L}$ is minimized. 
Well-known examples of loss functions are the mean-squared error (regression tasks) and binary cross-entropy (classification tasks). 
In this paper, we shall consider two different loss functions: (a) mean-squared error (MSE) in Eq.~(\ref{eq:MSE_loss}) and (b) heteroscedastic loss in Eq.~(\ref{eq:hetero_loss}).  
The mean-squared-error loss is simply the proportional to the Euclidean distance between the predictions and the target values, averaged over the number of samples, given by
\begin{equation}\label{eq:MSE_loss}
    \mathcal{L} = \frac{1}{np}\sum_{i=1}^n|{\bf y}_i - {\bf \hat{y}_i}|^2\ ,
\end{equation}
where ${\bf y}_i$ is the output corresponding to the $i$th input ${\bf x}_i$, while ${\bf \hat{y}}_i \equiv {\bf f({\bf x}_i, {\bf \Omega})}$ is the prediction made by the neural network for the same input. 
$p$ is the size of the output vector. 

The goal during training is to find the global minimum of the loss function in the $\boldsymbol{\Omega}$-space \footnote{This can be a high-dimensional space, since the number of IAPs range from a few thousand to tens of millions.}. 
The usual way involves utilizing a gradient descent algorithm which requires two ingredients: (a) an efficient way to calculate the gradient of the loss function w.r.t $\boldsymbol{\Omega}$, (b) a rule to update the weights and biases in the direction of the steepest descent. 

The former can be computed using the backpropagation algorithm \cite{Rumelhart_1986} which computes the gradient backwards from the last layer, using the chain rule and layered structure of a neural network to avoid redundant calculations. 
Once the gradient is calculated, it is scaled by a step size (also called the learning rate) and the IAPs are updated in the opposite direction. 
There are numerous algorithms to carry out the update, like stochastic gradient descent, nesterov, adaptive moment estimation (Adam), etc. (see \cite{Prince_book_2023} for a detailed discussion on backpropagation and gradient descent methods). 

Usually, during training the above procedure must be carried out many times, i.e. the same training data is passed through neural network and the parameters are updated multiple times to reduce the loss.  
Training is complete when the loss converges to a minimum. The true test of a neural network is how it deals with data it was not trained on, or unseen data. 
A well-trained neural network should generalizes well, i.e. it should make accurate predictions even for the inputs that are not present in the training data. 
If the training of a neural network goes on for too long, it can memorize the training data and start to perform worse on unseen data. This is called overfitting \cite{Prince_book_2023}, and one must stop the training before this.

We would like to point to our reader Refs.\cite{Strang_book_2018, Mehta_2019, Prince_book_2023} for excellent pedagogical discussions on the various aspects of machine learning and neural networks.
For our work in this chapter, the choice of the gradient descent algorithm, learning rate, and other parameters related to the optimization of the loss function for our problem are discussed in Section~\ref{sec:architecture}.

\subsection{Generating simulated rotation curves\label{sec:simulated_RCs}}

Neural networks usually require a sufficiently large number of known pairs of inputs ${\bf x}$ and target outputs ${\bf y}$ to train on. 
In this case it will be a set of velocities from a rotation curve as input and the parameter vector ${\bf P}$ that can generate the rotation curve. 
However, for a fixed galaxy, we only have a single set of observed velocities, i.e. one rotation curve.
Hence, to successfully train a network to learn the relationship between rotation curves and parameters, we have to rely on a set of simulated rotation curves as training data.  
Consider a galaxy with an observed rotation curve between $R_{min}$ and $R_{max}$ consisting of $N_{obs}$ data points (i.e. $N_{obs}$ values of radii for which there are observed velocities). 
To generate $I$ simulated rotation curves for this galaxy, we follow the procedure below: 

\begin{enumerate}
    \item First, we define a uniform distribution for each parameter.
    While any random combination of parameters can form a velocity curve, not all of them will be visually similar to the observed one. 
    Hence, choosing sensible ranges for the uniform distributions for each galaxy is important. 
    We do this by examining how the numerically generated curves vary with each parameter in comparison to the observed rotation curve and choose the lower and upper limits accordingly.
    \item We make a random draw from each parameter distribution, which will give a parameter vector {\bf P} (see Eq.~(\ref{eq:params})) in the $5$D parameter space.
    \item We use $\{m, s\}$ to obtain the ULDM density profile from Eq.~(\ref{eq:schive_profile}), and then use $\{r_t, r_s\}$ to obtain the profile of the NFW skirt.
    Finally, given the total density profile, we calculate the dark matter velocity profile $V_{DM}$ using Eq.~(\ref{eq:circ_vel_2}). 
    \item Using the randomly drawn mass-to-light ratio $\Upsilon_*$ along with the fixed $V_d$ and $V_g$ values provided by the SPARC catalog, we obtain the baryonic contribution to the velocity curve. 
    \item We finally construct the full velocity curve between $R_{min}$ and $R_{max}$ for all the $N_{obs}$ radius values using Eq.~(\ref{eq:vel_components}). 
\end{enumerate}

The ranges of the uniform distribution for each parameter are shown in Table~\ref{tab:param_space} for all galaxies. 
We employ the above procedure for $I = 5\times 10^5$ randomly chosen parameter vectors ${\bf P}$ from the above-mentioned ranges to obtain simulated rotation curves $v(r)\in \mathbb{R}^{N_{obs}}$.
These simulated rotation curves will serve as the training inputs in Fig.~\ref{fig:ANN_cartoon}, while ${\bf P}$ will be the target parameters that the neural networks attempts to predict for every input rotation curve.

\begin{table}[h!]
\resizebox{\columnwidth}{!}{
\begin{tabular}{@{}cccccc@{}}
\toprule
\multirow{2}{*}{\textbf{Galaxy}} & \multicolumn{5}{c}{\textbf{Parameter ranges}} \\ \cmidrule(l){2-6}
& $m$ ($10^{-23}$\ eV) & $s\ (10^3)$ & $r_t$ (kpc) & $r_s$ (kpc) & $\Upsilon_* (M_\odot/L_\odot)$ \\ \cmidrule(r){1-6}
DDO 154 & $\left[1, 10\right]$ & $\left[3, 9\right]$ & $\left[1, 5.99\right]$ & $\left[1, 15\right]$ & $\left[0.3, 0.8\right]$ \\ \cmidrule(l){1-6}
ESO444-G084 & $\left[1, 10\right]$ & $\left[2, 9\right]$ & $\left[1, 4.44\right]$ & $\left[1, 15\right]$ & $\left[0.3, 0.8\right]$\\ \cmidrule(l){1-6}
UGC 5721 & $\left[1, 10\right]$ & $\left[1.5, 5\right]$ & $\left[1, 6.74\right]$ & $\left[1, 15\right]$ & $\left[0.3, 0.8\right]$ \\ \cmidrule(l){1-6}
UGC 5764 & $\left[1, 10\right]$ & $\left[2, 9\right]$ & $\left[1, 3.62\right]$ & $\left[1, 15\right]$ & $\left[0.3, 0.8\right]$\\ \cmidrule(l){1-6}
UGC 7524 & $\left[1, 10\right]$ & $\left[1, 9\right]$ & $\left[1, 10.69\right]$ & $\left[1, 15\right]$ & $\left[0.3, 0.8\right]$\\ \cmidrule(l){1-6}
UGC 7603 & $\left[1, 10\right]$ & $\left[2, 7\right]$ & $\left[1, 4.11\right]$ & $\left[1, 15\right]$ & $\left[0.3, 0.8\right]$\\ \cmidrule(l){1-6}
UGC A444 & $\left[1, 10\right]$ & $\left[2, 9\right]$ & $\left[1, 2.55\right]$ & $\left[1, 15\right]$ & $\left[0.3, 0.8\right]$\\
\bottomrule
\end{tabular}
}
\caption[Uniform ranges of parameters explored for all galaxies]{\justifying Uniform ranges for all parameters. Note that ULDM mass is chosen to be such that the size of the core is $\sim \mathcal{O}(1\ \text{kpc})$. 
Similarly, requiring that ULDM describes the inner regions for all galaxies, the lower limit for the transition radius is $r_t \geq 1\ \text{kpc}$. The upper limit is fixed to be the largest radius bin for which there is an observation and hence is galaxy specific.}
\label{tab:param_space}
\end{table}

\subsection{\label{sec:preprocessing}Pre-processing}

Pre-processing of training data usually involves converting it to some more usable type or to normalize the input data such that all inputs have a similar range of values \cite{Eisert_2022}. 
This is required to ensure that no component of the input is considered to be more important if it has a larger absolute value or variation.
It also prevents the gradient descent algorithms from taking too small or too large steps based on the absolute value of the inputs. 

Before proceeding, we split our simulated data into three parts. For each galaxy considered in our analysis, we reserve $0.8I$ ($4\times 10^5$) examples for training, and $0.1I$ ($5\times 10^4$) examples each for validation and testing. 
The neural network will update its weights and biases only based on the examples in the training set. 
Hence, the validation set acts as unseen data for the neural network and will only be used to monitor if the network is overfitting. 
It can also be used to measure performance across various hyperparameters values. 
Finally, test data also acts as unseen data and will be used to characterize the performance of the final neural network once it has been trained. A well trained neural network will have a similar loss value across all three datasets. 
Note that for the remainder of this paper, training set or data will refer to the $80\%$ subsample of the simulated dataset. 

While there are various techniques to scale the components of the input, we use z-score normalization (also called standardization) on our input.
For each radius bin for which we have a velocity value, we subtract the velocity from the mean and divide by the standard deviation of the training data, 
\begin{equation}
    \Tilde{{\bf v}}_l = \frac{{\bf v}_l - \mu_l}{\sigma_l}\ .
\end{equation}
Here $\mu_l$ and $\sigma_l$ are the mean and standard deviation over the $l$th radius value of the training data (validation and test data are not included in this calculation). The boldface here implies a column vector the size of the training data, i.e. $4\times 10^5$. 
We then use the mean and standard deviation of the training set itself to scale the validation and test sets, as well as the actual observed rotation curve before feeding them to the neural network. 

We also scale output parameters to ensure that each component has values which are close to $\mathcal{O}(1)$. This leads to scaling the output parameter vector as: 
\begin{equation}
    {\bf P} = \{m/10^{-23},\  s/10^3,\  r_t,\  r_s,\  10\Upsilon_*\}\ .
\end{equation}
The output parameters (${\bf P}$) that the neural network predicts therefore need to be scaled back to familiar units during final predictions.

\subsection{\label{sec:architecture}Neural network architecture}

It is easy to see from the discussion in Section~\ref{sec:ANN_basics} that a larger number of neurons per layer as well as a larger number of layers in a neural network will enable it to approximate a complex function better. 
However, the performance of the network also depends on other parameters like batch size, learning rate of the optimizer, choice of the optimization algorithm and activation function as well as the loss function used, etc. 
These parameters that characterize a neural network are called hyperparameters and choosing their optimal values for a neural network is a difficult task. This is because they must be chosen empirically, i.e. by training the neural network multiple times with various combinations of hyperparameters and choosing those which perform best on some unseen data. 

Often a full grid search in the hyperparameter space is computationally expensive, which has lead to the use to some other methods like random search or genetic algorithms \cite{GarciaArroyo_2024}. 
We employ a grid search for only the base architecture of the neural network, i.e. number of layers and number of neurons per layer, in a small grid of parameters. 
The rest of the hyperparameters are chosen by trial and error across various training runs.

The fixed hyperparameters, except for the base architecture of the network, are the following: 
\begin{enumerate}
    \item Activation function: For every neuron in the hidden layers, we implement the ReLU (Rectified linear unit) activation function, given by $a(z) = \mathrm{max}(z, 0)$. 
    \item Loss function: We use the mean-squared-error (MSE) loss given by Eq.~(\ref{eq:MSE_loss}). Note that we shall also implement a different loss function in Section~\ref{sec:hetero_uncertainty} given by Eq.~(\ref{eq:hetero_loss}). 
    \item Optimization algorithm: We use a momentum-based stochastic gradient descent algorithm called ADAM \cite{Kingma_ADAM_2017}. Parameters of the algorithm save for the learning rate, are kept at their default values.
    \item Learning rate: The step-size used for updating IAPs; we fix it to $10^{-4}$. 
    \item Batch size: For a stochastic gradient descent method, IAPs are updated based on loss calculated for a small subset of size $\mathcal{B}$ of the total training set. 
    This small subset is called a batch (or a mini-batch), and is randomly drawn from the training set without replacement. 
    We use $\mathcal{B} = 32$, implying $4\times 10^5/32 = 12500$ updates (or iterations) of the internal adjustable parameters before all samples from the training set are fed to the neural network once. 
    \item No. of epochs: When the entire training set passes through the neural network once, it is called an epoch.
    Usually, multiple epochs are required to adequately train a neural network. 
    Note that, if the training goes on for too many epochs, the neural network can memorize the training set, which leads to a poor performance on unseen data, i.e. validation and test sets (this is called overfitting). 
    To prevent this, we also keep an eye on the validation loss while training, and find that by $250$ epochs, the validation loss stops decreasing appreciably or starts to increase. 
    \item Dropout: To prevent overfitting we also utilize dropout regularization, where for every sample from the training set, the output of each neuron is set to zero with probability $d$, so that each sample is passed through a different `thinned' neural network \cite{Srivastava_2014}. 
    We set $d=0.2$ for every hidden layer in the network. 
\end{enumerate}

To find the optimum architecture of the neural network, we first fix the above hyperparameters at the values mentioned. We then construct a grid that consists of the number of hidden layers $L \in [1, 2, 3, 4]$ and neurons per layer $N_j \in [100, 150, 200]$. 
For a galaxy, for each combination of $L$ and $N_j$ we train a neural network for 250 epochs.
The performance of these trained networks is then evaluated using the loss on validation data, (recall that the weights and biases are not updated based on validation data) and we choose the architecture for which the validation loss is the lowest.
Note that this grid search is carried out with noise-induced simulated rotation curves (see Section~\ref{sec:noisy_case} for details) with the mean-squared-error loss function, while the obtained optimal architecture is utilized for the case of noiseless training data as well as the case with a heteroscedastic loss function (as we shall discuss in Section~\ref{sec:hetero_uncertainty}).

We perform this grid search for $4$ galaxies, and find that validation loss was lowest for networks with at least $2$ hidden layers, while the number of neurons per layer varied. 
It is worth noting that the difference between validation loss for the best performing architectures for a given galaxy was very small ($\sim \mathcal{O}(10^{-3})$). 
Therefore for purposes of this paper, we fix the number of hidden layers to $2$ and neurons per layer to $200$ for all 7 galaxies in the sample.

Finally, using the index notation discussed in Section~\ref{sec:ANN_basics}, the neural network architecture will comprise of an input layer ($j=0$), whose size is the number of observed radius values for a particular galaxy, $2$ hidden layers ($j = 1,2$) with $N_j = 200$ each and an output layer $j = 3$ with $N_3 = 5$ outputs corresponding to a vector given by Eq.~(\ref{eq:params}) in parameter space.

\section{\label{sec:MSE_predictions}Inferring model parameters using mean-squared-error loss function}

Once the simulated rotation curves are obtained and the architecture is finalized, we can now train our neural networks.
In this section, we train two neural networks for each galaxy in the sample: one without noise included in the inputs of the training set (i.e. simulated rotation curves) and one with noise included. 

For both neural networks, we use the mean-squared-error loss defined in Eq.~(\ref{eq:MSE_loss}) during training, while all other hyperparameters including the architecture are fixed to what was discussed in Section~\ref{sec:architecture}.

\subsection{\label{sec:noiseless_case}Noiseless Case}

Let us start with the simpler case, where simulated rotation curves without any noise are used for training. 
We follow the pre-processing steps in Section~\ref{sec:preprocessing} and then train $7$ different neural networks, each for a galaxy in our sample of DM dominated dwarf galaxies. 
The neural networks are trained for $250$ epochs while both training and validation loss are monitored.
The loss as a function of epochs is plotted in Appendix~\ref{app:loss_epochs} in Figs.~\ref{fig:loss_1} and~\ref{fig:loss_2} for all galaxies.
We then test the performance of the network by calculating the loss on the test data.
This is done to ensure that test loss is not too different from training loss, i.e. the neural network does just as well on unseen data as it does on training data. 

For the final test, the trained neural networks are given the observed rotation curves - one for each neural network - as input. 
Note that only the central values of the observed rotation curves are fed to the network. 
The neural network, using its trained weights and biases predicts a parameter vector $\bf{\hat{P}}$. 

Since we do not have target parameters for the observed rotation curves to check how good these predictions are, we must utilize another metric to discern if the predicted parameters are good. 
One way to do so is to use the predicted parameter vector ${\bf \hat{P}}$ to construct a rotation curve using Eqs.~(\ref{eq:uldm_nfw}) - (\ref{eq:circ_vel_2}), which can then be compared to the observed one using a $\chi^2_{red}$ (reduced $\chi^2$) value. 

We note the following results, using the example of UGC 5721, which carry over to other galaxies as well: 

\begin{itemize}
    \item To make sure performance on test data is good, along with calculating loss, one can also compare constructed rotation curves using the target parameters to those obtained from the predicted parameters. We do this for three randomly selected examples from the test data in Fig.~\ref{fig:test_noiseless}. 
    It can be seen that, while the rotation curves don't match exactly (since the loss is not zero for parameter predictions), the predicted curves look visually similar and are close to the target rotation curves. 
    
    \begin{figure}[h!]
    \includegraphics[width = 0.8\textwidth]{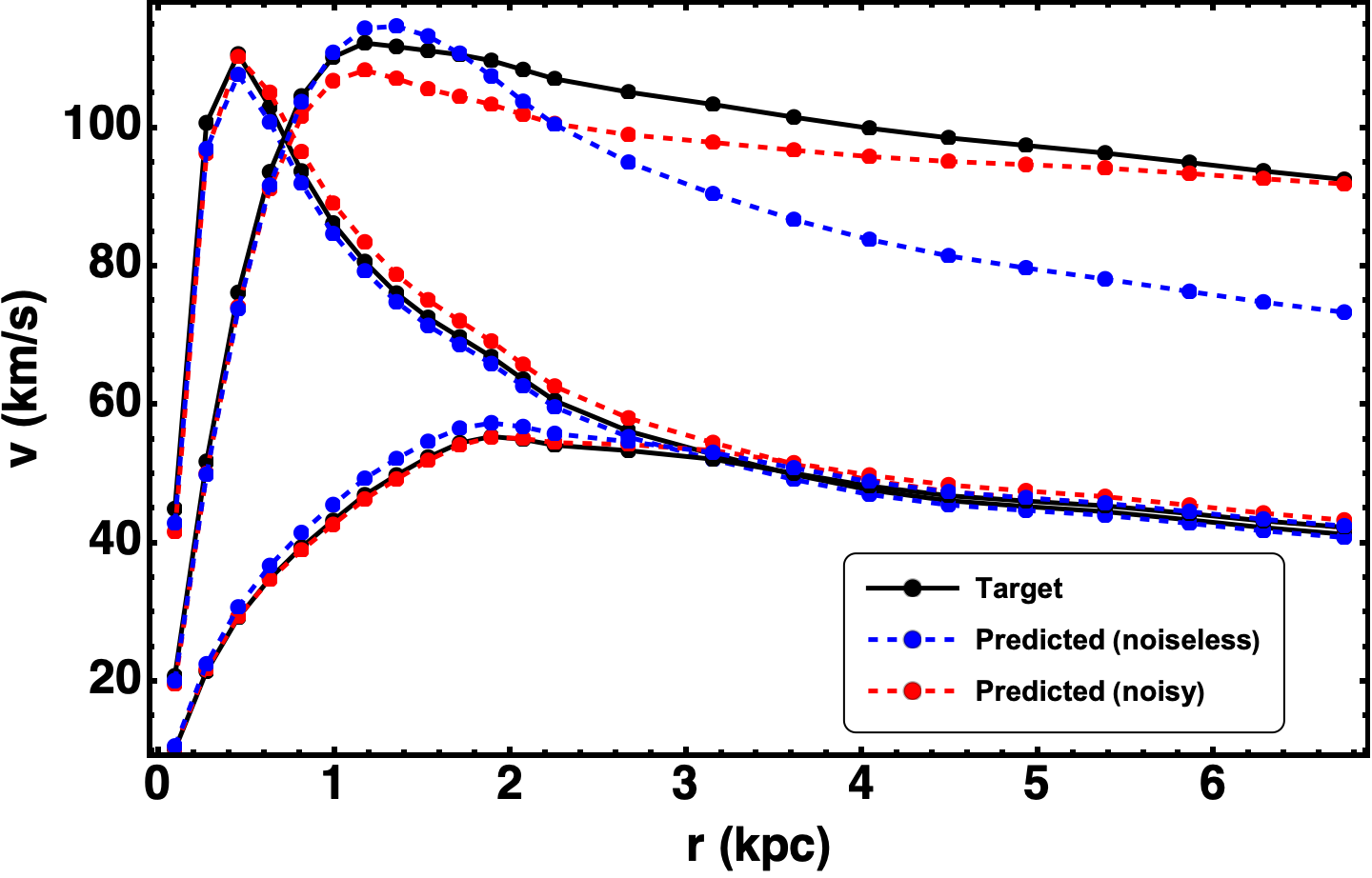}
    \caption[Performance of ANN on test data]{Target (black) and predicted rotation curves (blue for the noiseless case and red for the noisy case) constructed from randomly selected target parameters from the test dataset and the corresponding predicted parameters that the neural network infers respectively.}
    \label{fig:test_noiseless}
    \end{figure}
    
    \item However, when observed rotation curves are given as input, the predicted parameters are not able to reconstruct the observed rotation curve with a low $\chi^2_{red}$, which can be seen by the blue curve in Fig.~\ref{fig:UGC05721_compare}. One reason for this could be that even without the error bars, central values of the observed velocities do not form a smooth rotation curve (see for instance, the dip in the observed velocity at $\sim 1\ \text{kpc}$ in Fig.~\ref{fig:UGC05721_compare}). 
    Since the dark matter component for simulated data is a smooth curve, these bumps are not learned by the neural network.
    This can be seen in other galaxies as well, where in Fig.~\ref{fig:MSE_comparison}, parameters for observed velocities (without uncertainties) which are smoother are predicted better by the corresponding neural network. 

    \begin{figure}[h!]
    \centering
    \includegraphics[width=0.8\textwidth]{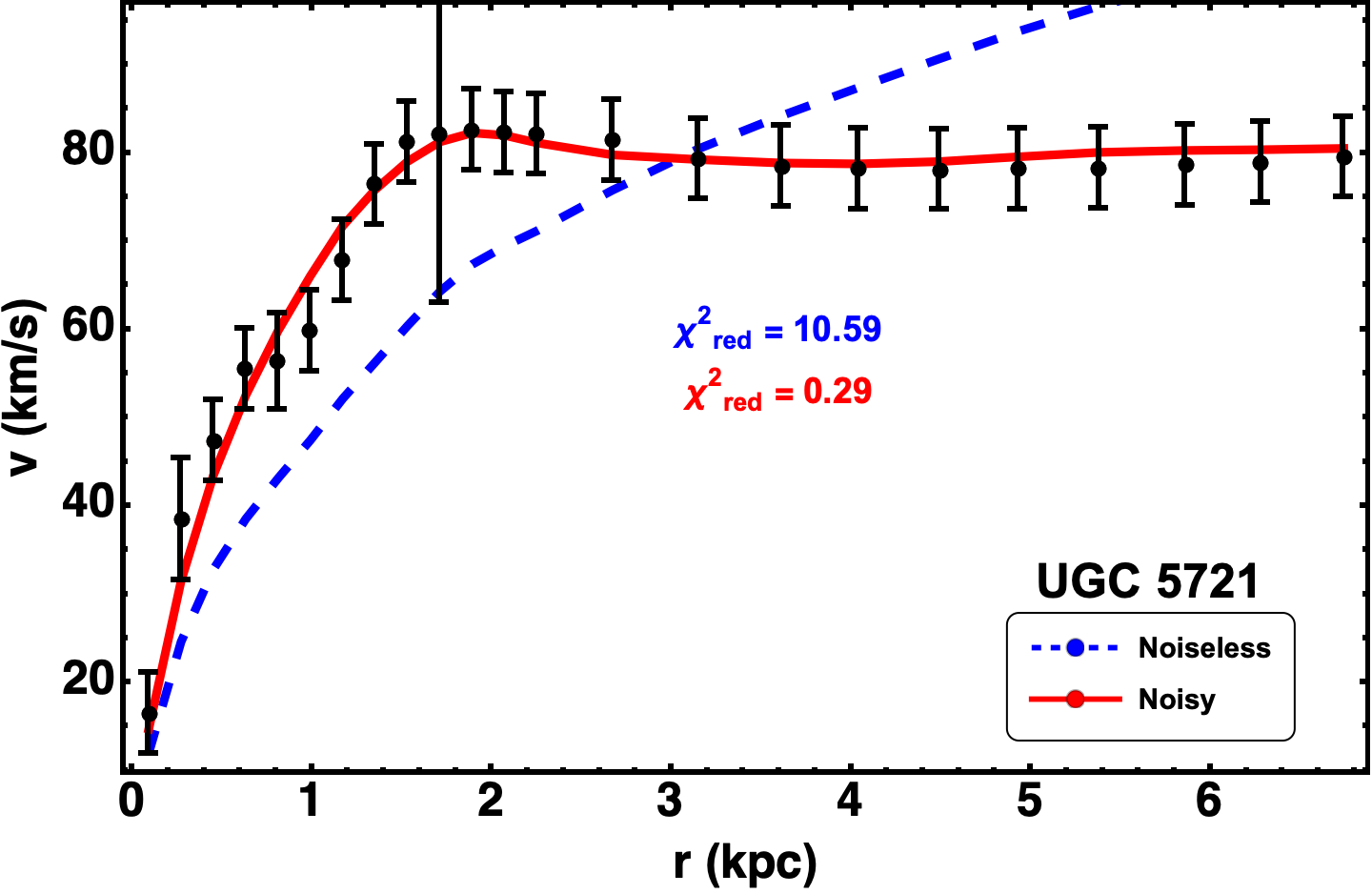}
    \caption[ANN (with MSE loss function) prediction for UGC 5721]{Performance of the neural networks trained on MSE loss with (red) and without (blue) noisy inputs, for UGC 5721.}
    \label{fig:UGC05721_compare}
\end{figure}
\end{itemize}

\subsection{Noisy case}\label{sec:noisy_case}

\begin{figure}[h!]
\begin{tabular}{ccc}
\includegraphics[width=0.3\textwidth]{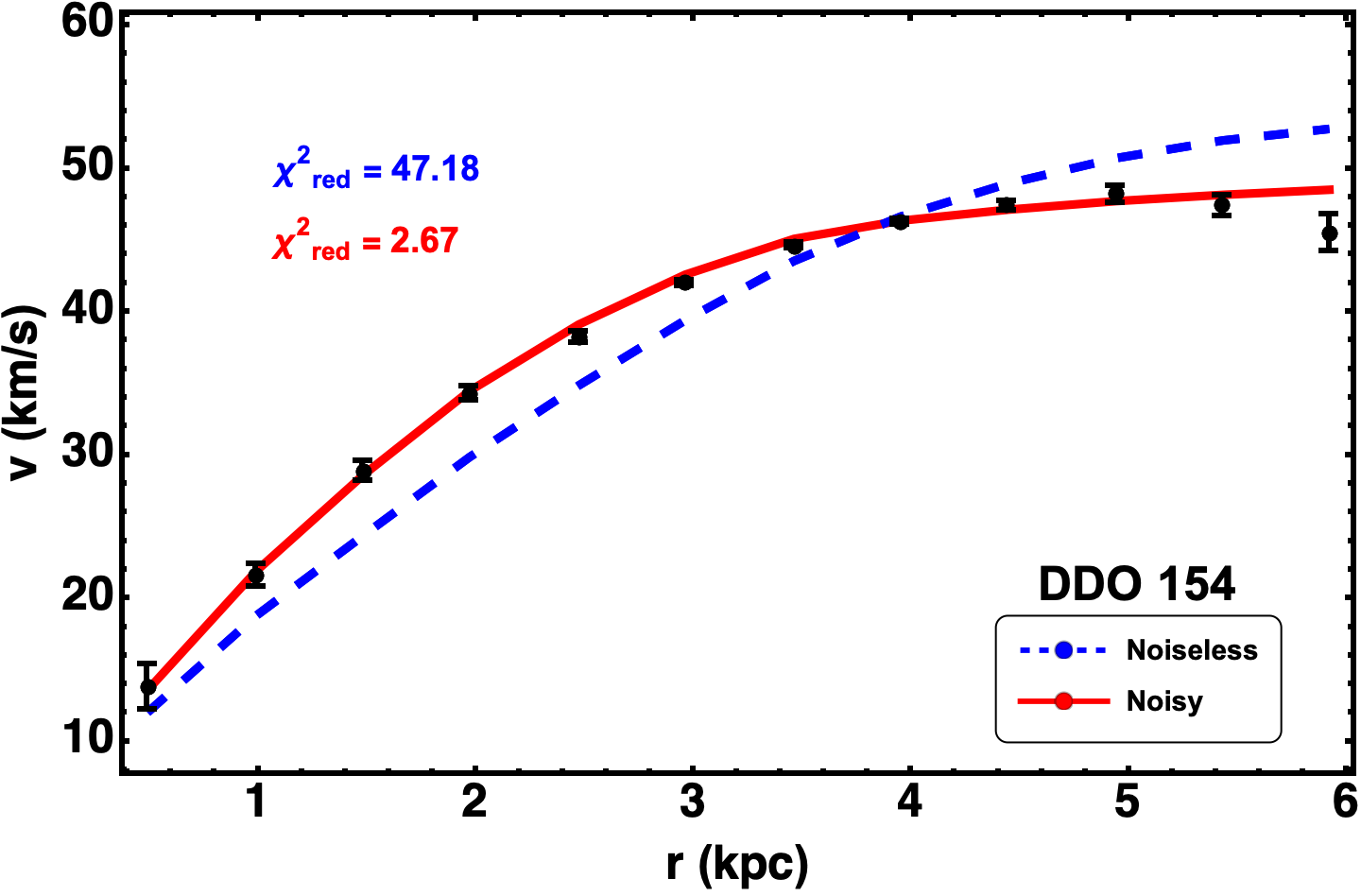} & 
\includegraphics[width=0.3\textwidth]{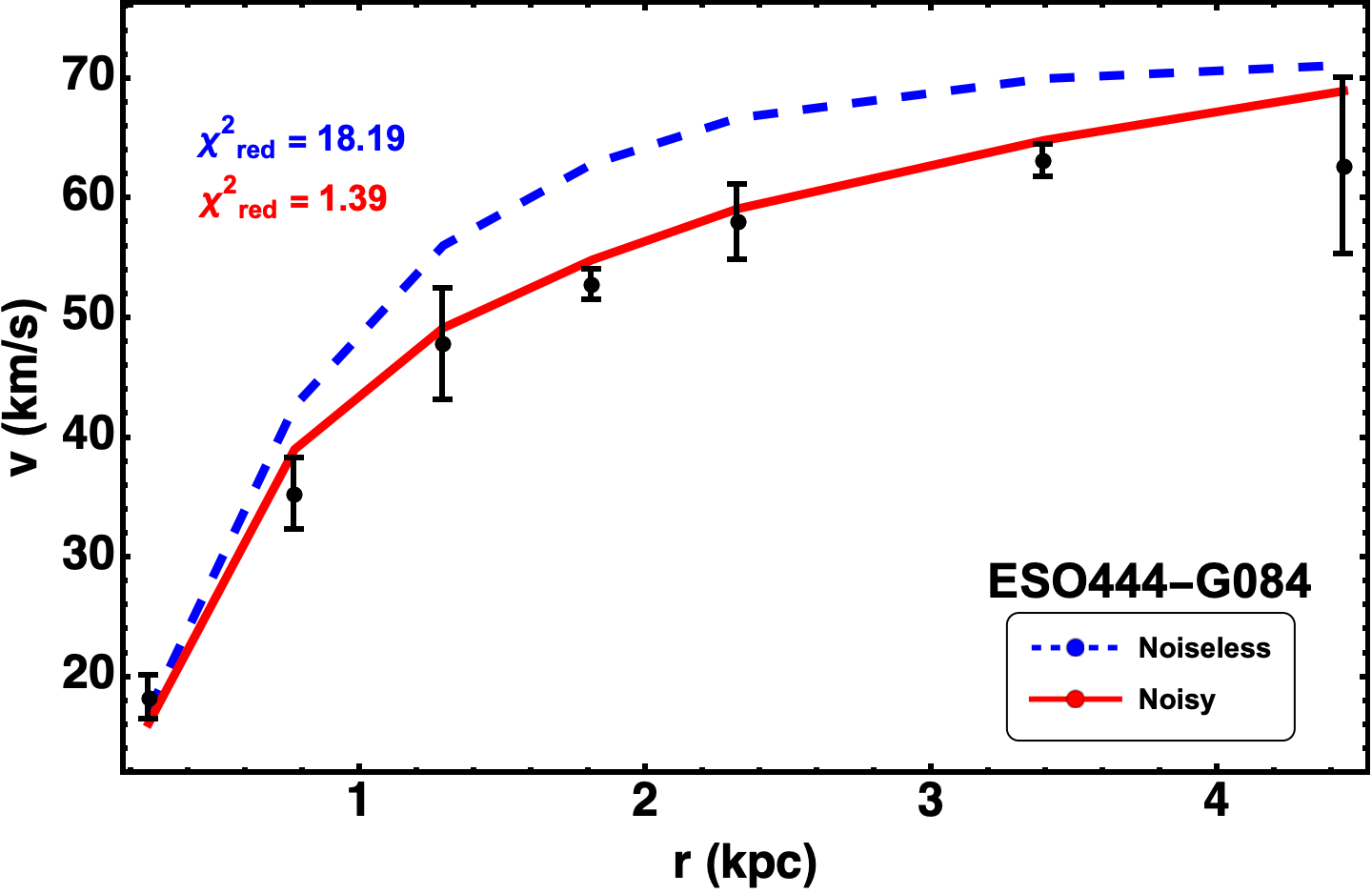} &
\includegraphics[width=0.3\textwidth]{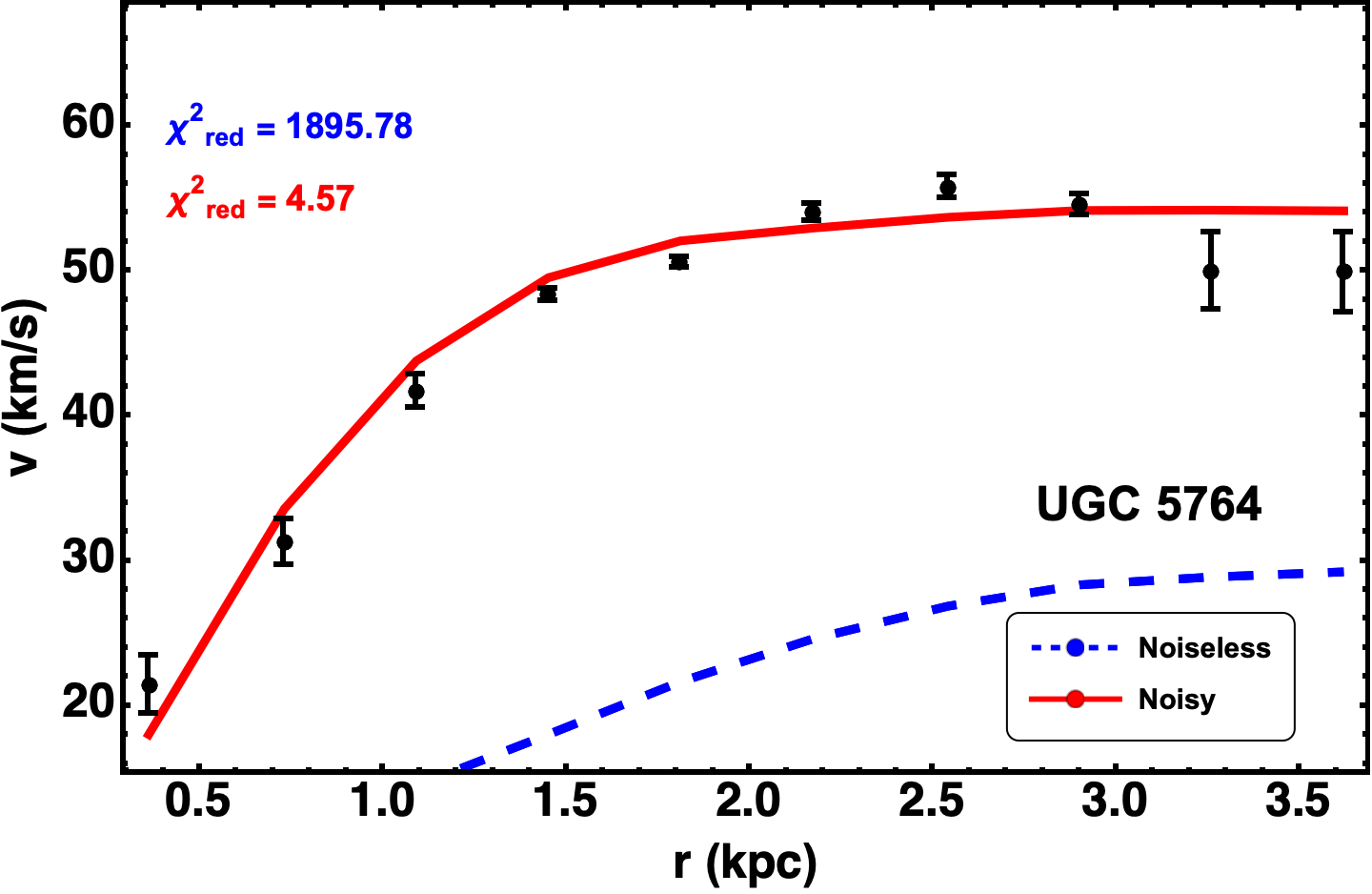} \\
(a) DDO154 & (b) ESO 444-G084 & (c) UGC 5764 \\[6pt]
\includegraphics[width=0.3\textwidth]{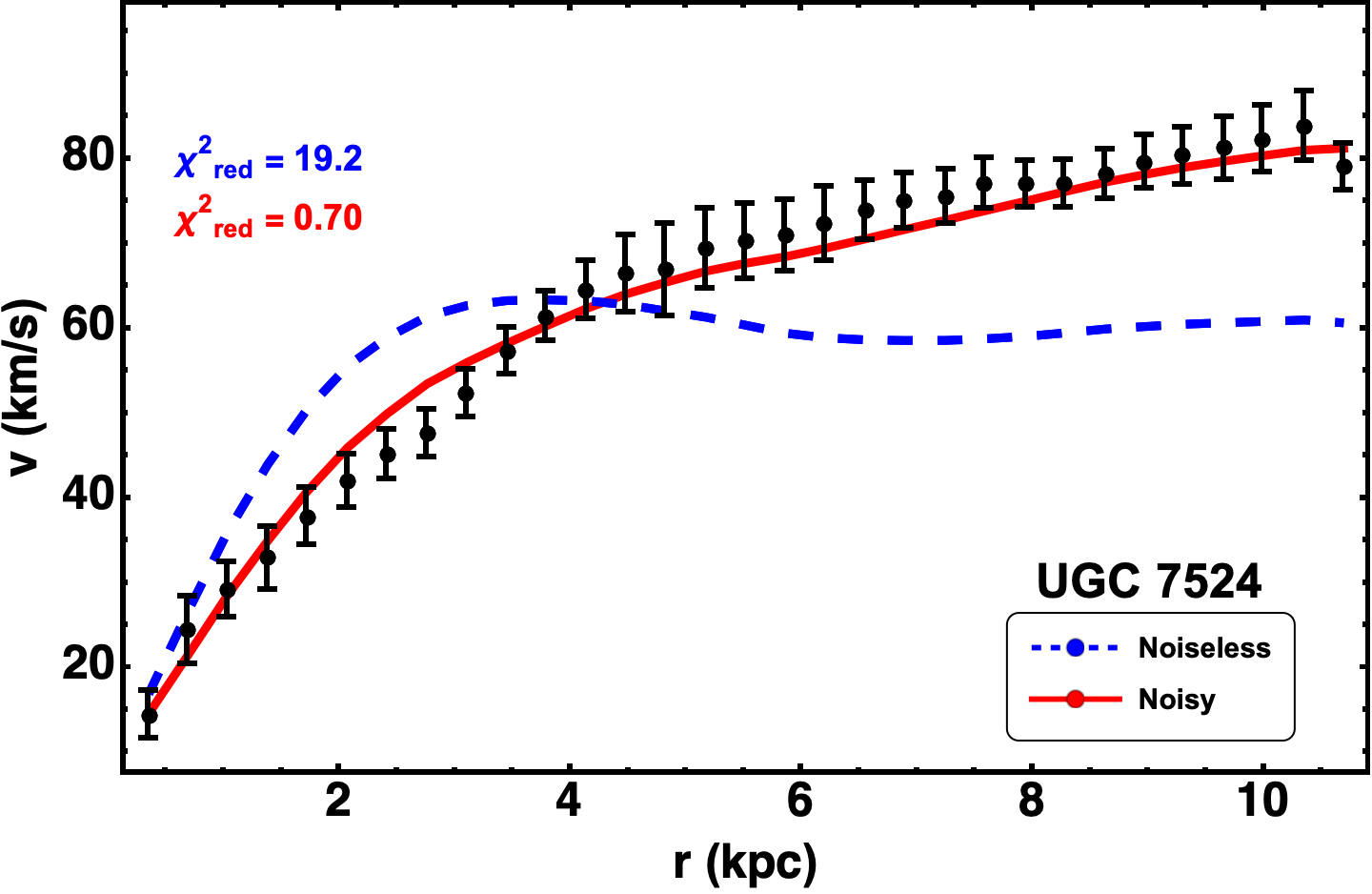} &   \includegraphics[width=0.3\textwidth]{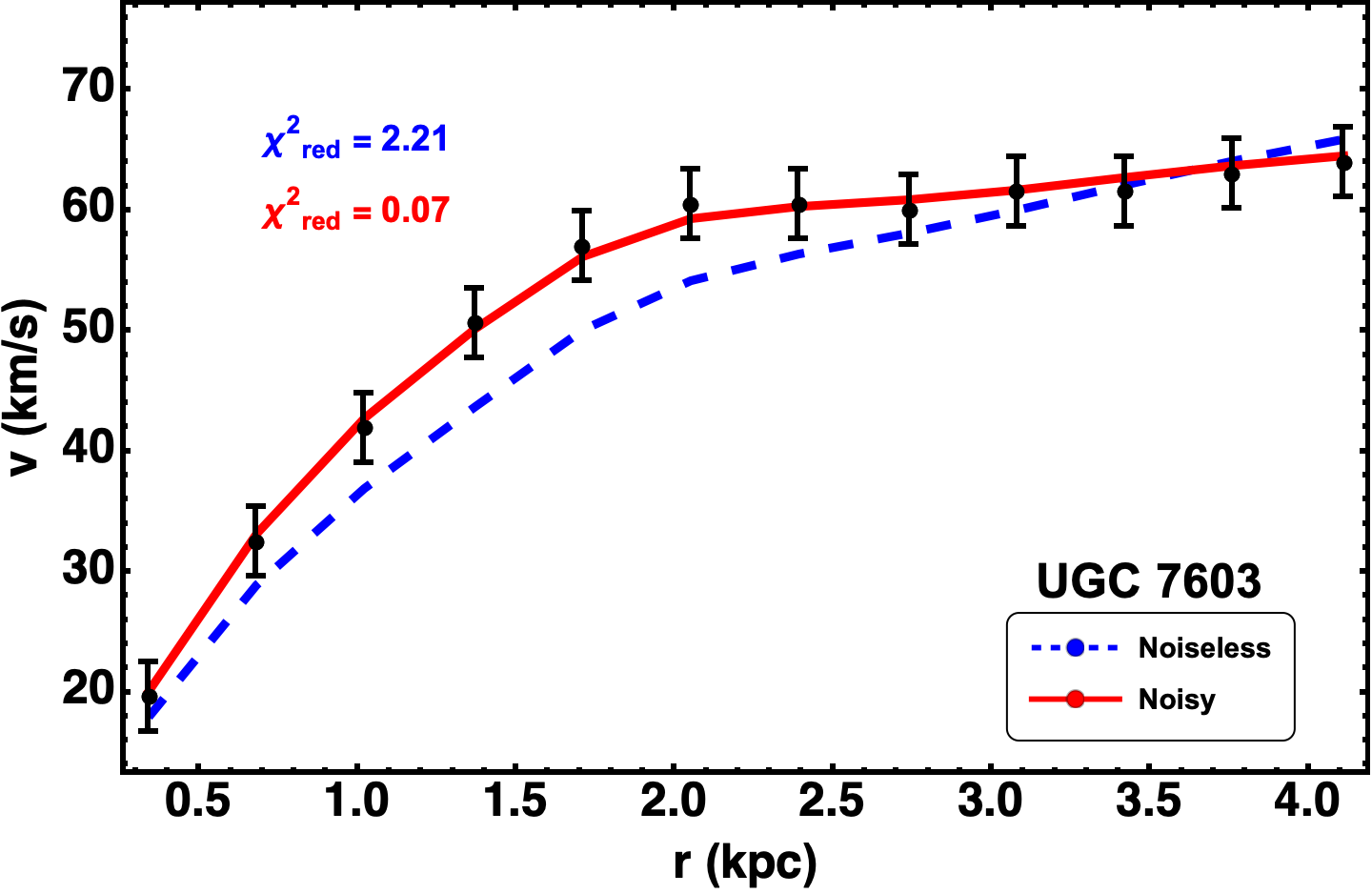} &
\includegraphics[width=0.3\textwidth]{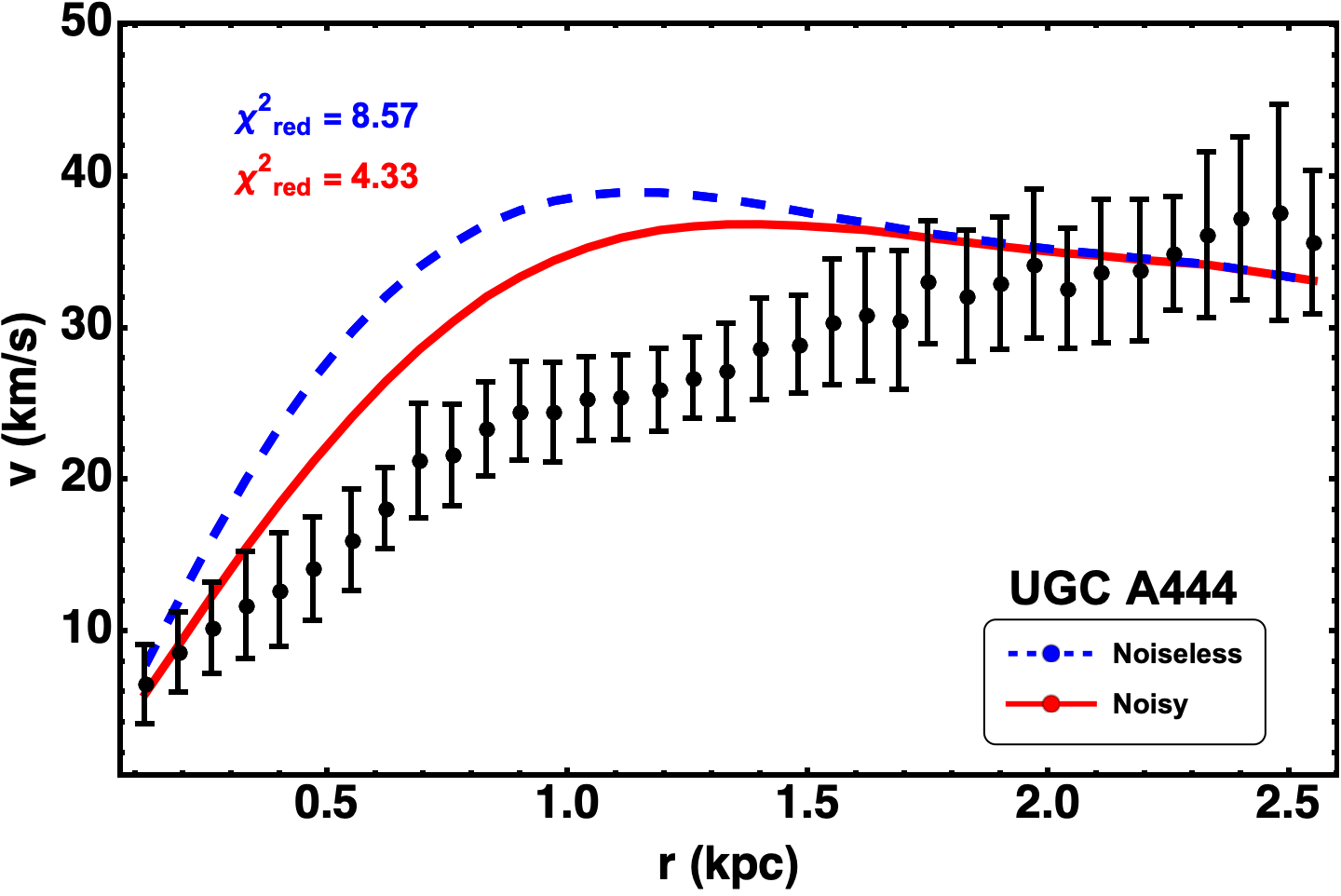} \\
(d) UGC 7603 & (e) UGC 7524 & (f) UGCA 444
\end{tabular}
\caption[ANN (trained with MSE loss fucntions) predictions for remaining 6 galaxies]{\justifying Curves corresponding to the predicted parameters for the neural networks trained on noisy (red) and noiseless (blue) input data with MSE loss function. Note that only the central value of the observed rotation curves were given as input.}
\label{fig:MSE_comparison}
\end{figure}

In this section, we consider a neural network with the same hyperparameters and training time as in the previous section, but with noise included in simulated rotation curves. 
We shall see how this improves performance of the network when confronted with observed rotation curves.

The SPARC catalog provides (for all galaxies) central values for the observed velocity at different radius values along with the uncertainty at that bin, i.e. $v(r) = v_{obs}(r) \pm \sigma(r)$. 
Hence, at some radius bin $r_i$ we denote the observed velocity as $v_{obs}(r_i) \pm \sigma(r_i)$. 
This noise, when incorporated during training, could possibly lead to better predictions \cite{Wang_2020, Pal_2023}. 
As we shall see in this section, that is precisely the case. 
 
In every simulated rotation curve, for each radius bin $r_i$, we add a random draw from the distribution $\mathcal{N}(0, \sigma(r_i)^2)$ as noise to the simulated velocity $v_{sim}(r_i)$.
Note that the target outputs, i.e. the known parameters, are not noisy and will not be affected by this process. 
The usual pre-processing steps are carried out as discussed in Section~\ref{sec:preprocessing}.

It is interesting to note that training with noisy inputs is equivalent to Tikhonov regularization \cite{Bishop_1995}, a strategy that is also employed in \cite{Wang_ECoPANN_2020} for cosmological parameter estimation. 
However, unlike the procedure in the above papers, we do not add noise after every epoch, and choose to include noise at the pre-processing stage itself. 

We train neural networks for $250$ epochs with the same hyperparameters except that inputs will now be noisy. In this case, validation loss does converge by $250$ epochs, after which it starts to increase, signaling overfitting. Similar to the previous section, using the example of UGC 5721, we note the following results: 

\begin{itemize}
    \item After training, we evaluate the performance of the neural network on test data, for which we know the target parameters. We find that the test loss is similar to the training and validation loss implying no overfitting. 
    It is worth noting that due to the regularization effect of the input noise, the MSE loss for all data sets is higher compared to the noiseless case. 
    However, the when rotation curves corresponding to target and predicted parameters are constructed, we find good agreement between the two as seen in Fig.~\ref{fig:test_noiseless}. 

    \item On the other hand, unlike the noiseless case, we find that the performance on observed rotation curves has improved drastically.
    We feed as inputs, the central values of the observed velocities and construct rotation curves from the output parameters predicted. 
    The corresponding curves and their $\chi^2_{red}$ are shown in Fig.~\ref{fig:MSE_comparison} by the red curves.
    Note that the $\chi^2_{red}$ values are far smaller compared to the noiseless case, with predicted curves fitting much better than before. 
\end{itemize}

Similar results are obtained for all the other galaxies in our sample, as seen in Fig.~\ref{fig:MSE_comparison}.
In conclusion, neural networks trained with noisy training data perform much better than those trained without noise. 
In the next section we shall discuss how one can also obtain the uncertainties associated with the parameters along with the point-estimates.

\section{Estimating uncertainties in model parameters}{\label{sec:uncertainties}}

As we discussed in the previous section, observed rotation curves have errors associated with every velocity observation. 
Given this uncertainty, it would be useful to obtain uncertainties in the parameter predictions as well.  

It is important to note that in machine learning uncertainty is usually separated into \cite{Kendall_HS_2017}: (a) Epistemic uncertainty viz., the uncertainty associated with the choice of the model and its weights, which can be reduced by using more training data and (b) Aleatoric uncertainty viz., the uncertainty that is inherent to the data itself (like errors in measurement or observations) and cannot be reduced by increasing the size of the training set.

In this section, following recent work \cite{Wang_ECoPANN_2020, Ho_2021, Pal_2023} we try to account for the aleatoric uncertainty in multiple ways, for neural networks trained on noisy inputs. 
First we explore how multiple realizations of the observed rotation curves can enable us to obtain multiple estimates of parameters in Section~\ref{sec:multiple_realizations}. 
These point-estimates can be used to construct `joint' and `marginalized' distributions for the predicted parameters. 
In Section~\ref{sec:hetero_uncertainty} we also explore the use of a heteroscedastic loss function to implicitly learn the uncertainty during training itself.

\subsection{Using multiple realizations of the observations}\label{sec:multiple_realizations}

A straightforward way to obtain uncertainties in parameter predictions - assuming that errors in observations are Gaussian - is to draw multiple samples from the Gaussian $\mathcal{N}(v_{obs}(r), \sigma_r^2)$ for each value of radius $r$ observed. 

Thus, for every galaxy in our sample, we now have multiple realizations of the observed rotation curves for that galaxy. 
We can use these realizations as inputs to the trained neural network corresponding to each galaxy. 
Each realization will lead to a slightly different parameter prediction ${\bf P}$. 
We generate $1000$ realizations and obtain $1000$ predictions for each galaxy in the sample. 

For every galaxy, we then consider the $50\%$ quantile (i.e., median) for each parameter as the point-estimate and $16\%$, $84\%$ quantiles (corresponding to a $1\sigma$ region of a $1$D normal distribution) as the uncertainties. 
This definition will help us in comparing our results to those obatined using MCMC in Section~\ref{sec:mcmc_comparison}.
Note that a similar procedure was used in \cite{Wang_ECoPANN_2020} to estimate cosmological parameters from CMB observations.
The resultant parameters and their uncertainties are shown in Table~\ref{tab:preds_hetero}. 
We also plot the rotation curves constructed from predictions and obtain the goodness of fit to the observations using $\chi^2_{red}$ (reduced $\chi^2$).
The rotation curves are shown in Fig.~\ref{fig:all_comparison} by the purple dashed curves. 
It is clear that similar to the case with a single realization of the observed rotation curve, the rotation curves corresponding to the mean value of parameters fit observations just as well, or even better, especially in the case of DDO 154 and UGCA 444.

\begin{table}
\centering
\resizebox{\linewidth}{!}{
\begin{tabular}{c|c|ccccc}
\toprule
\multirow{2}{*}{\textbf{Galaxy}} & \multirow{2}{*}{\textbf{Method}} & \multicolumn{5}{c}{\textbf{Predictions}} \\ \cmidrule(l){3-7}
& & $m$ ($10^{-23}$ eV) &  scale $s$ & $r_t$ (kpc) & $r_s$ (kpc) & $\Upsilon_*$ ($M_\odot/L_\odot$)\\ \cmidrule(r){1-7}

\multirow{3}{*}{DDO 154} & MSE & $1.96_{-0.16}^{+0.12}$ & $5380.39_{-103.57}^{+101.64}$ & $3.63_{-0.69}^{+0.85}$ & $6.37_{-0.57}^{+0.63}$ & $0.61_{-0.08}^{+0.05}$ \\ [0.2cm]

& Hetero & $1.95_{-0.30}^{+0.30}$ & $5350.71_{-292.16}^{+292.16}$ & $3.50_{-0.87}^{+0.87}$ & $6.87_{-3.90}^{+3.90}$ & $0.61_{-0.12}^{+0.12}$ \\ [0.2cm]

& MCMC & $1.79_{-0.10}^{+0.16}$ & $5328.85_{-90.48}^{+136.65}$ & $3.25_{-0.83}^{+1.61}$ & $4.01_{-1.56}^{+3.56}$ & $0.70_{-0.14}^{+0.08}$ \\ [0.1cm] \cmidrule(r){1-7}

\multirow{3}{*}{ESO 444-G084} & MSE & $5.06_{-0.98}^{+0.69}$ & $4954.22_{-157.20}^{+123.29}$ & $1.26_{-0.08}^{+0.2}$ & $9.91_{-1.06}^{+0.72}$ & $0.58_{-0.01}^{+0.01}$ \\ [0.2cm]

& Hetero & $5.26_{-1.07}^{+1.07}$ & $5055.20_{-363.85}^{+363.85}$ & $1.26_{-0.27}^{+0.27}$ & $9.53_{-3.38}^{+3.38}$ & $0.56_{-0.14}^{+0.14}$ \\ [0.2cm]

& MCMC & $5.29_{-1.02}^{+0.94}$ & $5181.16_{-324.19}^{+257.23}$ & $1.14_{-0.11}^{+0.21}$ & $10.39_{-3.63}^{+3.15}$ & $0.58_{-0.18}^{+0.16}$ \\ [0.1cm]  \cmidrule(r){1-7}

\multirow{3}{*}{UGC 5721} & MSE & $2.28_{-0.30}^{+0.37}$ & $3042.17_{-127.19}^{+155.87}$ & $2.44_{-0.34}^{+0.57}$ & $7.74_{-1.76}^{+0.94}$ & $0.65_{-0.04}^{+0.03}$ \\ [0.2cm]

& Hetero & $2.21_{-0.34}^{+0.34}$ & $3052.34_{-163.26}^{+163.26}$ & $2.55_{-0.72}^{+0.72}$ & $7.12_{-4.04}^{+4.04}$ & $0.66_{-0.13}^{+0.13}$ \\ [0.2cm]

& MCMC & $2.12_{-0.22}^{+0.30}$ & $3064.23_{-99.85}^{+124.44}$ & $2.56_{-0.41}^{+0.36}$ & $7.27_{-4.74}^{+5.20}$ & $0.70_{-0.13}^{+0.07}$ \\ [0.1cm]  \cmidrule(r){1-7}

\multirow{3}{*}{UGC 5764} & MSE & $3.97_{-0.42}^{+0.35}$ & $4658.97_{-119.27}^{+122.23}$ & $1.92_{-0.32}^{+0.64}$ & $4.18_{-1.00}^{+1.83}$ & $0.56_{-0.01}^{+0.01}$ \\ [0.2cm]

& Hetero & $3.87_{-0.43}^{+0.43}$ & $4769.11_{-192.48}^{+192.48}$ & $2.00_{-0.51}^{+0.51}$ & $4.75_{-3.74}^{+3.74}$ & $0.57_{-0.14}^{+0.14}$ \\ [0.2cm]

& MCMC & $4.25_{-0.35}^{+0.43}$ & $4879.48_{-106.92}^{+123.52}$ & $1.40_{-0.30}^{+0.28}$ & $1.42_{-0.32}^{+0.90}$ & $0.56_{-0.18}^{+0.17}$ \\ [0.1cm]  \cmidrule(r){1-7}

\multirow{3}{*}{UGC 7524} & MSE & $1.90_{-0.43}^{+0.62}$ & $4403.68_{-256.26}^{+376.14}$ & $3.02_{-0.75}^{+1.30}$ & $10.35_{-1.70}^{+1.29}$ & $0.57_{-0.03}^{+0.03}$ \\ [0.2cm]

& Hetero & $1.55_{-0.52}^{+0.52}$ & $4370.11_{-422.96}^{+422.96}$ & $3.02_{-1.43}^{+1.43}$ & $9.56_{-2.86}^{+2.86}$ & $0.57_{-0.14}^{+0.14}$ \\ [0.2cm]

& MCMC & $1.36_{-0.28}^{+0.70}$ & $4196.24_{-315.63}^{+609.70}$ & $2.48_{-0.89}^{+1.22}$ & $10.35_{-3.22}^{+2.92}$ & $0.64_{-0.17}^{+0.12}$ \\ [0.1cm] \cmidrule(r){1-7}

\multirow{3}{*}{UGC 7603} & MSE & $2.59_{-0.27}^{+0.47}$ & $4128.01_{-177.46}^{+326.71}$ & $2.29_{-0.42}^{+0.47}$ & $7.35_{-0.36}^{+0.68}$ & $0.54_{-0.04}^{+0.05}$ \\ [0.2cm]

& Hetero & $2.61_{-0.65}^{+0.65}$ & $4243.85_{-368.33}^{+368.33}$ & $2.20_{-0.77}^{+0.77}$ & $7.43_{-4.04}^{+4.04}$ & $0.56_{-0.14}^{+0.14}$ \\ [0.2cm]

& MCMC & $2.48_{-0.42}^{+0.62}$ & $4186.56_{-221.71}^{+345.33}$ & $2.18_{-0.83}^{+0.74}$ & $6.03_{-3.56}^{+5.78}$ & $0.57_{-0.17}^{+0.15}$ \\ [0.1cm] \cmidrule(r){1-7}

\multirow{3}{*}{UGCA 444} & MSE & $7.96_{-0.95}^{+1.05}$ & $7178.33_{-510.66}^{+384.62}$ & $1.74_{-0.01}^{+0.01}$ & $7.93_{-0.02}^{+0.02}$ & $0.55_{-0.001}^{+0.001}$ \\ [0.2cm]

& Hetero & $6.84_{-1.61}^{+1.61}$ & $7958.19_{-742.22}^{+742.22}$ & $1.47_{-0.35}^{+0.35}$ & $8.23_{-3.91}^{+3.91}$ & $0.57_{-0.14}^{+0.14}$ \\ [0.2cm]

& MCMC & $6.90_{-1.40}^{+1.13}$ & $8467.32_{-640.01}^{+386.14}$ & $1.34_{-0.22}^{+0.47}$ & $8.57_{-4.56}^{+4.38}$ & $0.57_{-0.18}^{+0.16}$ \\
\bottomrule
\end{tabular}
}
\caption[Parameter values and their uncertainties inferred from the ANN and MCMC approaches]{\justifying Parameters and their uncertainties obtained by feeding observed rotation curves as input to neural networks trained using: (a) the multiple realization method using MSE loss function (Section~\ref{sec:multiple_realizations}), (b) a heteroscedastic loss function (Section~\ref{sec:hetero_uncertainty}) and (c) using MCMC (Section~\ref{sec:mcmc_comparison}).}
\label{tab:preds_hetero}
\end{table}

The method of multiple realizations that we have discussed here gives us a sequence of parameters which can then be used to construct joint and marginal distributions. 
One could then compare these distributions to the posteriors obtained from a likelihood-based MCMC sampling approach.
In Section~\ref{sec:mcmc_comparison} we do this for UGC 5721 and compare the contours with those obtained using the approach described in this section.

\subsection{\label{sec:hetero_uncertainty}Employing a heteroscedastic loss function}

Another way to account for aleatoric uncertainty involves changing the loss function used for training the neural network. 
Consider the case where, for every output corresponding to the parameter prediction, one assigns an additional output that represents the Gaussian variance in that output. 
This implies that instead of learning a point estimate of the parameter, the neural network now learns the mean and variance of the Gaussian distribution to which the parameter belongs. 
Hence, in the output layer, for every $k$th parameter in ${\bf P}$, we assign an additional output to be its variance $\sigma_k^2$. 
For our neural network, we shall now have an output layer $j = 3$ with $N_3 = 10$ outputs, with the first five outputs corresponding to the parameter vector ${\bf \hat{P}}$ and the last five to the uncertainties of the parameters. 

Here, it is important to note that we do not have known values for $\sigma_k^2$ in the output of the simulated training set. To learn these uncertainties, one must introduce an uncertainty term in the loss function which is given by (for $n$ samples),
\begin{equation}\label{eq:hetero_loss}
    \mathcal{L}_{HS} = \frac{1}{n}\sum_{i=1}^{n}\left[\frac{1}{2p}\sum_{k=1}^p\left(e^{-s_{ik}}(y_{ik} - \hat{y}_{ik})^2 + s_{ik} \right)\right]\ ,
\end{equation}
where $y_{ik}$ is the $k$th parameter for the $i$th sample in the training set, while $\hat{y}_{ik} \equiv f_{k}({\bf x}_i, {\boldsymbol \Omega})$ is the neural network prediction for the same. This is called heteroscedastic loss \cite{Kendall_HS_2017}. 

Here, the actual outputs corresponding to the uncertainties have been redefined to $s_{ik} \equiv \log{\hat{\sigma}_{ik}^2}$ where $\hat{\sigma}_{ik}^2$ is the predicted variance for the predicted parameter $\hat{y}_{ik}$, ensuring a more stable loss \cite{Kendall_HS_2017}.
The first term (i.e. the squared difference between the predicted and target value) in the parenthesis is weighed by $\exp{(-s_{ik})}$, which penalizes too small values of $s_{ik}$ by increasing the loss. On the other hand, the second term ensures that too large values of $s_{ik}$ are also penalized.

An important caveat to note here is that this loss function assumes that the parameters are independent random variables drawn from a Gaussian distribution.

\subsubsection{\label{sec:hetero_inference}Training and inference using heteroscedastic loss}

Recent work \cite{Pal_2023} has suggested that cosmological parameters inferred from a neural network trained using heteroscedastic loss can infer uncertainties that are similar to those obtained using MCMC. 
We look at the UGC 5721 galaxy as an example. 
Utilizing the same architecture and the hyperparameter values as the previous case with noisy input data, we train the neural network for $250$ epochs. We find that the validation loss has converged by this epoch. 

\begin{figure}[h!]
\centering
\begin{tabular}{cccc}
\includegraphics[width=0.3\textwidth]{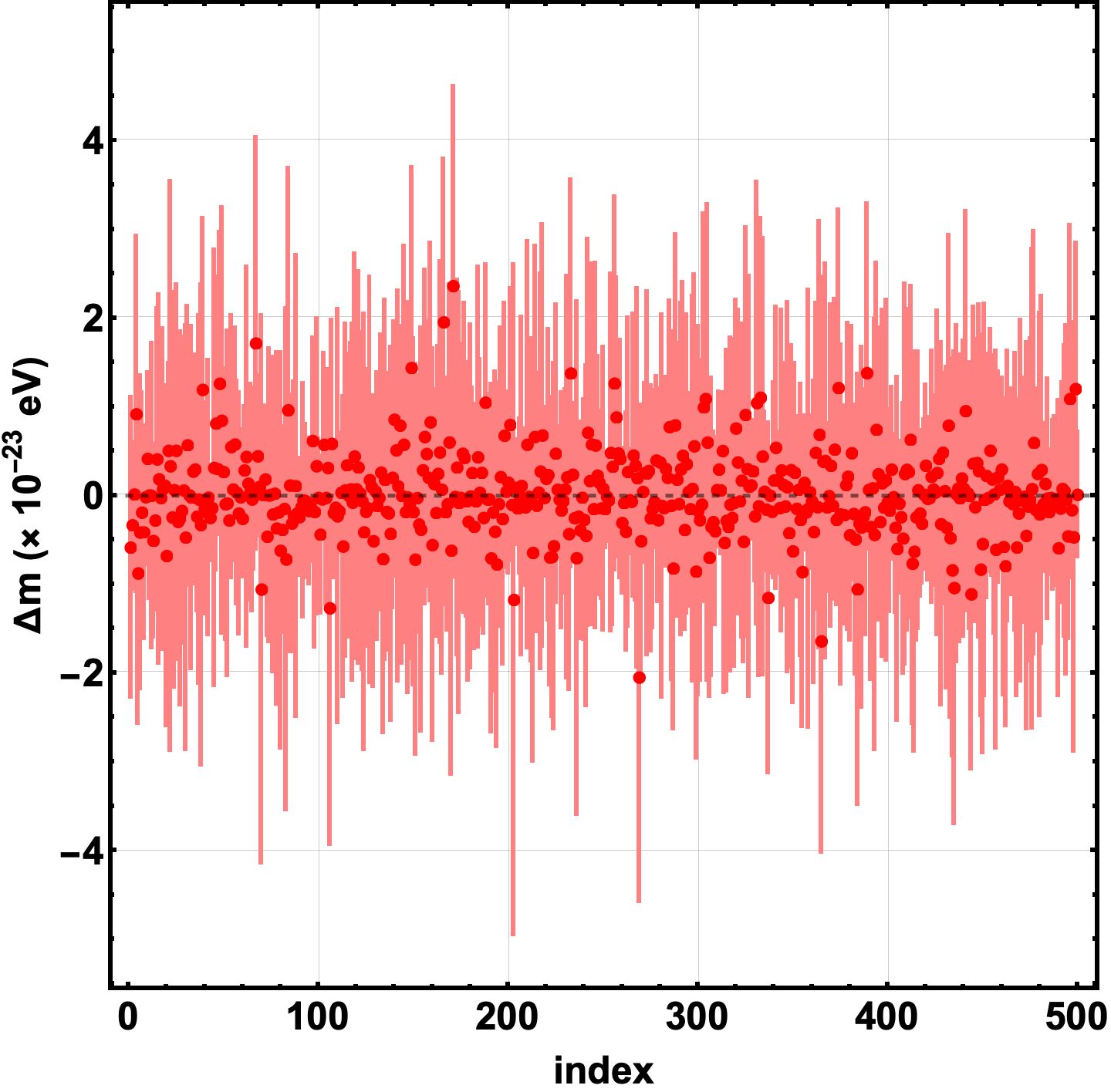} &
\includegraphics[width=0.3\textwidth]{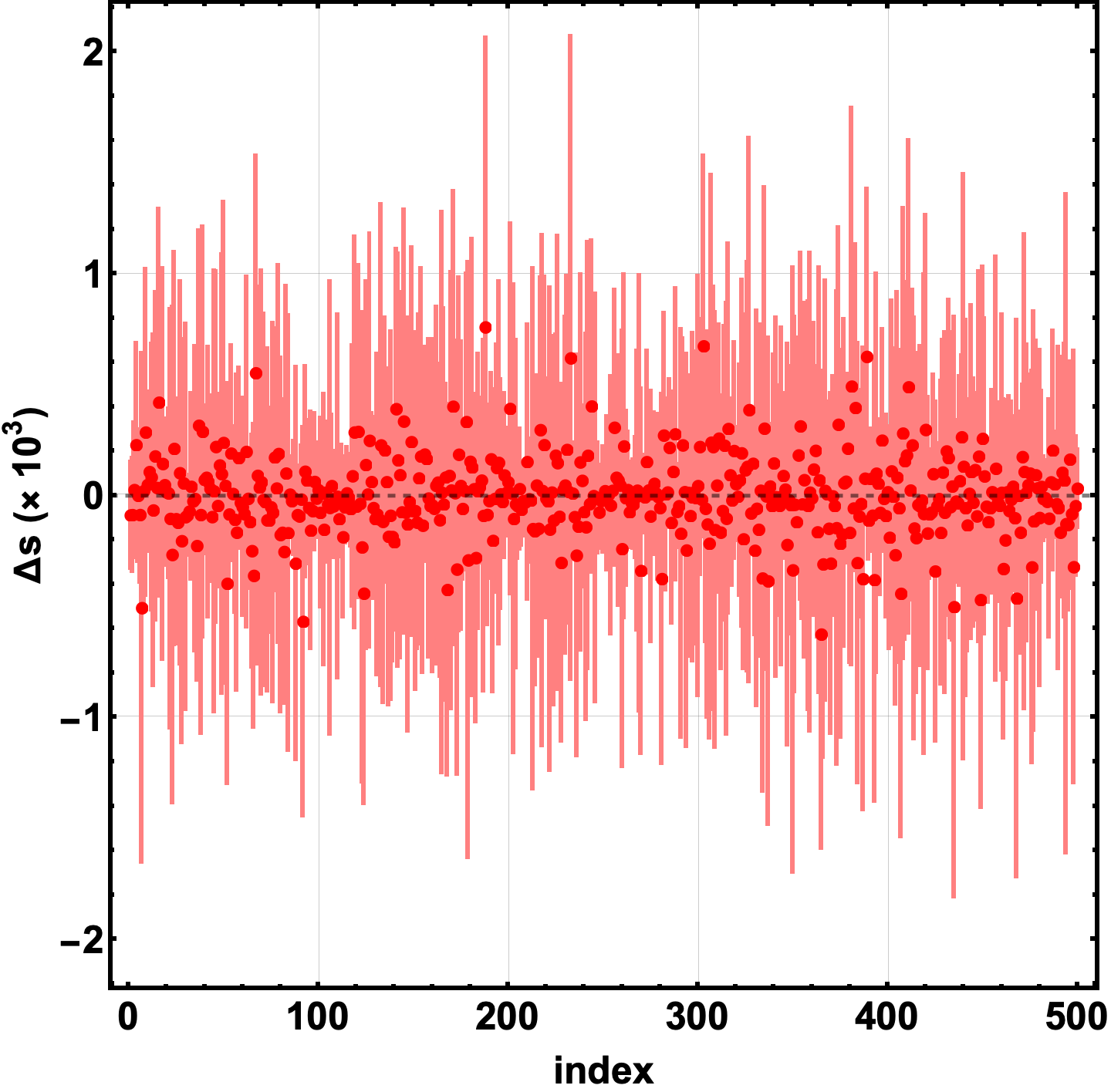} &
\includegraphics[width=0.3\textwidth]{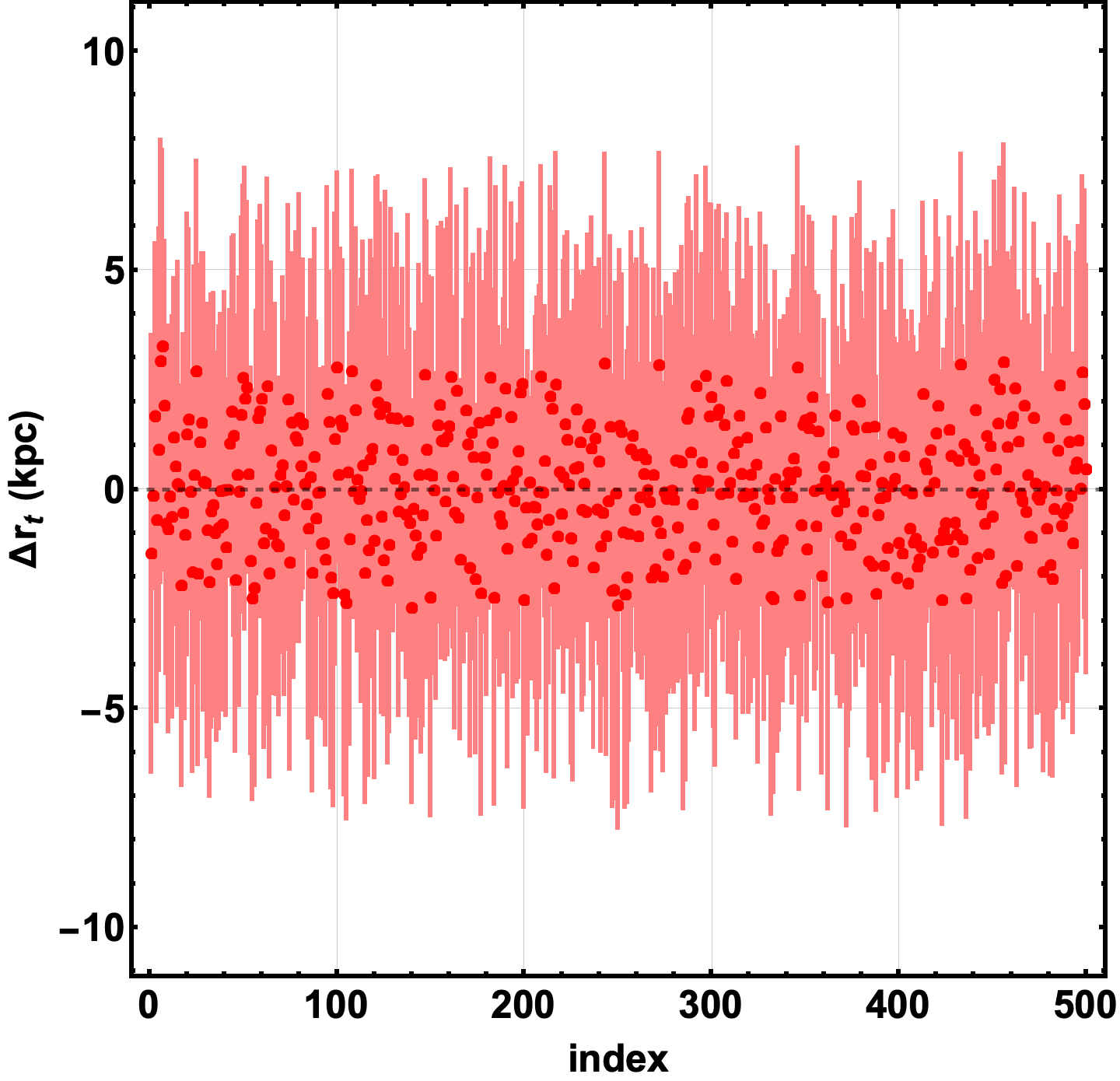} \\
(a) ULDM mass, $m$ & (b) scale parameter, $s$ & (c) transition radius, $r_t$  \\[6pt]
\end{tabular}
\begin{tabular}{cccc}
\includegraphics[width=0.3\textwidth]{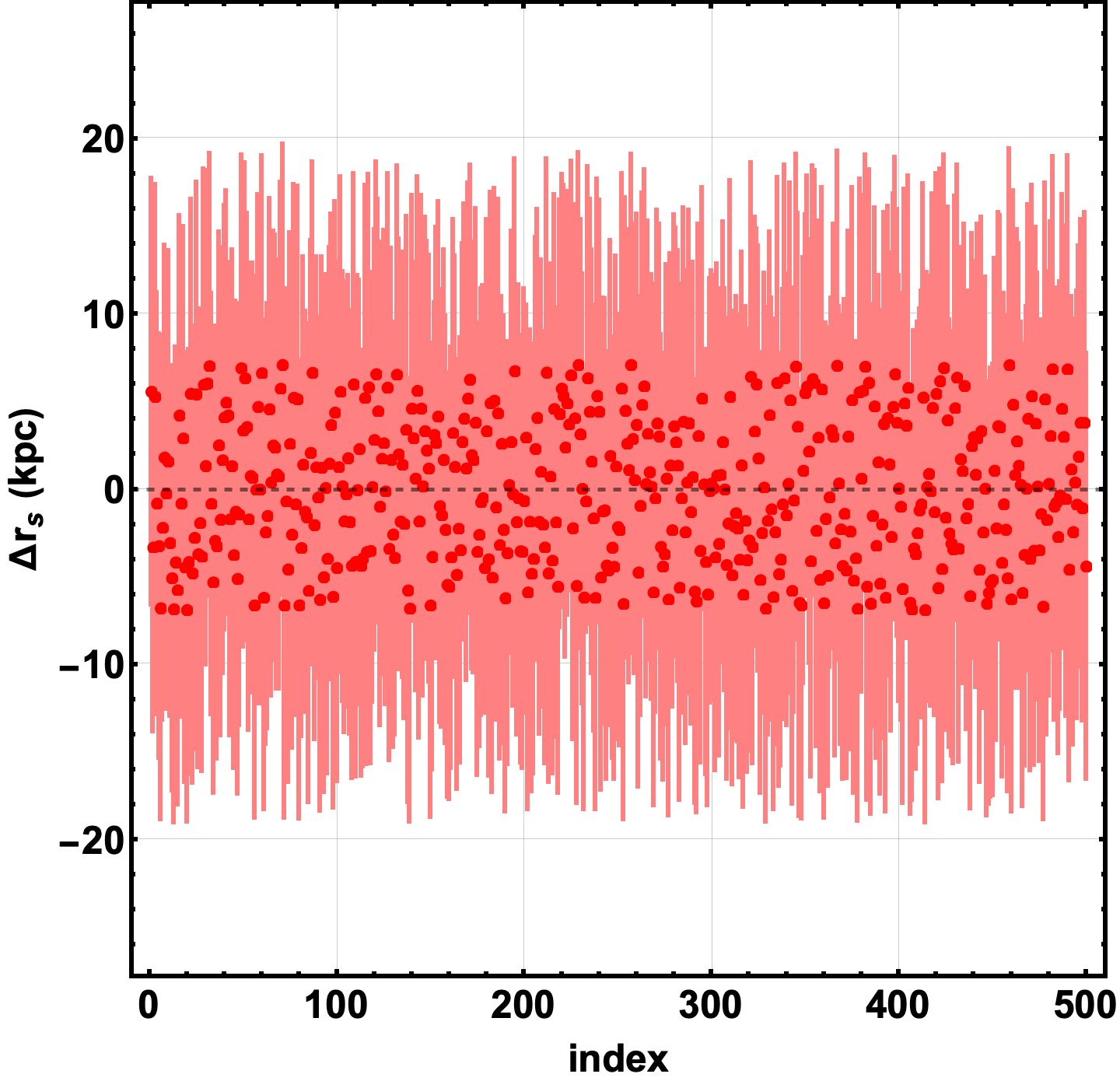} &
\includegraphics[width=0.3\textwidth]{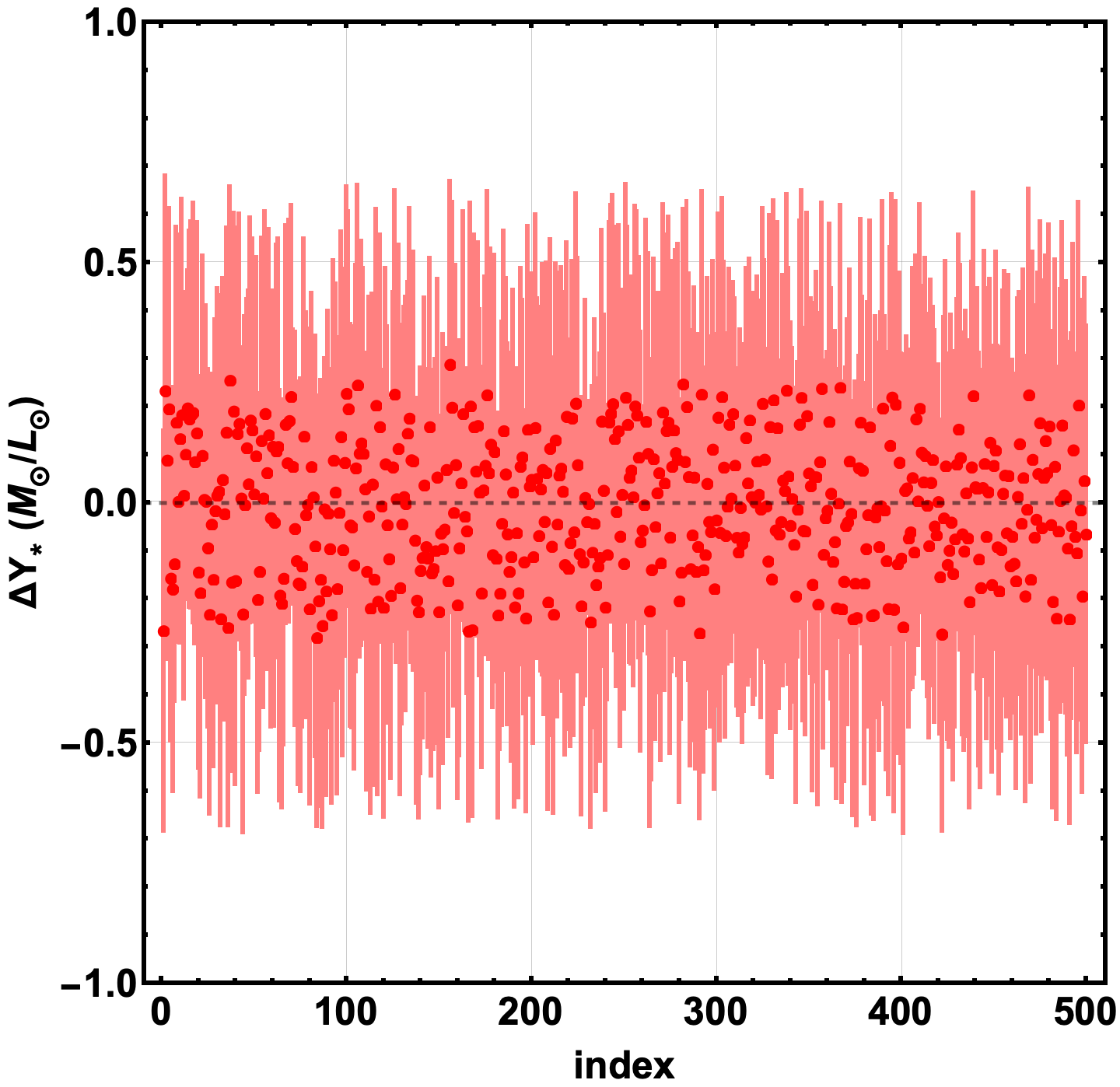} \\
(d) scale radius, $r_s$ & (e) stellar mass-to-light ratio, $\Upsilon_*$ \\[6pt]
\end{tabular}
\caption[Performance of ANN trained with heteroscedastic loss function on test data]{\justifying Performance of the neural network for UGC 5721 trained using a heteroscedastic loss function on test data. The red dots are the difference between inferred and target parameter values, e.g. $\Delta m = m_{\text{true}} - m_{\text{inferred}}$, while the pink lines correspond to three times the inferred uncertainty of prediction.}
\label{fig:hetero_uncert}
\end{figure}

Here, to test on unseen data, since the neural networks also predicts uncertainties along with the point estimates of the parameters, we employ a different method than previous. 
For $500$ samples from test data, we plot the difference between the predicted value and target value for each parameter. We also plot $3\sigma_p$ for each sample. 
The plot is shown in Fig.~\ref{fig:hetero_uncert}, where the pink lines are $3\sigma_p$ values while the red dots are differences between predicted and target parameters. 
Similar to \cite{Pal_2023}, for most samples, the difference between predicted and target value of each parameter lies within three times the uncertainty values.
Hence, these uncertainties appear to capture the ability of the neural network (with the chosen hyperparameters) to learn the parameters from the training data. 

Note that larger differences in $r_t$, $r_s$ and $\Upsilon_*$ and correspondingly larger uncertainties. 
The larger uncertainties in the $r_s$ and $\Upsilon_*$ are expected, since rotation curves of dark matter dominated dwarf galaxies are not sensitive to these parameters. 
For the case of $r_s$, since we force $r_t\geq 1\ \text{kpc}$ and rotation curves extend only till $\mathcal{O}(5)\ \text{kpc}$ for most galaxies in our sample, a change in the NFW scale radius does not alter the rotation curve significantly. 
Similarly, because dark matter dominated galaxies have small baryonic contribution to begin with, changing $\Upsilon_*$ does not affect the rotation curve much.
We shall see in Section~\ref{sec:mcmc_comparison}, that this quantification of the uncertainty agrees well with the uncertainty in parameters obtained from MCMC methods. 

For the final test, we feed the central values of the rotation curve to the trained neural network as input, giving us an output parameter vector as well as the uncertainty associated with each parameter. 
The parameters and uncertainties are shown in Table~\ref{tab:preds_hetero}. 
Notice again, the large uncertainties (i.e., the same order of magnitude as the inferred values) obtained for $r_s$ and $\Upsilon_*$. 

\begin{figure}[h]
    \centering
    \includegraphics[width=0.75\textwidth]{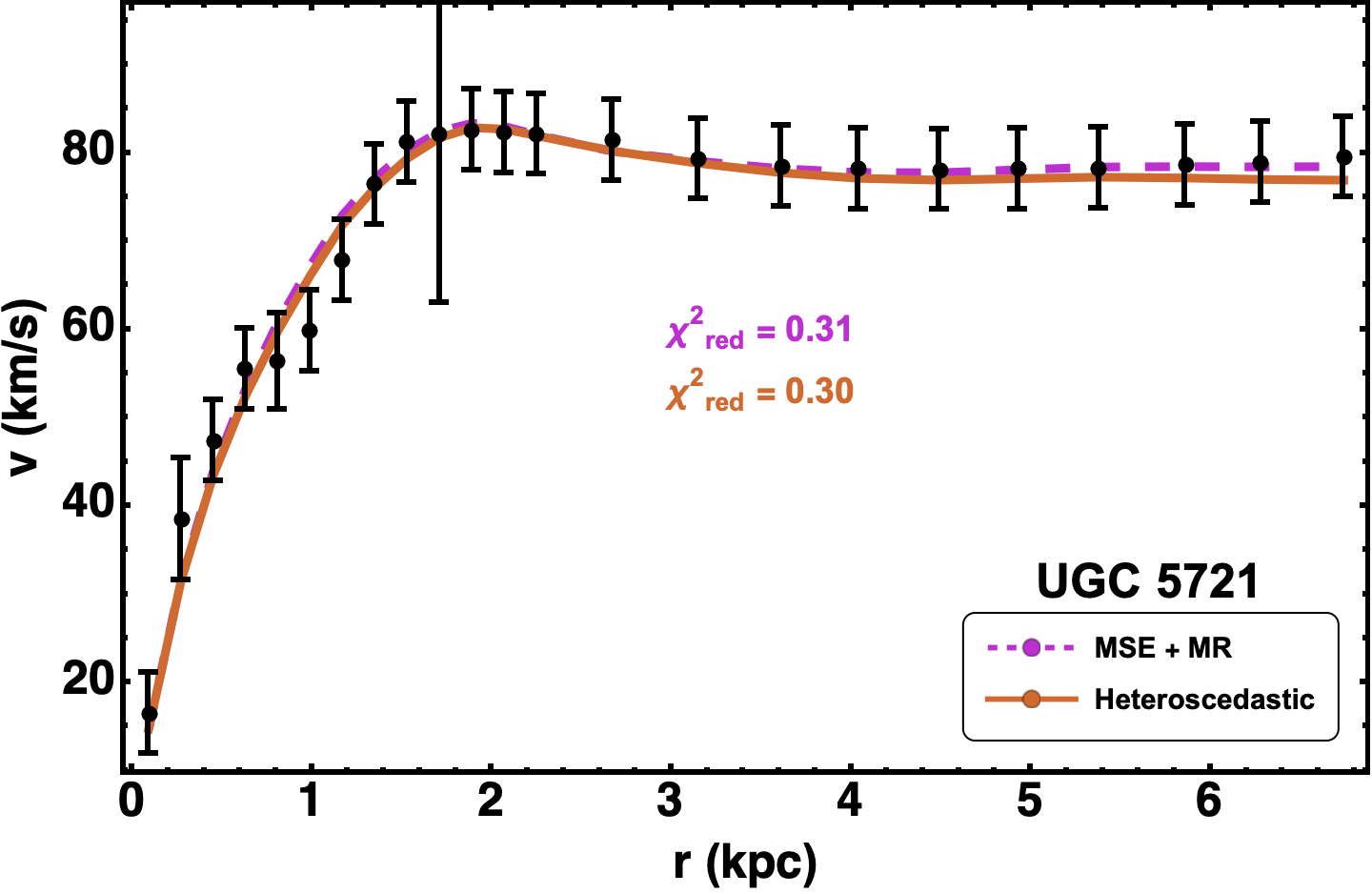}
    \caption[ANN (with heteroscedastic loss function) prediction for UGC 5721]{\justifying Rotation curves corresponding to the predicted parameters for the neural network trained for UGC 5271 on noisy input data with: (a) MSE loss and mean of multiple parameter predictions (purple dashed) and (b) heteroscedastic loss (orange).}
    \label{fig:UGC05721_hetero}
\end{figure}

The rotation curve corresponding to the above inferred parameters agrees well with observations for UGC 5721 which can be seen in Fig.~\ref{fig:UGC05721_hetero} by the orange curve.
We reiterate that only the central value of observed rotation curve is given to the neural network during inference, hence, there is only one prediction for the parameters and their uncertainties.
Then network has learned the uncertainty implicitly from the noisy training data. 

\begin{figure}[h]
\begin{tabular}{ccc}
\includegraphics[width=0.3\textwidth]{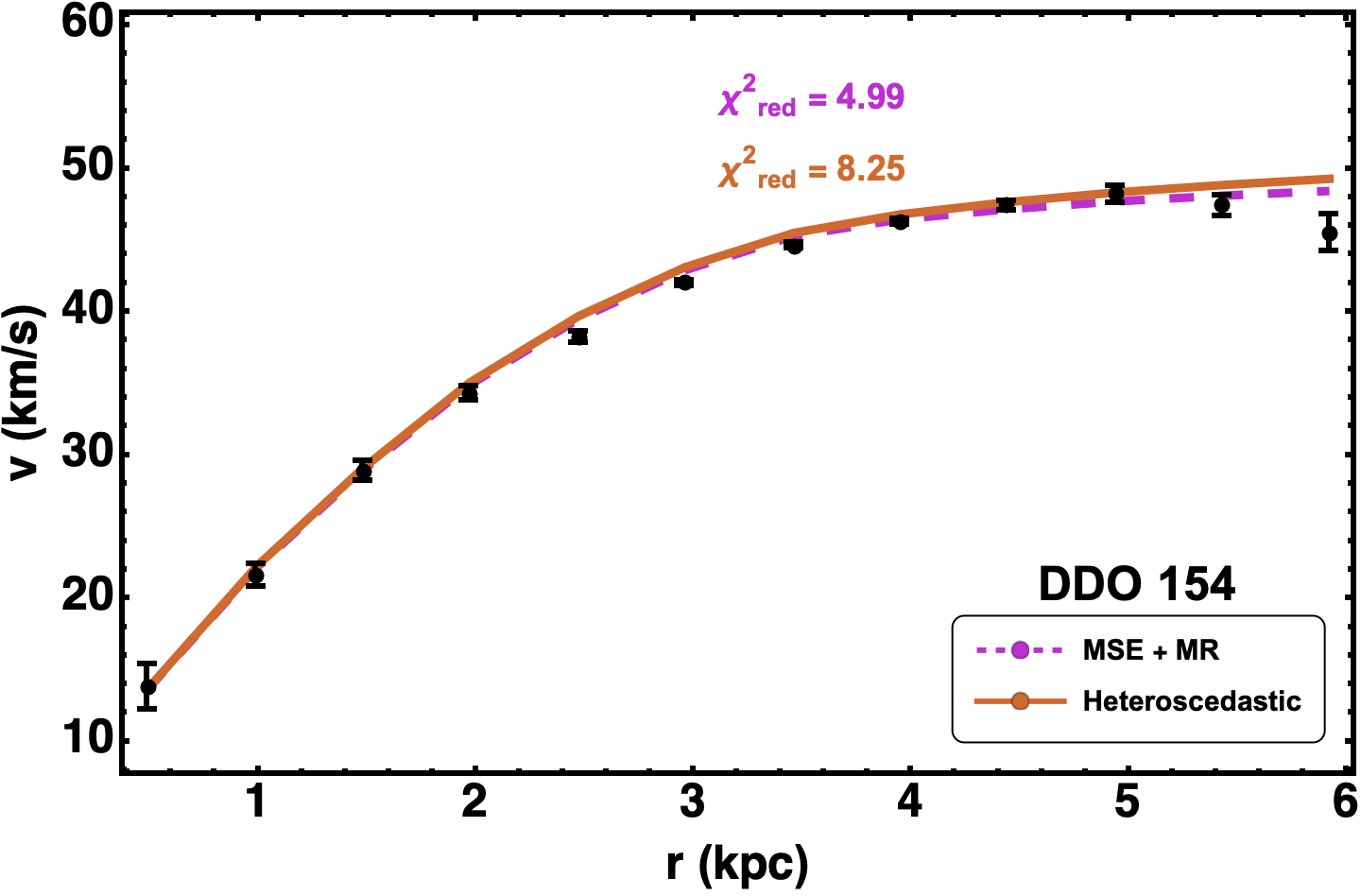} & 
\includegraphics[width=0.3\textwidth]{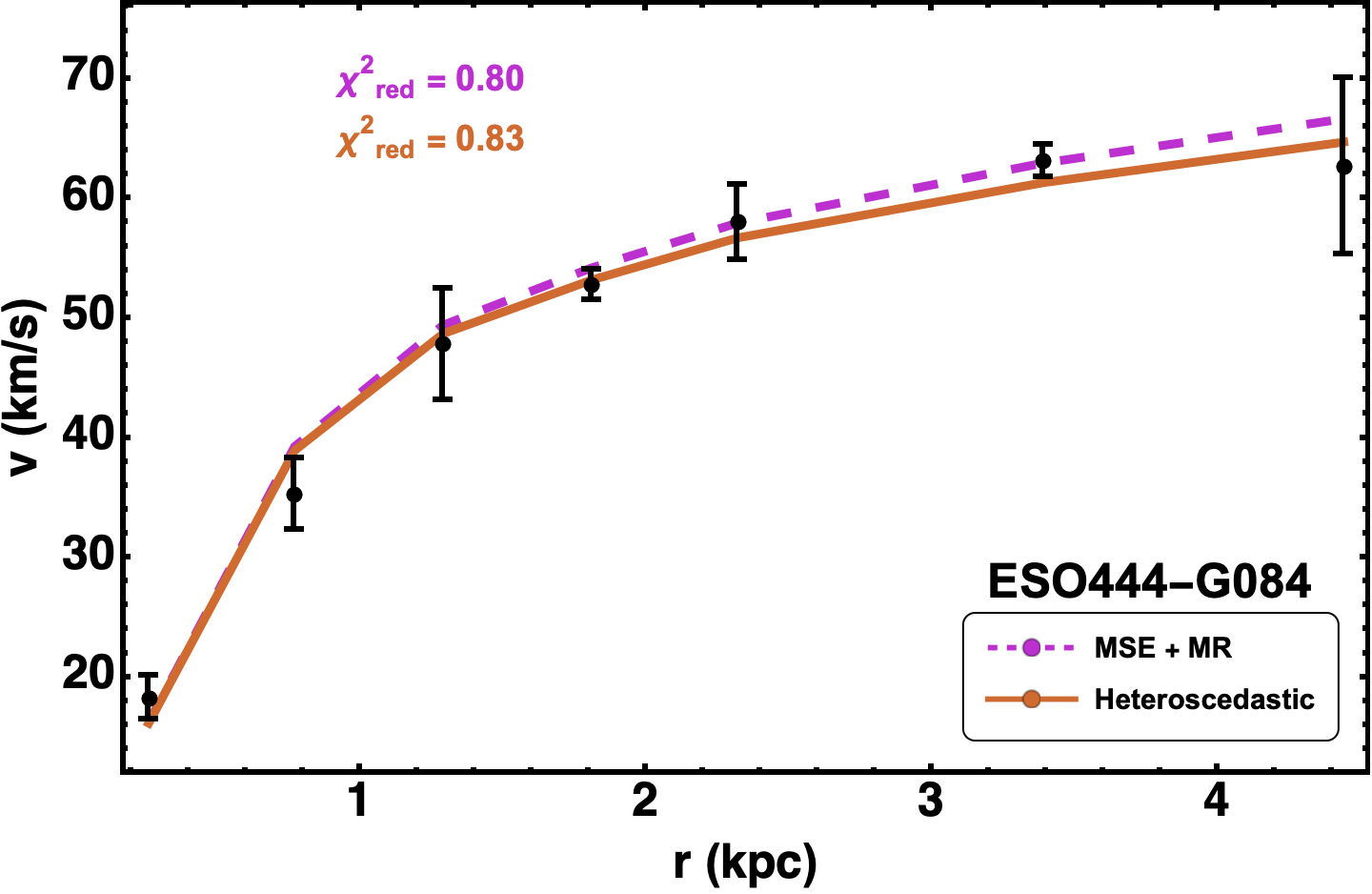} &
\includegraphics[width=0.3\textwidth]{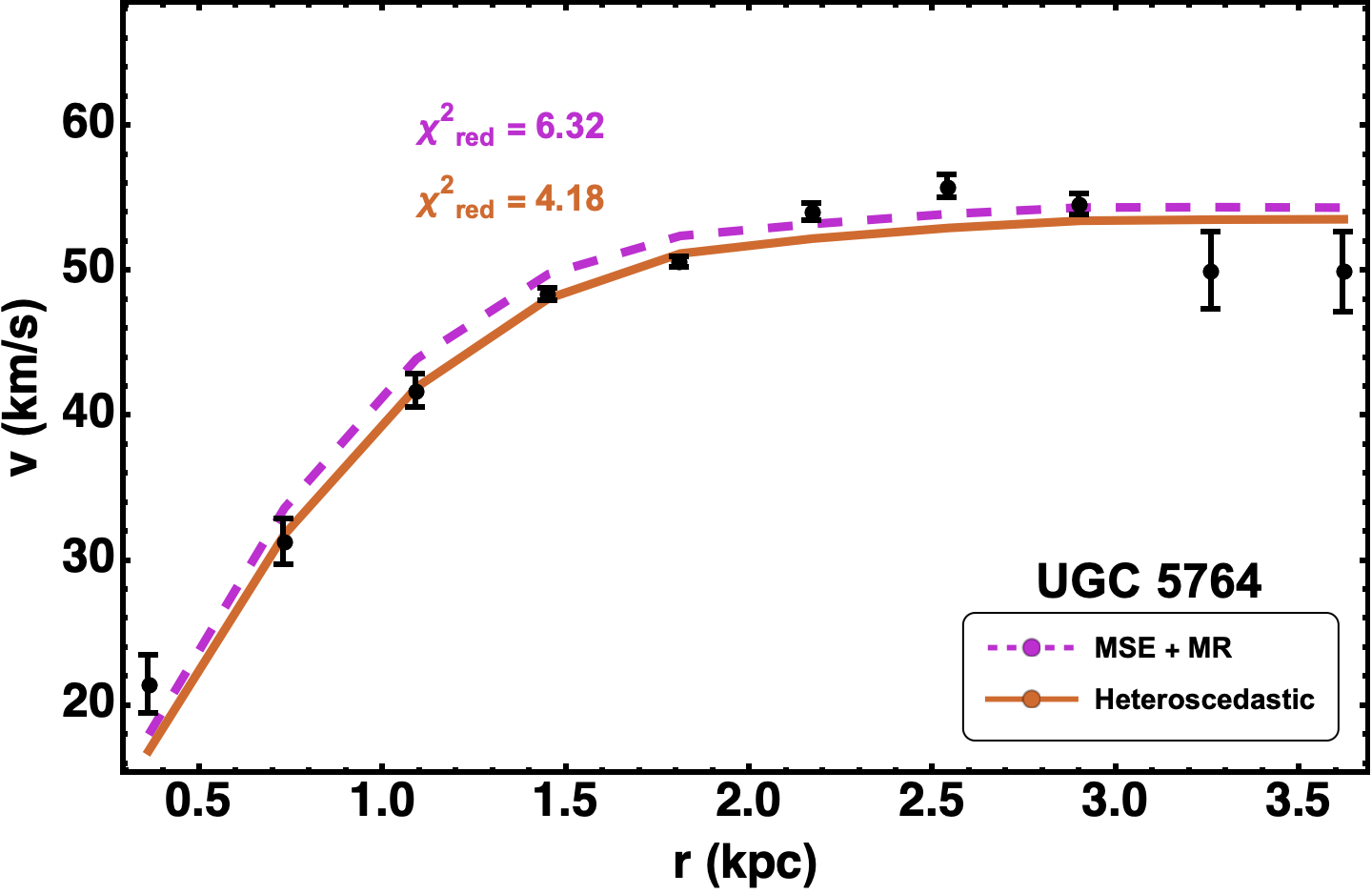} \\
(a) DDO154 & (b) ESO 444-G084 & (c) UGC 5764 \\[6pt]
\includegraphics[width=0.3\textwidth]{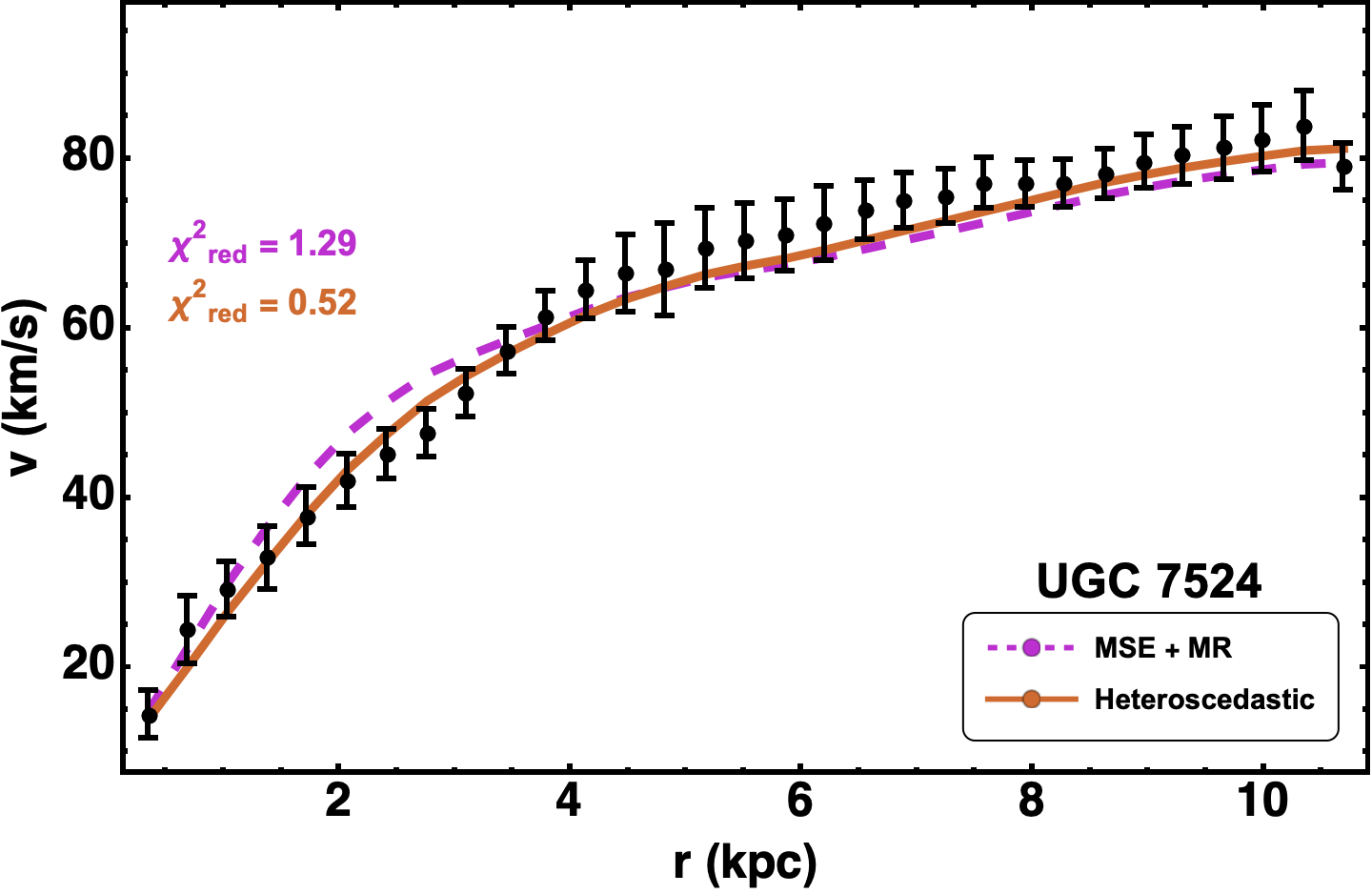} &   \includegraphics[width=0.3\textwidth]{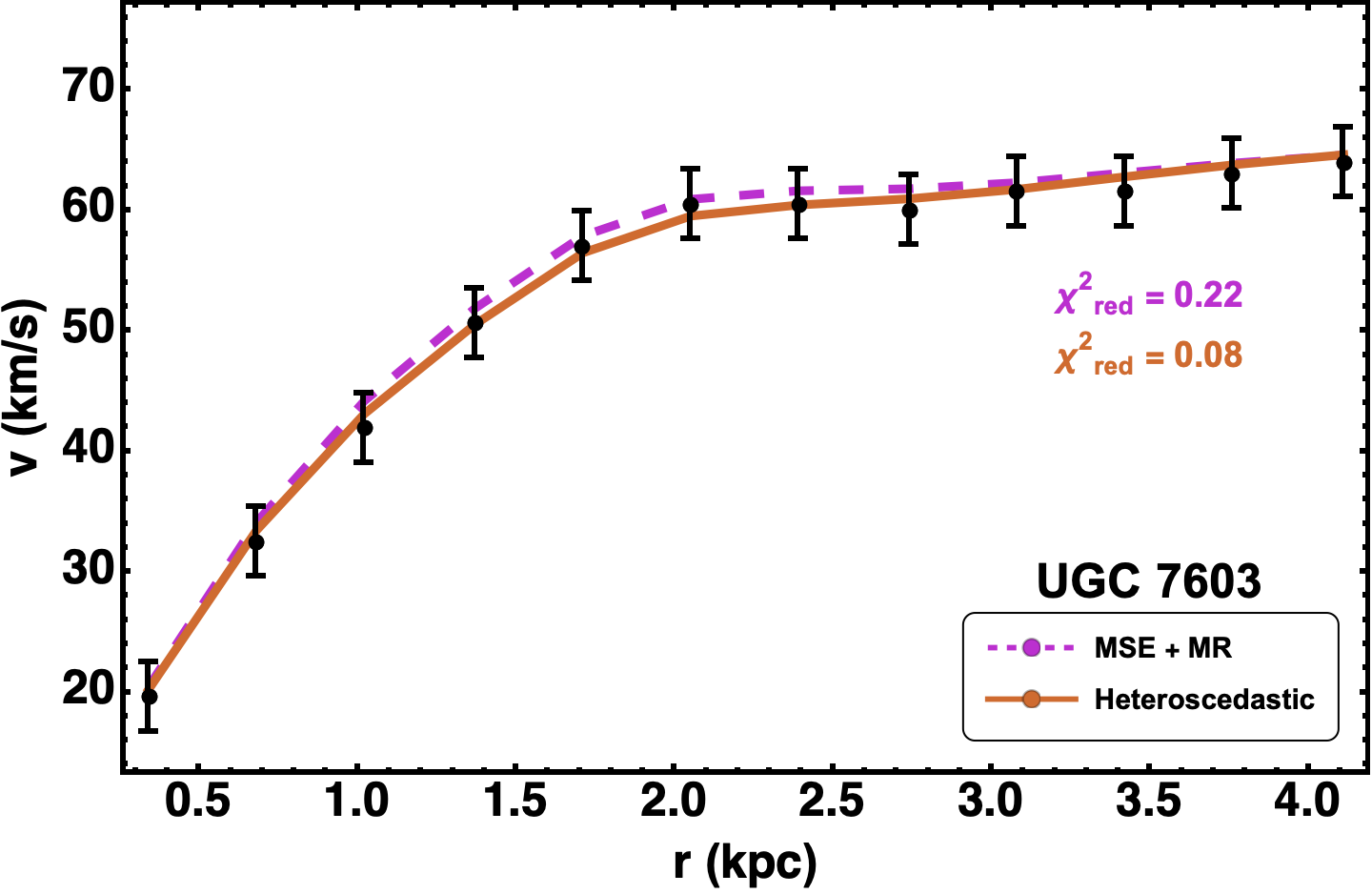} &
\includegraphics[width=0.3\textwidth]{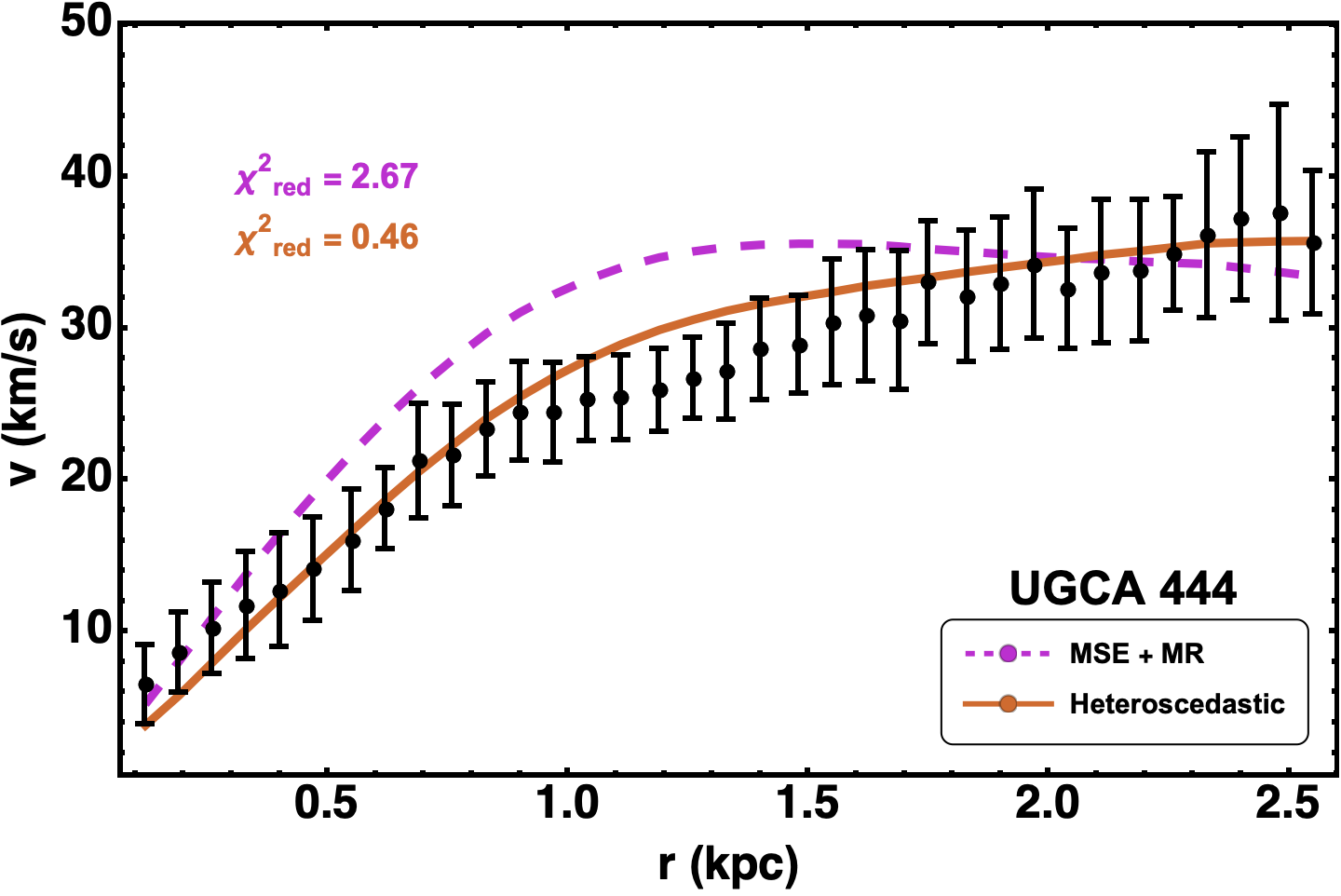} \\
(d) UGC 7603 & (e) UGC 7524 & (f) UGCA 444
\end{tabular}
\caption[ANN (with heteroscedastic loss function) prediction for remaining 6 galaxies]{\justifying Rotation curves corresponding to the predicted parameters for each of the three cases studied:(a) MSE loss function with noisy input data with multiple realizations of the observed rotation curves from Section~\ref{sec:multiple_realizations} (purple dashed curves) , and (b)  Heteroscedastic loss function with noisy inputs from Section~\ref{sec:hetero_uncertainty} (orange curves). The reduced $\chi^2$ are also shown for each case.}
\label{fig:all_comparison}
\end{figure}

The story remains the same for all galaxies in the samples, rotation curves for which are shown in Fig.~\ref{fig:all_comparison}, while the parameter values and their uncertainties are listed in Table~\ref{tab:preds_hetero}.
Notice the larger uncertainties associated with $r_s$ and $\Upsilon_*$ for the other galaxies as well, which is remarkably consistent with the expectation that the rotation curves for DM dominated galaxies with $r_t \geq 1\ \text{kpc}$ are not very sensitive to $r_s$ and $\Upsilon_*$ values. 
This implies that uncertainty in the heteroscedastic loss function captures, in some sense, the effect a parameter will have on a rotation curve. 
This is interesting since the neural network has only approximated the functional dependence between parameters and rotation curves by looking at various examples. 

Therefore we see that without having to define a likelihood or plotting a posterior distribution, one can use heteroscedastic loss function obtain accurate parameter predictions as well as uncertainties that capture, to some extent, the sensitivity of the data to the parameters.
It is worth noting that this uncertainty is based on the loss function of the ANN that compares the output of the neural network to the true output (or the ground truth). 
On the other hand, the uncertainty in the Bayesian inference approach stems from the likelihood function, which compares the theoretical and the observed data assuming a distribution, often a Gaussian, for errors.

\section{Comparison with MCMC}\label{sec:mcmc_comparison}

In this section, we compare the multiple realizations approach of obtaining a sequence of parameters as well as the uncertainties obtained using the heteroscedastic loss function with a likelihood-based MCMC approach. 

Recall from Section~\ref{sec:multiple_realizations}, that for a fixed galaxy, we sample from the Gaussian $\mathcal{N}(v_{obs}(r), \sigma_r^2)$ for all radius values to generate multiple realizations of the observed rotation curve. 
One can think of this as sampling curves from the vicinity of the observed rotation curve while being consistent with observations. 
For a sufficiently accurate neural network, one can then think of the multiple parameter predictions (using the multiple realizations as input) as a `chain' of parameters that satisfy the observed rotation curve of that galaxy. 
One can then plot 2D and 1D projections of the parameter values for this chain in a corner plot to visualize how the parameters are distributed. 

The above procedure appears qualitatively similar to the approach used in Bayesian inference using MCMC.
Indeed, authors in \cite{Wang_ECoPANN_2020} employed such a method to obtain chains of parameters and found that the corresponding 2D contour plots along with 1D projections using ANNs agreed well with the 2D joint distributions and 1D marginalized distribution obtained using MCMC sampling. 

\begin{figure}[h!]
    \includegraphics[width=
    0.75\textwidth]{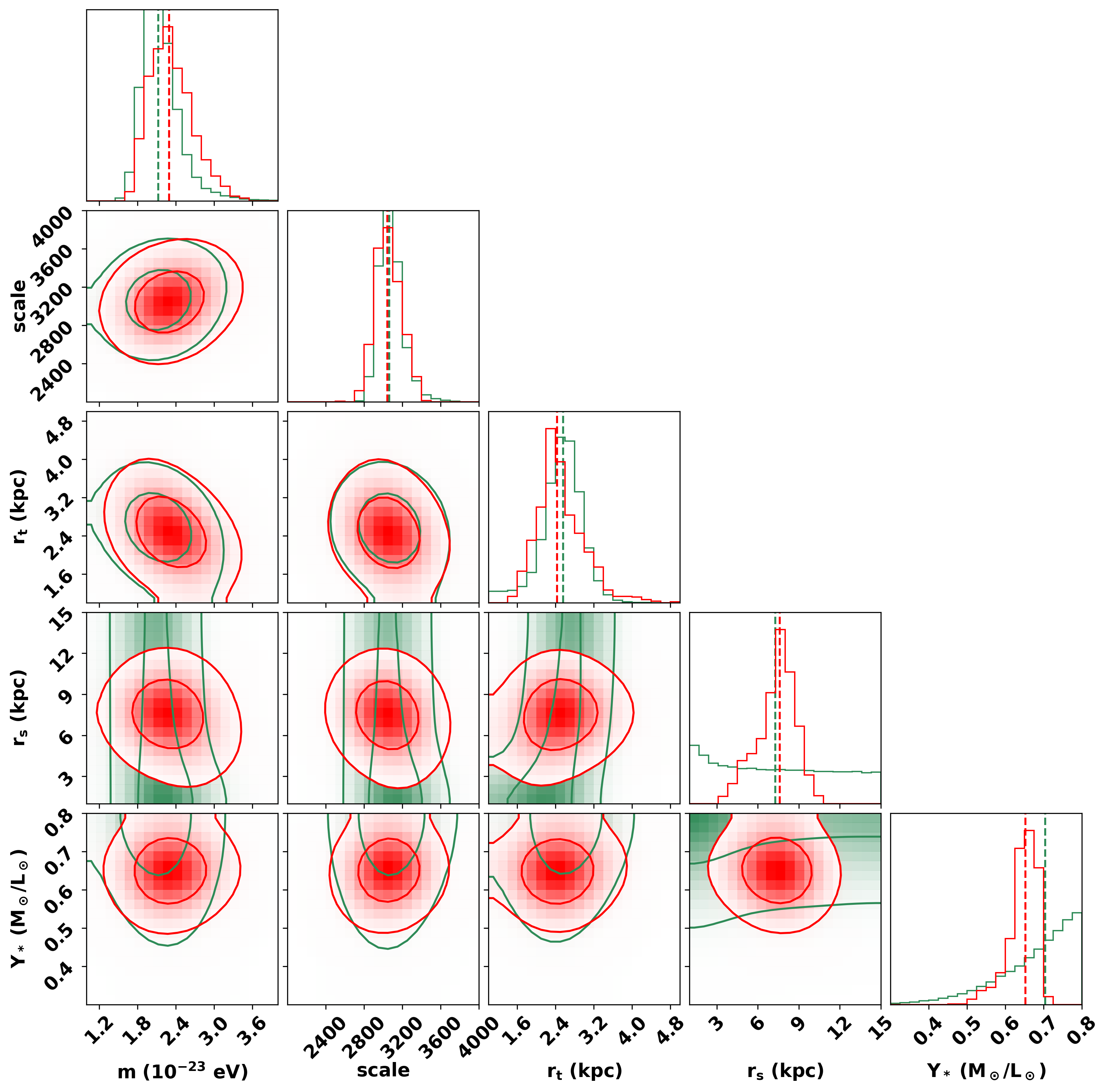}
    \caption[Comparing contour plots from the ANN (MSE loss) and MCMC approach]{\justifying $2$D and $1$D distributions of parameters obtained using our approach as described in Section~\ref{sec:multiple_realizations} (in red), plotted over joint and marginal distributions obtained using MCMC for UGC 5721 (in green). The vertical dashed lines (red and green) in the $1$D histograms are the $50\%$ quantile values for both approaches. The $2$D contours show $\sim 39.3\%$ and $86.4\%$ confidence regions.}
    \label{fig:ANN_MCMC_contour_plot}
\end{figure}

Therefore, we proceed to use the `chain' of $1000$ inferred parameters, to obtain the median and $1\sigma$ confidence interval for each parameter for every galaxy in our sample as shown by the purple points with error bars in Fig.~\ref{fig:MSE_hetero_MCMC}. 
One can also use the chain of parameters to plot contours of the 2D joint and 1D marginalized distributions in a corner plot. 
Contours obtained for the case of UGC 5721, as an example, are shown in red in Fig.~\ref{fig:ANN_MCMC_contour_plot}. 

To compare with Bayesian inference, we sample the posterior using an ensemble sampler incorporated in the emcee\footnote{\href{https://emcee.readthedocs.io/en/v2.2.1/}{https://emcee.readthedocs.io/en/v2.2.1/}} python package, for all galaxies in the sample. 
We use the same model described in Section~\ref{sec:model_and_data}, where the theory rotation curve is parameterized by $5$ parameters in Eq.~(\ref{eq:params}).
We use the same uniform priors as those defined in Table~\ref{tab:param_space}.
Here, the log-likelihood is written as
\begin{equation}
    \mathrm{ln}\mathcal{L}_{\text{MCMC}} = -\frac{1}{2}\sum_i\left(\frac{v_{obs}(r_i) - v_{th}(r_i; {\bf P})}{\sigma_i}\right)^2\ .
\end{equation}
We then run the emcee sampler with 48 walkers (at different random initial points in the parameter space) for $20,000$ steps, out of which the initial $10,000$ are removed to account for the burn-in period. 
The median and $1\sigma$ confidence intervals are shown in Fig.~\ref{fig:MSE_hetero_MCMC} by the green points with error bars for all parameters for every galaxy in our sample (they are also listed in Table~\ref{tab:preds_hetero}).  
These chains can also be used to plot $2$D and $1$D posterior distributions, which are shown for the case of UGC 5721 in Fig.~\ref{fig:ANN_MCMC_contour_plot} by the green contours.

Note that, for the neural network trained using the heteroscedastic loss function, one does not get multiple parameter estimates.
Here, the uncertainty is learned implicitly during training, and hence one cannot plot the joint distributions.
We therefore plot the point-estimate and uncertainties obtained using this method in Fig.~\ref{fig:MSE_hetero_MCMC}, denoted by the orange points with error bars. 

We note the following observations:
\begin{itemize}
    \item The median values obtained from the multiple realizations approach as well as the point-estimates obtained from neural networks trained using heteroscedastic loss function lie within $1\sigma$ intervals obtained from the MCMC approach for most parameters and galaxies.
    
    \item As seen from the case of UGC 5721 in Fig.~\ref{fig:ANN_MCMC_contour_plot}, it is also clear that contours obtained using the ANN approach seem to capture correct correlations between some parameters, for instance, the positive correlation between $m$ and $s$, the negative correlation between $r_t$ and $m$ as well as $r_t$ and $s$.
    
    \item Predicted uncertainties are also similar for all three methods for $m$, $s$ and $r_t$.
    However, the multiple realizations approach severely under-estimates the uncertainties for $r_s$ and $\Upsilon_*$, which, as we discussed in Section~\ref{sec:multiple_realizations}, do not affect the rotation curves significantly. 
    
    This is captured correctly by the neural networks trained using heteroscedastic loss, which gives large uncertainties for $r_s$ and $\Upsilon_*$ for all galaxies in Fig.~\ref{fig:MSE_hetero_MCMC}, agreeing well with the MCMC approach. 
\end{itemize}

\section{\label{sec:discussion}Summary and discussion}

We live in the era of data-driven cosmology and astrophysics, where ever larger and more accurate data sets are becoming available \cite{Amin_whitepaper_Snowmass2021}. 
Extracting information from such data sets is essential to constrain models of new physics.
While the usual method for doing this involves a likelihood-based approach using Markov Chain Monte Carlo (MCMC), with advancement in computer hardware, it is worthwhile to explore machine learning and in particular, deep learning to develop novel and complementary approaches to tackle the same problem.

With that in mind, in this chapter, we have explored the use of artificial neural networks in learning model parameters from observed galactic rotation curves.
Neural networks are powerful tools that can be used to approximate a wide-range of complex functions \cite{Hornik_1990}, which is particularly useful when the relationship between some input vector ${\bf x}$ and output vector ${\bf y}$ is highly non-linear and complex.

In this chapter, unlike a likelihood-based approach, we train neural networks with rotation curves as input and the model parameters as output using a large sample of simulated rotation curves whose parameter values are already known\footnote{There are various other approaches that can be used here to carry out parameter estimation along with uncertainty quantification using neural networks including Bayesian neural networks (BNNs) and normalizing flows (\cite{Hortua_2020, Hortua_2023, Stachurski_2023, Kolmus_2024} for recent work).}. 
By looking at samples in the training data, the neural network 
updates its internal adjustable parameters (called weights and biases) to approximate a function $f: \mathbb{R}^{N_{obs}}\rightarrow\mathbb{R}^P$ where $N_{obs}$ is the number of observed velocities in a rotation curve and $P = 5$ is the number of parameters, which are:  
ULDM particle mass $m$, the scale parameter $s$ associated with the mass of the core, transition radius $r_t$, i.e. the radius at which the core profile transitions to a NFW profile, the scale radius of the NFW profile $r_s$, and $\Upsilon_*$, the stellar mass-to-light ratio which tunes the baryonic contribution to the rotation curves (See Section~\ref{sec:model_and_data} for a detailed discussion on the model and dataset used). 

\begin{figure}[h!]
\centering
\begin{tabular}{cc}
\includegraphics[width=0.45\textwidth]{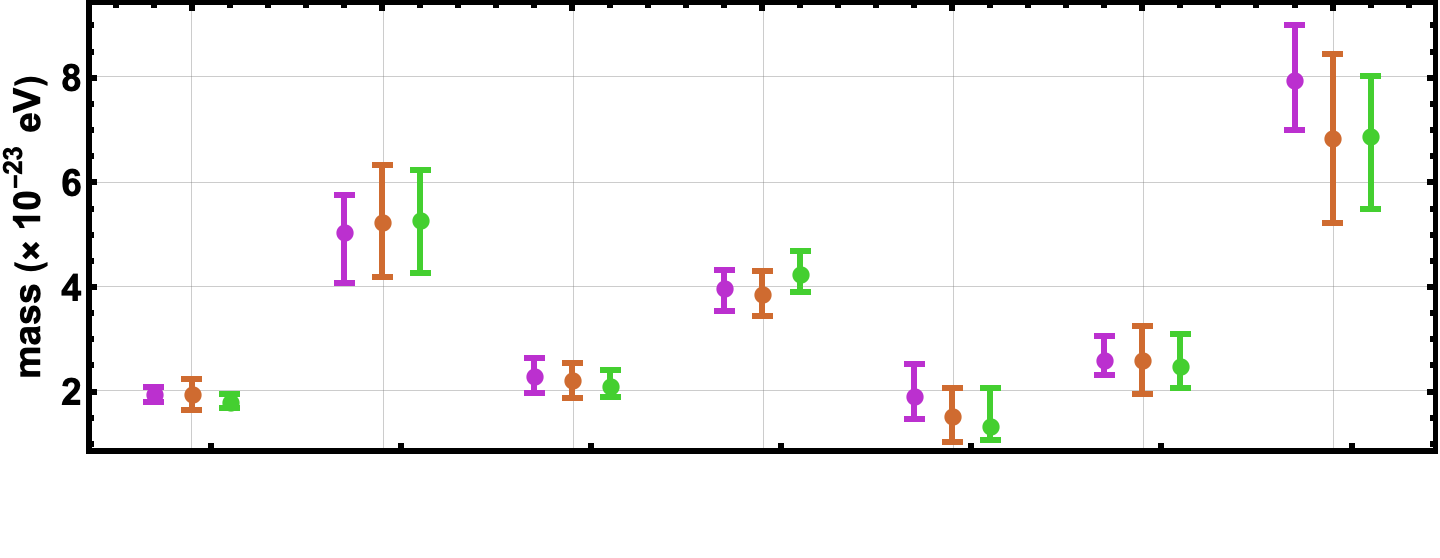} &
~ \\
(a) ULDM mass, $m$ & ~ \\[6pt]
\includegraphics[width=0.45\textwidth]{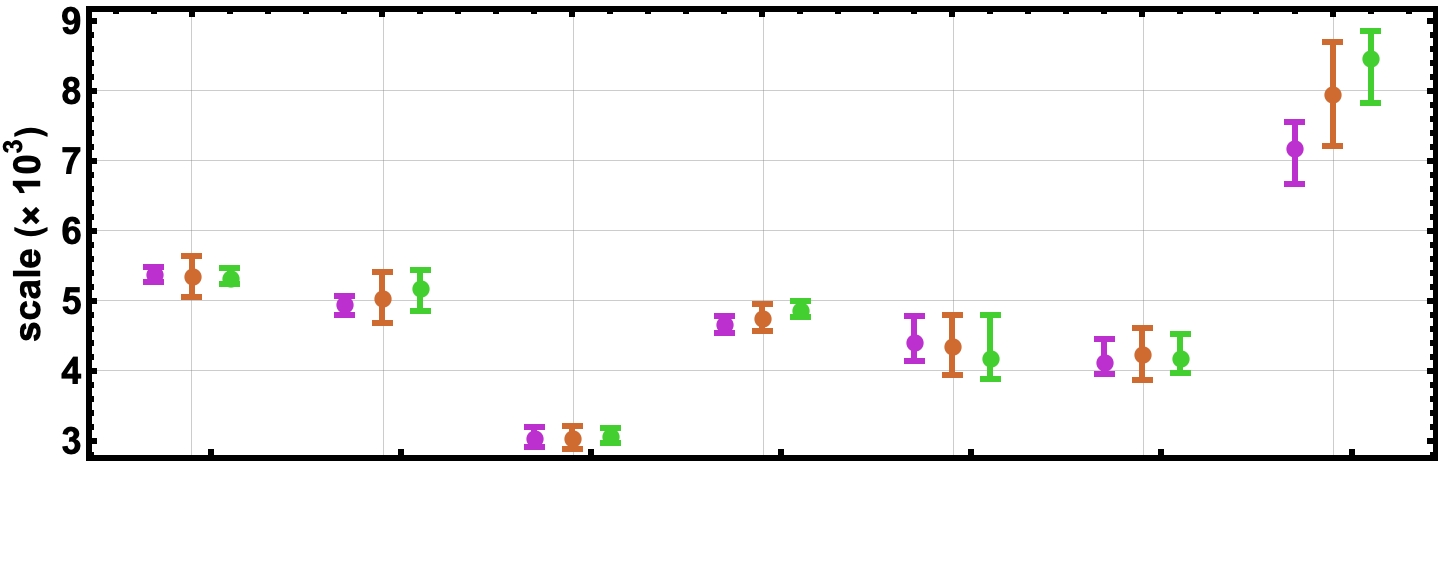} &
\includegraphics[width=0.45\textwidth]{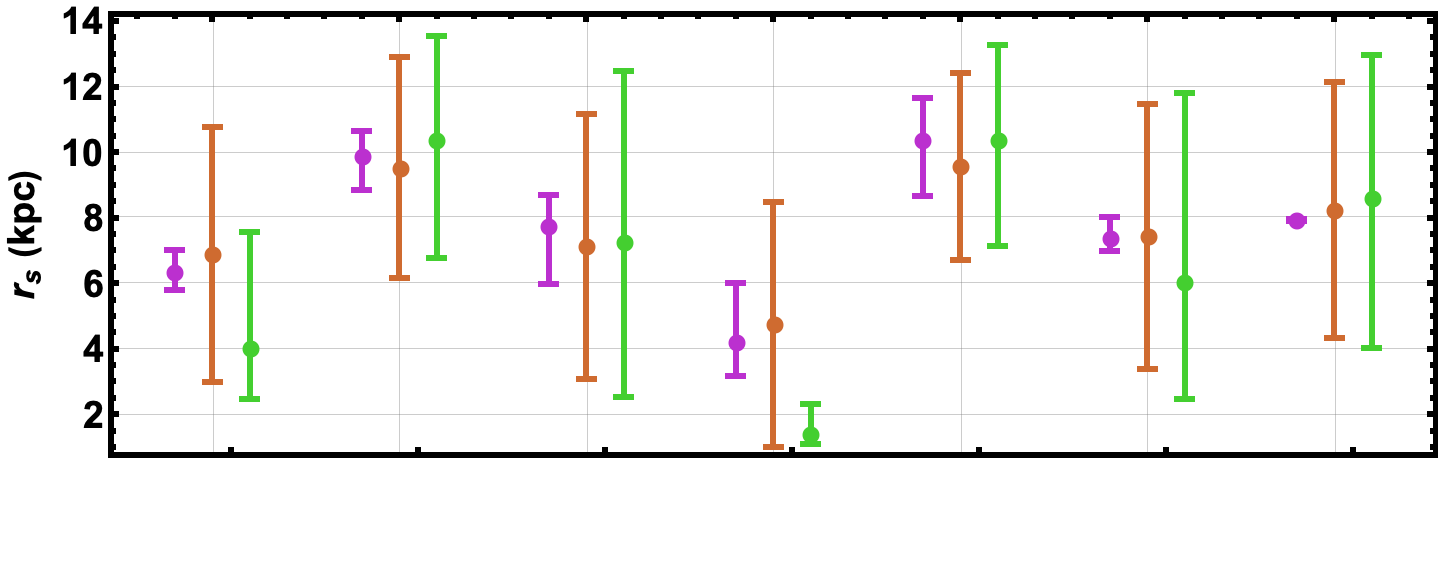}  \\
(b) scale, $s$ & (d) scale radius, $r_s$ \\[6pt]
\includegraphics[width=0.45\textwidth]{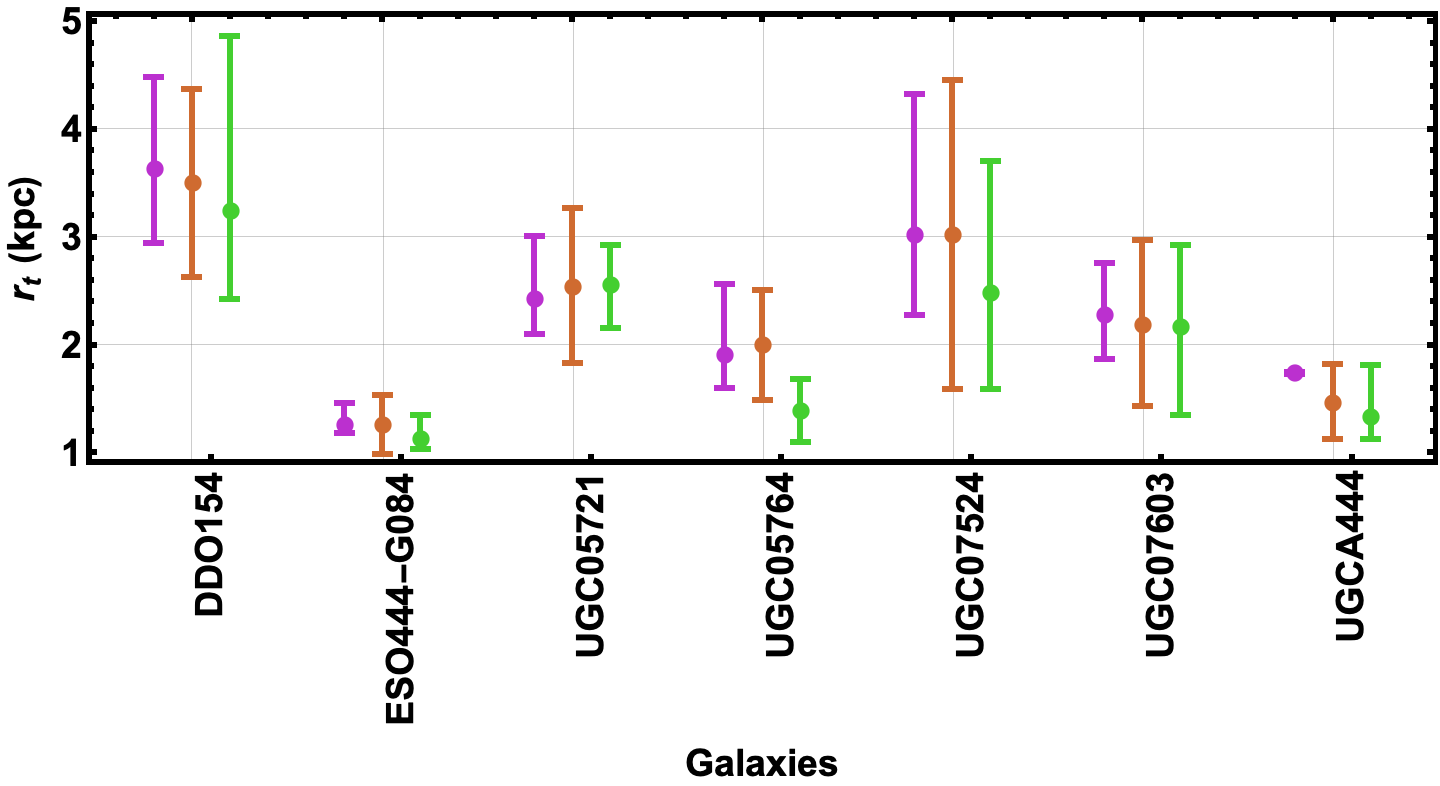} & \includegraphics[width=0.45\textwidth]{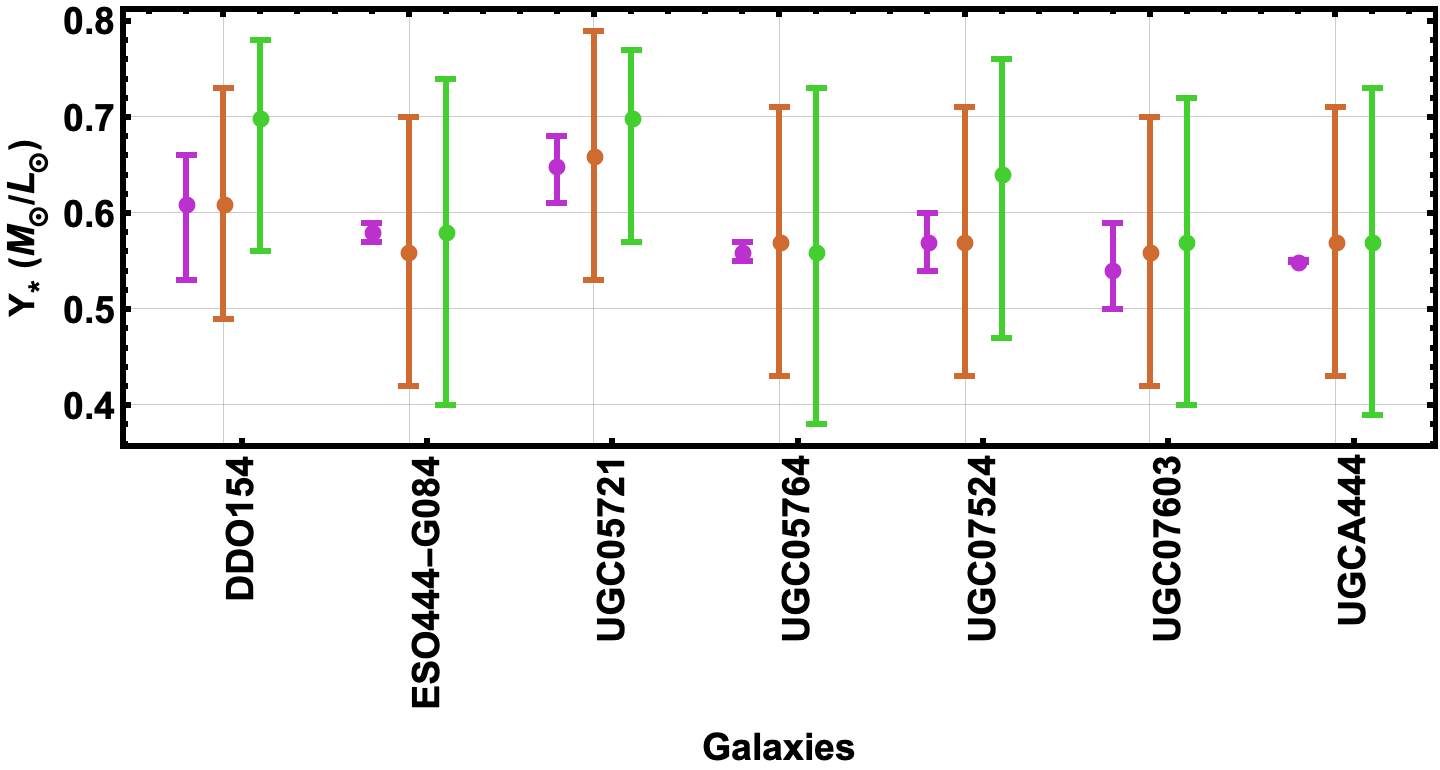} \\
(c) transition radius, $r_t$ & (e) stellar mass-to-light ratio, $\Upsilon_*$\\
\end{tabular}
\caption[Parameter values and $1\sigma$ uncertainties for ANN and MCMC approaches]{\justifying Inferred parameters and their uncertainties using multiple realization method (purple), heteroscedastic loss function (orange) and MCMC runs using emcee (green).}
\label{fig:MSE_hetero_MCMC}
\end{figure}

The observed rotation curve is then given as input to the trained neural networks and we get inferred point-estimates of parameters as output. 
Here is a short summary of our findings:

\begin{enumerate}
    \item Neural networks trained on noisy simulated rotation curves perform much better than the noiseless case when confronted with observed rotation curves. 
    The noise included during training appears to capture the fact that observed rotation curves are not ``smooth'' functions (see Fig.~\ref{fig:MSE_comparison} and the discussion around it).

    \item To quantify the uncertainty associated with the inferred parameters based on the uncertainty in the observations, we generated multiple realizations of the observations as described in Section~\ref{sec:multiple_realizations}, which lead to multiple parameter estimations using observations, resembling a chain of parameters. 
    The median values obtained using this method lead to rotation curves which agree well with observed rotation curves, as shown in Fig.~\ref{fig:all_comparison} by the purple curves.

    \item However, comparison with parameter estimation using MCMC in Section~\ref{sec:mcmc_comparison} suggests that while the multiple realizations method can capture some of the correlations between a few parameters, namely $m$, $s$ and $r_t$, it struggles when confronted with hard-to-constrain parameters like $r_s$ and $\Upsilon_*$ which don't affect the rotation curves in our sample significantly (see Figs.~\ref{fig:ANN_MCMC_contour_plot} and~\ref{fig:MSE_hetero_MCMC}).

    \item Instead of mean-squared-error (MSE) loss function, using a heteroscedastic loss function in Eq.~(\ref{eq:hetero_loss}) to train the network enables us to learn uncertainty associated with the parameter during training itself. 
    When confronted with observed rotation curves, the ANNs perform just as well as the networks trained using MSE loss function (see Fig.\ref{fig:all_comparison}).

    \item For the case of neural networks trained using the heteroscedastic loss function, while one cannot obtain $2$D and $1$D distribution of parameters, the ANNs are able to capture the amount of uncertainty better for the hard-to-constrain parameters ($r_s$ and $\Upsilon_*$). 
    Indeed, as shown in Fig.~\ref{fig:MSE_hetero_MCMC}, for almost all galaxies, the parameters and the uncertainties inferred by neural networks with the heteroscedastic loss function agree well with those obtained from MCMC analysis. 
\end{enumerate}

In summary, we have shown that with the chosen architecture and hyperparameters, neural networks can prove to be a useful tool in obtaining parameter estimates that can describe observed rotation curves well, i.e. with a small $\chi^2_{red}$. 
Our exploration, along with other recent works \cite{Wang_2020, Wang_ECoPANN_2020, Pal_2023, Hagimoto_2024, Artola_2024} demonstrates that, the use of neural networks for this class of problems can be a useful complementary approach to standard likelihood-based approaches.
In particular, our work in this chapter agrees well with the recent results obtained in \cite{Pal_2023}, where model parameters as well as their uncertainties obtained using an ANN trained with a heteroscedastic loss function agreed well with those obtained using MCMC. 

Before closing we would like to note some caveats and future prospects: 
\begin{inparaenum}[(a)]
 \item In the case of Bayesian inference, the propagation of uncertainty from observed data to the parameters is well-understood. 
 On the other hand, while the parameter uncertainties obtained using the heteroscedastic loss function agree, in magnitude, with those obtained via the MCMC approach, their interpretation remains a challenge.
 This must be explored further, and we leave it to a future paper. 
 \item
 The true values of each parameter are assumed to be independently and normally distributed in the heteroscedastic loss function, which may not be the case. 
 \item
 We have not carried out a full hyperparameter exploration to obtain an optimal neural network. It is likely that there exists a better choice of hyperparameters that perform better by giving better estimates of the parameter values and the associated uncertainties for this dataset.
 \item
 Information about the radius values for which velocities (observed or simulated) are obtained has not been used in this analysis, since the input vector is just a list of velocities. Therefore, the neural networks we have trained cannot differentiate between two rotation curves with the same velocities but different observed radii. 
 An interesting direction would be to train a neural network capable of handling input rotation curves from multiple galaxies with a varied range of radii and velocities to infer a single parameter value for the fundamental parameter of our model, i.e., ULDM particle mass $m$, while inferring different galaxy specific parameters.
 We leave this exploration to future work.
 \item In this chapter, we deal with the vanilla fuzzy dark matter model, i.e., a scalar field with negligible self-interactions.
 In chapter~\ref{chpt:paper_2}, we have demonstrated that repulsive self-interactions can satisfy observed rotation curves while also satisfying an empirical core-halo mass relation, unlike the case of no self-interactions \cite{Bar_2022}.
 It could then be interesting to include the effect of $\lambda\varphi^4$ kind of self-interactions for the ultralight scalar field, and infer $\lambda$ along with the parameters we considered in this chapter.  
\end{inparaenum}
\justifying
\chapter{Summary and future prospects}\label{chpt:conclusion}

\section{Summary}
The physical nature of dark matter is one of the most important unsolved mysteries in cosmology. 
Evidence for its existence can be found across various epochs since the big bang as well as from various length scales all the way from galactic scales to cosmological scales~\cite{Profumo_2019, Cirelli_DM_review_2024}.
However, the microscopic nature of dark matter, which includes its mass, spin, couplings, etc. are yet unknown.  
The goal of this thesis was to take a signature-driven approach and explore what observational data can say about these properties, in particular mass and self-coupling (of the $\lambda\varphi^4$ kind) of the DM particle, assuming a spin-zero ultra light boson (called Ultra Light Dark Matter (ULDM)). 
We did this by describing the central regions of galactic halos by self-gravitating field configurations whose properties depend on particle mass $m$ and self-interaction strength $\lambda$. 
Of course, for such an approach to work, the mass and self-coupling of the dark matter candidate should be in the range that can be probed by observed data. Here, ULDM is an ideal candidate, since with $m\sim 10^{-22}$, its deBroglie wavelength is large $\sim\mathcal{O}(\text{kpc})$, ensuring novel signatures at galactic scales. 
Our approach involves studying the implications of ULDM self-coupling on already existing constraints as well as exploring novel techniques like neural networks to learn parameters from observed data. 

To that effect we have looked at the following observations to explore and uncover the mass and self-coupling of ULDM: (a) amount of mass contained within a region around the centre of galactic halos (see chapter~\ref{chpt:paper_1}), (b) galactic rotation curves of dark matter dominated dwarf galaxies (see chapters~\ref{chpt:paper_2} and~\ref{chpt:paper_4}) as well as (c) the existence of satellite dwarf galaxies orbiting around the centres of a larger host halo (see chapter~\ref{chpt:paper_3}). 
As discussed in chapter~\ref{chpt:numerical_solutions}, we have used stationary and quasi-stationary solutions of the equations of motion of ULDM, the Gross-Pitaevskii-Poisson equations eqs.~(\ref{eq:GrossPitaevskii}) and~(\ref{eq:Poisson}), to describe the density profiles in the inner regions of galactic halos. 
We have found that, the above observations not only probe particle masses as small as $m\sim 10^{-22}\ \text{eV}$ 
they also probe small self-couplings in the range $|\lambda|\sim 10^{-90}-10^{-95}$.
It is important to note that while there has been recent interest in exploring the role of self-interactions of ULDM on observations, many stringent constraints on ULDM work in the regime where the self-interaction strength $\lambda = 0$ (see \cite{Irsic_2017, Hlozek_2018, Rogers_2021, Bar_2022, Dalal_2022, Khelashvili_2023} for a few recent examples).
We summarise our main results as follows: 

\begin{itemize}
    \item \textbf{Chapter~\ref{chpt:paper_1}}:
    
    In this chapter, based on \cite{Chakrabarti_2022}, we find that even in the presence of a supermassive black hole at the centre of a halo, 
    the density profile of the soliton around it squeezes (expands) in the presence of attractive (repulsive) self-interactions.
    Here, using the observed halo mass and the observed mass of the SMBH at the centre of the M87 halo, self-couplings as small as $\lambda\sim \pm 10^{-95}-10^{-96}$ can alter the density profile to an extent where the amount of mass contained within a spherical region of radius $10\ \text{pc}$ is different for different $\lambda$ values for a fixed $m$. 
    We also allowed $m$ to vary and obtained constraints in the $\lambda-m$ plane for M87 (see figures~\ref{Excluded_-} and~\ref{Excluded_+}). 

    \item \textbf{Chapter~\ref{chpt:paper_2}}:
    
    In this chapter, we considered observed rotation curves of dwarf galaxies. 
    Since including self-interactions alters the shape of the density profile of the solitonic solutions of the GPP system, it is expected that the corresponding velocity curve also changes, as seen by eq.~(\ref{eq:circ_vel}). 
    We therefore looked at the observed rotation curves of dark matter dominated low surface brightness galaxies from the SPARC catalogue \cite{Lelli_2016} in \cite{Dave_2023} and found that: if ULDM must simultaneously satisfy observed rotation curves as well as a Soliton-Halo relation (see eq.~(\ref{eq:modified_SH}) and the discussion around it) between the masses of the ULDM core and the surrounding halo, then for $m\sim 10^{-22}\ \text{eV}$ one requires repulsive self-interactions of strength $\lambda \gtrsim 10^{-90}$ (see figure~\ref{fig:LSB_with_inter}). 
    This highlights the importance of taking into account the effect of self-interactions, since if $\lambda = 0$, Ref.~\cite{Bar_2022} demonstrated that the above condition cannot be satisfied by any $m\in \left[10^{-24}\ \text{eV}, 10^{-20}\ \text{eV}\right]$. 

    It is important to note that too large attractive self-interactions can lead to unstable solutions that can collapse \cite{Chavanis_2011, Levkov_2017, Schiappacasse_2018}. 
    Hence, stability of the cores requires that $\hat{\lambda}_\text{ini}$ (dimensionless self-coupling as defined in chapter~\ref{chpt:numerical_solutions}) not be too negative.  
    We found that for values of $\hat{\lambda}_\text{ini}$ that are allowed, the resultant deviation from the $\lambda = 0$ case is not enough to evade the constraints for $m= 10^{-22}\ \text{eV}$.

    \item \textbf{Chapter~\ref{chpt:paper_3}}:
    
    In chapter~\ref{chpt:paper_3}, having noted that astrophysical observations can be sensitive to both the sign and strength of self-interactions, we look at a slightly different astrophysical system: a satellite dwarf galaxy orbiting in the gravitational potential of a larger host halo. 
    In this scenario, using quasi-stationary solutions to model the loss of ULDM mass from the satellite dwarf, we find that the presence of very small self-interactions can have a drastic impact on the timescales over which the satellite loses most of its mass \cite{Dave_2024}. 
    Hence, the mere existence of a satellite dwarf with some observed properties can be used to probe the mass and self-coupling of ULDM. 

    In particular, using the quasi-stationary formalism, for a satellite galaxy with a known orbital period and core mass, the presence of attractive self-interactions (leading to a higher central density) extends the potential barrier as seen in figure~\ref{fig:SI_barrier}, thereby increasing the timescale over which the satellite is tidally disrupted. 
    On the other hand, repulsive self-interactions spread the inner core, lower the central density of the satellite dwarf galaxy and squeeze the potential barrier as seen in figure~\ref{fig:SI_barrier}, and hence decrease its lifetime. 
    The effects described above are particularly important for the values of $m$ that are allowed by observations, since it was suggested in \cite{Hertzberg_2023} that, ULDM with $m = 10^{-22}\ \text{eV}$ is not compatible with the existence of the Fornax dwarf spheroidal.  
    We found in \cite{Dave_2024}, that for the same $m = 10^{-22} \ \text{eV}$, if one includes attractive self-interactions with $\lambda \lesssim -2.12\times 10^{-91}$, the lifetime of Fornax is extended enough for it to survive until present day (see figure~\ref{fig:saving_fornax}). 

    \item \textbf{Chapter~\ref{chpt:paper_4}}:
    
    Finally, noting that it is through analysing observed data that one can learn and put constraints on various models of new physics, we attempted to explore a new way to infer parameters from data, i.e. using neural networks. 
    The parameters of interest were those describing the dark matter density profile along with a Baryonic parameter, and the observational data we used were rotation curves of 7 dark matter dominated galaxies from the SPARC catalogue \cite{Dave_2025}.
    For numerical ease, we carried out this work in the $\lambda = 0$ limit, leaving us with $5$ free parameters: $\{m, s, r_t, r_s, \Upsilon_*\}$ (see section~\ref{sec:model_and_data}). 
    The usual approach, explored for instance in Refs~\cite{Bernal_2017, Khelashvili_2023, Banares_Hernandez_2023} involves  Bayesian inference.
    We, on the other hand, find that neural networks trained on a larger number of simulated rotation curves, when fed the observed rotation curve are able to infer parameters that describe observed rotation curves well (see figure~\ref{fig:all_comparison}).
    We also compared the parameter uncertainties obtained from the ANN approach to those obtained from the Bayesian approach. 
\end{itemize}

Therefore, we have demonstrated that 
(a) astrophysical data we consider is quite sensitive to both ULDM mass $m$ and self-coupling $\lambda$, 
(b) for $m\sim 10^{-22}\ \text{eV}$, the value of the self-coupling that can have non-negligible impact on observations we consider can be as small as $|\lambda| \gtrsim 10^{-91}$, and 
(c) such a small self-coupling can, in some cases, evade stringent constraints on ULDM with $\lambda = 0$. 
In-fact, in the last few years, there has been an increase in interest in the effect of self-interactions of ULDM on various observations \cite{Foidl_2022, Mocz_2023, Winch_2023, Budker_2023, Ravanal_2023, Indjin_2023, Kadota_2024, Indjin_2025, Lee_2025, GalazoGarcia_2025}.

\section{Future prospects}

Here we briefly discuss some future prospects that can be explored given our work in this thesis.

\begin{itemize}
    \item The exploration in this thesis raises a number of interesting questions. For instance, broadly, what other constraints on $m$ will be altered if the effects of self-interactions are taken into account? 
    For instance, the very stringent bounds on Ultra-Light Axion (ULA) mass $m > 2\times 10^{-20}\ \text{eV}$ \cite{Irsic_2017, Rogers_2021} can be evaded if one allows for a larger initial axion field value, which will necessarily require one to consider the full cosine potential \cite{Desjacques_2018, Leong_2019, Winch_2023}, with all its higher order terms. 

    \item Similarly, it is worth asking how the constraints ($m > 3\times 10^{-19}\ \text{eV}$) in \cite{Dalal_2022}, - obtained by comparing observed velocity dispersion and the half-light radius of Segue 1 and Segue 2 ultra-faint dwarfs to those expected in the FDM paradigm from stellar heating - are altered in the presence of self-interactions. 
    While the authors claim that this effect is dependent solely on the deBroglie wavelength, i.e. mass of the FDM particle, it is worth investigating how much of an effect self-interactions will have on the final constraints.
    Another interesting question to ask here is what fraction of the total DM can ULDM with self-interactions occupy compared to ULDM with no self-interactions.

    \item Further, a recent paper \cite{Zimmermann_2024} argues that the observed density profile of the Milky Way satellite galaxy Leo II requires $m > 2.2\times 10^{-21}\ \text{eV}$ (the argument relies on the reconstruction of the observed density profile using a superposition of ground and excited state solutions of a linearized Schrödinger-Poisson system).
    As we have seen in chapter~\ref{chpt:numerical_solutions} the ground state solution of the GPP system features a smaller and a centrally denser core in the presence of attractive self-interactions. 
    It would then be interesting to figure out if and by how much the presence of a $\lambda < 0$ self-coupling alters this constraint as well. 

    \item Finally, an important point to note is that the value of $\lambda$ required to alter the constraints for the observations in chapters~\ref{chpt:paper_2} and~\ref{chpt:paper_3}, is a few orders of magnitude higher than the fiducial value $\lambda \sim 10^{-96}$ expected for ULAs \cite{Hui_2017} if the density fraction of axions has to satisfy $\Omega_{a} = \Omega_{DM} \sim 0.1$. 
    It would be interesting to explore how this enhancement of $\lambda$ can be achieved for ULA, while satisfying the current relic abundance constraints (see discussion in appendix~E of \cite{Fox_2023}).

    \item In chapter~\ref{chpt:paper_4}, we have not included the self-interactions to maintain numerical ease while generating training data for the neural network. 
    The obvious next step is to include the dimensionless self-interaction strength $\hat{\lambda}_\text{ini}$ as a free parameter. 
    Since $\hat{\lambda}_\text{ini}$ is discreet, one has to modify the neural network architecture to predict one discreet parameter and 5 continuous ones. 
    Additionally, the MCMC sampling algorithm itself needs to be altered, since the standard Metropolis Hastings or even the ensemble sampling algorithm employed by the python package `emcee' \cite{Foreman_Mackey_2013} do not handle discreet parameters by default. 
    As we have mentioned at the end of chapter~\ref{chpt:paper_4}, since we train a different neural network for each galaxy in our sample, the values of the fundamental parameter $m$ that we obtain are different. While it possible to conduct a joint analysis of all galaxies using the Bayesian approach such that one obtains a single best- fit value of ULDM mass $m$ across all galaxies, it would be interesting to explore whether that can be done for the case of neural networks as well.

    \item In this work, we have only considered a single scalar field with self-interactions. 
    However, string theory predicts the existence of multiple light scalar fields~\cite{Arvanitaki_2010}.
    It is then interesting to consider a dark matter model consisting of multiple scalar fields. 
    Such a model can potentially solve the tension arising from the ULDM description of different astrophysical observations, like dwarf spheroidals and ultrafaint dwarfs~\cite{Pozo_2023}, or to describe rotation curves of galaxies in the SPARC catalogue~\cite{Street_2022}. 
    There has been a growing interest in studying the formation of solitons in such a scenario as well~\cite{Luu_2018, Glennon_2023, Jain_2023, Luu_2023, Mirasola_2024}. 
    It will be interesting to explore this avenue further by including the effect of self-interactions~\cite{Glennon_2023, Mirasola_2024} as well as confronting various observations.
\end{itemize}


Efforts of the scientific community in the attempt to resolve the mystery of dark matter have only multiplied over the last few decades, as newer models and detectors have been built as well as novel astrophysical and cosmological probes have been explored.
As we exclude more and more of the parameter space of the WIMP~\cite{Billard_2021} class of dark matter models, several alternatives, like the one considered in this thesis, have now started garnering a lot of interest.
Particularly, for the case of axions, axion-like particles, or ultra light scalars, new detection strategies have been used or developed to constrain the parameter space further in the ultra light regime~\cite{Xia_2023, Kaplan_2022, Arakawa_2023, Filzinger_2023, Kobayashi_2022, Afzal_2023}.  
We hope that our work here, highlighting the important role of self-interactions of ULDM, motivates further study of the observational consequences of a self-interacting ultra light dark matter candidate.

\newpage
\phantomsection

\thispagestyle{publicationpage}

\addcontentsline{toc}{chapter}{List of Publications}
\begin{center}\textbf{\huge List of Publications}\end{center}
\justifying
\begin{itemize}
    \item \textbf{\large Published}:
\begin{enumerate}
    \item S.~Chakrabarti, \textbf{B.~Dave}, K.~Dutta and G.~Goswami, \emph{Constraints on the mass and self-coupling of ultra-light scalar field dark matter using observational limits on galactic central mass}, \href{https://doi.org/10.1088/1475-7516/2022/09/074}{\emph{J. Cosmol. Astropart. Phys.} \textbf{09} (2022) 074} [\href{https://arxiv.org/abs/2202.11081}{arXiv:2202.11081 [astro-ph.CO]}]
    \item \textbf{B.~Dave} and G.~Goswami, \emph{Self-interactions of ULDM to the rescue?}, \href{https://doi.org/10.1088/1475-7516/2023/07/015}{\emph{J. Cosmol. Astropart. Phys.} \textbf{07} (2023) 015} [\href{https://arxiv.org/abs/2304.04463}{arXiv:2304.04463 [astro-ph.CO]}]
    \item \textbf{B.~Dave} and G.~Goswami, \emph{ULDM self-interactions, tidal effects and tunnelling out of satellite galaxies}, \href{https://doi.org/10.1088/1475-7516/2024/02/044}{\emph{J. Cosmol. Astropart. Phys.} \textbf{02} (2024) 044} [\href{https://arxiv.org/abs/2310.19664}{arXiv:2310.19664 [astro-ph.CO]}]
\end{enumerate}

    \item \textbf{\large Under Review}:

    \begin{enumerate}
        \item \textbf{B.~Dave} and G.~Goswami, \emph{Learning from galactic rotation curves: a neural network approach} [\href{https://arxiv.org/abs/2412.03547}{arXiv:2412.03547 [astro-ph.CO]}]\\(Under review) 
    \end{enumerate}

\end{itemize}

\cleardoublepage 
\phantomsection  
\renewcommand*{\bibname}{References}

\addcontentsline{toc}{chapter}{\textbf{References}}

\printbibliography


\appendix
\chapter*{APPENDICES}
\addcontentsline{toc}{chapter}{APPENDICES}
\chapter{Details of scaling, and shapes of the curves in chapter~\ref{chpt:paper_1}}\label{AppendixA}

\section{Empirical correlations}
\label{sec:scaling}
For any given value of mass of halo $M_{\text{halo}}$, there exist empirical relations such as the soliton mass $-$ halo mass relation (see \cite{Schive_PRL_2014} and the discussion in section 3.1 of \cite{Davies_2020}), which can be used to obtain the observed value of dimensionless soliton mass ${\hat M}_{\text{emp}}$ from the halo mass $M_{\text{halo}}$

 \begin{equation} \label{eq:memp}
 \hat{M}_\text{emp} = 5.45\times 10^{-3} \left(\frac{M_\text{halo}}{2\times 10^{14}\ M_\odot}\right)^{1/3} \; .
\end{equation}

It is important to note that, as pointed out in ref.~\cite{Bar_2019}, the soliton mass - halo mass relation was obtained from simulations \cite{Schive_Nature_2014, Schive_PRL_2014} in the absence of any SMBHs. Furthermore, the mass range of halos for which the simulations are carried out is $M_\text{halo} \sim (10^9 - 5\times 10^{11})\ M_\odot$ for $m \approx 10^{-22}\ \text{eV}$. Thus, for a galaxy with SMBH at the centre such that $M_\text{halo} \gtrsim 10^{12}\ M_\odot$ and for dark matter mass $m$ values much different from $10^{-22}\ \text{eV}$, strictly speaking, this relation is not applicable. Despite this, we proceed with the use of this relation in the rest of the work since our main focus is only on illustrating the method we present. 

Similarly, there exists an empirical relation which relates halo mass $M_{\text{halo}}$ to the mass of the central BH $M_{\bullet}$ (see \cite{Bandara_2009} and the discussion in section 3.2 of \cite{Davies_2020}). Using eq.~(\ref{eq:alpha}), one can find the corresponding observed value of the parameter $\hat \alpha$ i.e. ${\hat \alpha}_{\text{emp}} (m, M_{\text{halo}})$:

 \begin{equation} \label{eq:alphaemp}
\hat{\alpha}_\text{emp} = 1.18\times 10^{-2}\left(\frac{M_\text{halo}}{2\times 10^{14} \ M_\odot}\right)^{1.55}\left(\frac{m}{10^{-22}\ \text{eV}}\right) \; .
\end{equation}

In summary, given the halo mass $M_{\text{halo}}$ one can use such correlations to determine the observed dimensionless soliton mass ${\hat M}_{\text{emp}}$ using eq.~(\ref{eq:memp}). Similarly, given the halo mass $M_{\text{halo}}$ and the ULDM mass $m$ of interest, one can determine the observed value of the quantity ${\hat \alpha}$ from eq.~(\ref{eq:alphaemp}).

In some cases e.g. for the Milky Way galaxy or M87 galaxy, both $M_\bullet$ and $M_{\text{halo}}$  are well measured and hence one can use this measured value of $M_\bullet$ (along with $m$ of interest) to determine ${\hat \alpha}_{\text{emp}}$ from eq.~(\ref{eq:alpha}).

\section{Determination of the scaling parameter ``$s$"}
\label{sec:parameter}
 
One can use the following strategy to find the amount of scaling: 

\begin{itemize}
 \item If a particular object such as a galactic halo is to be used to constrain the parameters of ULDM, its own mass, $M_{\text{halo}}$, is already fixed. In order to proceed, let us also fix the mass of ULDM particle i.e. $m$.
 \item Start with some initial value of ${\hat \lambda}$ called ${\hat \lambda}_{\text{ini}}$ (note that this value will correspond to some value of self-coupling $\lambda$ as defined by eq.~(\ref{eq:dimensions2}) in terms of $m$).
 \item Determine ${\hat M}_{\text{emp}}$ using eq.~(\ref{eq:memp}).
 Similarly, determine ${\hat \alpha}_{\text{emp}}$ from either eq.~(\ref{eq:alphaemp}), or, if the mass of central black hole is already known, directly using eq.~(\ref{eq:alpha}).
 \item Choose an initial value for $\hat \alpha$, call this ${\hat \alpha}_{\text{ini}}$ and solve eqs.~(\ref{eq:GP_dimless_SMBH}) and (\ref{eq:P_dimless_SMBH}) to find the corresponding ${\hat M}$ and call it ${\hat M}_{\text{ini}}$.
 \item Find the value of scaling parameter such that the mass of the soliton after scaling is equal to the empirical mass of the soliton (using eq.~(\ref{eq:scaling_mass})) i.e. find $s$ such that
\begin{equation}
s = \frac{{\hat M}_{\text{ini}}}{{\hat M}_{\text{emp}}} \; .
\end{equation}
 \item This scaling should also transform $\hat \alpha$, 
 so, transform $\hat \alpha$ using this scaling parameter, 
 i.e. find ${\hat \alpha}_{\text{trans}} = s^{-1} {\hat \alpha}_{\text{ini}}$. 
We then ask ourselves: is ${\hat \alpha}_{\text{trans}} = {\hat \alpha}_{\text{emp}}$? 
In practice, we have to ask whether they are equal to some desired accuracy i.e. we have to ask whether the following is true:
\begin{equation}
 \left[ \frac{ {\hat \alpha}_{\text{trans}} - {\hat \alpha}_{\text{emp}} }{ {\hat \alpha}_{\text{emp}} } \right] < 0.05 \; ,
\end{equation}
(assuming $5 \%$ accuracy).
\item If the answer is `yes', we have found the correct ${\hat \alpha}_{\text{ini}}$ and the corresponding $s$ is the desired scaling parameter. On the other hand, if it is `no', we try another value of ${\hat \alpha}_{\text{ini}}$. 
\item Once the desired ``$s$" is obtained, the value of self-coupling probed using the object of mass $M_{\text{halo}}$, and for ULDM of mass $m$, is given by (eq.~(\ref{eq:scaling_lambda}))  
\begin{equation}\label{eq:lambda}
 {\hat \lambda} = s^2 {\hat \lambda}_{\text{ini}} \; ,
\end{equation}
which can be used to determine the corresponding value of probed $\lambda$ using eq.~(\ref{eq:dimensions2}).
\end{itemize}
By following this procedure, given $M_{\text{halo}}$, $m$ and ${\hat \lambda}_{\text{ini}}$, we can find the initial $\hat \alpha$ i.e. ${\hat \alpha}_{\text{ini}}$ which we must start with, such that, after scaling by an amount $s$, the dimensionless soliton mass is 
${\hat M}_{\text{emp}}$ and after scaling, the $\hat \alpha$ parameter takes up its desired empirical value ${\hat \alpha}_{\text{emp}}$.
Thus, $m$, ${\hat \lambda}_{\text{ini}}$ and $M_{\text{halo}}$ are the final free variables.

This explains how the various curves in figure~\ref{fig:densityprofile} have been obtained. The halo mass is fixed to $M_{\text{halo}} = 2 \times 10^{14} M_{\odot}$ while the ULDM mass is $m = 10^{-22} ~ {\rm eV}$. Different curves correspond to various values of ${\hat \lambda}_{\text{ini}}$ chosen which imply different values of the self coupling $\lambda$. Note that each curve corresponds to a different value of the scaling parameter ``$s$". 
 In particular, as one can see from figure~\ref{s_m} in appendix~\ref{sec:m&lambda}, the value of $s$ for ${\hat \lambda}_{\text{ini}} = +1$ is $s_+ = 181.5$, while the value of $s$ for ${\hat \lambda}_{\text{ini}} = -1$ is $s_- = 66.6$. The square of the ratio of these numbers gives the ratio of $\lambda$ corresponding to ${\hat \lambda}_{\text{ini}} = +1$ and ${\hat \lambda}_{\text{ini}} = -1$ (which is mentioned in the caption of figure~\ref{fig:densityprofile}). The reasoning in appendix~\ref{sec:m&lambda} explains why one expects $s_+$ to be always larger than $s_-$.

\section{The shapes of the curves in figure~\ref{M87WithBH}}
\label{sec:Shape1}

In this section, we try to understand why the shapes of the various curves in figure \ref{M87WithBH} are what they are.

Let us first think about the case of a theoretical soliton without self-interactions and without a central black hole. It is worth recalling that the size of theoretical soliton is $R \sim {\cal O}(m^{-1})$ and its mass is $M \sim {\cal O}(M_{pl}^2\cdot  m^{-1})$. Thus, as we increase $m$, the soliton gets smaller and lighter. Let us start with DM of extremely small mass, the corresponding soliton shall be very big (and very heavy) and only a very small fraction of the soliton will be inside a sphere of radius $r_*$ of the centre of the soliton. Within this small central region of the soliton, if we assume that its density is constant, $\rho_0$, the mass within $r_*$ will be $M_{r<r_*} \sim \rho_0 r_*^3$. Now, since $\rho_0 \sim M R^{-3}$, the density itself goes as $m^{2}$, so, $M_{r<r_*} \sim m^2 r_*^3$. This is the reason why the left part of a plot of $\ln M_{r<r_*}$ against $\ln m$ will appear to be a straight line with slope 2. As we increase $m$, the soliton keeps getting smaller, at some point, the entire soliton is inside $r_*$, so, $M_{r<r_*}$ becomes equal to $M$ which we know goes as $m^{-1}$ - this explains why the right part of a plot of $\ln M_{r<r_*}$ against $\ln m$ will be a straight line with slope -1. 

When we need to go from theoretical soliton to real soliton, we need to use the scaling parameter. It turns out that in the absence of a central black hole and in the absence of scalar self-interactions, the scaling parameter $s$ itself does not depend on $m$. Thus, the argument in the last paragraph is applicable to real solitons also. In the presence of central black hole or self-interactions (or both), $s$ will be $m$ dependent and strictly speaking, the arguments of the last paragraph will not apply.
The presence of central black hole squeezes the soliton. In addition, the presence of attractive or repulsive self-interactions can further compress or expand it. Finally, the density of the soliton is not exactly the same at all radii. These additional effects slightly modify the shapes of the curves in figure \ref{M87WithBH} but the basic explanation remains the same.

Now let us understand the distinction between the curve corresponding to ${\hat \lambda} = 0$, the solid curve labelled ``0" in figure \ref{M87WithBH}, and the various curves corresponding to the presence of interactions, ${\hat \lambda} \neq 0$. 
In the rightmost part of the figure~\ref{M87WithBH}, irrespective of the sign of the interactions or the presence of a central black hole, the behaviour is identical to the behaviour in ${\hat \lambda} = 0$ case. This is because, as we saw above, the rightmost part of figure~\ref{M87WithBH} corresponds to large $m$ i.e. large ${\hat r}_*$ for which the condition ${\hat r}_* < s {\hat r}_{\text{sol}} $ is satisfied and the entire soliton is inside the region of interest. Thus, the integral in eq.~(\ref{eq:M_full}) will be the same for all the cases as the upper limit of the integral can be replaced by $\infty$.

To understand the left part of figure~\ref{M87WithBH} in the presence of interactions, first recall that ${\hat \lambda} > 0$ corresponds to repulsive self interactions while ${\hat \lambda} < 0$ corresponds to attractive self-interactions.
For ${\hat \lambda} > 0$, the soliton is bigger and hence for every value of $m$ to the left of the peak, less mass will be enclosed within the region of interest compared to the case of no interactions. This causes the corresponding curve to be lower than the curve for ${\hat \lambda} = 0$ case. 
On the other hand, for ${\hat \lambda} < 0$ or in the presence of a central black hole, the soliton will be smaller and hence for every value of $m$ to the left of the peak, more mass will be enclosed within the region of interest compared to the case of no interactions and this will cause the corresponding curve to be higher than the curve for ${\hat \lambda} = 0$ case. 

It is useful to understand in a different way how this works out. For the unscaled (i.e. theoretical soliton), the integrand ${\hat \phi}^2 {\hat r}^2$ in eq.~(\ref{eq:mass_dimless}) can be plotted against ${\hat r}$.  One finds that the curve corresponding to ${\hat \lambda} > 0$ lies above the curve corresponding to ${\hat \lambda} < 0$. The dimensionless soliton mass ${\hat M}$ will be the area under this curve. Thus, we expect ${\hat M}_{\text{ini}}^+ > {\hat M}_{\text{ini}}^-$, i.e. the dimensionless soliton mass for unscaled soliton for the case of repulsive self-interactions is greater than that in the case of attractive self-interactions. This particular point can also be understood from figure~\ref{Mhatinivsalphahatini} and the discussion in section~\ref{sec:m&lambda} below.
Consequently, eq.~(\ref{eq:s}) implies that $s_+ > s_-$ i.e. the scaling parameter for ${\hat \lambda} > 0$ case is more than that in ${\hat \lambda} < 0$ case. This will cause the integrand in eq.~(\ref{eq:M_limited}) for the two cases to scale as $1 / s_+^2$ and $1 / s_-^2$ i.e. the integrand for ${\hat \lambda} > 0$ case will decrease more than the integrand for the ${\hat \lambda} < 0$ case. This causes the curves for repulsive interactions in figure~\ref{M87WithBH} to be below the curves for attractive interactions.

\section{Probed $\lambda$ for a chosen $m$ and for a given object}
\label{sec:m&lambda}

It is very useful to understand the shapes of the curves and excluded regions in $\lambda-m$ plane using the formalism we have developed. Similarly, one might wonder, for a given object (fixed $M_{\text{halo}}$) and a chosen range of $m$ values, what typical values of $\lambda$ one can probe or test using the method presented here. In particular, one might wonder whether there are any regions in the $\lambda-m$ plane which can not be probed using the present method.
In this section, we clarify all of these issues.

\begin{figure}[ht]
  \includegraphics[width = 0.9\textwidth]{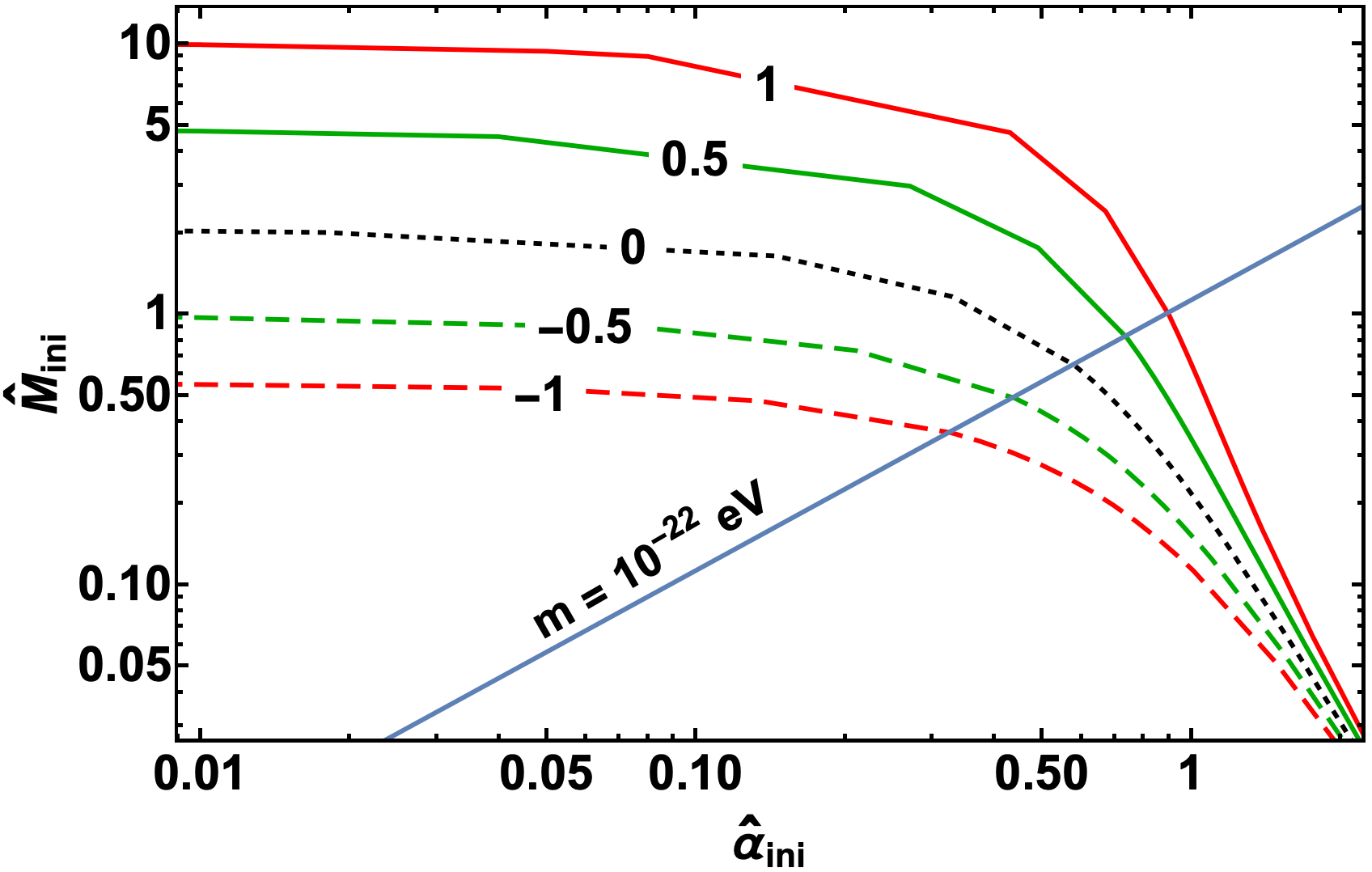}
  \caption[The dependence of $\ln {\hat M}_{\text{ini}}$ on $\ln {\hat \alpha}_{\text{ini}}$ for various choices of ${\hat \lambda}_{\text{ini}}$]{The dependence of $\ln {\hat M}_{\text{ini}}$ (unscaled dimensionless soliton mass as defined in eq.~(\ref{eq:mass_dimless})) on $\ln {\hat \alpha}_{\text{ini}}$ (unscaled dimensionless black hole `strength' as defined in eq.~(\ref{eq:alpha})) for various choices of ${\hat \lambda}_{\text{ini}}$ for M87 galaxy. From this figure, it is easy to see that ${\hat M}_{\text{ini}}^+ > {\hat M}_{\text{ini}}^-$. Given this, eq.~(\ref{eq:s}) implies that $s_+ > s_-$, thus, for a given value of $|{\hat \lambda}_{\text{ini}}|$, the probed $\lambda$ obtained from eq.~(\ref{eq:lambda0}) for attractive self-interactions will be more than the probed $\lambda$ for negative interactions. The straight line is the one obtained from the right hand side of eq \ref{eq:LHS=RHS}, its slope is 1 and intercept is given by eq (\ref{eq:obs}).
}
  \label{Mhatinivsalphahatini}
\end{figure}

Once the values of $\hat \alpha$ and ${\hat \lambda}$ in eqs.~(\ref{eq:GP_dimless_SMBH}) and (\ref{eq:P_dimless_SMBH}) are known, let us say ${\hat \alpha} = {\hat \alpha}_{\text{ini}}$ and ${\hat \lambda} = {\hat \lambda}_{\text{ini}}$, we can solve for ${\hat \phi}({\hat r})$ and find the dimensionless mass of the soliton i.e. ${\hat M} = {\hat M}_{\text{ini}}$. In figure~\ref{Mhatinivsalphahatini}, we show the dependence of ${\hat M}_{\text{ini}}$ on ${\hat \alpha}_{\text{ini}}$ for a few possible values of ${\hat \lambda}_{\text{ini}}$. This plot will help us understand many issues in this section. 

From eq.~(\ref{eq:dimensions2}) and eq.~(\ref{eq:lambda}), we can conclude that, for a given object (i.e. fixed $M_{\text{halo}}$), the probed self coupling value $\lambda$ depends on $m$, ${\hat \lambda}_{\text{ini}}$ and the value of scaling parameter $s$ since

\begin{equation}\label{eq:lambda0}
 \lambda = 8 s^2 {\hat \lambda}_{\text{ini}} \left( \frac{m}{M_{pl}} \right)^2 \; .
\end{equation}
If $s$ were a constant, for a given ${\hat \lambda}_{\text{ini}}$ (and a given object i.e. given $M_{\text{halo}}$), the plot of $\ln \lambda$ against $\ln m$ shall be a straight line with slope 2. In reality, as we shall now argue, the scaling parameter $s$ itself depends on all the three independent variables $M_{\text{halo}}$, $m$ and ${\hat \lambda}_{\text{ini}}$.

We saw in section~\ref{sec:parameter} that, in order to turn an unrealistic soliton into a realistic model of the core of DM halo, we must satisfy the requirement, eq.~(\ref{eq:s}),
\begin{equation}\label{eq:LHS=RHS}
 {\hat M}_{\text{ini}} ({\hat \alpha}_{\text{ini}} , {\hat \lambda}_{\text{ini}}) = \left( \frac{ {\hat M}_{\text{emp}} }{{\hat \alpha}_{\text{emp}}} \right) {\hat \alpha}_{\text{ini}} \; ,
\end{equation}
where, because of eqs.~(\ref{eq:memp}) and (\ref{eq:alphaemp}), the ratio in the brackets on the RHS goes as $M_{\text{halo}}^{-1.22} \cdot m^{-1}$.
The LHS of this equation is what is shown as the family of curves in figure \ref{Mhatinivsalphahatini}. 
The right hand side of this equation suggests that, in figure~\ref{Mhatinivsalphahatini}, for a given value of $m$, the observationally consistent values of $\ln {\hat M}_{\text{ini}}$, as a function of $\ln {\hat \alpha}_{\text{ini}}$, should be straight lines with slope 1 and intercept which is roughly given by (using eqs.~(\ref{eq:memp}) and (\ref{eq:alphaemp})),

\begin{equation}\label{eq:obs}
\text{Intercept} = \ln\left[0.47\left(\frac{M_\text{halo}}{2\times 10^{14} \ M_\odot}\right)^{-1.22}\left(\frac{m}{10^{-22}\ \text{eV}}\right)^{-1}\right]
\end{equation}
For M87, for DM of mass $m = 10^{-22}~{\rm eV}$, the corresponding line is shown in figure~\ref{Mhatinivsalphahatini}.
Thus, for a chosen value of ${\hat \lambda}_{\text{ini}}$, the correct value of ${\hat M} = {\hat M}_{\text{ini}}$ and ${\hat \alpha}_{\text{ini}}$ will be the one at the intersection of the curves in figure~\ref{Mhatinivsalphahatini} and straight lines with slope 1 and intercept given by eq.~(\ref{eq:obs}).

\subsection{The shapes of curves in figures~\ref{Excluded_-} and \ref{Excluded_+}} 

\begin{figure}[ht]
  \includegraphics[width = 0.95\textwidth]{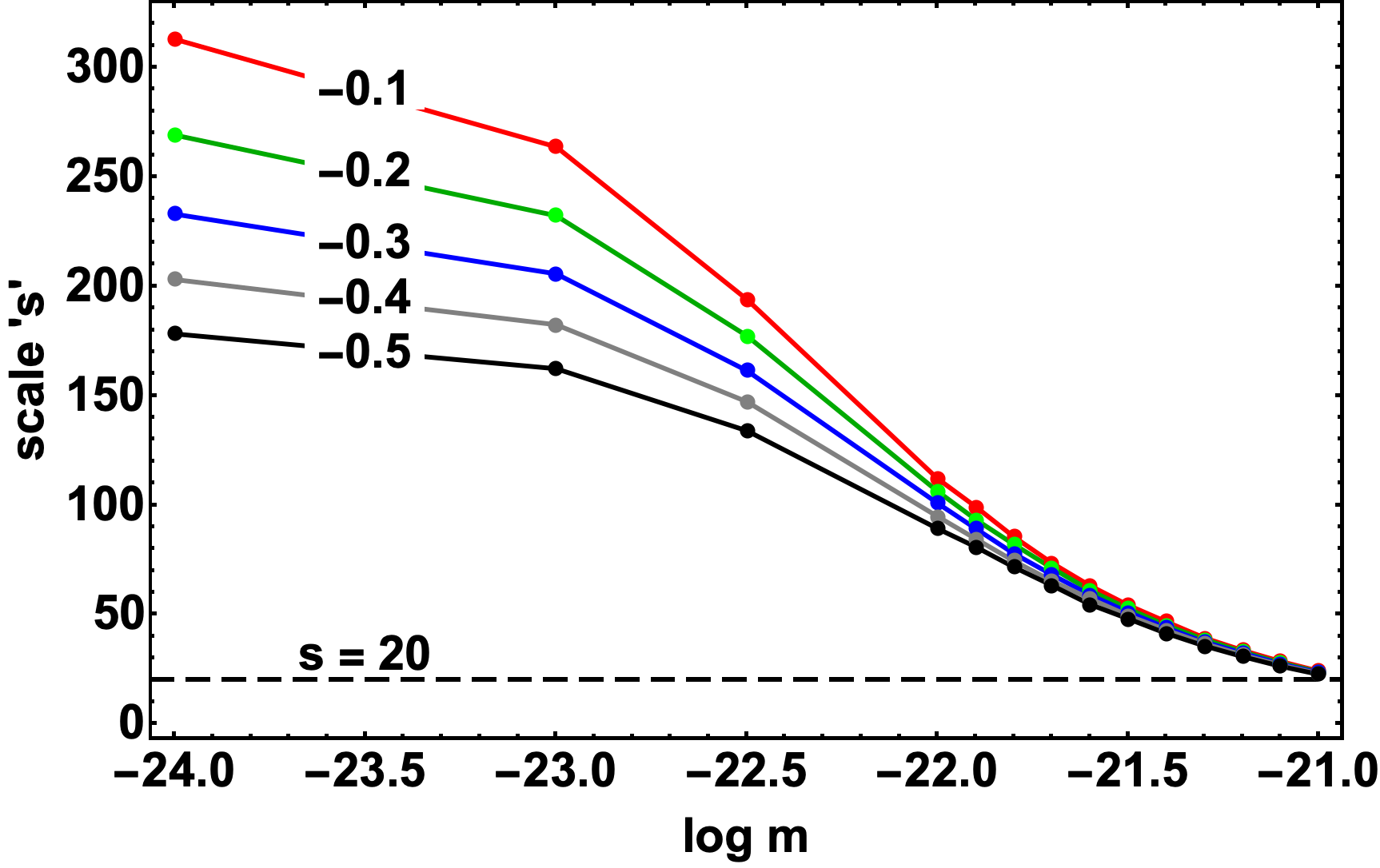}
  \caption[Scaling parameter $s$ against $m$ for various values of ${\hat \lambda}_{\text{ini}}$]{The plot of scaling parameter $s$ against $m$ for various values of ${\hat \lambda}_{\text{ini}}$ for M87 galaxy. As long $s$ is sufficiently large compared to 1, the formalism in this paper is trustworthy.
}
  \label{s_m}
\end{figure}

As we increase $m$, eq~(\ref{eq:obs}) informs us that the intercept of the line representing the RHS of eq.~(\ref{eq:LHS=RHS}) decreases, this means that in figure~\ref{Mhatinivsalphahatini} we end up exploring the rightmost regions of the plot - the corresponding ${\hat \alpha}_{\text{ini}}$ is larger and ${\hat M}_{\text{ini}}$ is smaller. Since $M_{\text{halo}}$ is fixed, ${\hat M}_{\text{emp}}$ remains the same, eq.~(\ref{eq:s}) then implies that $s$ must decrease. This can be verified from the curves in figure~\ref{s_m}.

For very small values of $m$, the intercept of the line (using eq.~(\ref{eq:obs})) becomes too large and the line intersects the curves in figure~\ref{Mhatinivsalphahatini} in the leftmost, flatter part. Under this circumstance, if we increase $m$ slightly, the intercept decreases and the line shifts only slightly downwards, causing the intersection point of the line and the curve to be shifted only slightly to the right. Since we are in the leftmost, flatter part of the curves in figure~\ref{Mhatinivsalphahatini}, this means that the corresponding value of ${\hat M}_{\text{ini}}$ doesn't decrease appreciably. Thus, from eq.~(\ref{eq:s}), we find that $s$ shall not change much - something easy to verify from figure~\ref{s_m}. 

Thus, the quantity $\partial s / \partial m$ has less negative values for smaller values of $m$ and more negative values for larger values of $m$. Now, using eq.~(\ref{eq:lambda0}), one can easily show that
\begin{equation}
 \frac{\partial \lambda}{\partial m} = 16 {\hat \lambda}_{\text{ini}} \left[  \left( \frac{m s^2 }{M_{pl}^2} \right) + s  \left( \frac{\partial s}{\partial m} \right) \left( \frac{m}{M_{pl}} \right)^2  \right] \; ,
\end{equation}
i.e.
\begin{equation}
 \frac{\partial \ln | \lambda| }{\partial \ln m} = 2 \left[  1 +  \frac{\partial \ln s}{\partial \ln m} \right] \; .
\end{equation}
Thus, the plot of $\ln | \lambda| $ against $\ln m$ shall be a straight line with slope 2 for smaller values of $m$ while the slope will decrease from 2 for larger values of $m$. This explains the basic shape of the curves in figures~\ref{Excluded_-} and \ref{Excluded_+}.

\subsection{Change in probed $\lambda$ as ${\hat \lambda}_{\text{ini}}$ changes (for fixed $M_{\text{halo}}$ and $m$)}

Let us first focus our attention on the case with ${\hat \lambda}_{\text{ini}} < 0$ i.e. attractive self-interactions.

\begin{figure}[ht]
  \includegraphics[width = 0.95\textwidth]{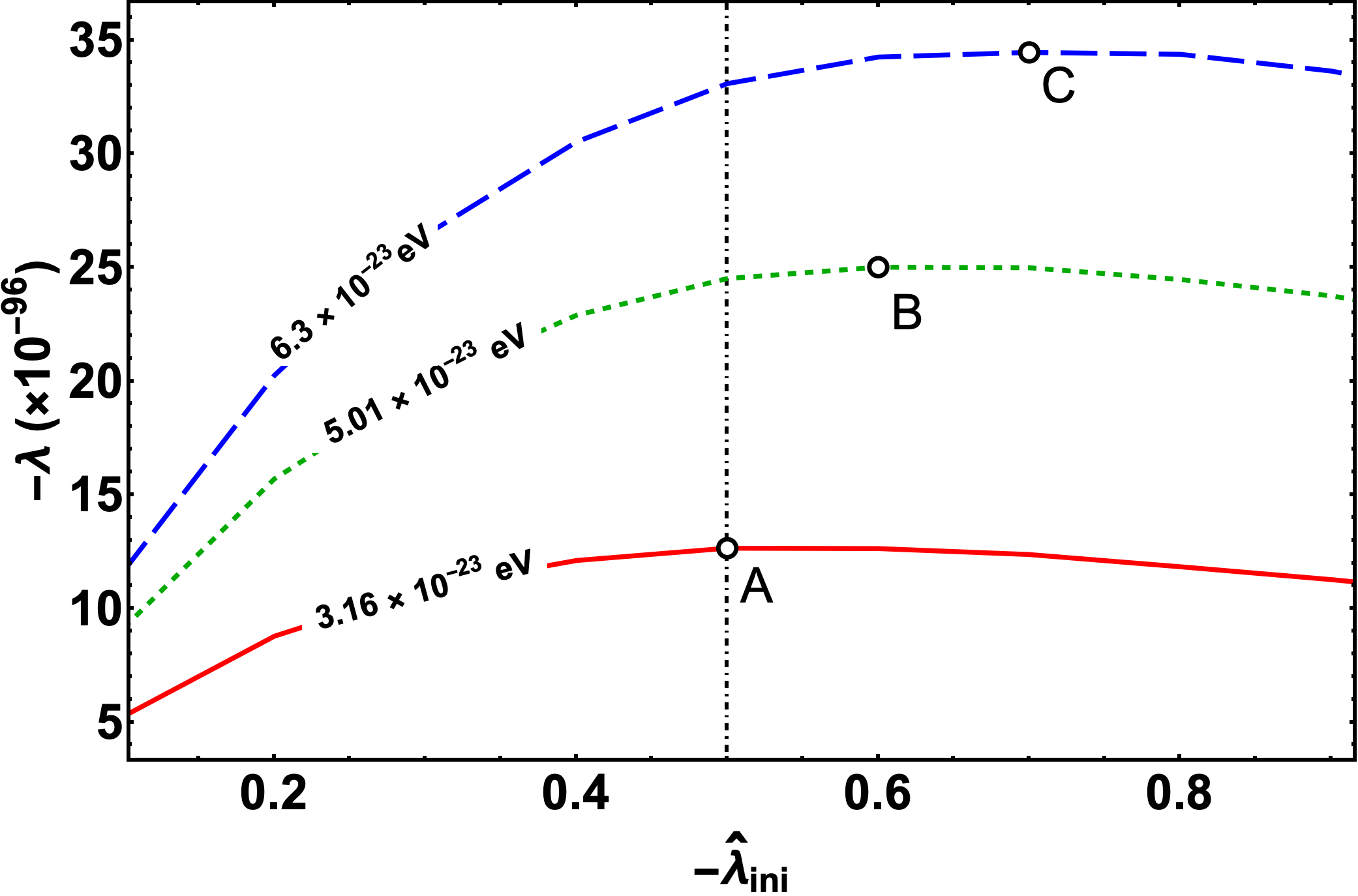}
  \caption[The plot of $\lambda$ against $\hat{\lambda}_{\text{ini}}$ for three different values of $m$]{The plot of $\lambda$ against $\hat{\lambda}_{\text{ini}}$ for three different values of $m$ as marked on each curve. Points A, B and C correspond to the local maximum for each curve. The vertical dot-dashed line corresponds to $\hat{\lambda}_{\text{ini}} = -0.5$.}
  \label{lambda0-lambdac}
\end{figure}

When ${\hat \lambda}_{\text{ini}}$ becomes more negative, from figure~\ref{Mhatinivsalphahatini}, this leads to a lower value of ${\hat M}_{\text{ini}}$, thus, eq.~(\ref{eq:s}) then implies that $s$ must decrease - this is true for all values of $m$ but the decrease in $s$ is more for smaller $m$ and less for larger $m$. 
Thus, for negative self-interactions, in eq (\ref{eq:lambda0}), the function $s^2$, which depends on ${\hat \lambda}_{\text{ini}}$ falls very quickly at smaller $m$ and less quickly at larger $m$.
Since eq.~(\ref{eq:lambda0}) involves a product of a quickly decreasing function $s^2 ({\hat \lambda}_{\text{ini}})$ and an increasing function $|{\hat \lambda}_{\text{ini}}|$, this implies that, as we increase $| {\hat \lambda}_{\text{ini}} |$, the probed values of $\ln | \lambda| $ will first increase and then decrease. This means that, for a given $m$, curves corresponding to ever increasing values of $| {\hat \lambda}_{\text{ini}} |$ will have a $\ln | \lambda| $ value which will first increase and then decrease, this can be seen from figure~\ref{lambda0-lambdac}.  

Thus, for a given object (i.e. fixed $M_{\text{halo}}$), $\lambda$ depends on $m$ and ${\hat \lambda}_{\text{ini}}$. The dependence of $\lambda$ on ${\hat \lambda}_{\text{ini}}$ for different values of $m$ is shown in figure~\ref{lambda0-lambdac} and is completely expected given the above explanation.
What we have argued is that for any given object (such as M87 galaxy), for every choice of $m$, there is always a local maximum in the plot of probed $\lambda$ against ${\hat \lambda}_{\text{ini}}$. If the value of $\lambda$ corresponding to the local maximum is called $\lambda_{\rm max}$, we can find such a value for every $m$. If, for a given object with a fixed $M_{\text{halo}}$, one plots $\lambda_{\rm max}$ against $m$, in such a plot, for a given $m$, any value of $\lambda > \lambda_{\rm max}$ for that $m$ can not be probed for any choice of ${\hat \lambda}_{\text{ini}}$ by the method we have presented in this paper. This is the origin of the inaccessible region in $\lambda-m$ plane for attractive self-interactions. From figure~\ref{Mhatinivsalphahatini} and eq.~(\ref{eq:lambda0}), it is easy to see that this will not happen for positive self-interactions.

\subsection{Maximum soliton mass in the presence of a central black hole}\label{app:max_mass_smbh}

Another way to understand the existence of a maximum $\lambda$ is to extend the argument for the existence of a maximum soliton mass in the case of attractive self-interactions (as discussed in section~\ref{sec:max_mass}) to include a SMBH at the centre. 

\begin{figure}[ht]
    \centering
    \begin{subfigure}[b]{0.45\textwidth}
        \centering
        \includegraphics[width = \textwidth]{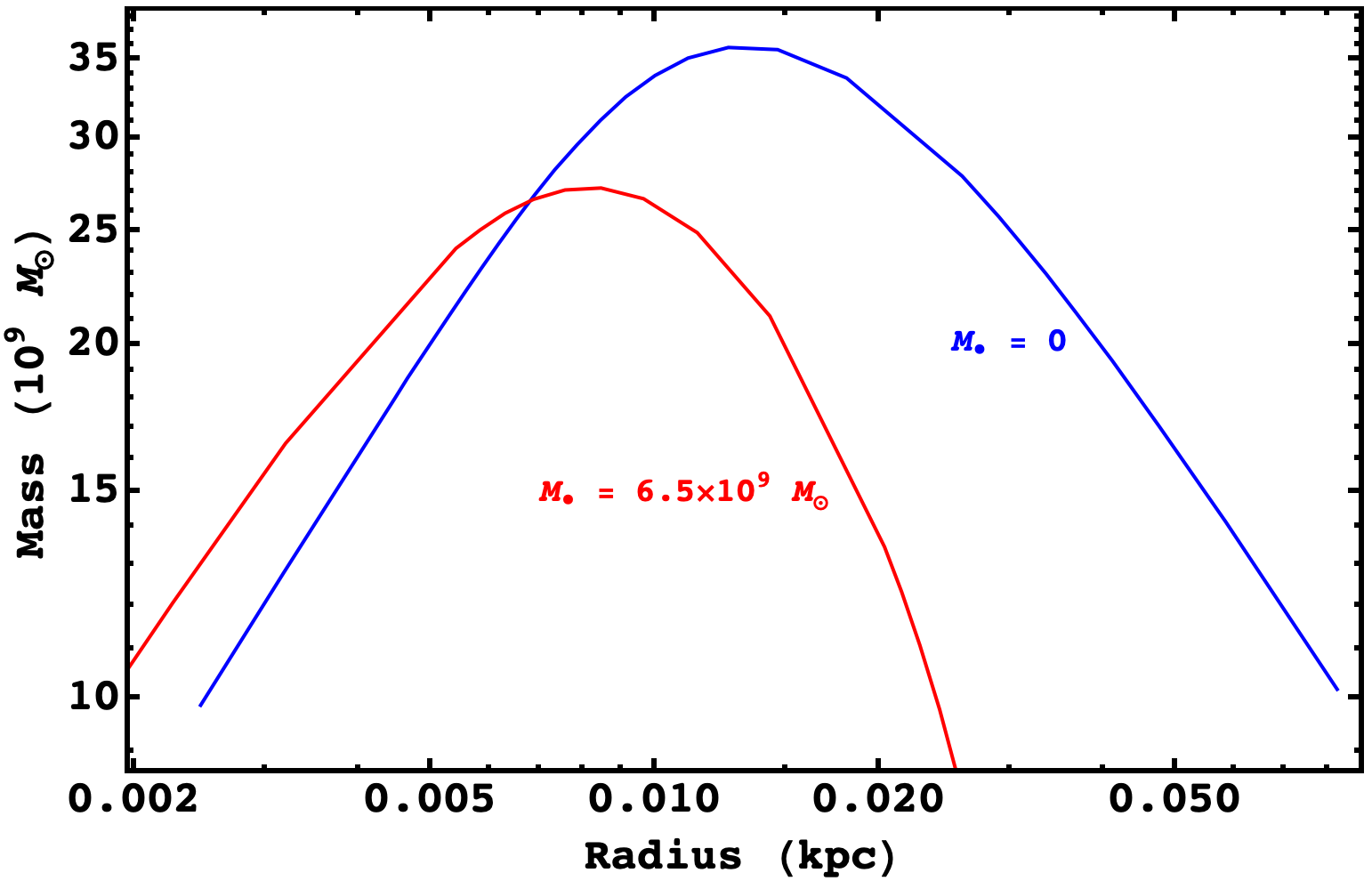}
        \caption{Effect of SMBH on maximum mass and the overall mass-radius curve for a fixed $m = 10^{-22}\ \text{eV}$ and $\lambda = 10^{-95}$.}
        \label{fig:mass_radius_smbh}
    \end{subfigure}
    \hfill
    \begin{subfigure}[b]{0.45\textwidth}
        \centering
        \includegraphics[width = \textwidth]{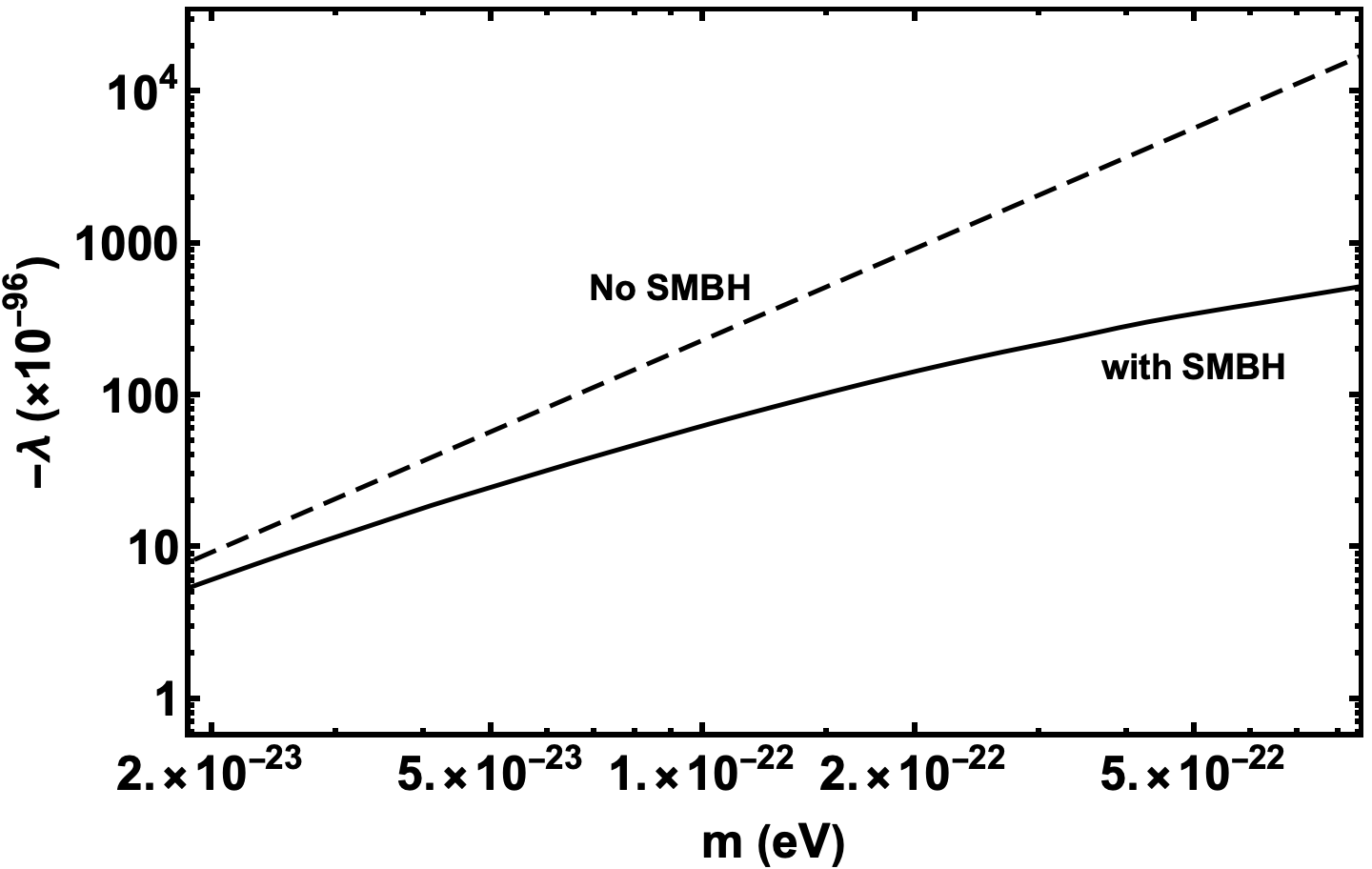}
        \caption{The maximum $\lambda$ corresponding to every $M_{\text{fin}}$ that satisfies eq.~(\ref{eq:Schive_SH}) are plotted with (solid) and without (dashed) SMBH.}
        \label{fig:lambda_m_SMBH}
    \end{subfigure}
    \caption[Impact of the M87* SMBH ($M_\bullet = 6.5\times 10^9\ M_\odot$) on maximum mass in the $M-R$ plane and $\lambda-m$]{Impact of the M87* SMBH ($M_\bullet = 6.5\times 10^9\ M_\odot$) on maximum mass in the $M-R$ plane and $\lambda-m$ is shown. Halo mass is taken to be $M_h = 2\times 10^{14}\ M_\odot$.}
    \label{fig:effect_of_SMBH}
\end{figure}

Similar to the discussion in section~\ref{sec:mass_radius_curves}, to determine maximum mass in the presence of attractive self-interactions, we first fix $\hat{\lambda}_{\text{fin}}$ to determine scale $s$. 
In this case since we have a SMBH with a fixed physical mass $M_\bullet$, we also fix some $\hat{\alpha}_{\text{fin}}$ (see eqs.~(\ref{eq:alpha}) and~(\ref{eq:alpha_scaling})) such that $\hat{\alpha}_{\text{ini}}/s = \hat{\alpha}_{\text{fin}}$ for every $\hat{\lambda}_{\text{ini}}$. We incorporate this in the following manner: 

\begin{enumerate}
    \item For a fixed $m$ and $M_\bullet$ find the desired $\hat{\alpha}_{\text{fin}}$.
    \item Fix $\hat{\lambda}_{\text{fin}}$ such that for every $\hat{\lambda}_{\text{ini}}$ one gets the required scale $s$ from eq.~(\ref{eq:lambda_scale}).
    \item Using scaling symmetry, find $\hat{\alpha}_{\text{ini}} = s\hat{\alpha}_{\text{fin}}$, for every $\hat{\lambda}_{\text{ini}}$.
    \item Solve the GPP system to obtain $\hat{M}_{\text{ini}}$ and $\hat{R}_{\text{ini}}$ for the given $\hat{\lambda}_{\text{ini}}$ and $\hat{\alpha}_{\text{ini}}$. 
\end{enumerate}
Because of the squeezing effect of $\hat{\alpha}_{\text{ini}}$ on the density profile, the corresponding $\hat{M}_{\text{ini}}$ and $\hat{R}_{\text{ini}}$ will be different compared to $\hat{\alpha}_{\text{ini}} = 0$. It is then expected that $M_{max}$ will be altered in the presence of a black hole. Using the same argument as in section~\ref{sec:max_mass} and eq.(\ref{eq:max_mass}) maximum mass for a fixed $\lambda_{\text{fin}}$ in the presence of a black hole can be determined by $(\hat{M}_{\text{ini}}|\hat{\lambda}_{\text{ini}}|^{1/2})_{max}$. As an example, we consider the M87 halo with $M_h = 2\times 10^{14}\ M_\odot$ and $M_\bullet = 6.5\times 10^9\ M_\odot$. For $m = 10^{-22}\ \text{ev}$ the desired scaled $\hat{\alpha}_{\text{fin}} = 0.0048$. We plot the mass-radius curve for both $\hat{\alpha} = 0$ (blue curve) and $\hat{\alpha}_{\text{fin}} = 0.0048$ (see red curve) in figure.~\ref{fig:mass_radius_smbh}. We find that $M_{max}$ is smaller in the presence of a black hole, which is in agreement with the results in \cite{Chavanis_2019}.

From the above discussion note that $\hat{\alpha}_{\text{ini}}$ is dependent on $m$ through $\hat{\alpha}_{\text{fin}}$. One can therefore get a corresponding curve in the $\lambda-m$ plane by fixing the scaled soliton mass $\hat{M}_{\text{fin}}$ instead of $\hat{\lambda}_{\text{fin}}$, setting the desired scale to be $s = \hat{M}_{\text{ini}}/\hat{M}_{\text{fin}}$ for every value of $\hat{\lambda}_{\text{ini}}$. Similar to eq.~(\ref{lambda_constraint}), it transforms eq.~(\ref{eq:max_mass}) into

\begin{equation}
    |\lambda_{max}| = 64\pi \left(\hat{M}_{\text{ini}}|\hat{\lambda}_{\text{ini}}|^{1/2}\right)_{max}^2\frac{m_{pl}^2}{M_{\text{fin}}^2}\ .
\end{equation}
Now for every soliton mass $M_{\text{fin}}$ there is a maximum allowed $\lambda_{max}$ such that $M_{\text{fin}} = M_{max}(\lambda_{max})$. In the presence of black hole, it is then clear that if $M_{\text{fin}}$ remains unchanged, $\lambda_{max}$ will be smaller. One can fix the scaled soliton mass by assuming that $M_{\text{fin}}$ must satisfy some soliton-halo relation (e.g. eq.~(\ref{eq:Schive_SH})) for a fixed $M_h$. The resultant curve in the $\log{\lambda}-\log{m}$ plane when $\hat{\alpha}_{\text{fin}} = 0$ is a straight line (dashed curve in figure~\ref{fig:lambda_m_SMBH}). However, $\hat{\alpha}_{\text{fin}} \neq 0$, for every $\hat{\lambda}_{\text{ini}}$ and $m$ one must find some $\hat{\alpha}_{\text{ini}}$ such that $\hat{\alpha}_{\text{fin}}$ as well as $\hat{M}_{\text{fin}}$ correspond to the desired $M_\bullet$ and soliton mass $M_{\text{fin}}$ respectively. Hence, for every $m$ the corresponding $\lambda_{max}$ will be smaller than the case with $\hat{\alpha}_{\text{fin}} = 0$. as shown by the solid curve in figure~\ref{fig:lambda_m_SMBH}.

Therefore, in figure~\ref{Excluded_-}, the light gray region labelled `Can't be Probed' is the same as the `Not Allowed' region (where $\lambda > \lambda_{max}$ and no solitonic solutions exist) that has been distorted and pushed down due to the presence of a black hole at the centre.


\subsection{Change in probed $\lambda$ as $M_{\text{halo}}$ increases (for fixed ${\hat \lambda}_{\text{ini}}$  and $m$)}

As halo mass is increased, from eq.~(\ref{eq:obs}), the intercept of the line of interest will decrease, causing ${\hat M}_{\text{ini}}$ to decrease and ${\hat M}_{\text{emp}}$ to increase. Thus, $s$ must decrease too quickly causing the probed value of $\lambda$ to decrease too. Thus, one can probe smaller values of self-coupling using heavier halos provided the corresponding value of $s$ doesn't become too small for the formalism to be applicable.

In summary, $s$ decreases for increasing values of $M_{\text{halo}}$, $m$ and $| {\hat \lambda}_{\text{ini}} |$ for negative self-interactions.  If $s$ becomes too small, the formalism we are working with, in which the GR effects have been ignored, will not be valid. This imposes another practical restriction on the region of $\ln \lambda - \ln m$ plane that we can probe using our methods.
For the purpose of illustrating the basic idea, we restrict our attention to the range of parameter values for which the scaling parameter always stays bigger than 20, as is shown in figure~\ref{s_m}.

\chapter{An approximate approach to the tunnelling of ULDM}\label{app:approx_appraoch}

In chapter~\ref{chpt:paper_3}, in order to solve GPP equations modified by the presence of tidal potential, we took the approach of quasi-stationary states involving complex energy and outgoing wave boundary conditions.
In standard textbooks, one often deals with tunneling problems (to study e.g. $\alpha$-decay of nuclei) by using WKB approximation which involves real energy (see e.g. \cite{Griffiths_Schroeter_2018}).
It is a good idea to attempt to analyse the current problem in the usual textbook approach, this is done in this section.
Before proceeding, we must state the basic assumption for this method: We do not solve the system numerically in the presence of the tidal-potential. Instead, we solve with GPP system with $\hat{\omega}_{\text{\text{ini}}} = 0$ to obtain the stationary-state eigenvalue $\hat{\gamma}$ (corresponding to a node-less solution), $\hat{\phi}(\hat{r})$ and $\hat{\Phi}_{SG}(\hat{r})$. Now we construct an approximate effective potential 
    \begin{equation}\label{eq:eff_pot}
        \hat{V}_{\text{eff}} = \hat{\Phi}_{SG} + \hat{\Phi}_H + \hat{\Phi}_{SI}\ ,
    \end{equation}
where $\hat{\Phi}_{SI} \equiv 2\hat{\lambda}_{\text{\text{ini}}}|\hat{\phi}(\hat{r})|^2$ and $\hat{\Phi}_H \equiv -\frac{3}{2}\hat{\omega}_{\text{\text{ini}}}^2\hat{r}^2$ with some $\hat{\omega}_{\text{\text{ini}}} > 0$. It is important to note that in reality the presence of a tidal term in the Scrh\"{o}dinger equation should alter the density profile $|\hat{\phi}(\hat{r})|^2$ as well as the ground state energy $\hat{\gamma}$ (see figure~\ref{fig:example_sol}). This in-turn will alter $\hat{\Phi}_{SG}$. However, for small values of $\hat{\omega}$, $\hat{V}_{\text{eff}}$ as we have defined it, should serve as a good enough approximation. 
    
To understand the effect of $\hat{\omega}_{\text{\text{ini}}}$, let us construct $\hat{V}_\text{eff}$ for $\hat{\lambda}_{\text{\text{ini}}} = 0$ and different values of $\hat{\omega}_{\text{\text{ini}}}$. The resultant potential barriers are shown in figure~\ref{fig:approx_pot_barrier}. It is clear that a larger $\hat{\omega}_{\text{\text{ini}}}$ will lead to a narrower barrier, making it easier to tunnel through, while a smaller $\hat{\omega}_{\text{\text{ini}}}$ will widen the barrier until $\hat{\omega}_{\text{\text{ini}}}= 0$ for which there will no tunnelling. Figure~\ref{fig:example_sol} also illustrates that a small enough $\hat{\omega}_{\text{\text{ini}}}$ will have a negligible effect on the total potential in the inner region. Treating the soliton as a particle with energy $\hat{\gamma}$, for every $\hat{\omega}_{\text{\text{ini}}}$, the width of the potential barrier experienced by the particle will be $\hat{r}_2 - \hat{r}_1$ where $\hat{\gamma} = \hat{V}_{\text{eff}}$ at $\hat{r}_1$ and $\hat{r}_2$. The transmission probability $\hat{T}_p$ at $\hat{r}_2$ following \cite{Griffiths_Schroeter_2018} is then simply 

\begin{equation}\label{eq:transmission}
    \hat{T}_p = e^{-2\hat{\sigma}}\ ,
\end{equation}
where $\hat{\sigma}$ is the integral

\begin{equation}
\hat{\sigma} = \int_{\hat{r}_1}^{\hat{r}_2}\sqrt{2\left(\hat{V}_{\text{eff}}(\hat{r}) - \hat{\gamma}\right)}d\hat{r}\ .
\end{equation}
It is worth noting that $\hat{\sigma}$ is invariant under scaling (this can be seen by applying scaling relations from section~(\ref{sec:quasi_numerical})). Now, assuming that the particle is moving around in the potential well hitting the barrier at $\hat{r}_1$ repeatedly, the frequency of collisions can be written as $\hat{v}/2\hat{r}_1$, where $\hat{v}$ is the velocity of the particle. A product of the transmission probability and the collision frequency is the probability of tunnelling per unit time i.e. the decay rate $\hat{\Gamma}. $\footnote{Note that $\hat{\Gamma} = \frac{\hbar}{mc^2}\Gamma$.} Since $\hat{\Gamma}$ is the inverse of time, it scales as $\Gamma \rightarrow s^{-2}\Gamma$ (from eq.~(\ref{eq:scaling_omega})). Since any change in the exponent will dominate over any change in the velocity, one can approximate the decay rate as 

\begin{equation}\label{eq:wkb_decay}
    \hat{\Gamma} = s^{-2}\frac{\hat{v}}{2\hat{r}_1}e^{-2\hat{\sigma}} \approx s^{-2}e^{-2\hat{\sigma}}\ .
\end{equation}
Here we have ignored the $\frac{\hat{v}}{2\hat{r}_1}$ factor since we do not have an estimate for the velocity of the 'particle' in this case. Moreover, the exponential will have the dominant contribution to the decay rate.

\section{Negligible self-interactions} 

\begin{figure}[ht]
    \centering
    \includegraphics[width = 0.65\textwidth]{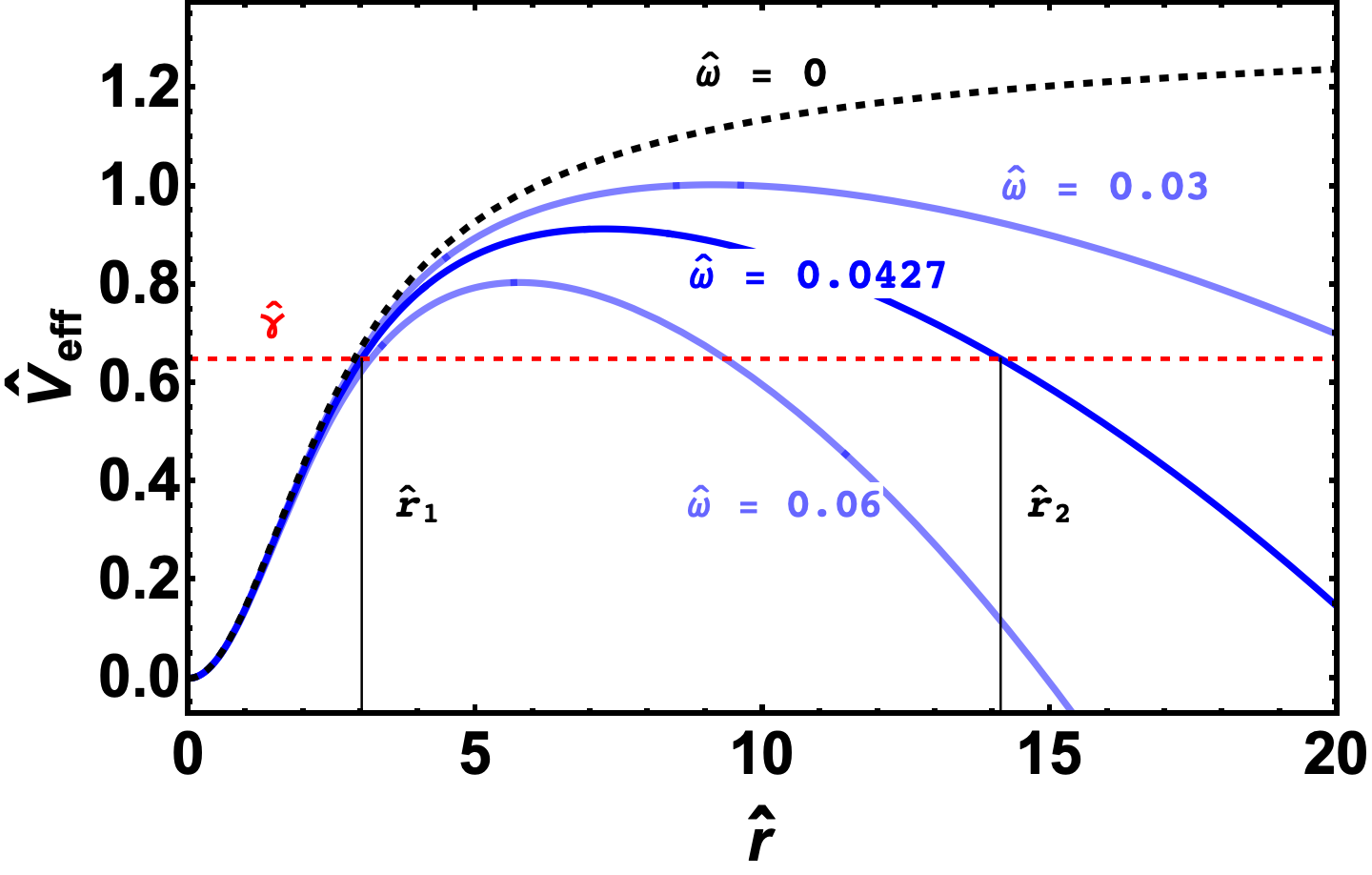}
    \caption[Approximate effective potentials for various values of $\hat{\omega}_\text{\text{ini}}$]{Approximate dimensionless effective potentials ($\hat{V}_\text{eff} = V_{\text{eff}}/c^2$) for various values of unscaled dimensionless orbital frequency $\hat{\omega}_\text{\text{ini}}$ (as defined in eq.~(\ref{eq:omega})) are shown. For the UMi spheroidal, required value of $\hat{\omega}_{\text{\text{ini}}} = 0.0427$ is shown by the dark blue curve. $\hat{d} = \hat{r}_2 - \hat{r}_1$ is the width of the barrier while the red line is the ground state energy eigenvalue in the absence of the tidal potential.}
    \label{fig:approx_pot_barrier}
\end{figure}

Let us apply this approach to a real system, that of the `UMi' spheroidal, whose central density is $\rho_c \approx 0.15\ M_\odot \text{pc}^{-3}$ and the orbital period is $T \approx 1.6\ \text{Gyr}$, where $\omega = 2\pi/T$ \cite{Hertzberg_2023}. Following the discussion in section~\ref{sec:fixed_rhoC}, the scale value $s = 7225$ and $\hat{\omega}_{\text{\text{ini}}} = 0.0427$ are obtained. The resultant approximate effective potential is shown in figure~\ref{fig:approx_pot_barrier}. The decay rate using eq.~(\ref{eq:wkb_decay}) (after scaling) is then estimated to be 

\begin{equation}
    \Gamma \sim 10^{-3}/\text{t}_{\text{uni}}\ ,
\end{equation}
which is an order of magnitude larger than the one obtained in \cite{Hertzberg_2023}. As discussed there are several sources of uncertainty: (a) The self-gravitational potential $\hat{\Phi}_{SG}$ will be altered in the presence of the tidal potential, (b) The extra term in the Schr\"{o}dinger equation will also impact the eigenvalue $\hat{\gamma}$. If either of these two quantities change even a little, the width of the barrier experienced by the soliton might change which will lead to a drastically different decay rate because of the exponential dependence on $\hat{\sigma}$. This implies that this approximation can only work when $\hat{\omega}_{\text{\text{ini}}}$ is small enough to not appreciably affect the solution of the SP system. Further, we have also ignored $\hat{v}/2\hat{r}_1$ factor, which can lead to further alterations to the decay rate. 

However, as we have now seen, even a crude toy-approach using the WKB approximation is useful for qualitatively understanding the impact of $\hat{\omega}$. This is also true for the effects of self-interactions (which we shall discuss in section~\ref{sec:SI_toy}).

\section{Non-negligible self-interactions}\label{sec:SI_toy}

In this section, similar to what we did in the main paper we shall look at two cases: (a) fixed scaled core density and (b) fixed scaled core mass. For both cases we look at three values of $\hat{\lambda}_\text{\text{ini}} = \{-0.05, 0, 0.1\}$. We follow the same procedure for obtaining the appropriate scale value as we did in sections~\ref{sec:fixed_rhoC} and~\ref{sec:fixed_Mc} and find:

\begin{figure}[ht]
    \centering
    \begin{subfigure}[t]{0.45\textwidth}
        \centering
        \includegraphics[width = \textwidth, height = 2in]{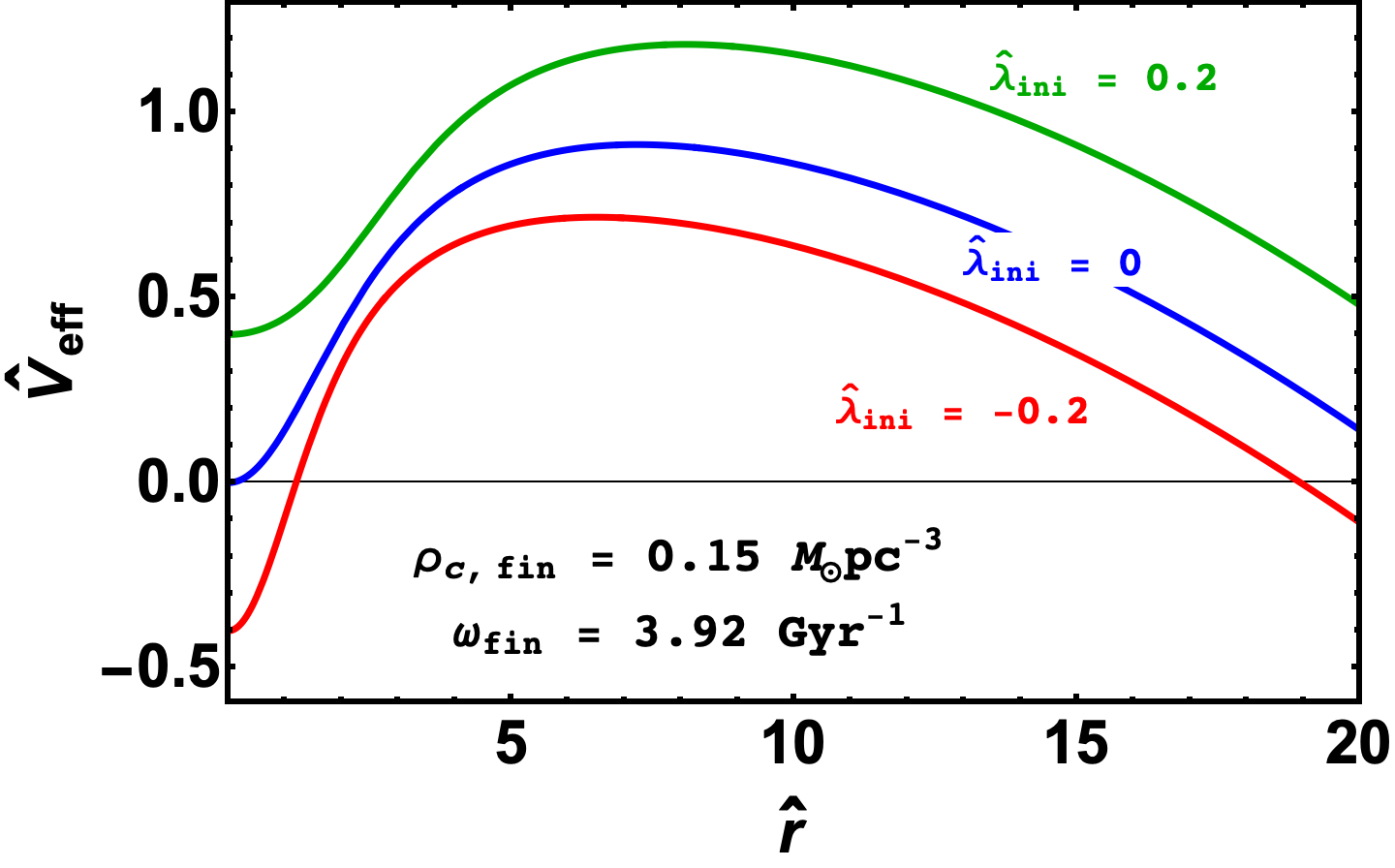}
        \caption{Approximate effective potentials for three values of $\hat{\lambda}_\text{\text{ini}}$ are shown, for a fixed $\hat{\omega}_\text{\text{ini}} = 0.0427$ and $s = 7225$ which corresponds to a scaled $\rho_{c,\text{\text{fin}}}= 0.15\ M_\odot\text{pc}^{-3}$ and $\omega_\text{\text{fin}} = 3.92\ \text{Gyr}^{-1}$.}
        \label{fig:wkb_fixed_rhoc}
    \end{subfigure}
    \hfill
    \begin{subfigure}[t]{0.45\textwidth}
        \centering
        \includegraphics[width = \textwidth, height = 2in]{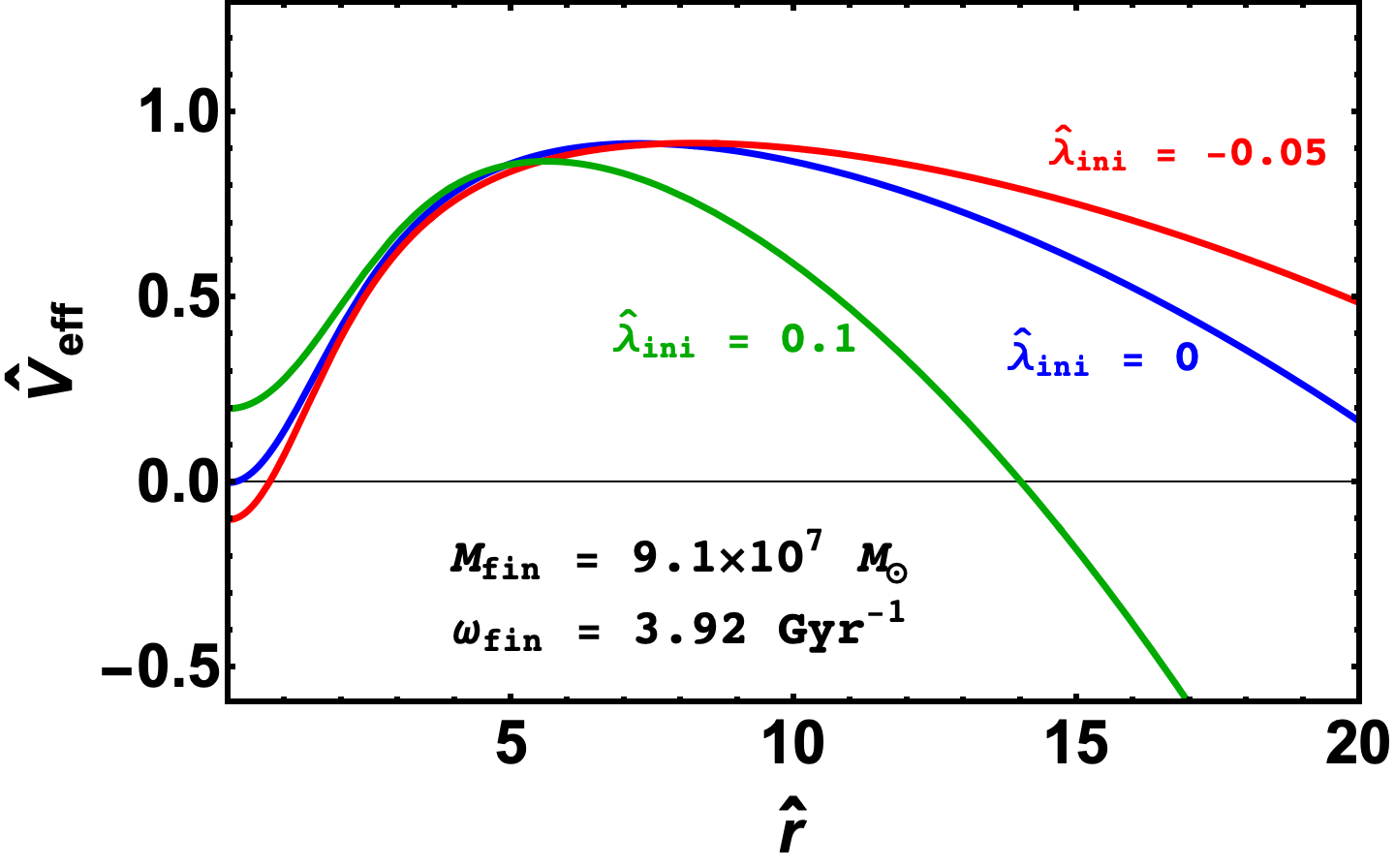}
        \caption{Approximate effective potential $\hat{V}_\text{eff}$ for different values of $\hat{\omega}_{\text{\text{ini}}}$ and $\hat{\lambda}_\text{\text{ini}}$ such that scaled $M_\text{\text{fin}} = 9.1\times 10^7\ M_\odot$ and $\omega_\text{\text{fin}} = 3.92\ \text{Gyr}^{-1}$. The barriers are similar to what we get from numerical solutions in figure~\ref{fig:SI_barrier}.}
        \label{fig:wkb_fixed_core_mass}
    \end{subfigure}
    \caption[Comparing effective potentials for fixed scaled density and fixed scaled core mass approaches]{Left panel illustrates the effects of self-interactions, as defined in eq.~(\ref{eq:dimensions2}), on the dimensionless effective potential for objects with the same central density and orbital period. In the right panel, we consider different self-interaction strengths for objects with the same core and orbital period.}
    \label{fig:approx_SI_barriers}
\end{figure}

\begin{itemize}
    \item The scaled central density and orbital frequency are fixed to be $\rho_{c, \text{\text{fin}}} = 0.15\ M_\odot\text{pc}^{-3}$ and $\omega_\text{\text{fin}} = 3.92\ \text{Gyr}^{-1}$. The corresponding scale value is fixed for all $\hat{\lambda}_\text{\text{ini}}$ to be $s = 7225$ while $\hat{\omega}_\text{\text{ini}} = 0.0427$. For $\hat{\lambda}_\text{\text{ini}} = 0$ the scaled lifetime is $\tau \approx 235\ t_\text{uni}$. For $\hat{\lambda}_\text{\text{ini}} = \pm0.2$ the corresponding scaled $\lambda = \pm 1.41\times 10^{-91}$ while the scaled lifetimes are found to be $\tau \approx 6387\ t_\text{uni}$ and $\tau \approx 36.60\ t_\text{uni}$ for repulsive and attractive self-interactions respectively.  

    \item For $M_\text{\text{fin}} = 9.1\times 10^7\ M_\odot$ and $\omega_\text{\text{fin}} = 3.92\ \text{Gyr}^{-1}$, we find that for $\hat{\lambda}_\text{\text{ini}} = 0$ the required unscaled dimensionless $\hat{\omega}_\text{\text{ini}} \approx 0.042$. The scaled lifetime of $\tau \approx 337.44\ t_\text{uni}$. For $\hat{\lambda}_\text{\text{ini}} = -0.05$, the corresponding self-coupling turns out to be $\lambda \approx -2.78\times 10^{-92}$ with a lifetime of $\tau \approx 8.74\times 10^4\ t_\text{uni}$. On the other hand, for $\hat{\lambda}_\text{\text{ini}} = 0.1$, the lifetime turns out to be $\tau \approx 5.9\times 10^{-2}\ t_\text{uni}$ with a scaled self-coupling $\lambda \approx 1.11\times 10^{-91}$. 
\end{itemize}

The impact of self-interactions we observe above agrees with the full numerical approach qualitatively, albeit the actual values of lifetimes are far from the ones we get from a numerical implementation. This is due to the approximate nature of this method, as discussed in the previous section. However, the intuition regarding the effect of scaling relations on the potential barrier experienced by the soliton remains intact, as seen in figure~\ref{fig:approx_SI_barriers} for both the fixed central density and fixed core mass cases.

\chapter{Monitoring the loss function}\label{app:loss_epochs}

In this section, we plot the loss functions for each of the three neural networks, i.e., with noiseless and noisy training data for MSE loss function and noisy training data for heteroscedastic loss function - trained for each of the $7$ galaxies in our sample. 
\begin{figure*}
\begin{tabular}{ccc}
\includegraphics[width=0.28\linewidth]{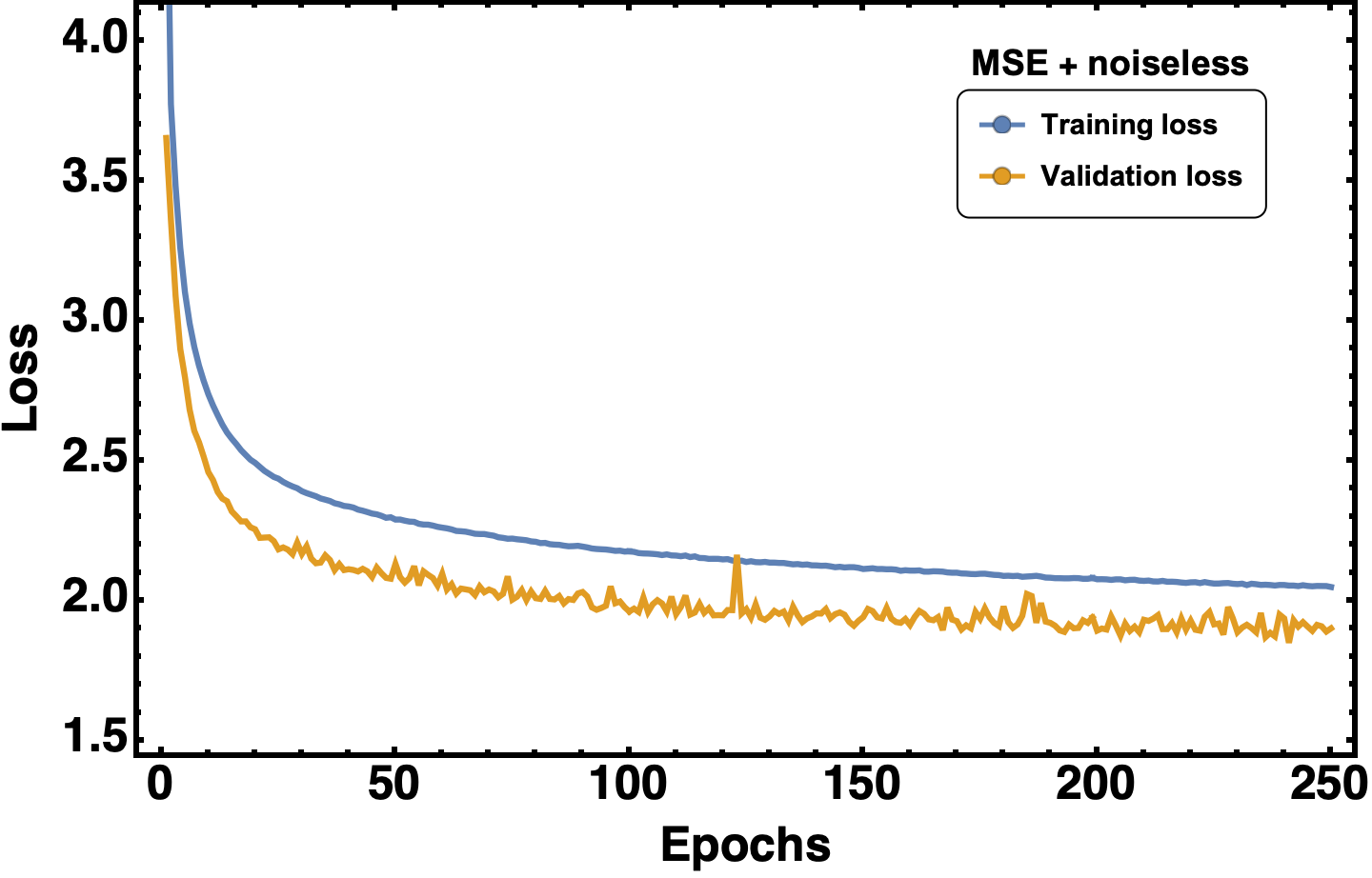} &   \includegraphics[width=0.28\linewidth]{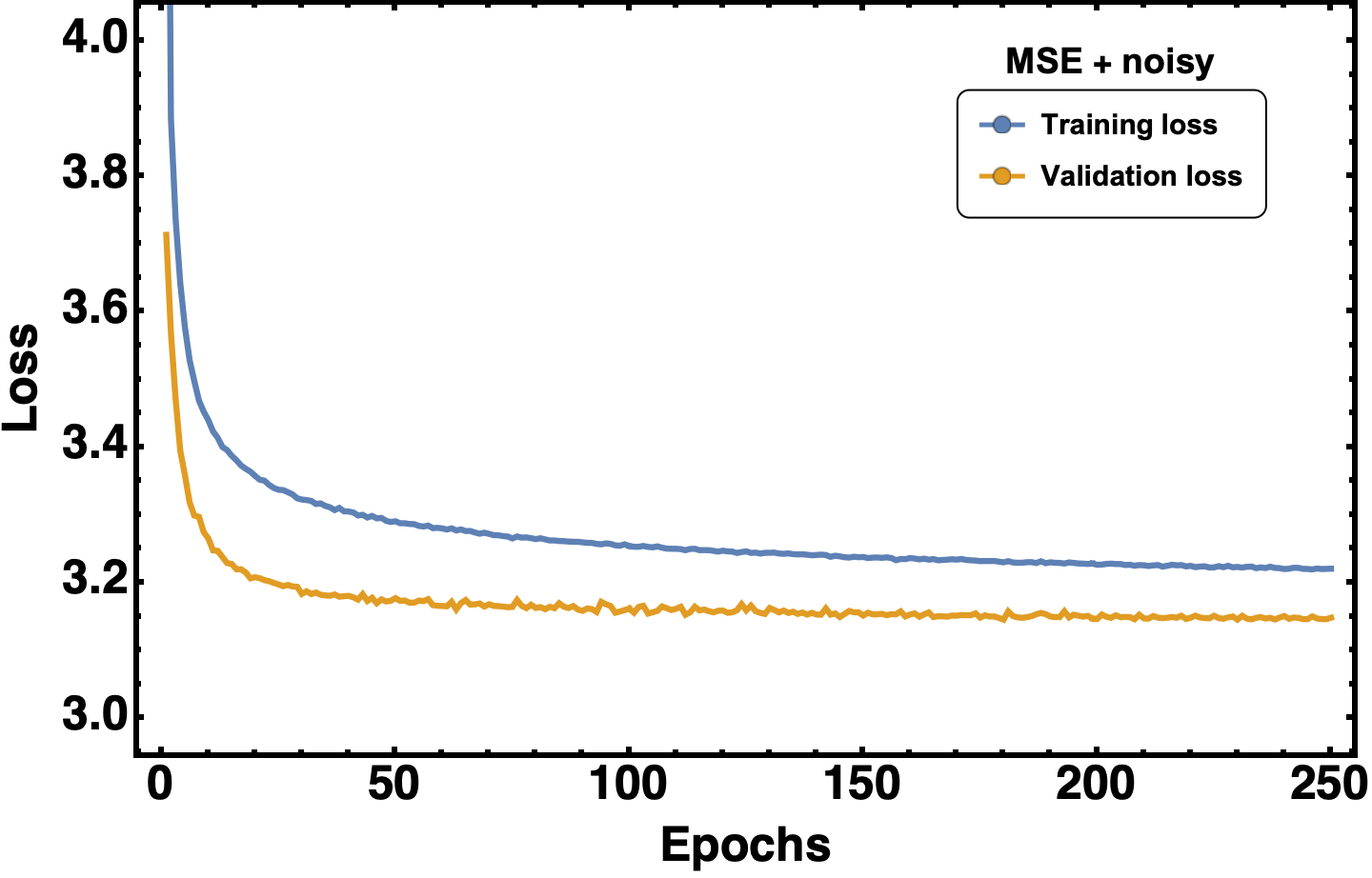} &
\includegraphics[width=0.28\linewidth]{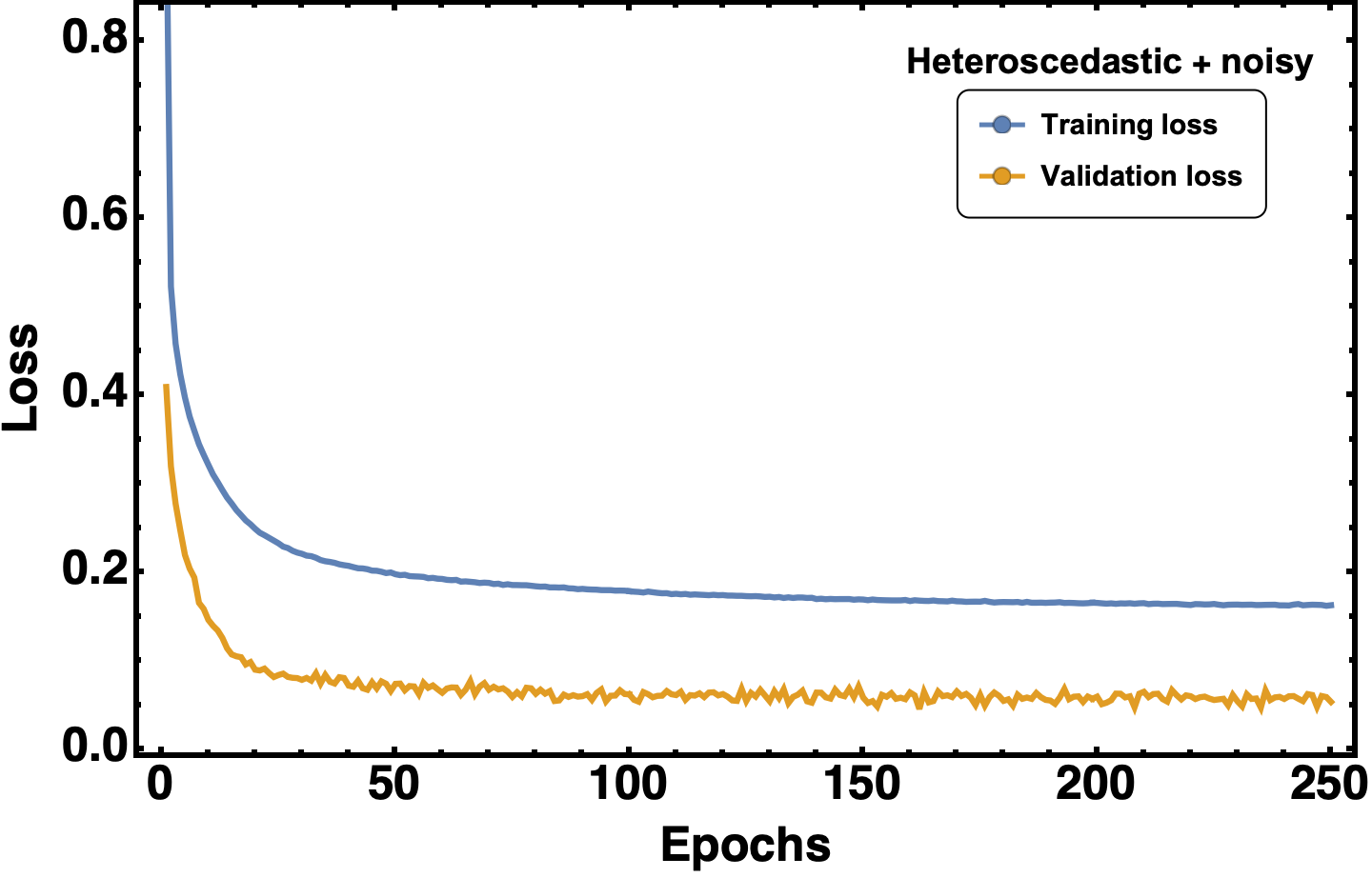} \\
& (a) DDO 154 & \\[5pt]
\includegraphics[width=0.28\linewidth]{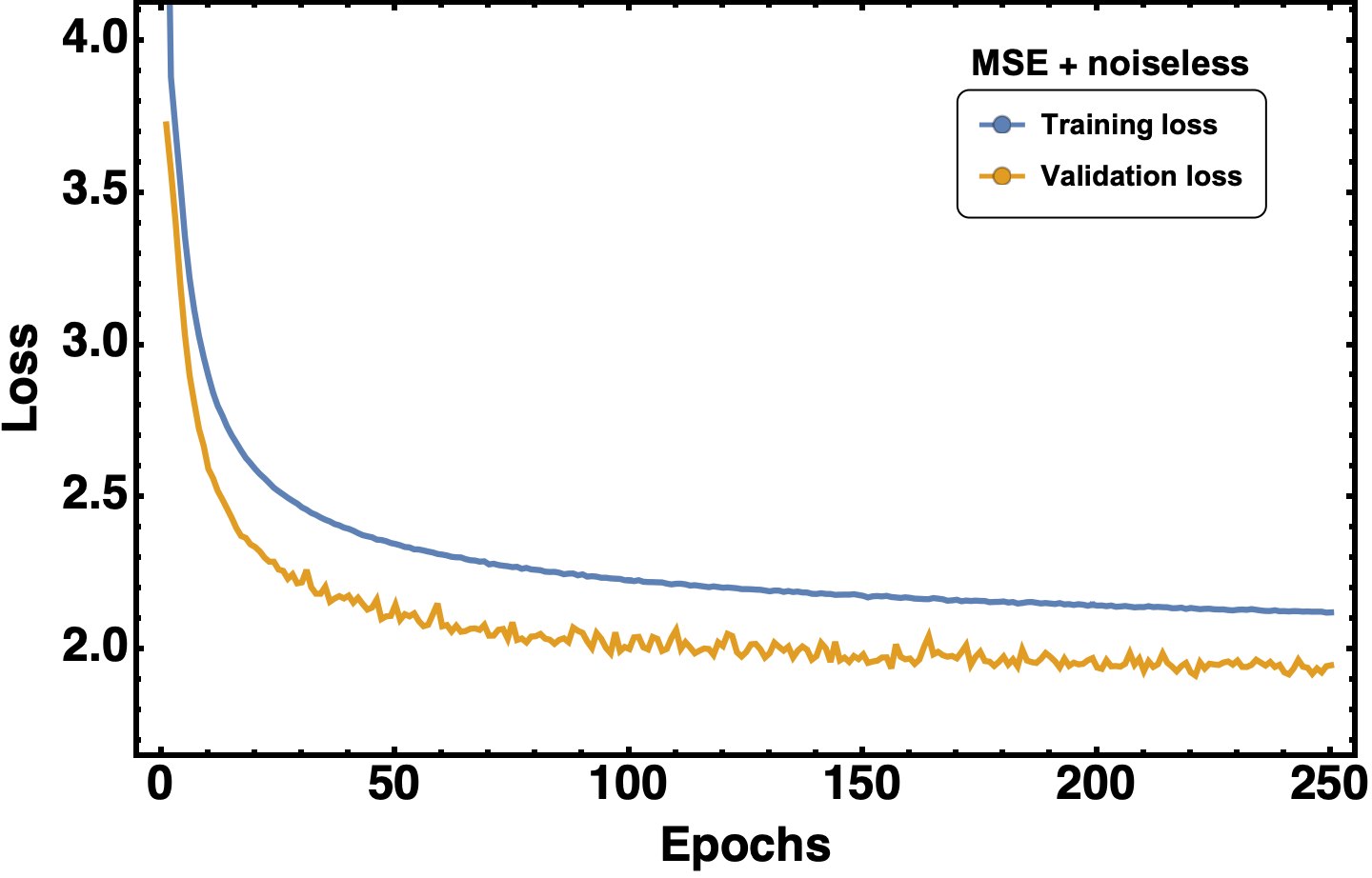} &   \includegraphics[width=0.28\linewidth]{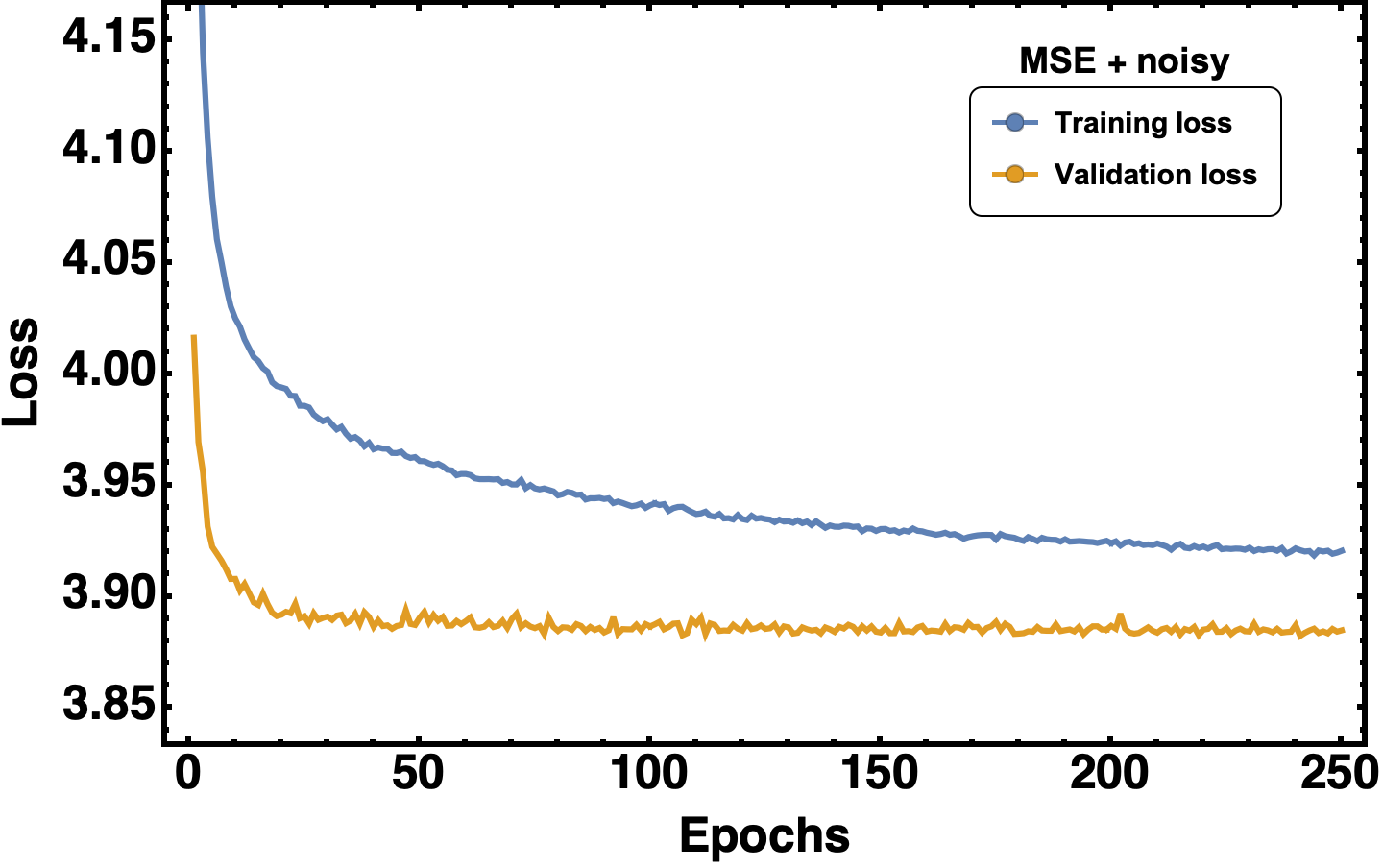} &
\includegraphics[width=0.28\linewidth]{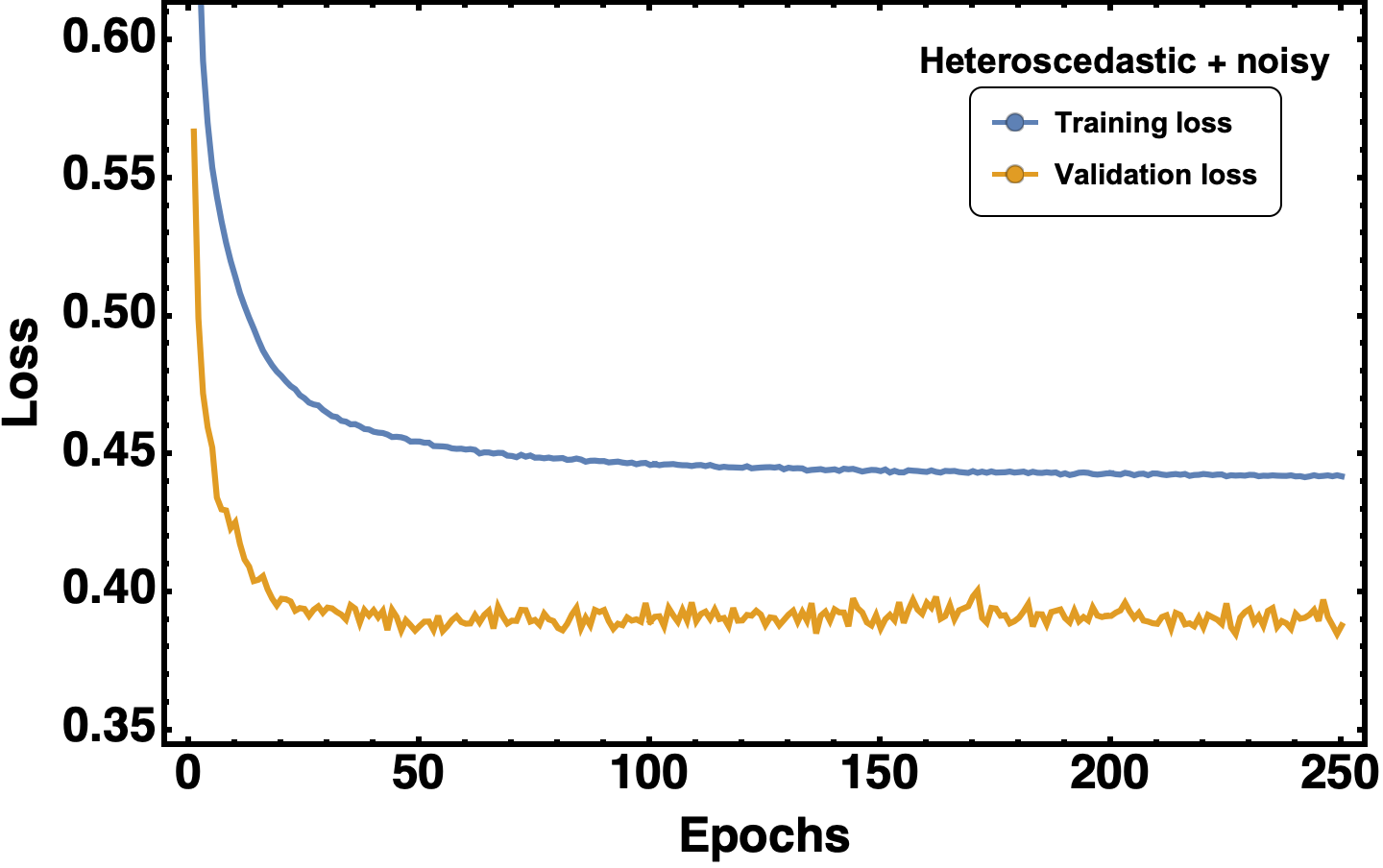} \\
& (b) ESO 444-G084 & \\[5pt]
\includegraphics[width=0.28\linewidth]{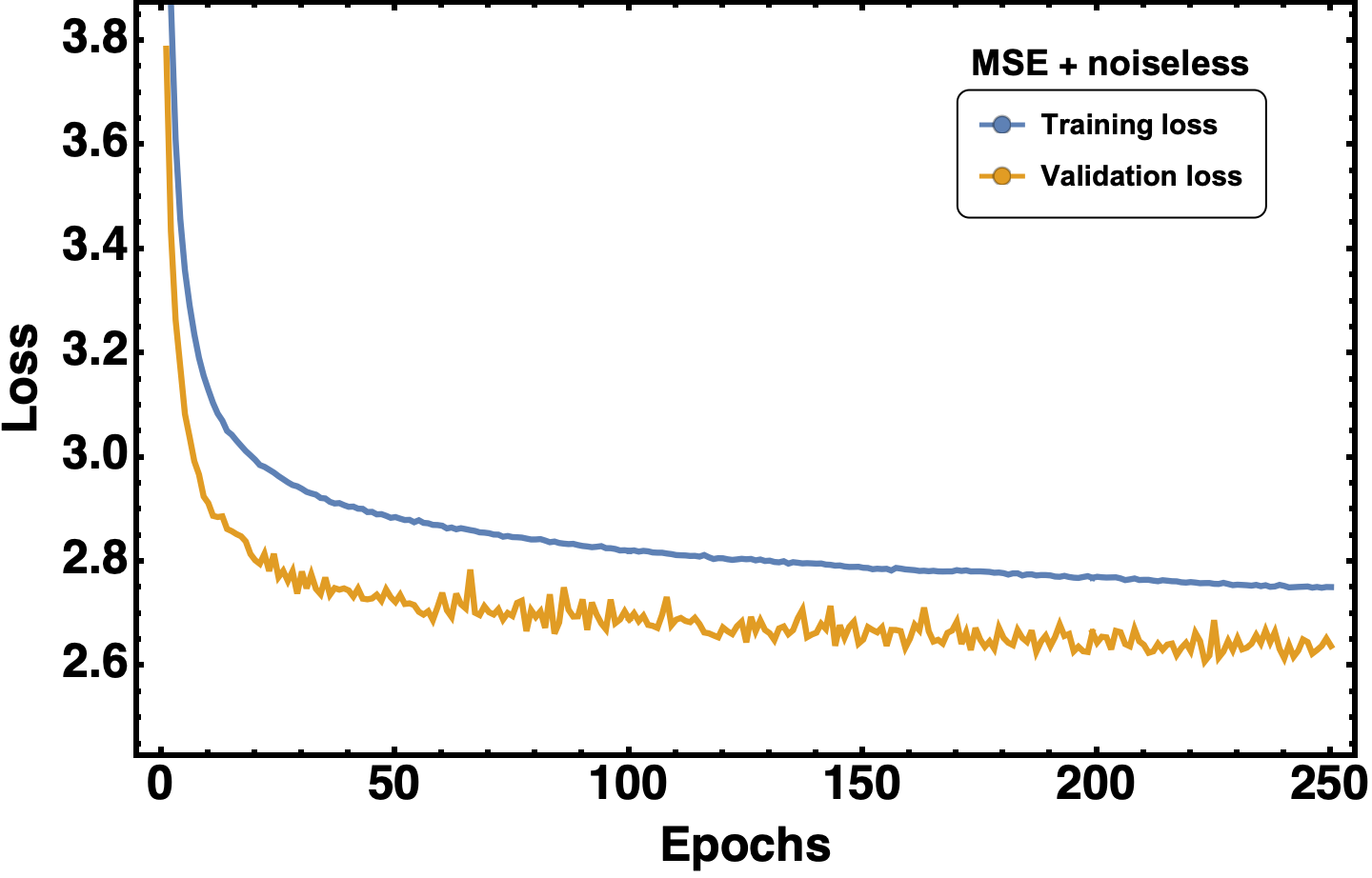} &   \includegraphics[width=0.28\linewidth]{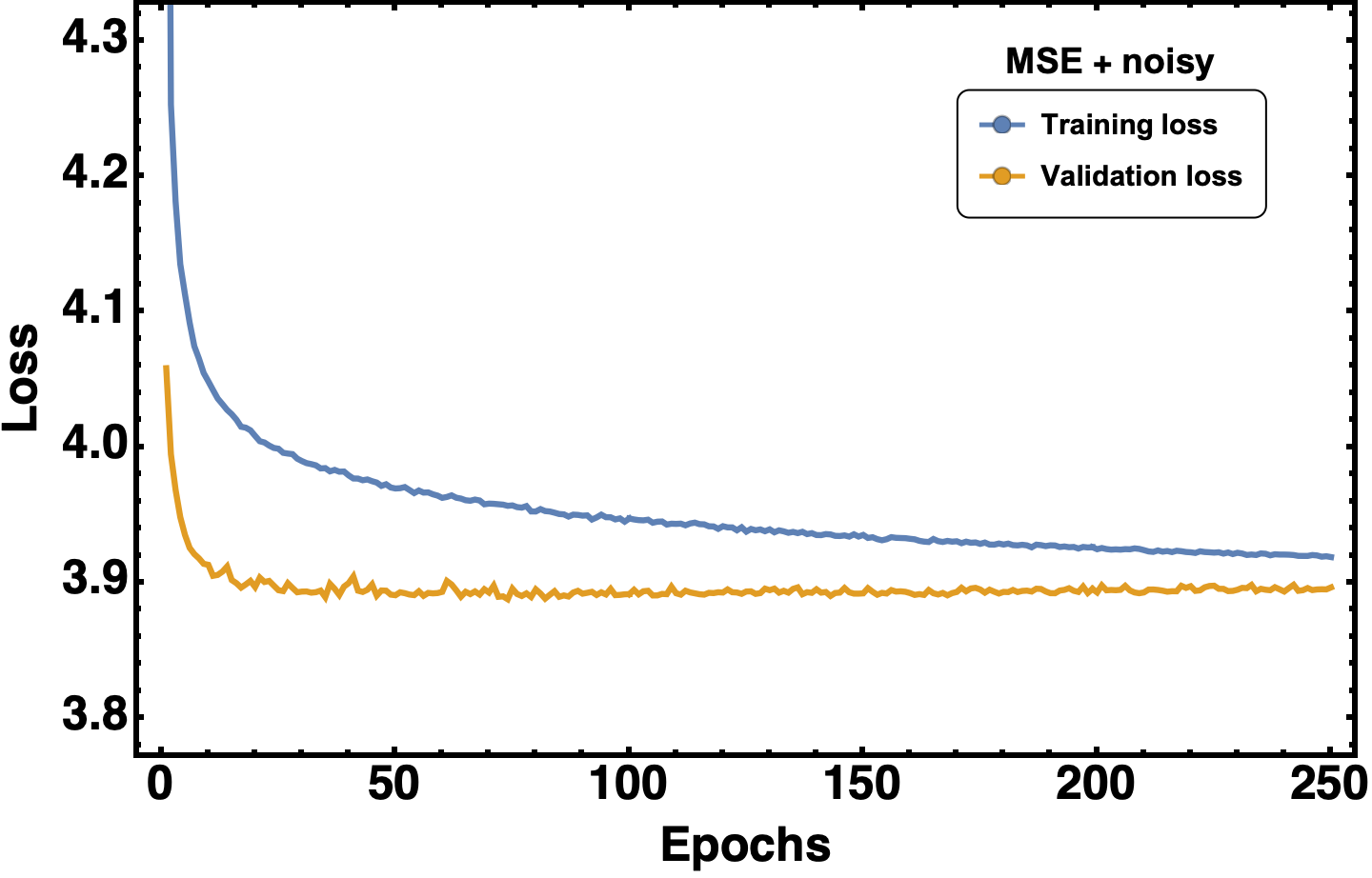} &
\includegraphics[width=0.28\linewidth]{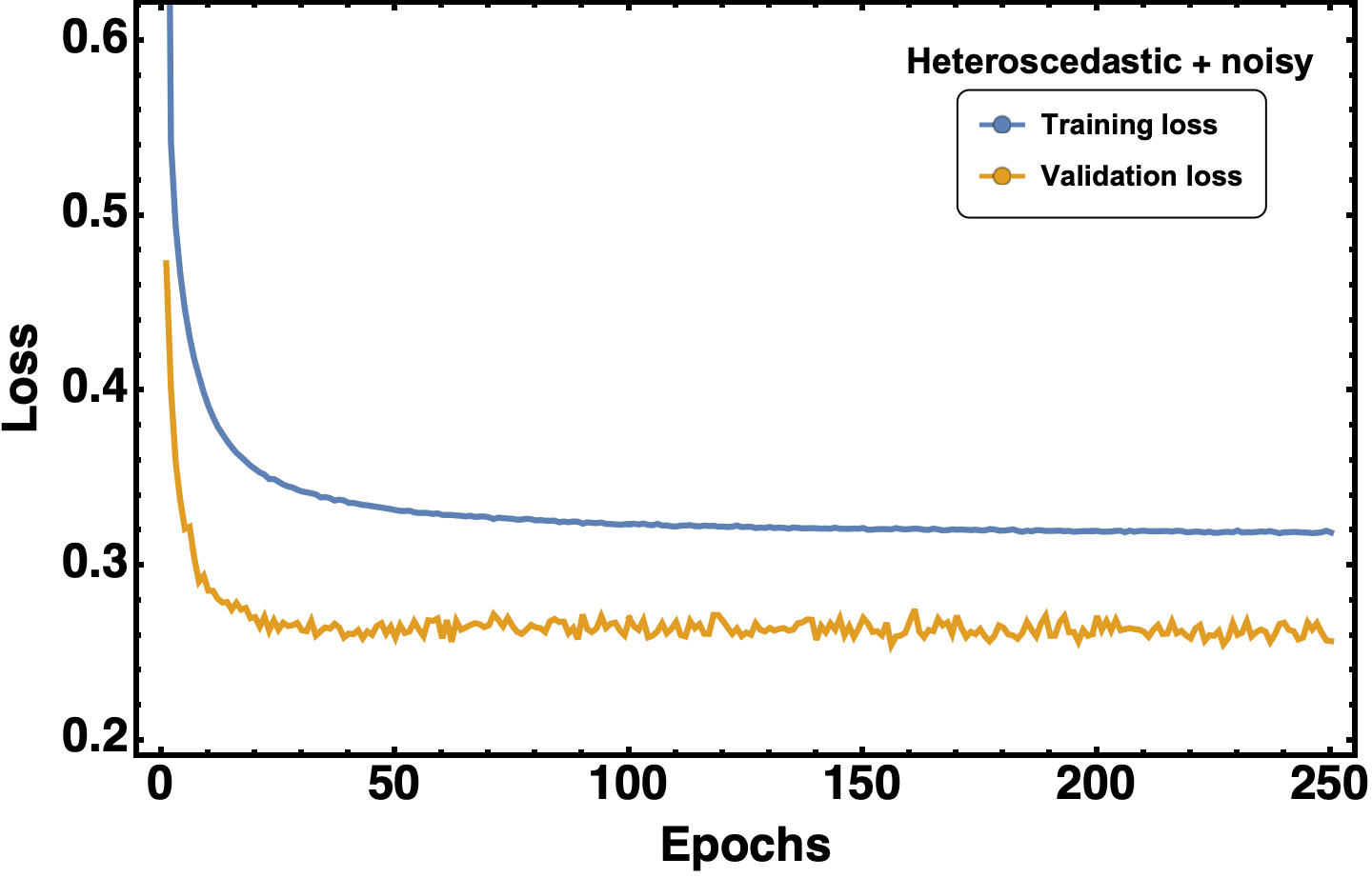} \\
& (c) UGC 5721 & \\[5pt]
\includegraphics[width=0.28\linewidth]{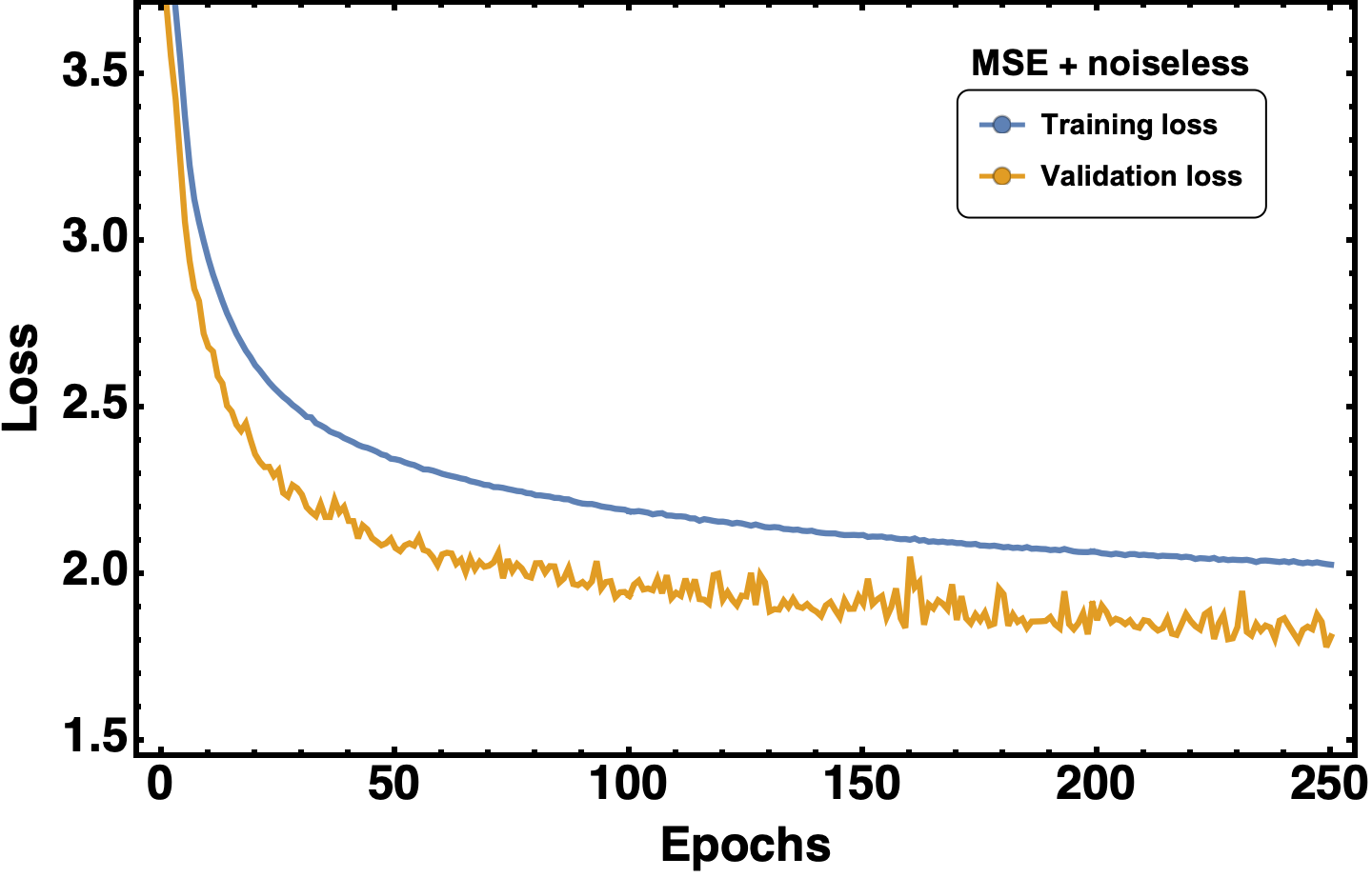} & 
\includegraphics[width=0.28\linewidth]{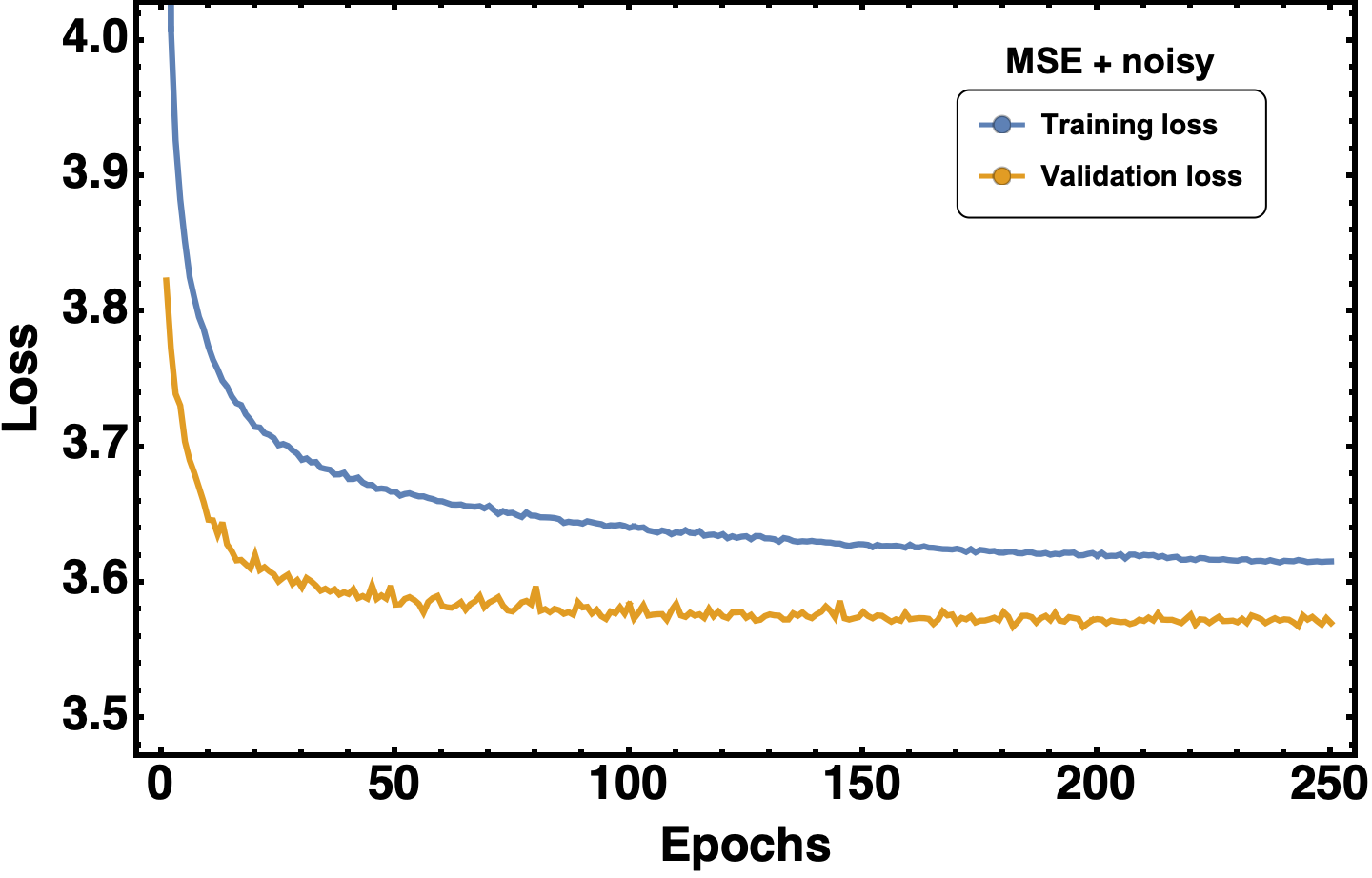} &
\includegraphics[width=0.28\linewidth]{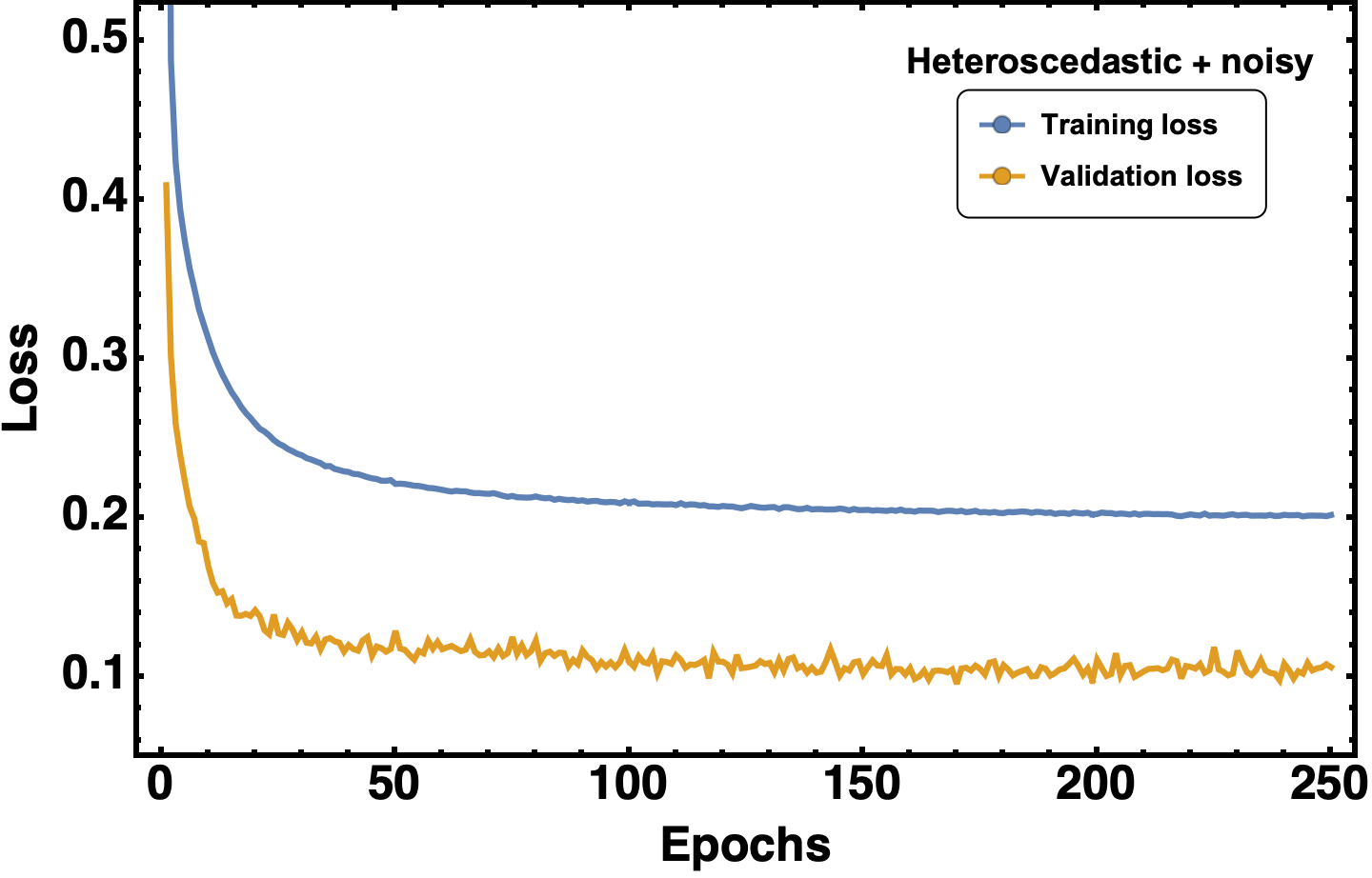} \\
& (a) UGC 5764 & 
\end{tabular}
\caption[Loss function as a function of epochs for the first 4 galaxies in our sample]{\justifying Loss plotted with respect to epochs for the three cases we have considered as mentioned in the text as well as the top right corner of each plot. 
Training loss is denoted by blue while validation loss is denoted by orange for three galaxies from our sample.}
\label{fig:loss_1}
\end{figure*}

\begin{figure*}
\begin{tabular}{ccc}
\includegraphics[width=0.28\linewidth]{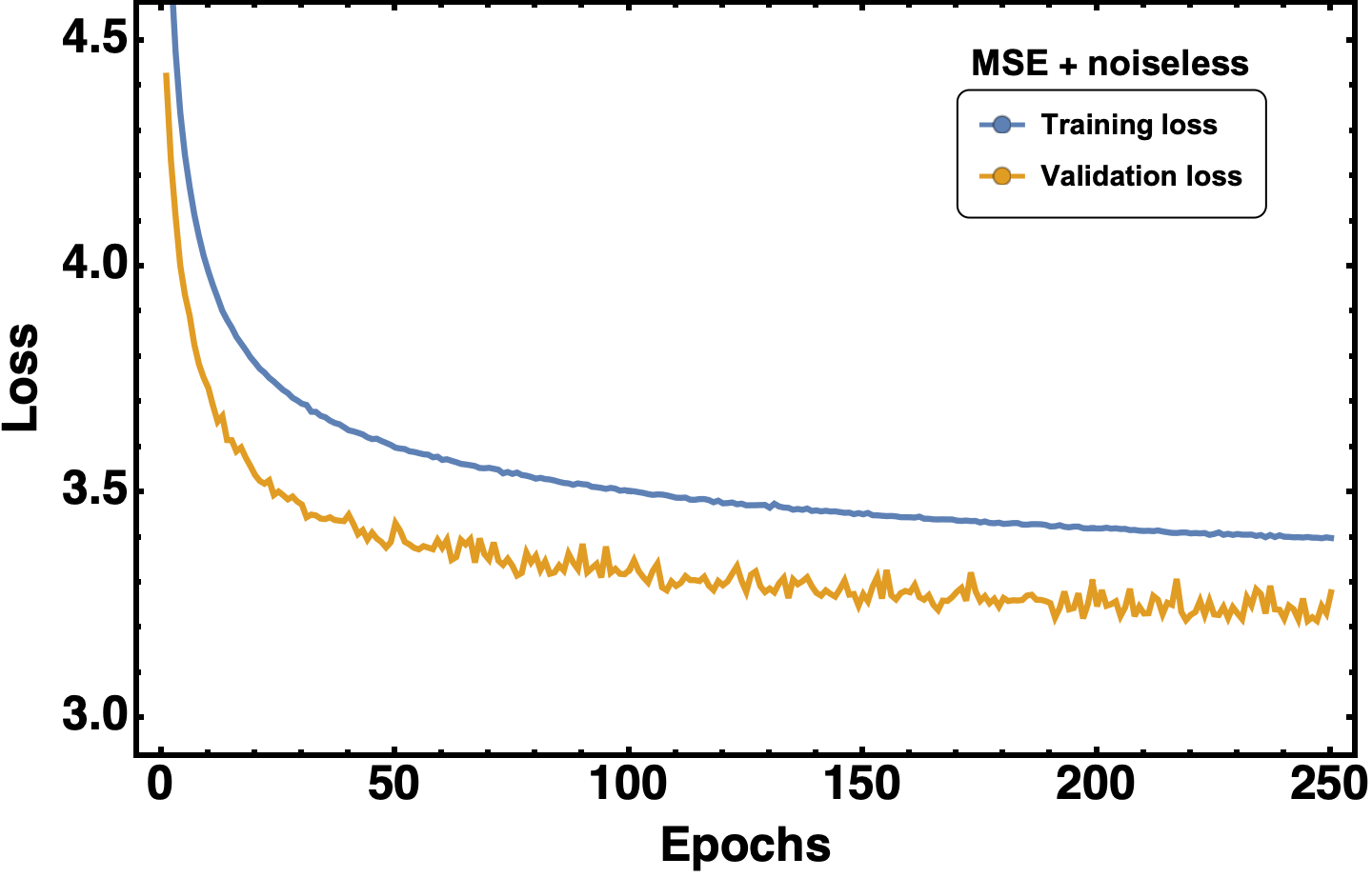} &   \includegraphics[width=0.28\linewidth]{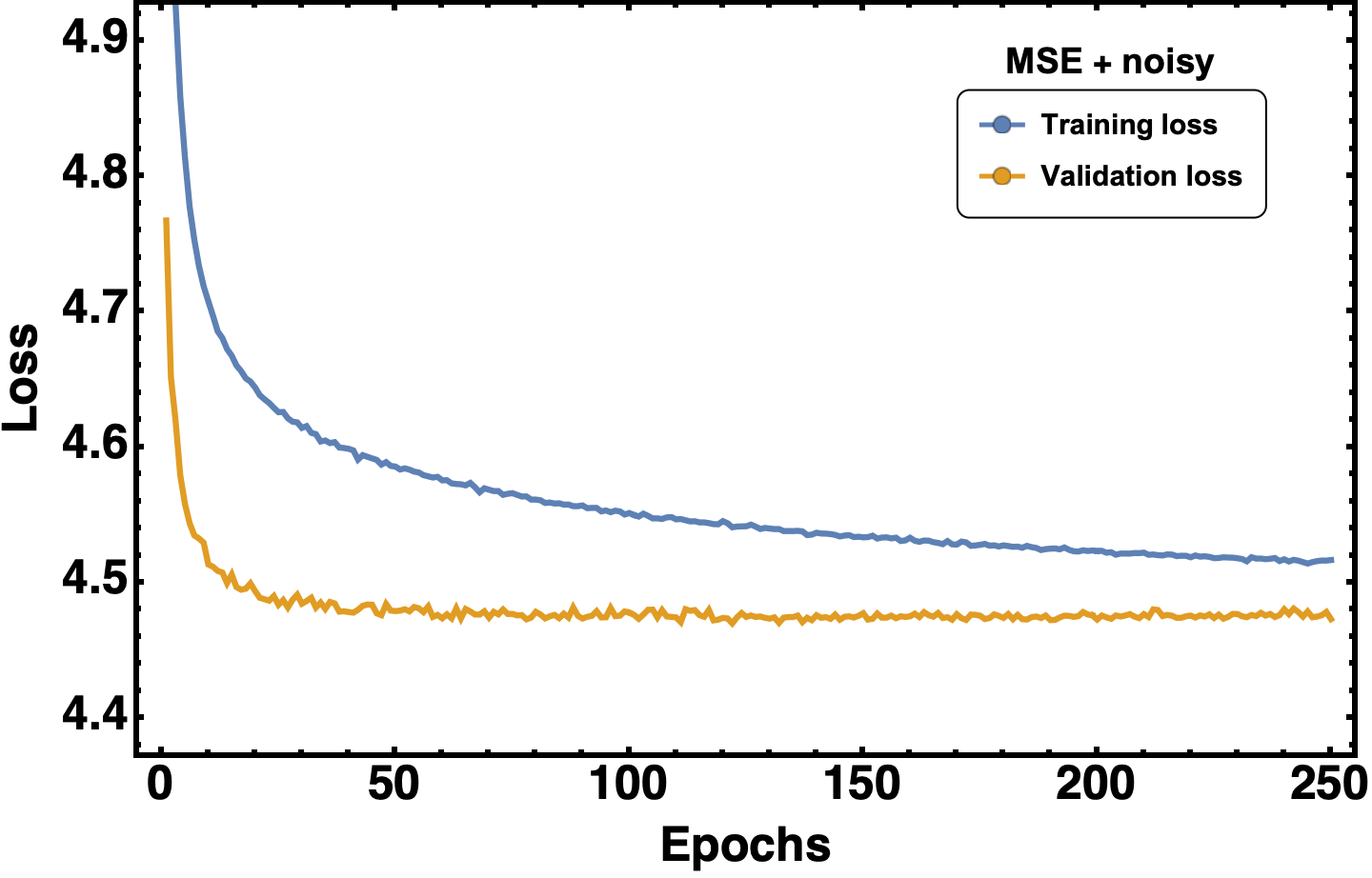} &
\includegraphics[width=0.28\linewidth]{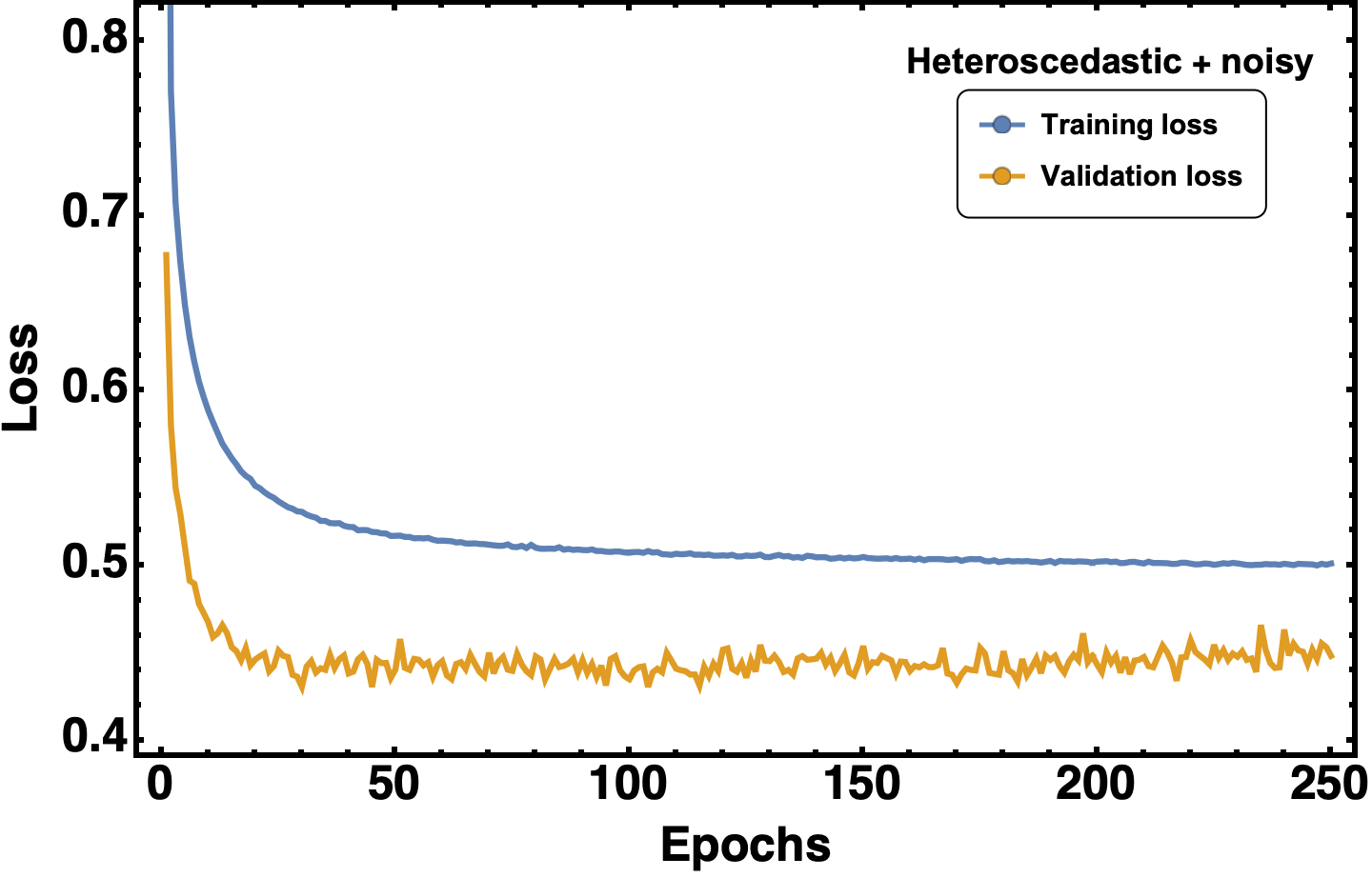} \\
& (b) UGC 7524 & \\[5pt]
\includegraphics[width=0.28\linewidth]{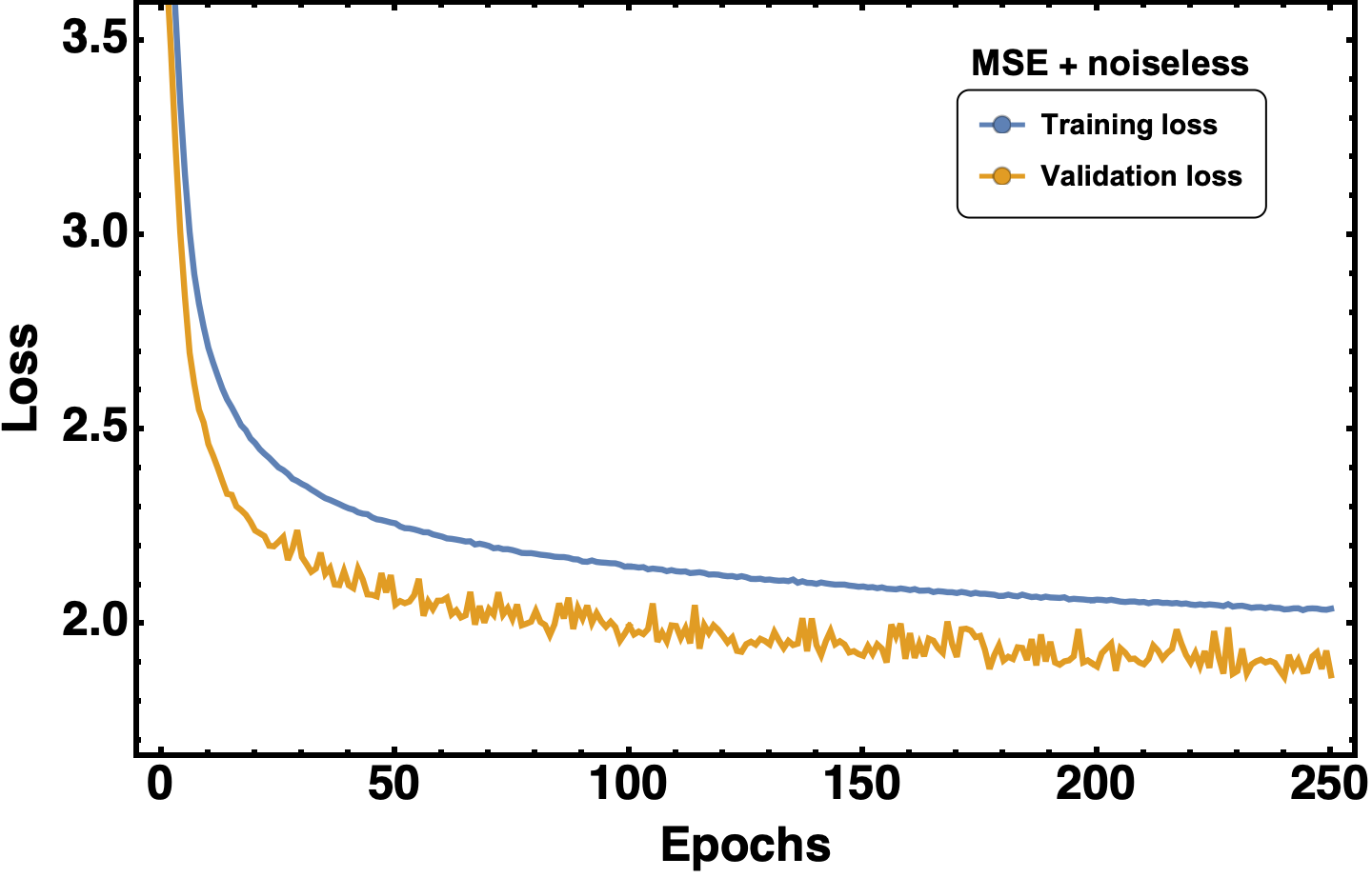} & 
\includegraphics[width=0.28\linewidth]{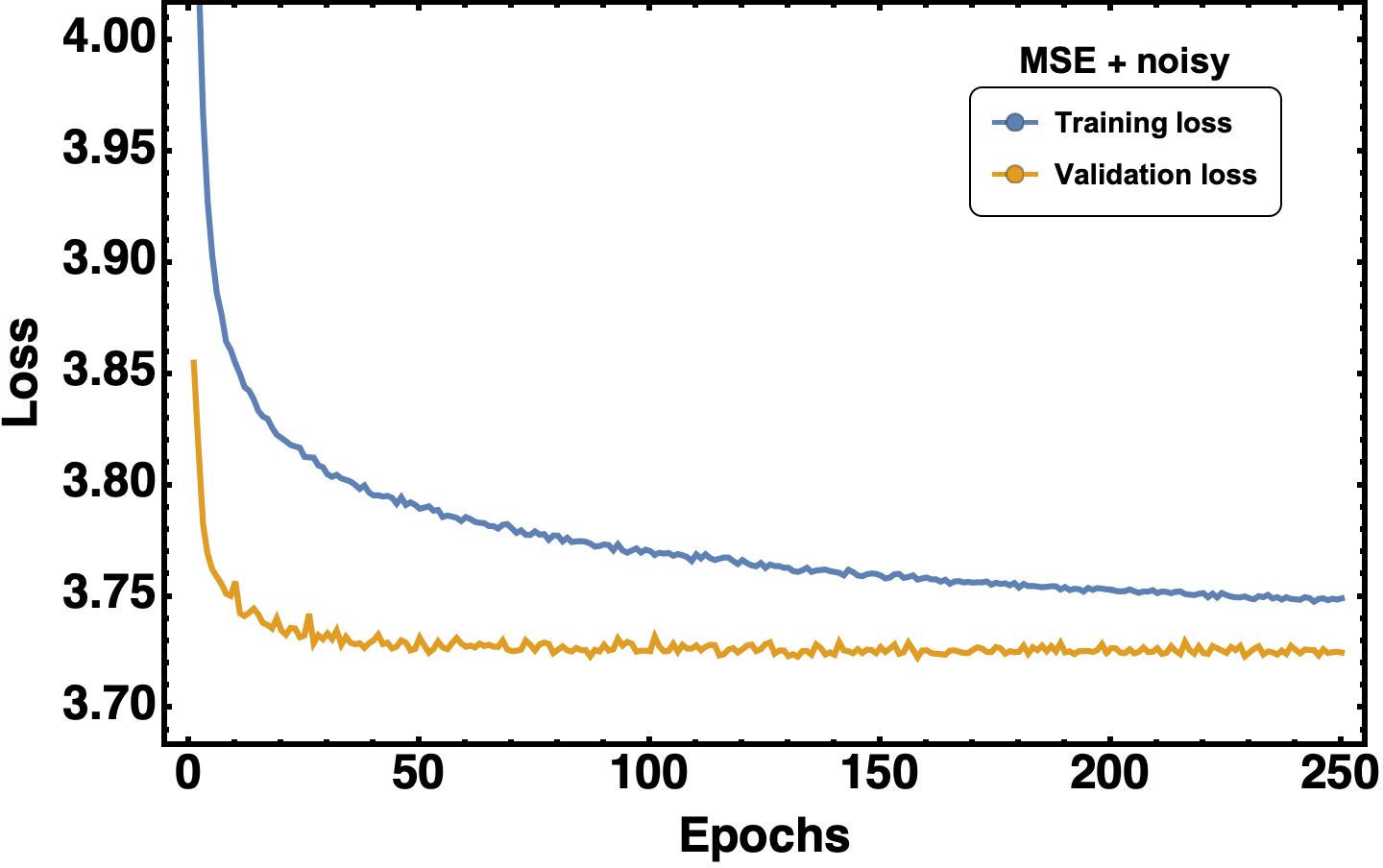} &
\includegraphics[width=0.28\linewidth]{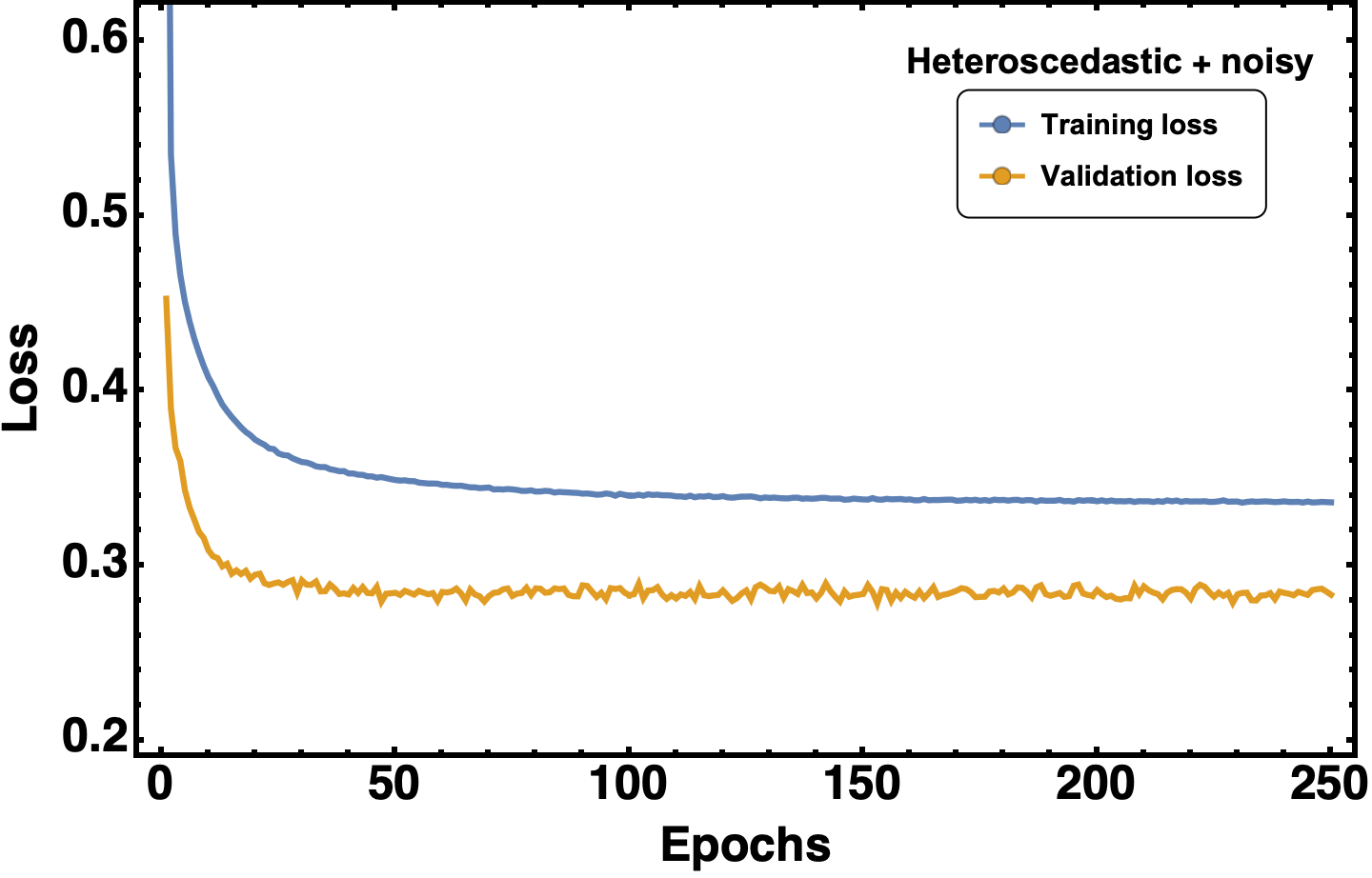} \\
& (c) UGC 7603 & \\[5pt]
\includegraphics[width=0.28\linewidth]{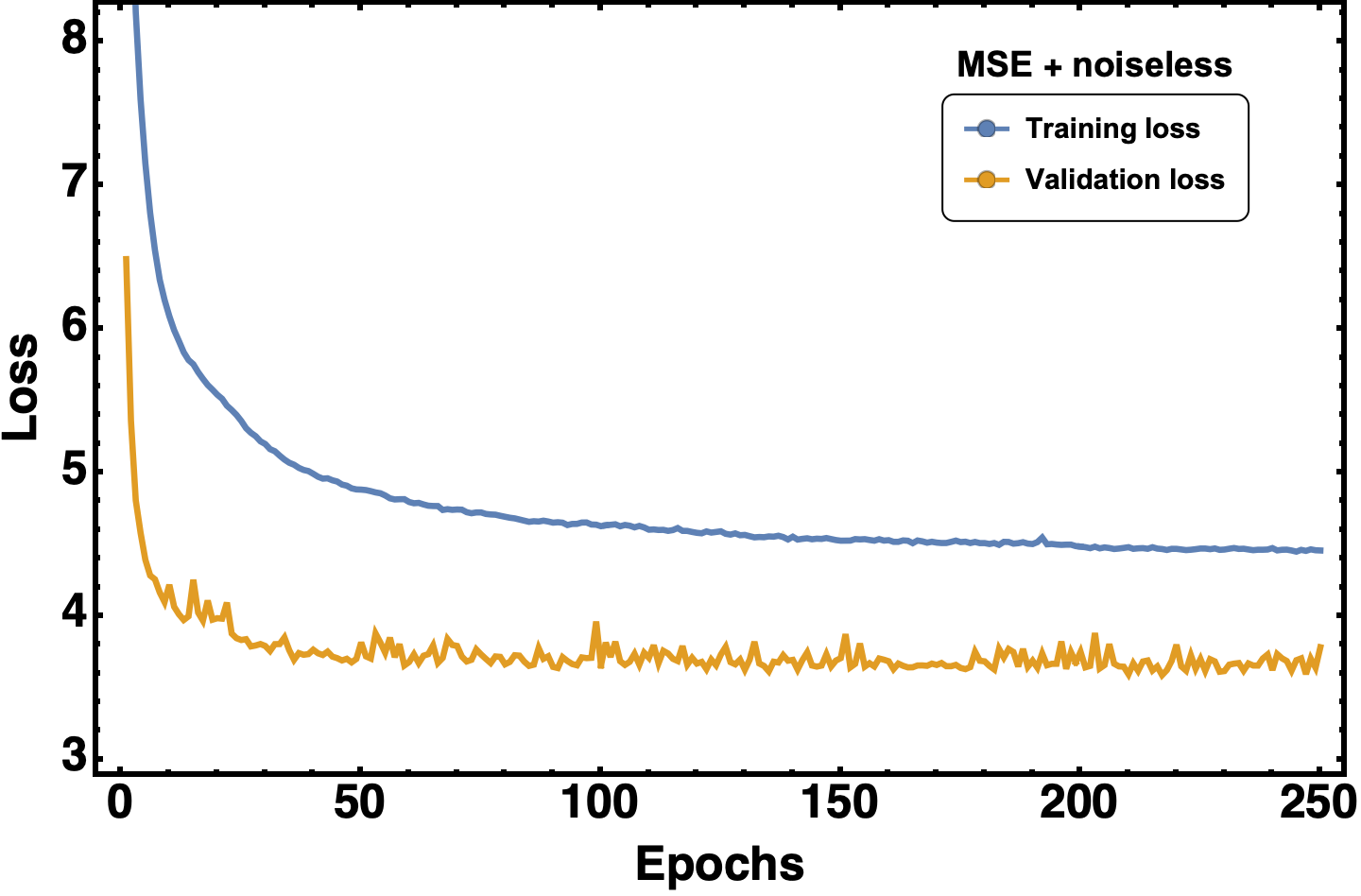} &   \includegraphics[width=0.28\linewidth]{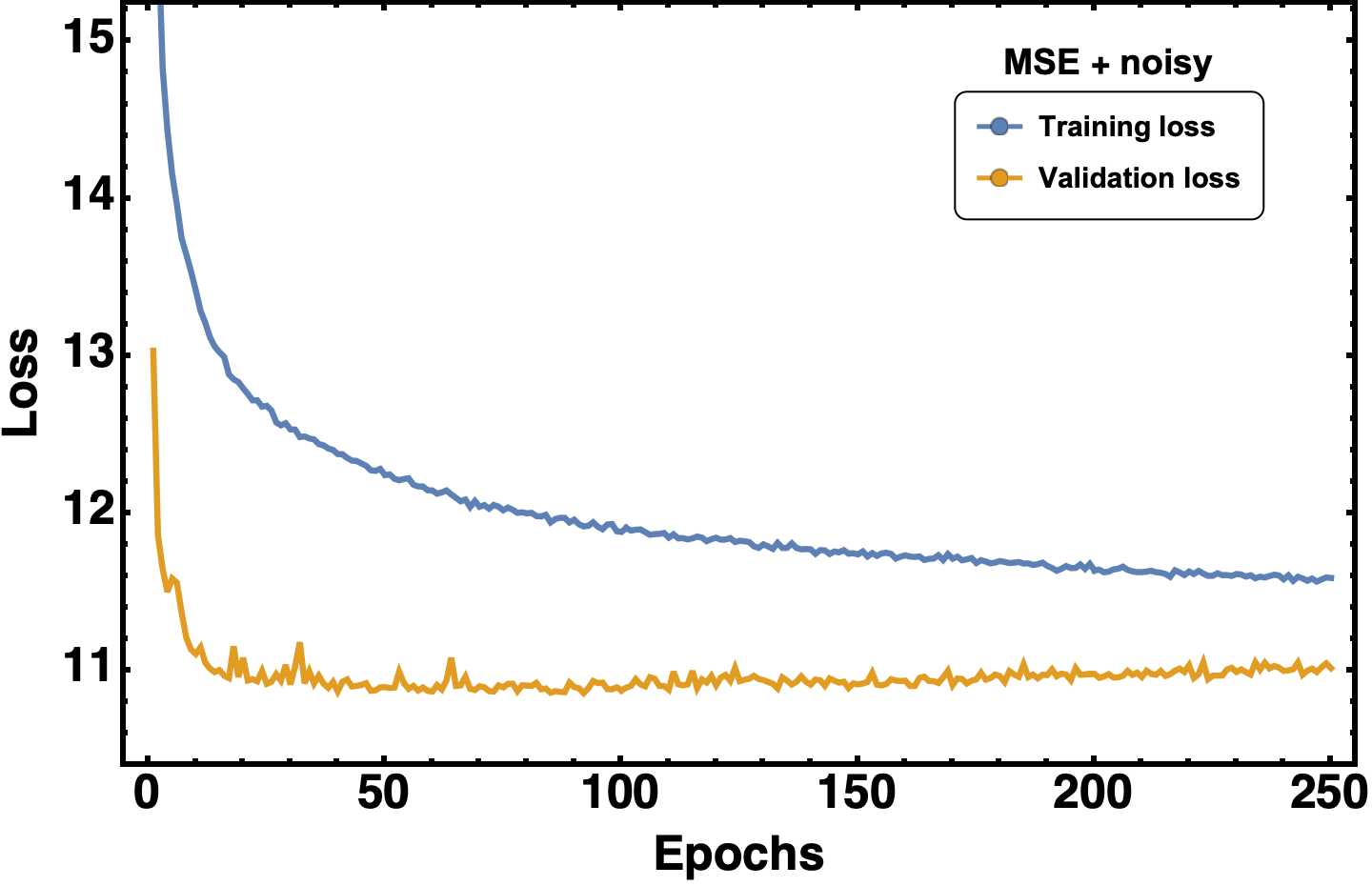} &
\includegraphics[width=0.28\linewidth]{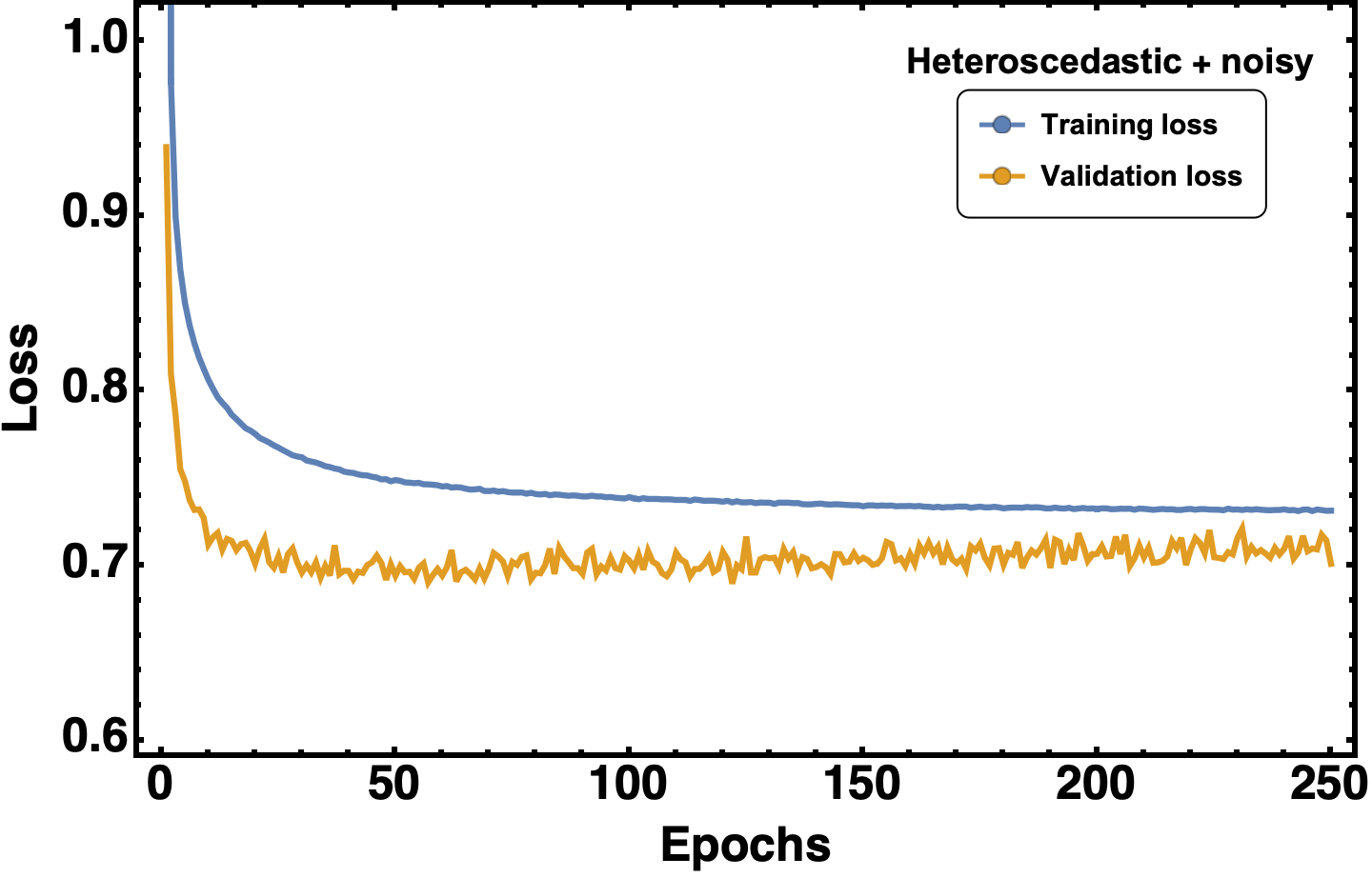} \\
& (d) UGCA 444 & 
\end{tabular}
\caption[Loss function as a function of epochs for the remaining 3 galaxies in our sample]{Same as Fig.~\ref{fig:loss_1} for the remaining three galaxies in our sample. }
\label{fig:loss_2}
\end{figure*}

The plots are show in Figs.~\ref{fig:loss_1} and~\ref{fig:loss_2}. Note that for all galaxies, by $250$ epochs, validation loss for every case has stopped changing noticeably. 
Moreover, for the case of noisy training data with MSE loss function, validation loss has started increasing slightly for some galaxies (see the case for UGC 5721 and UGCA 444).
This implies that the neural network is beginning to overfit and further training will lead to worse performance on unseen data. 

\end{document}